\documentclass[12pt,titlepage]{report}
\usepackage{times}
\usepackage{setspace}
\usepackage{acronym}
\usepackage{amsmath}
\usepackage{amssymb}
\usepackage{graphicx}
\usepackage[round]{natbib}
\usepackage{hyperref}
\usepackage{multirow}
\usepackage{epstopdf}
\usepackage{subfigure}
\usepackage{rotating}
\DeclareGraphicsExtensions{.ps,.pdf,.eps,.jpg}

\newcommand{\be}{\begin{equation}}
\newcommand{\ee}{\end{equation}}

\newcommand{\magsim}
{\ \lower2pt\hbox{$\sim $}\mkern-14mu \raise2pt\hbox{$>$}\ }

\newcommand{\s}{{\rm \, s}}

\newcommand{\aj}{\textit{Astron.\ J.}}

\newcommand{\msun}{\mbox{$M_\odot$}}      % mass sun
\newcommand{\mse}{\mbox{m s$^{-1}$}}       % meters per second
\newcommand{\cmse}{\mbox{cm s$^{-1}$}}     % centimeters per second
\newcommand{\ms}{\mbox{m s$^{-1}$~}}       % meters per second
\newcommand{\cms}{\mbox{cm s$^{-1}$~}}     % centimeters per second
\newcommand{\etal}{\mbox{\rm et al.}}     % et al.
\newcommand{\muas}{\mbox{$\mu as$}}       % microarcseconds
\newcommand{\mjup}{M$_{\rm JUP}$}         % mass jupiter
\newcommand{\mearth}{M$_{\rm Earth}~$}     % mass earth
\newcommand{\mearthe}{M$_{\rm Earth}$}     % mass earth
\newcommand{\gtsima}{\stackrel{\textstyle >}{\sim}}
\newcommand{\simgt}{\scriptsize{\raisebox{-2pt}{$\gtsima$}}\normalsize}

\newcommand{\ti}{$\,\times\,$}
\newcommand{\etae}{$\eta_{\oplus}$}

\begin{document}

\pagenumbering{roman} % START WITH ROMAN NUMERALS FOR INTRODUCTORY MATERIAL

\begin{titlepage}

\begin{center}

{\Large \bf Worlds Beyond: A Strategy for the Detection and Characterization of Exoplanets }\\

\vspace{.25in}

{\Large Report of the ExoPlanet Task Force: with ERRATUM}\\

\vspace {.2in}

{\Large Astronomy and Astrophysics Advisory Committee}

\vskip  4truein

{\large Washington, D.C.}

\vspace{0.3in}

\vspace{0.05in}

{\large April 22, 2009}

\vfill

\clearpage

\vskip 0.5truein
\baselineskip16pt

The Exoplanet Task Force

\end{center}

\begin{description} 
\item Debra Fischer, San Francisco State University
\item Heidi Hammel, Space Science Institute
\item Thomas Henning, Max Planck Institute \emph{ESA Liason}
\item Lynne Hillenbrand, California Institute of Technology
\item James Kasting, Penn State University
\item Greg Laughlin, University of California, Santa Cruz
\item Jonathan I. Lunine, The University of Arizona \emph{Chair}
\item Bruce Macintosh, Lawrence Livermore Laboratories
\item Mark Marley, NASA Ames Research Center
\item Gary Melnick, Harvard-Smithsonian Center for Astrophysics
\item David Monet, United States Naval Observatory
\item Charley Noecker, Ball Aerospace and Technologies Corp.
\item Stan Peale, University of California, Santa Barbara
\item Andreas Quirrenbach, Landessternwarte Heidelberg
\item Sara Seager, Massachusetts Institute of Technology
\item Josh Winn, Massachusetts Institute of Technology

\end{description}

\clearpage

\vspace{2.5 in}

\emph{\\Round and round, round and round\\
Round and round the Sun\\
The wandering stars go on forever\\
Can you name them one by one?\\}

\indent\indent{A Child's Introduction to Outer Space, 1958}

\newpage

\vspace{1.0 in}

\begin{center}

\bf{Acknowledgements}

\end{center}

\noindent {The Task Force is grateful to the following reviewers for reading and providing comments:}

\begin{description} 

\item Reta Beebe, {\it New Mexico State University}
\item Andrew Gould, {\it Ohio State University}
\item Marc Kuchner, {\it  NASA Goddard Spaceflight Center}
\item David Latham, {\it Harvard Smithsonian Center for Astrophysics}
\item Eugene Levy, {\it Rice University }
\item Victoria Meadows, {\it University of Washington}
\item Didier Queloz, {\it University of Geneva}

\end{description}

\vspace{0.2 in}

\noindent{The Task Force is deeply grateful for the expertise and sustained efforts of Dana Lehr and Michael Briley from the NSF, and Stephen Ridgway and Zlatan Tsvetanov of NASA, who together greatly improved the quality of the process and the product. }

\vspace{0.2 in}

\begin{center}

\noindent {\bf {Government disclaimer}}

\end{center}

Portions of this work were performed under the auspices of the
U.S. Department of Energy by Lawrence Livermore National Laboratory in
part under Contract W-7405-Eng-48 and in part under Contract
DE-AC52-07NA27344.

\clearpage

\centerline{{\bf Erratum}}
\vskip 0.2in

The calculations described on pages 85-92 of Chapter 9 (``Depth of
search comparisons'') employed an incorrect mass-radius relation for
iron-rock planets, due to the misapplication of the fitting formulas
of Fortney et al.~(2007) (specifically, log$_e$ was used instead of
log$_{10}$). Fig.~1 shows the original and corrected mass-radius
relations.

The most important effect of this error was to make the prospects for
using JWST/NIRSpec to characterize the atmospheres of transiting
planets appear more favorable than they are in reality. Furthermore,
some of the assumptions made in the JWST/NIRSpec calculation may be
regarded as overly optimistic. The selectively-absorbing portion of
the planetary disk was taken to be an annulus with thickness $10H$,
where $H=kT/\mu g$ is the pressure scale height. A thickness of $10H$
would require unusually high atomic or molecular absorption. Also, the
calculations used $\mu=18$~amu, the value for water vapor, but the
pressure is likely to be dominated by heavier species. Both of these
choices cause the signal to be larger than expected for an Earthlike
planet.

In addition, due to an editing error, the last sentence of the last
full paragraph on page 88 erroneously referred to ``secondary eclipse
events'' instead of transit events. This editing error had no effect
on the calculation, which was indeed performed for transits.

We have repeated the calculations using the corrected mass-radius
relation and an annulus of thickness $5H$ with $H=kT/\mu g$ and
$\mu=28$~amu. Fig.~2 is the revised version of Fig.~9.3(b) from the
original report, showing the depth of search for a hypothetical
all-sky transit mission. Fig.~3(a) is the revised version of
Fig.~9.3(c), showing the subset of the systems from the all-sky survey
that are characterizable with JWST/NIRSpec. The sharp boundary at
10~$M_\oplus$ reflects the threshold in our calculation between
iron-rock planets and gas giants. Transmission spectroscopy of
iron-rock planets with NIRSpec using our criteria ($R=100$ at 2$\mu$m,
SNR~$>10$ per resolution element after observing 10 transits) will be
extremely challenging or impossible.\footnote{It should be noted that
  even the choice of $5H$ may be considered optimistic. On Earth,
  almost all the water is in the troposphere, which extends to a
  height of about 17~km or approximately $2H$.}

The results are very sensitive to the values of the planetary radius
and its atmospheric scale height. Fig.~3(b) shows the results for
ice-rock planets (or ``water worlds'') in which the ice-mass fraction
is 90\%, using the mass-radius relation from Fortney et
al.~(2007). Many more such planets are characterizable, including some
planets in the habitable zone.

The results also depend sensitively on the criteria for
characterizability. As another example, we consider NIRCam photometry
using a filter centered at 2~$\mu$m with bandwidth $0.5$~$\mu$m. We
assume that 1~e$^-$/s is recorded at AB~28.0~mag (based on the
aperture of 25~m$^2$ and an overall throughput of 50\%) and that the
noise is dominated by photon noise from the star. We require SNR~$>5$
for observation of a single transit. (In practice, the atmospheric
absorption feature probably will not fill the entire band, and
multiple transit observations would be needed.)  As above, the
selectively-absorbing annulus has a width $5H$ with $H=kT/\mu g$ and
$\mu=28$~amu. The results are shown in Figs.~3(c) and 3(d), for
iron-rock and ice-rock planets, respectively. Unsurprisingly, the
prospects are better with this more forgiving definition of
characterizability.

In light of these revisions, the prospects for JWST to characterize
the atmospheres of transiting planets are more limited in both the
range of planetary types and the range of observing
modes. Nevertheless, the ExoPTF remains convinced that the M-dwarf
track is a fruitful and worthwhile parallel track to the more
expensive and prolonged G-dwarf track, by providing near-term
opportunities for revealing studies of planetary properties through
radial-velocity surveys, transit surveys, and observations with
Spitzer and JWST.

The error in the mass-radius relationship also affected the
depth-of-search calculations regarding the detectability of planets
with coronagraphs (pages 89-92), although the effect was comparatively
minor. Fig.~4 shows the corrected versions of the depth-of-search
figures for 2.5m and 4.0m space coronagraphs, and for a 30m
ground-based telescope with adaptive optics.

\vskip 0.1in

We thank Tom Greene for bringing these issues to our attention. We
also note that since the publication of our report, two other
groups\footnote{Kaltenegger \& Traub~2009, ApJ, in press
  [arxiv:0903.3371]; Deming et al.~2009, PASP, submitted
  [arxiv:0903.4880]} have undertaken more detailed calculations
regarding the characterization of transiting planets with JWST.

\clearpage

\begin{figure}[ht]
\centering
    \label{fig:mass-radius}
    \includegraphics[width=11cm]{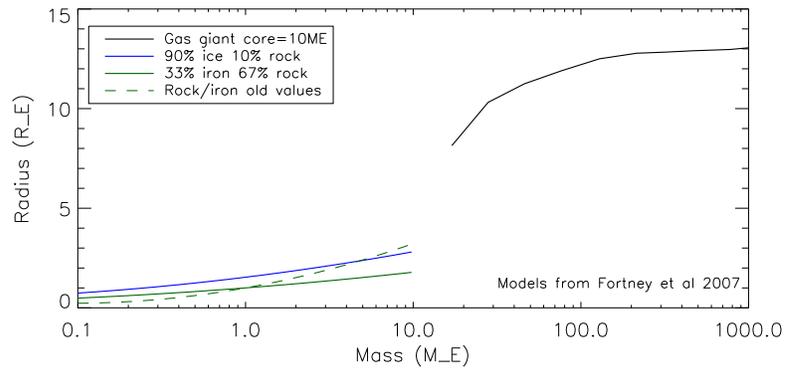}
    \caption{ The original (dashed line) and corrected (solid green
      line) theoretical mass-radius relations for iron-rock planets.
      Also shown is the relation for ice-rock planets (solid blue
      line) and gas giants (black line).}
\end{figure}

\begin{figure}[]
\centering
    \label{fig:allsky}
    \includegraphics[width=7cm]{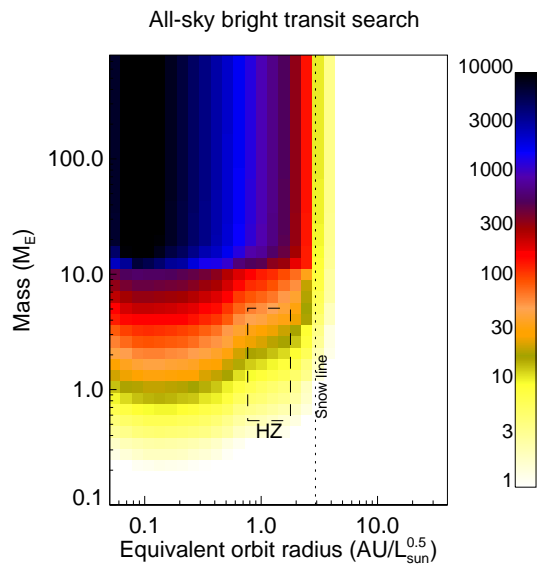}
    \caption{Revised Fig.~9.3(b) of the ExoPTF Final Report, showing
      the depth-of-search of a hypothetical all-sky transit survey.}
\end{figure}

\begin{figure}[ht]
\centering
\subfigure[] % caption for subfigure a
{
    \label{fig:nirspec-iron}
    \includegraphics[width=7cm]{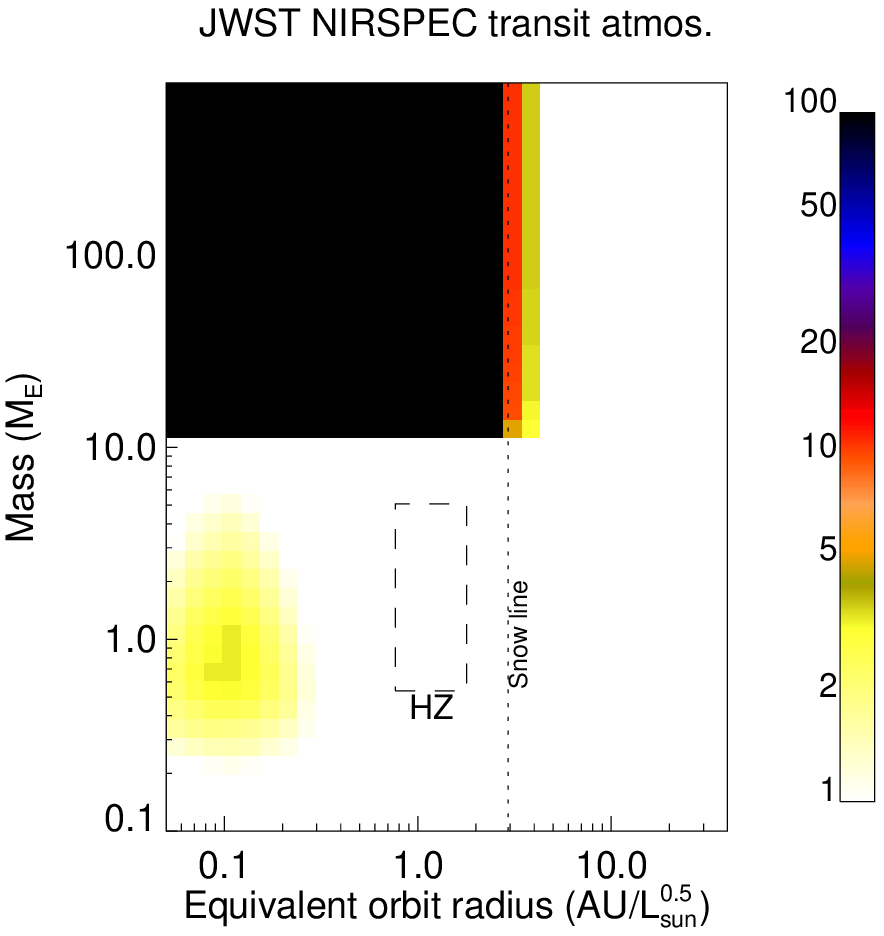}}
\hspace{1cm}
\subfigure[] % caption for subfigure b
{
    \label{fig:nirspec-ice}
    \includegraphics[width=7cm]{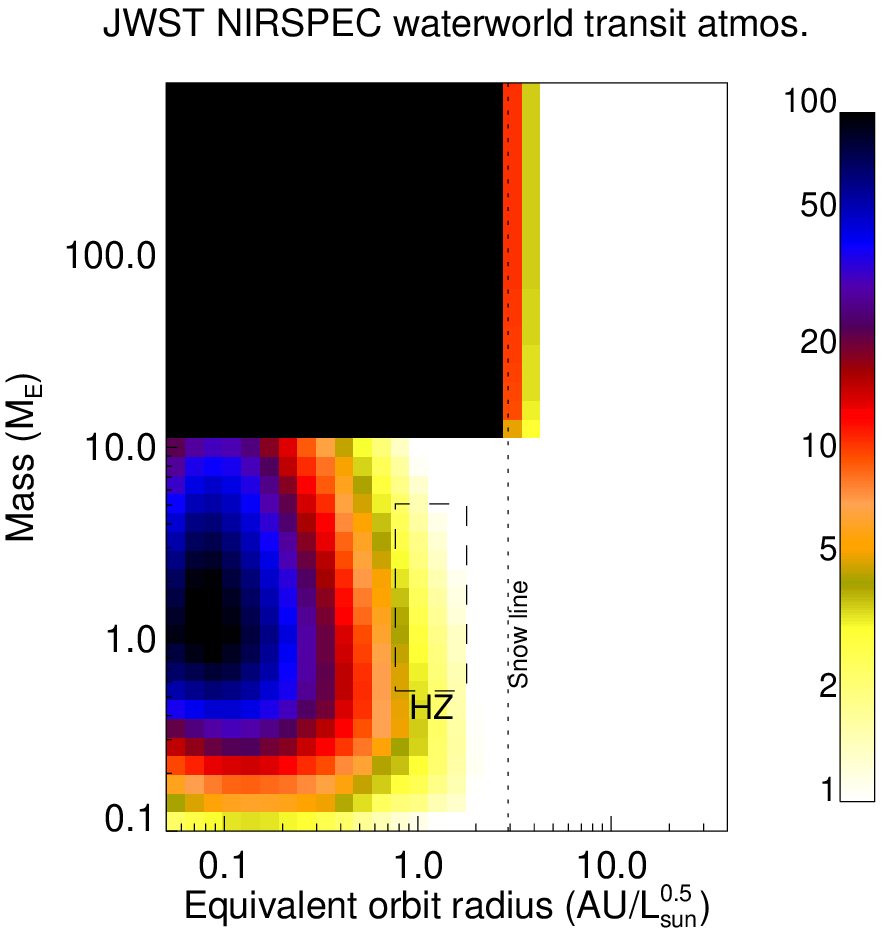}}    
\hspace{1cm}
\subfigure[] % caption for subfigure c
{
    \label{fig:nircam-iron}
    \includegraphics[width=7cm]{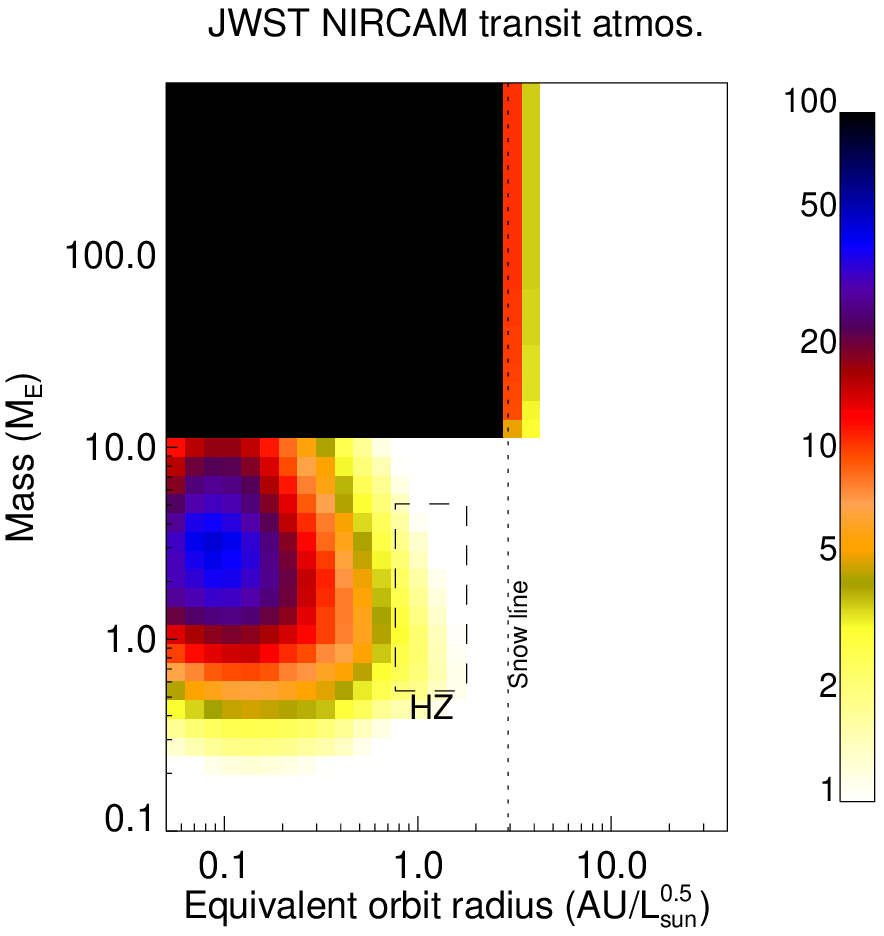}}
\hspace{1cm}
\subfigure[] % caption for subfigure d
{
    \label{fig:nircam-ice}
    \includegraphics[width=7cm]{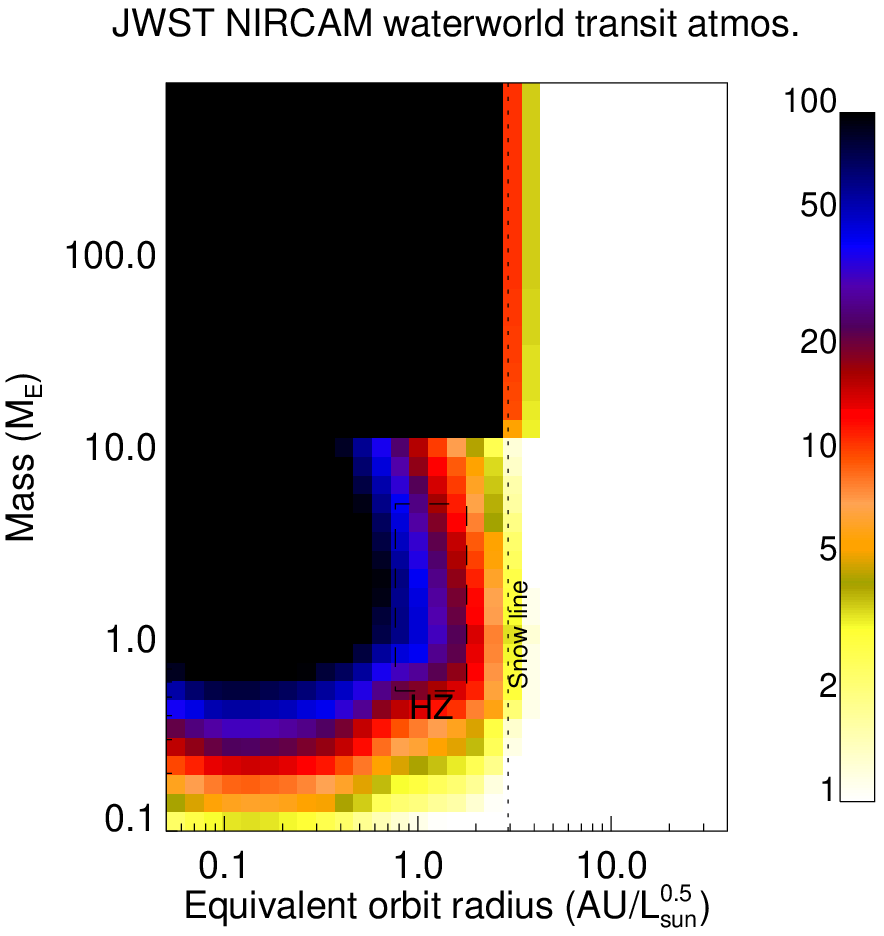}}    
\caption{(a) Revised Fig.~9.3(c) of the ExoPTF Final Report. Plotted
  is the number of targets from a hypothetical all-sky transit survey
  that are characterizable by transmission spectroscopy with JWST/NIRspec,
  according to the criterion given on the previous page.  (b) Same,
  but with the mass-radius relation for low-mass planets changed from
  iron-rock (33\% iron) to ice-rock (90\% ice). (c) and (d): Same as (a) and (b), but
  for JWST/NIRcam filter photometry using criteria described on
  the previous page.
}
\label{fig:jwst} % caption for the whole figure
\end{figure}

\begin{figure}[ht]
\centering
\subfigure[] % caption for subfigure a
{
    \label{fig:exao}
    \includegraphics[width=7cm]{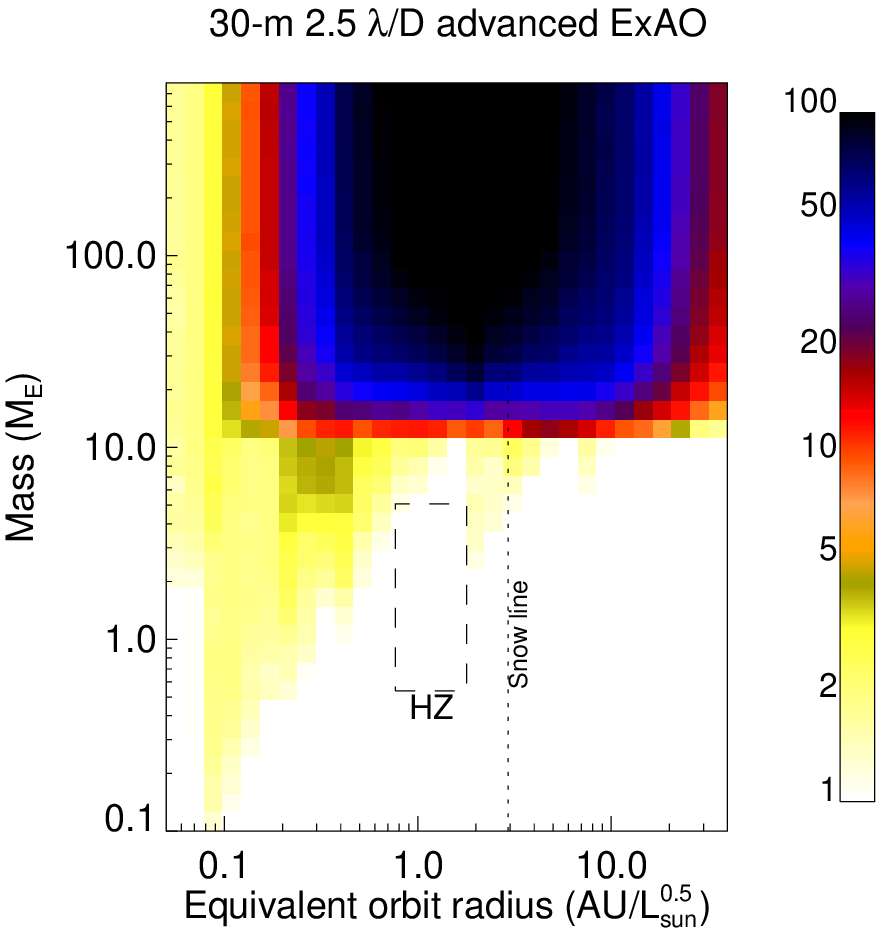}}
\hspace{1cm}
\subfigure[] % caption for subfigure b
{
    \label{fig:tpf25}
    \includegraphics[width=7cm]{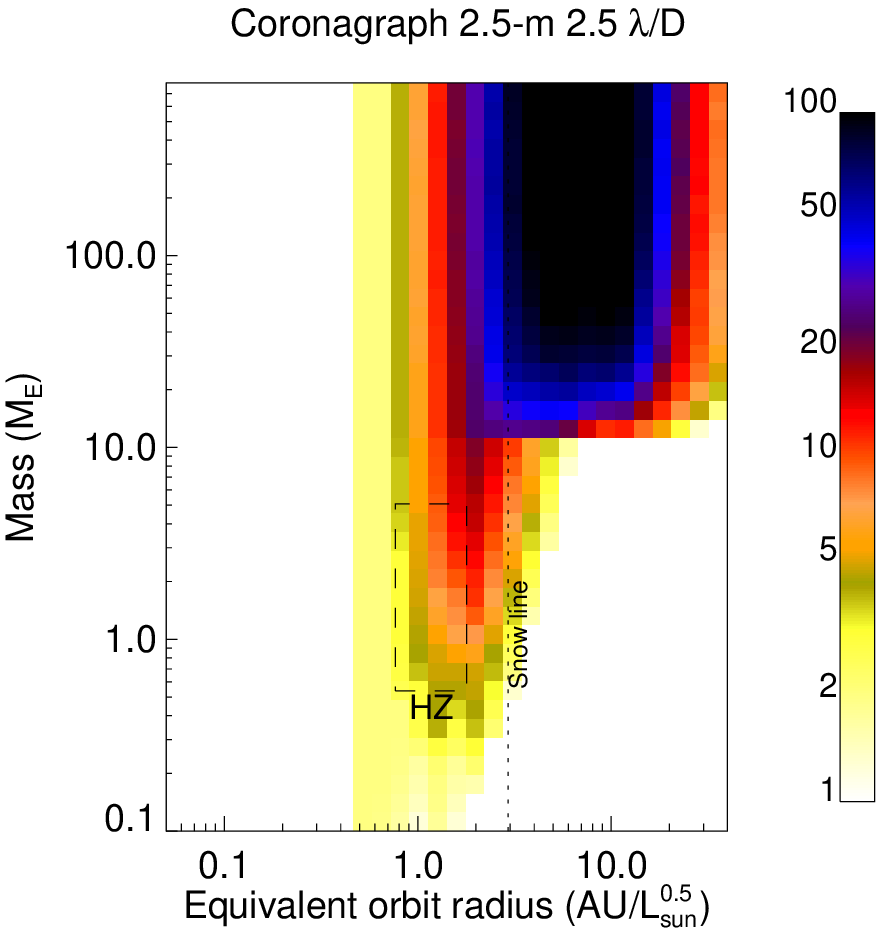}}
\hspace{1cm}
\subfigure[] % caption for subfigure c
{
    \label{fig:tpf40}
    \includegraphics[width=7cm]{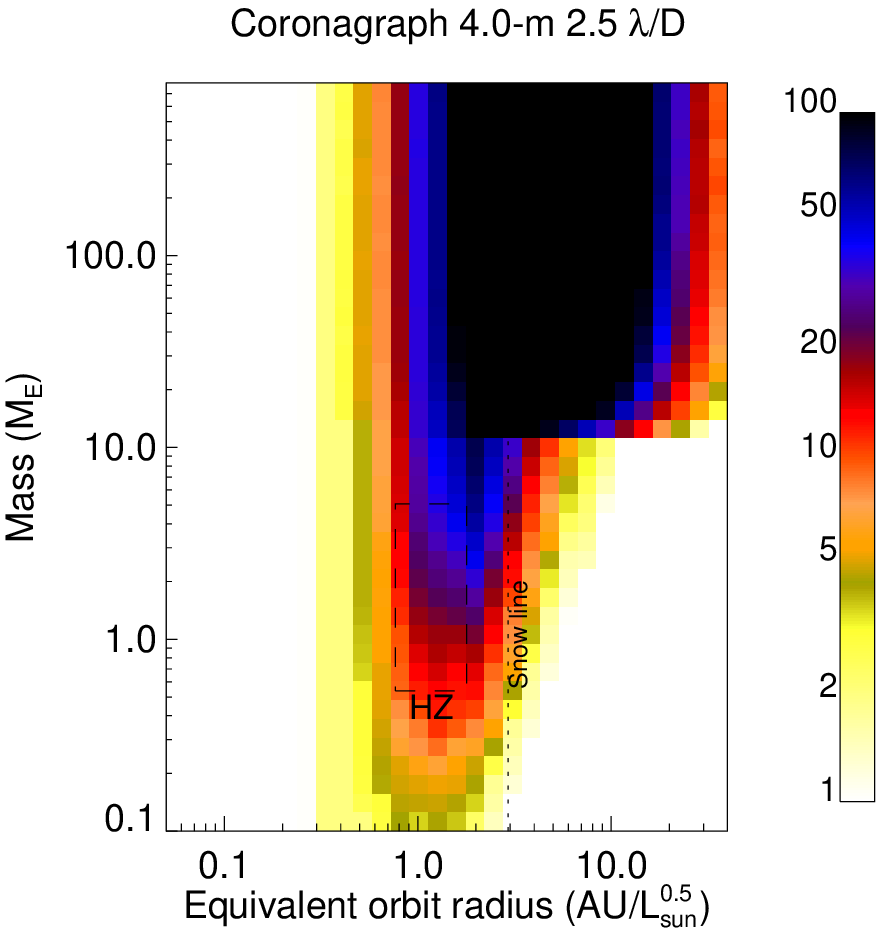}}
\hspace{1cm}
%\subfigure[] % caption for subfigure d
%{
%    \label{fig:nircam-ice}
%}
%    \includegraphics[width=7cm]{}}
\caption{(a) Revised Fig.~9.4(b) of the ExoPTF Final Report, showing
  the sensitivity of a 30-m ground based telescope equipped with
  advanced adaptive optics and a 2.5$\lambda/D$ coronagraph.
  (b) Revised Fig.~9.4(c) of the ExoPTF Final Report, showing
  the depth-of-search of a space coronagraph mission with a 2.5-m
  primary mirror and an inner working angle of 2.5$\lambda/D$.
  (c) Revised Fig.~9.4(d) of the ExoPTF Final Report, showing
  the depth-of-search of a space coronagraph mission with a 4.0-m
  primary mirror and an inner working angle of 2.5$\lambda/D$.
}
\label{fig:coronagraph} % caption for the whole figure
\end{figure}

\clearpage

\section{Executive summary} 

\subsection{Introduction}

We stand on a great divide in the detection and study of
exoplanets. On one side of this divide are the hundreds of known
massive exoplanets, with measured densities and atmospheric
temperatures for a handful of the hottest exoplanets. On the other
side of the divide lies the possibility, as yet unrealized, of
detecting and characterizing a true Earth analog--an ``Earth-like" planet (a planet of one Earth mass or Earth radius
orbiting a sun-like star at a distance of roughly one astronomical unit). This Exoplanet
Task Force Report describes how to bridge this divide. The
recommendations emphasize immediate investment in technology and space
mission development that will lead to discovering and characterizing
Earth analogs. 

We recognize that setting a goal of detecting planets like the Earth sets the bar high.  It is important that the program target such objects, if we are to determine whether the conditions we find on our own world are a common outcome of planetary evolution. The only example of a habitable world we have is our own one-Earth-mass planet, and indeed our nearest neighbor, Venus, is nearly the same mass but uninhabitable by virtue of closer proximity to the Sun. Searching for planets, for example, five times the mass of Earth is easier, but should they turn out to lack habitable atmospheres we would not know whether this is by chance or a systematic effect of the higher mass.  The  connection of the strategy described here to the big questions that motivate astronomical endeavors lies in understanding whether our home world is a common or rare outcome of cosmic evolution. 

The discovery of Earth analogs will change the way we humans view our
place in the cosmos. Just 500 years ago the standard western view of the cosmos was that Earth stood
at the center with all the planets and the Sun moving around it.  Copernicus's bold treatise of 1543 brought about a
reluctant paradigm change that Earth and the other planets orbited the
Sun. Later our Sun was recognized to be but one of 100 billion stars
in the Milky Way Galaxy and today astronomers believe the Milky Way is
but one of hundreds of billions of galaxies in the Universe. Earth remains special
as the only planet we know of with life. Finding Earth analogs that
show signs of habitability or atmospheric indicators of life
will bring about a new paradigm shift---one that completes
the Copernican Revolution.

Discovering an Earth analog is one of the most challenging feats for
any planet finding technique. While an Earth analog only 30 light
years distant is not fainter than the faintest objects (galaxies) ever
observed by the Hubble Space Telescope, the adjacent massive, huge,
and overwhelmingly bright parent sun-like star makes planet detection
extremely challenging. The Sun is 100 times larger, 300,000 times more
massive and 10 million to 10 billion times brighter than
Earth. Detecting Earth in reflected light is like searching for a
firefly 6 feet from a searchlight that is 2400 miles distant.  Each of
the five different exoplanet finding techniques are most sensitive to
planet-star combinations that are very different from the Earth-sun
system. At the same time---for the first time in human history--
several different exoplanet discovery techniques are close to finding Earth
analogs. The Exoplanet Task Force Report recommends using the radial
velocity technique to push to find Earth-mass planets around nearby
sun-like stars, for the handful of stars that are bright enough for
monitoring. The only technique appropriate to survey the nearest
hundred or so bright sun-like stars in the mid-term is space-based
astrometry, and this is one cornerstone of the Task Force
recommendations. To study the planet atmosphere for signs of
habitability or life, direct imaging is required, and the Task Force
Report recommends investment in direct imaging technology development
across different wavelengths and techniques. Once Earth-mass planets
are known to orbit nearby sun-like stars, the Task Force recommends
launching a direct imaging space mission for habitability
characterization.

The hierarchy of pressing questions in the search for Earth analogs are ``Do Earth-like planets exist?Ó ``Are they common?Ó and ``Do they show signs of habitability or biosignatures?".  NASA's Kepler mission aims to answer the first two questions by monitoring a large number of faint stars to find the frequency of Earth analogs.  We call this frequency $\eta_{\oplus}$ (``eta sub Earth"), the fraction of sun-like stars that have at least one planet in the habitable zone (where the habitable zone is the region around the star with temperatures suitable for surface liquid water). In order to assess whether an individual Earth analog has signs of habitability we must find Earth analogs around the nearest and brightest stars. Only the bright stars host planets bright enough to take spectra---i.e., to ``fingerprint" the planet's atmosphere in order to identify molecular features and biosignatures. The Task Force strategy focuses on  the Earth analogs around the bright, nearby sun-like stars (called F, G, and K stars), provided that $\eta_{\oplus}$ is high enough. 

There is an exciting possibility of a fast track to finding and
characterizing habitable exoplanets. This is the search for big Earths
(super Earths) orbiting small, cool stars (called M-dwarf stars or M-dwarfs).  M-dwarfs are
much less luminous than the sun so that the locations amenable to
surface liquid water to support life as we know it are very close to
the star. The relative size and mass of the planet and star are more
favorable for planet detection than the Earth-Sun analog. Two different yet complementary
planet searching techniques (transits and radial velocities) are very sensitive to super Earths 
orbiting with close separations to M-dwarfs---and indeed a handful of super
Earths orbiting small stars have already been discovered. A set of
transiting planets (those that go in front of their stars as seen from
Earth) will enable average density measurements and hence
identification of terrestrial-type planets. The Task Force report
recommends bolstering support for both radial velocity and transit
discovery and characterization of super Earths orbiting in the
habitable zones of M-dwarfs. Suitable transiting planets can have their
atmospheres characterized by the  James Webb Space Telescope, now under development,
building on the current Spitzer Space Telescope studies of hot
Jupiters.

Beyond the search for habitable exoplanets, the Task Force has
identified the observation of ``planetary architectures" (orbital arrangements and types of planets around a star) as key to determining the diversity of planetary systems (and hence how common or uncommon is our own solar system's geography) and the long-term habitability
of inner terrestrial planets as affected by overall system architecture. A third major area of planetary
formation completes the picture for understanding the origin and
evolution of exoplanetary systems.

\subsection{Background to the Recommendations}
The Task Force developed a 15 year strategy for the detection and characterization of
extrasolar planets (``exoplanets") and planetary systems, as requested by
NASA and the NSF to the Astronomy and Astrophysics Advisory
Committee. The charge to the Task Force is given in the Appendix. The
strategy is an outgrowth of the efforts underway for two decades to
detect and characterize extrasolar planets-- from which over 260 planets
and dozens of multiple planet systems have been found and studied. It
is informed by a variety of technological studies within the
astronomical community, industry, NASA centers and NSF-funded
facilities that point the way toward techniques and approaches for
detection and characterization of Earth-sized (0.5--2 times Earth's
radius) and Earth-mass (0.1--10 times the mass of the Earth) planets
in the solar neighborhood.  The raw material for the strategy was
provided in the form of invited briefings and 85 white papers received
from the community.

The strategy we developed is intended to address the following questions, given in
priority order:

\begin{enumerate}

{\bf \item What are the physical characteristics of planets in the
habitable zones around bright, nearby stars?}

{\bf \item What is the architecture of planetary systems?}

{\bf \item When, how and in what environments are planets formed?}

\end{enumerate}

The time horizon for the strategy is divided into three epochs, 1--5,
6--10 and 11--15 years.  A two-pronged effort toward answering the
first question is recommended, one focussed on ultimately detecting
and characterizing Earth-size/Earth-mass planets around M-dwarfs
(``M-dwarf planets") using ground-based and space assets in place or
under development today, and the second with the ultimate goal of
detecting and characterizing Earth-size/Earth-mass planets around
stars like our Sun with new capabilities whose technologies are under
development today  (Fig. ~\ref{fig:two-pronged}). The first of these uses optical and near-infrared
doppler spectroscopic (radial velocity) and transit surveys from the
ground, possibly supplemented by a space-based transit survey if
needed, to find M-dwarf planets approaching the size/mass of the Earth
that can be characterized with Spitzer (even after its cryogen is
depleted) and/or the James Webb Space Telescope. This track is expected to be
completed within the second time epoch (6--10 years).

\begin{figure}[h]
\begin{center}
\includegraphics*[scale=0.5]{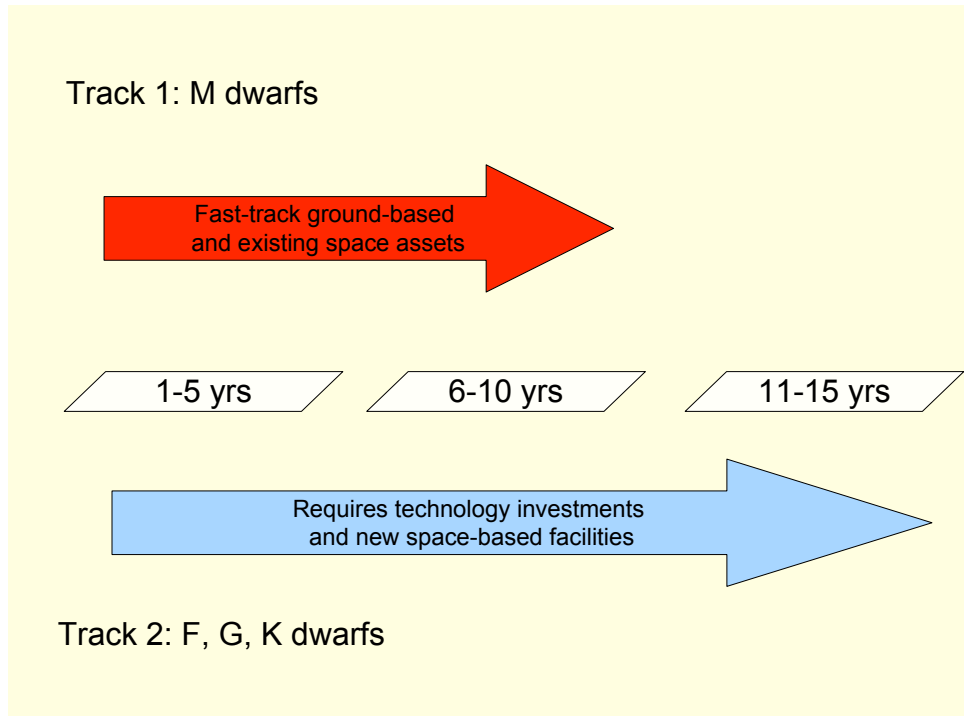}
\caption{The recommended strategy is shown conceptually as consisting of parallel tracks. The top track for M-dwarfs relies on ground-based and space assets that exist or are under development. The bottom track, for Sun-like stars, requires technological investments and new space-based facilities, but is flexible and builds on previous findings. \label{fig:two-pronged}}
\end{center}
\end{figure} 

The other track is focussed on stars more similar to the Sun, what are
called F, G, and K dwarfs. Its key feature is deployment in the second
epoch of an astrometric facility in space with the sensitivity to
survey for one Earth-mass planets around up to 100 solar-type stars in our
cosmic neighborhood, providing a target list for a direct-detection
mission to be deployed in the final epoch. We envision this to be an
astrometry mission that heavily emphasizes sub-microarcsecond
planet-finding science, thereby limiting its scope and hence
cost when compared with mission architectures such as the Space Interferometry Mission (SIM) that would have broader goals across astrophysics. Following this, in the third epoch, a direct-detection mission is
deployed capable of doing a spectroscopic examination of Earth-mass
planets identified by the astrometric mission. The scope of the direct
detection mission depends on the findings of the astrometry
mission. If, for some reason, an astrometry mission is not executed,
then an assessment of whether a direct detection program should proceed will
depend upon measured values of  $\eta_\oplus$,  and on the dust in the system (the
``exo-zodiacal environment") around target stars, among other factors. The
COROT and Kepler space-based transit searches will constrain
$\eta_\oplus$ early enough (within the first five years of this
strategy) to scope the size of the direct-detection mission and,
together with results of technology studies, determine whether it
should be based on coronagraphic/occulter or nulling interferometric
technologies. The astrometric mission itself does not require transit
survey results since it will provide high value data on planetary
system architectures even if specifically Earth-mass planets turn out
to be non-existent or so rare that no candidates are found within a volume accessible
to direct detection.

This approach is adaptable to new discoveries and surprises along the
way. It provides the potential for early results on Earth-sized
planets around M-dwarfs. The strategy moves beyond M-dwarf Earths, on
which life may be limited due to potential habitability issues, to
find Earth analogs around stars similar to our own. Because the
astrometric survey approach is insensitive to confusion from
background objects and zodiacal dust clouds around candidate stars,
its use in the second epoch ensures significant results on planetary
system architectures and the existence of Earth-mass planets even in
the unhappy circumstance that most systems are much dustier than our
own. By using the astrometric mission to survey for targets for the
direct detection mission, the strategy allows maximally efficient
design of direct detection approaches unburdened by the need to
conduct surveys.  Decision points contingent on the exo-zodiacal and
$\eta_\oplus$ surveys occur early enough in the strategy to permit a
shift in focus in the later part of the strategy.

Throughout the 15 year period of the strategy, surveys of the architectures of planetary systems
 with ground- and space-based
capabilities, and work toward ground-based direct detection and
examination of larger (Neptune-class) planets, would be pursued to
complement the search for planets like our own. A sensitive survey of
the distribution of extra-solar (exo-) zodiacal emission in solar-type
stars should be conducted in the first time epoch to enable key decisions on
the direct detection mission.

Although the Task Force believes this strategy to be 
reasonably paced it is also flexible in the sense that individual
elements can be delayed or stretched out, while the overall program
continues to provide exciting discoveries.

Complementary to the planet search and characterization strategy is
the use of existing and anticipated ground- and space-based resources
to study circumstellar disks in their planet-forming and
post-formation phases, with the goal of understanding planet formation
and inferring the presence of planets in young systems.

Theroetical investigations are critical for interpreting and guiding
key observations. Because theory results span all time epochs, the
recommendations are described in Section 1.4 covering all years.  This
placement does not reflect the priority of theory compared to the
observational and technological recommendations in Section 1.1--1.3. The interplay between theory and observation is an ongoing activity that is essential to returning the full value of our science missions. 

Exoplanet science and technology is a rapidly changing field. There is potential for new, transformational technologies that may significantly impact planet discovery. One such example, that of a new radial velocity technique, arose just as this report was being completed. The new radial velocity technique enables precise wavelength calibration of high-resolution spectrographs using laser frequency combs. Demonstrated in the lab, the frequency comb spectrograph still needs to be tested on a telescope. If the significant challenges of long-term instrument stability and intrinsic stellar variability can be met, the precise calibration could enable discovery of Earth analogs around hundreds of sun-like stars. This development serves as a 
reminder that transformational techniques will arise, and may call for reevaluation of the relevant decision points in the strategy. 

Recommendations of the Task Force to implement the strategy are below, with the detailed programmatic elements depicted in figure  ~\ref{fig:recommended_program}.
The recommendations are divided into different time epochs, the 1--5
year, 6--10 year, and 11-15 year time frames.  While we do not
strictly prioritize the recommendations, we do order them---within
each time epoch---according to the priority order of the compelling
science questions introduced above.  Where appropriate, we further
organize the recommendations into the F, G, K (i.e., sun-like)
strategy and the M-dwarf strategy.

We expect the report and our recommendations, based as they are
on input from 85 {\it white papers} from the community and a year-long
intensive study process, to be a
key part of the community input to the panel(s) dealing with
Exoplanets and to the Survey committee during the upcoming 2010
Decadal Survey of Astronomy and Astrophysics. Our recommendations for detailed assessment 
of critical techology developments should be implemented as soon as possible, ideally well within the time frame of the Decadal Survey process, to enable the strategy proposed here to move forward in a timely fashion.

\subsection{A. Recommendations for 1--5 Years}                                                                                                                                                                                                                                

\bigskip

{\bf  A. I. a.  What are the physical characteristics of planets in the
habitable zones around bright, nearby {\it F, G, K} stars?}

\bigskip

{\bf \emph{Recommendation A. I. a. 1}} {\it Sufficient
investment in ground-based telescope time for radial velocity
measurements to enable the discovery of low-mass exoplanets down to
Earth mass planets orbiting bright stars.  The required precision for the detection of Earth analogs is substantially
better than 1 m/s.  In the first time period, we recommend feasibility
studies for extreme Doppler precision (down to several cm/s) for bright
star targets.  It is also critical to continue surveys for planets of all detectable masses
with a target list well in excess of 1000 stars, with 3000 observations
per year total.}

\bigskip 

The investment could be NASA time on the Keck telescope, including
programs such as the NASA/Keck $\eta_\oplus$ program, as well as the
dedication of underutilized and existing 3-to-4 m class telescopes
with high-precision and high-throughput spectrographs.  A focus on
bright stars including the brightest M-dwarfs is the fast track to
finding a potentially habitable exoplanet.

\bigskip

In order to implement two key spaceborne capabilities---astrometry and
direct detection---within the 15 year timeframe of the strategy, technological
studies and mission development are required early on. NASA has
already invested substantially in technologies for both, and
particularly for spaceborne astrometry, so that there is a strong
foundation already for completing the additional technology
developments needed to conduct astrometric and direct detection
missions in space.

\bigskip

{\bf \emph{Recommendation \hspace{0.05in} A. I. a. 2}}
{\it Preparations should begin for a space-borne astrometric mission
capable of surveying between 60 and 100 nearby main sequence stars
with the goal of finding planets down to the mass of the Earth orbiting
their parent star within the habitable zone - i.e., approximately
0.8 to 1.6 AU, scaled appropriately for stellar luminosity.  Achieving
this goal will require the capability to measure convincingly wobble
semi-amplitudes down to 0.2 microarcseconds (``$\muas$") integrated over the mission lifetime.
Space-borne astrometry is currently the only technique that can distinguish
masses in the range of 1-10 earth mass. To this end, a rigorous 
technical feasibility
assessment should be undertaken immediately,  and any additional
required technology development beyond what has been accomplished to
date should be completed promptly, leading to an implementation phase
start in the intermediate 6-10 year time period. The feasibility assessment should include a critical analysis of stellar systematic effects, such as starspots.}

\bigskip

{\bf \emph{Recommendation \hspace{0.05in} A. I. a. 3}} 
{\it Technological development of space-borne
direct detection capabilities to ultimately find and characterize Earth-sized
planets should be undertaken at the start of the strategy. This
includes visible wavelength internal and external coronagraphs and IR
nulling interferometers.  A key enabling technology for internal
coronagraphs is advanced wavefront sensing and control; support must
be sufficient to assess the viability of internal coronagraphs
operating at an inner working angle IWA $< 3.5\lambda$/D. (The IWA is
the minimum angular separation from the star at which the observatory
can detect a planet.) Additional technologies that need attention
include, but are not limited to, next-generation deformable mirrors,
low-noise detectors, coronagraphic masks, and ultraprecise optical
surfaces. A key enabling technology for external occulters is
validated diffraction modeling. Support must be sufficient to complete
demonstration scalable to flight dimensions. Additional technologies
that need attention include, but are not limited to, alignment
sensors, deployment methods, high-specific-impulse thrusters, and
studies of plume effects. Also needed are mission models assessing the
science harvest as it depends on system size. }

\bigskip

Because both the astrometry and direct detection technologies are so
crucial to the strategy, ongoing technological development should be
supplemented with in-depth reviews by experts in the various fields.

{\bf \emph{Recommendation \hspace{0.05in} A. I. a. 4 }}{\it NASA should
establish a blue-ribbon panel, consisting largely of physicists or
optical scientists with expertise in wave optics, to evaluate various
coronagraph and wavefront control concepts and ensure that no
fundamental physical effect has been overlooked in planning for an
optical wavelength, direct detection mission. An equivalent panel
should be established for direct detection by interferometry. }

\bigskip

Sizing of direct detection systems, indeed their feasibility, for
studying planets around sun-like stars depends on how typical is our
solar system's dust emission--that is, what is the distribution of  zodiacal emission around other stars.

{\bf \emph{Recommendation \hspace{0.05in} A. I. a. 5 }}
{ \it Invest in a census of exo-zodi systems
around solar-type stars that might be targets for exoplanet searches.}

\bigskip

\noindent {\bf  A. I. b.  What are the physical characteristics of planets in the
habitable zones around bright, nearby {\it M } stars?}

\bigskip
This parallels Recommendation A. I. a. 1, but for the smaller M-dwarfs.

\bigskip

{\bf \emph{Recommendation \hspace{0.05in} A. I. b. 1 }}{ \it In view of
the fact that M-dwarfs might harbor the most detectable Earth-sized
planets, search the nearest thousand M-dwarfs (J $\leq 10$) for
transiting low-mass exoplanets with radial-velocity measured masses. }

\bigskip

Near-IR spectrographs are potentially the best way
to find terrestrial mass planets in the habitable zones of main
sequence stars later than M4 (a relatively warm or ``early" M-dwarf); optical spectroscopy is
competitive and already available for earlier spectral types.
Develop IR spectrographs with a target precision of
1 m/s for radial velocity surveys of late M-dwarfs.  A near-term
demonstration of 10 m/s is critical to validate this technique.

{\bf \emph{Recommendation \hspace{0.05in} A. I. b. 2 }} {\it Develop
near-IR spectrographs with 1 m/s precision for radial velocity planet
surveys of late M-dwarfs, once feasibility at 10 m/s precision has been
demonstrated.}

\bigskip

Given the possibility that some M-dwarfs might harbor planets whose
properties could be studied with warm-Spitzer and/or JWST, these space
assets form an important part of the strategy.

{\bf \emph{Recommendation \hspace{0.05in} A. I. b. 3}} {\it Continue to operate Spitzer as a warm observatory for
characterizing low-mass transiting planets around
main sequence stars, particularly around M-dwarfs.}

\bigskip
\noindent {\bf A. II. What is the architecture of planetary systems?}

\bigskip

In this near-term period, progress toward a complete understanding of
planetary systems requires further development of two ground-based
techniques: microlensing, which will provide a census of planetary
mass and orbital separation as a function of stellar type and Galactic
environment; and extreme AO systems, which will enable the direct
detection of young giant planets.  Ground-based extreme-AO
coronagraphy on 8-m class telescopes represents the next major
scientific and technical step beyond the current generation of HST and
ground instruments.  Similar instruments on future Extremely Large
Telescopes (ELTs) will further advance the state-of-the-art and
provide a proving ground for technology that may ultimately be part of
an optical wavelength, direct detection approach.  Maintenance of US involvement in facilities that utilize or advance these technologies is key to addressing all three key scientific questions and to obtain full
benefit from the data.

\bigskip

{\bf \emph{Recommendation \hspace{0.01in} A. II. 1}}{ \it Increase
dramatically the efficiency of a ground-based microlensing network by
adding a single 2 meter telescope.}

\bigskip

{\bf \emph{Recommendation \hspace{0.05in} A. II. 2}}{ \it Continue
technological developments and implementation of key ground-based
capabilities such as extreme AO in the laboratory and on 8-m class
telescopes. Future ELT's should be designed with the capability to do
extreme-AO coronagraphy in mind. In the very near-term, establish a blue-ribbon panel to determine in detail the requirements, costs and opportunities associated with ground-based work on exo-planets.}

\bigskip

\noindent {\bf A. III. When, how and in what environments are planets formed?}

\bigskip

{\bf \emph{Recommendation \hspace{0.1in} A. III. 1}}{ \it Maintain
U.S. involvement in key facilities including Herschel and ALMA for
studies of disks, and continue support for archival analysis of
relevant Spitzer, Chandra, Hubble, and ground-based data.  }

\bigskip

{\bf \emph{Recommendation \hspace{0.05in} A. III. 2}} {{\it Sustain a healthy
level of support for ground-based, space-based, and theoretical
(see Section D) investigations of star and planet formation. }\bigskip

\bigskip

\subsection{B. Recommendations for 6--10 Years}

The recommendations here follow from the accomplishments of the first
time period and the logic of the strategy summarized above.

\bigskip

\noindent {\bf B. I. a. What are the physical characteristics of
planets in the habitable zones around bright, nearby {\it F, G, K}
stars?}

\bigskip

{\bf \emph{Recommendation \hspace{0.05in} B. I. a. 1}} {\it Launch and
operate a space based astrometric mission capable of detecting planets down to the mass of the Earth around 60--100 nearby
stars, with due consideration to minimizing the width of any blind
spot associated with Earth's parallax motion. (This requires a mission precision, over many visits to a given star, as small as 0.2 microarcseconds.)  }

\bigskip

{\bf \emph{Recommendation \hspace{0.05in} B. I. a. 2}}{ \it Contingent on
the latest knowledge of $\eta_\oplus$ and exo-zodi brightness for
potential target stars, move spaceborne direct detection into the
advanced formulation phase to enable a mission launching in the 11--15
year time frame.}

\bigskip

Note: if Kepler suffered a sufficiently serious mission failure,
development of the direct detection mission should proceed forward
based on COROT and ground-based results if they indicate a likelihood of
high $\eta_\oplus$. Likewise, if the astrometric mission fails or
turns out to be infeasible, pursuit of space-based direct detection in
the final time period would require a sufficiently large
$\eta_\oplus$ based on COROT and Kepler to give a reasonable
probability of mission success.

\bigskip

\noindent {\bf B. I. b. What are the physical characteristics of
planets in the habitable zones around bright, nearby {\it M} stars?}

\bigskip
Given the possibility that some M-dwarfs might harbor planets whose
properties could be studied with warm-Spitzer and/or JWST, these space
assets form an important part of the strategy.

{\bf \emph{Recommendation \hspace{0.05in} B. I. b. 1}} 
{ \it Use JWST to characterize Earth-sized
transiting planets around M-dwarfs.}

\bigskip

\noindent {\bf B. II. What is the architecture of planetary systems?}

\bigskip

{\bf \emph {Recommendation \hspace{0.05in} B. II. 1}} {\it Move planetary
system architecture studies and multiple-planet statistics beyond the
3 to 5 AU ``ice-line" (also called the ``snow-line") boundary for G-type stars by continuing
long-time-baseline Doppler spectroscopic studies.}

\bigskip

Section 4.2.4 and its associated appendix show that a spacebased microlensing mission has significant advantages over a ground-based network in being able to collect complete statistics on planetary masses and separations, including free-floating planets, as a function of stellar type and location in the Galaxy. 

{\bf \emph{Recommendation \hspace{0.05in} B. II. 2}} {\it  Without
impacting the launch schedule of the astrometric mission cited above,
launch a Discovery-class space-based microlensing mission to determine
the statistics of planetary mass and the separation of planets from their
host stars as a function of stellar type and location in the galaxy,
and to derive $\eta_\oplus$ over a very large sample. }

\bigskip

{\bf \emph{Recommendation \hspace{0.05in} B. II. 3}} {\it Begin
construction of a 30 meter telescope to do optical direct detection of
giant planets to understand planetary system architecture and planet
formation, and invest in appropriate instrumentation for planet
detection, characterization, and disk studies.}

\bigskip

\noindent {\bf B. III. When, how and in what environments are planets formed?}

\bigskip

{\bf \emph{Recommendation \hspace{0.05in} B. III. 1}} {\it Implement
next-generation high spatial resolution imaging techniques on
ground-based telescopes (AO for direct detection of young low mass
companions and interferometry for disk science).}

\bigskip

\begin{figure}[h]
\centering
\includegraphics[scale=0.4]{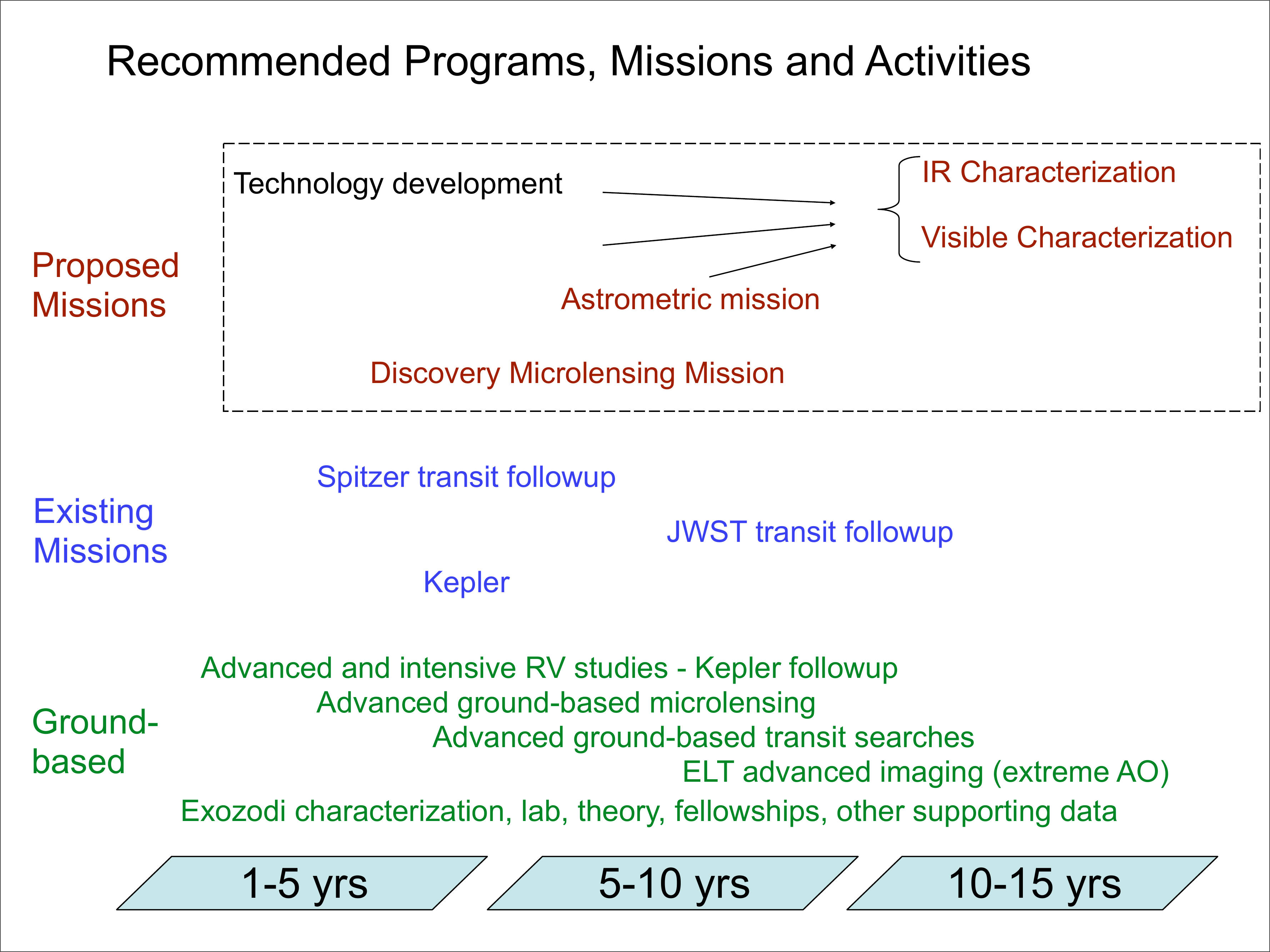}
\vspace{1mm}
\caption{Graphical depiction of the recommended strategy parsed in time periods and according to ground-based, space-based or existing assets.}
\label{fig:recommended_program}
\end{figure}

\subsection{C. Recommendations for 11--15 Years}

\bigskip

\noindent {\bf  C. I. What are the physical characteristics of planets in the
habitable zones around bright, nearby {\it F, G, K} stars?}

\bigskip

{\bf \emph{Recommendation \hspace{0.05in} C. I. 1}} {\it Provided
$\eta_\oplus$ is high and typical exo-zodi emission sufficiently low,
conduct space-based direct detection and characterization for
Earth-mass planets found by astrometry, in the habitable zone of
nearby solar-type stars.}

\bigskip

Given the current state of development of direct detection approaches,
the coronagraph/occulter appears to be more mature and less costly
than interferometric techniques. However, before a choice of which
direct detection approach to pursue is made, additional technological
studies as well as better constraints on $\eta_\oplus$ and exo-zodi
emission should be obtained.  

\bigskip

Should any of the following occur, namely $\eta_\oplus$ low, exo-zodi
emission high, or no astrometric candidates are found which are
suitable for direct detection missions, then technology studies of facilities capable of direct detection of Earth-sized planets around more
distant or more difficult targets are warranted.

{\bf \emph{Recommendation \hspace{0.05in} C. I. 2}} {\it Begin development
of a more ambitious space-based direct detection system, with
international collaboration where appropriate, to be launched beyond
the report's time horizon. Such a mission would either follow up on
the successful direct detection work begun in this period of the
strategy, or be used to overcome technological or observational (e.g.,
low $\eta_\oplus$) impediments that prevented such detection and
characterization in the report's 15-year time horizon. }

%{\bf \emph{Recommendation}} {Begin development of a more ambitious
%space-based direct detection system, with international collaboration
%where appropriate, to be launched beyond the report's time
%horizon. Such a mission would either follow up on the successful
%direct detection work begun in this period of the strategy, or be used
%to overcome technological or observational (e.g., low $\eta_\oplus$)
%impediments that prevented such detection and characterization in the
%report's time horizon. }

\bigskip

\noindent {\bf C. II. What is the architecture of planetary systems?}

\bigskip

{\bf \emph{Recommendation \hspace{0.05in} C. II. 1}}} {\it Should any of the
following occur, namely $\eta_\oplus$ low, exo-zodi emission high, or
no astrometric candidates are found which are suitable for direct
detection missions, then pursue with stronger emphasis studies of larger
planets via ground-based direct detection, and studies of the architecture of planetary systems with
ground- and space-based tools including microlensing.}

\bigskip

\noindent {\bf C. III. When, how and in what environments are planets formed?}

\bigskip

{\bf \emph{Recommendation \hspace{0.05in} C. III. 1}}
 {\it  Invest in technology for the
next-generation observations of planet-forming disks and disks where
young planets reside. These capabilities should include, for example,  sensitive
(equivalent to Spitzer and Herschel) far-infrared interferometric
observations from space to achieve the resolution of ALMA closer to
the peak of the dust spectral energy distribution. A particular need for the far-infrared is technology investment to increase detector performance. }

\subsection{D. Ongoing All Years}

\bigskip

Theoretical work is the tool to interpret the findings resulting from
the observational and technological recommendations in sections
1.1 through 1.4 above and to put them into a broader context.

\bigskip

{\bf \emph{Recommendation \hspace{0.05in} D. 1}} {\it A strong theory
program is essential to address all three compelling questions. Theory
programs include, planet atmosphere and interior studies, laboratory
astrophysics, n--body and hydrodynamic codes with large computational
demands to study planet formation, evolution, and dynamical evolution,
and stellar astrophysics (e.g. nearby young star samples, stellar
ages).}

\bigskip

{\bf \emph{Recommendation \hspace{0.05in} D. 2}} {\it NASA and NSF should
provide support for activities that maximize the knowledge return from
the data and train new scientists in the field, including theoretical studies
(See D. 1.), stellar properties surveys, and competitive fellowships
for young researchers.}

\bigskip

\bigskip

The implementation of this strategy--resulting potentially in the
detection and preliminary characterization of an Earth-like planet in
the habitable zone of another star--will be of profound scientific and
philosophical significance.  We assert here that the goals are
attainable within decades, but its accomplishment will require a
sustained commitment of fiscal resources. We expect, given the
universal nature of the questions being addressed, that this strategy
will be pursued as a collaboration among international partners. But
the United States' pivotal role in large space science endeavors over
the history of space exploration suggests that the nation must lead
the effort if it is to be completed in a timely fashion.

\bigskip

\end{titlepage}
\clearpage
\setcounter{page}{4}

% ============================================================================
% TABLE OF CONTENTS
% ============================================================================

% NOTE:  You have to run Typeset TWICE to use the TOC command!!!!!
%
%              The FIRST time just defines all the pagination but does NOT update the TOC.
%
%              The SECOND time actually updates and inserts the TOC.

% \setcounter{tocdepth}{0}   % shows only highest level in outline, i.e., the "Parts"
% \setcounter{tocdepth}{1}   % shows to next level in outline, i.e., the "Sections" under the "Parts"
\setcounter{tocdepth}{2}   % shows to next level in outline, i.e., the "subsections under the "Sections"
 
\tableofcontents
\pagebreak

\pagenumbering{arabic} % CHANGE TO ARABIC NUMERALS FOR MAIN DOCUMENT

% ============================================================================
% Begin New PART - INTRODUCTION
% ============================================================================

\chapter{Introduction}\label{intro}

This is a scientific strategy for the detection and characterization of extrasolar planets; that is, planets orbiting other stars. As such, it maps out--over a 15-year horizon--the techniques and capabilities required to detect and measure the properties of planets as small as Earth around stars as large as our own Sun. It shows how the technology pieces and their development fit together to achieve the strategy's primary goal: if planets like Earth exist around stars within some tens of light years of our own Solar System, those planets will be found and their basic properties characterized. Essential to this strategy is not only the search for and examination of individual planets, but also a knowledge of the arrangement, or \emph{architecture}, of planetary systems around as large a number of stars as possible; this is the second goal of the strategy. The final goal of the strategy is the study of disks around stars, important both to understand the implications of the variety of exoplanet systems for planet formation, and to determine how many nearby stars have environments around them clean enough of debris that planets may be sought and, if found, characterized. 

It is important that the program target planets as small as the Earth, if we are to determine whether the conditions we find on our own world are a common outcome of planetary evolution. The only example of a habitable world we have is our own one-Earth-mass planet, and indeed our nearest neighbor, Venus, is nearly the same mass but uninhabitable by virtue of closer proximity to the Sun. Searching for planets, for example, five times the mass of Earth is easier but should they turn out to lack habitable atmospheres we would not know whether this is by chance or a systematic effect of the higher mass.   This goal by no means excludes detection and study of larger planets, which themselves will provide a wealth of discoveries and new knowledge, but the Task Force strongly believes that public interest ultimately lies in knowing whether planets like our own are a common outcome of cosmic evolution. For that reason, we have set the bar high. 

The charge to the Exoplanet Task Force (``ExoPTF"), which conducted the study for the Astronomy and Astrophysics Advisors Committee (``AAAC"), is reproduced after the bibliography. The ExoPTF membership, listed above, was chosen from the astronomical commu"nity to represent the range of techniques and expertise involved in exoplanet research today: in instrumentation, observation, and theory. The Task Force met five times from February to September 2007. Representatives of NASA and the NSF were present as \emph{ex officio} members, and a number of presentations from outside experts were solicited to provide the Task Force with the most up to date information available. 

Early in the process, ``white papers" were solicited on the Task Force's public website, inviting interested individuals to submit ideas for techniques, missions, and theoretical investigations in the area of planet search and detection. Eighty-five white papers were received, read, and discussed by the Task Force. A number of the white papers provided important information that, when verified with outside experts, influenced the ultimate strategy; the sum total of the white papers provided an impressive indication of the breadth and depth of thinking in the astronomical community. 

Periodic briefings on the Task Force's progress were provided to the AAAC committee, and general reports were provided to the community at selected scientific conferences and workshops. A draft report was reviewed by the AAAC in October 2007 and a final report delivered in March 2008.

The principal goal of this strategy--the detection of habitable planets--is perhaps among the most challenging and at the same time most rewarding goals in modern science. In a sense, habitable planets are in an ``anti-sweet spot", wherein the position they occupy around nearby Sun-like stars does not correspond to the highest senstivity zone for any of the techniques discussed here. Further, Earth-sized planets will have extremely low signal-to-noise, making analysis of the detected signatures of these objects extremely difficult. The Task Force was encouraged, however,  by the extraordinary progress made in the last half-decade in detecting planets down to less than ten times the mass of the Earth, in characterizing extrasolar giant planets spectroscopically, in technological progress accomplished in key approaches that will be needed to detect and study Earth-sized planets, and in the innovative and entrepeneurial character of the community in proposing and in some cases executing novel approaches to detection and characterization. It is the vigor of the field and its practitioners that ensures ultimate success in this difficult endeavor. 

The Task Force firmly believes that the strategy presented herein is reasonably paced, technically possible, and scientifically compelling. It is robust against surprises both in the areas of science ({\it e.g.,} frequency of occurrence of planets) and technological development. But the strategy requires a level of stamina not typical of the Federal enterprise. Important and exciting results will come from the near-term portions of the strategy (\emph{i.e.,} within 5 years), but the ultimate goal of finding and characterizing a planet like our own, around a star like our Sun, will require the full 15-year time horizon for its realization. 

\pagebreak

% ============================================================================
% Begin New PART - SCIENTIFIC AND PHILOSOPHICAL SIGNIFICANCE
% ============================================================================

\chapter{Scientific and Philosophical Significance of Detecting other Earths}

The idea that humankind's home in the cosmos is just one of countless worlds has been favored by many cultures throughout human history.  The thread of western thought on this matter goes back to antiquity, but the view of the cosmos promulgated in medieval Europe put the Earth in a special place and discouraged speculation on the possibility of other inhabited worlds. Nonetheless, the German philosopher and theologian Albertus Magnus wrote in the 13th century: ``Do there exist many worlds, or is there but a single world? This is one of the most noble and exalted questions in the study of Nature." Two centuries later, Nicolas of Cusa asserted on a philosophical basis that the heavens were teeming with inhabited worlds.  The notion of a plurality of worlds was put on a scientific footing in the mid-16th century by Copernicus' cosmology that displaced the Earth from the center of the Solar System, so that it became possible to imagine an abundance of unseen worlds in the form of other solar systems, as Giordano Bruno wrote: \\

\begin{quote} 
\emph{``There are countless suns and countless earths all rotating around their suns in exactly the same way as the seven planets of our system. We see only the suns because they are the largest bodies and are luminous, but their planets remain invisible to us because they are smaller and non-luminous. The countless worlds in the universe are no worse and no less inhabited than our Earth."}\\
\end{quote}

These ideas were not popular in the chilly counter-Reformation environment of the times, and indeed Bruno was burned at the stake in 1600. Less than a decade later, however, Galileo Galilei saw through his telescope first that our Moon--rather than being a perfect sphere--was in fact a mountainous world in its own right, and then that planet-like bodies (moons) orbited Jupiter just as our Moon orbits the Earth. The progressive displacement of humanity's position from the cosmic center became irreversible. Today we know the Earth is one of countless bodies that orbit the Sun; the Sun orbits the center of the Milky Way Galaxy along with $\simeq 200$ billion other stars; there are tens of billions of galaxies like the Milky Way in the cosmos.

And yet, one special aspect of the Earth remains: of all the planets in the solar system, only Earth demonstrably is teeming with life. If life exists on other planets or moons of the Solar System, it does not have the profound impact on the properties of those bodies that life has on Earth, and for all we know Earth is the only abode of life around the Sun. No world in the solar system is remotely suitable for life as we know it in the way that Earth is; we must look beyond our own Solar System to the stars if we are to determine whether other habitable worlds like the Earth exist. To complete the Copernican revolution, we must detect and characterize ``extra-solar" planets the size of our home world.

Beginning in the early 1990s, planets around other stars began to be detected, first by their indirect effects on the motion of their parent stars, then directly. Today, over 260 planets are known to exist around other stars, about 10\% of which can be studied directly. All of these, with just a couple of exceptions, are more than ten times more massive than the Earth, and we could not today detect our own home world in orbit around a Sun-like star a mere 4 light years away--the distance to the closest star Proxima Centauri. But we are close to detecting and characterizing Earth-sized planets around smaller stars, and the means required to do so around a star like the Sun are so well understood that we argue here that the required technologies could be fielded within fifteen years. A strategy that maximizes the chances for success while minimizing risk forms the core of this report. 

Complementary to detecting and characterizing individual planets the size of the Earth is determining whether the architecture of our own solar system--rocky planets huddled close to the parent star and multiple giant planets in more distant orbits--is typical. We have enough information today to say that upwards of 15\% of Sun-like stars harbor giant planets, and that planets down to the size of Neptune also seem to be common, but have no information on the occurrence of planets the mass or size of the Earth. Because the formation of rocky planets like Earth may have been quite different from that of giant planets, extrapolation is risky. The scientific importance of understanding both the statistical occurrence of Earth-sized planets and the architectures of planetary systems is high, because these are crucial to understanding the processes by which planetary systems form. They therefore form a part of our strategy in conjunction with direct detection and characterization, as well as the study of planet forming disks and disk debris. Together, these elements will ensure that, no matter what the ultimate commonality or rarity of Earth-sized planets might be, the end result of the strategy will be a deep understanding of how planets fit into the overall scheme of cosmic evolution. 

Humankind stands today at the threshold of answering one of humankind's most ancient questions: is our home world the only suitable abode for life like us in the cosmos? Most humans would answer no, but only because the Copernican mindset is so firmly a part of our modern culture. What if we were rare enough that no such other Earth existed within the reach of modern astronomical telescopes, on the ground or in space? How would we react to the confirmed emptiness the universe presented to us? Would our specialness lead to a profound sense of faith, or a deep loneliness? Would it stiffen humankind's resolve to survive, to solve its political and environmental problems? Would our solitary state lead us to be more mindful of the precious uniqueness of our intellect and awareness?

Conversely, were we to find a planet like the Earth--rocky, similar in size to our world, possessed of an atmosphere-- around nearby stars, our perception of the cosmos around us would change in an equally profound fashion.  We would look up still with wonder, but a handful of stars in our night sky will forever after hold a special place in our imagination, tempting us with wild dreams of flight. Surely that too would make us refocus our energies to hasten the day when our descendants might dare to try to bridge the gulf between two inhabited worlds. \\

\begin{quote}
{\emph{``Which one is it, Mommy?" asked the older of her two children. They had walked away from the campfire, and gazing now at the familiar pattern of stars in the night sky, a question far different from any ever asked by thousands of generations of human beings drifted off in the cool night air. ``Look at the bright star over there," the woman responded to them ``now move your eyes a little to the right, and you'll see that slightly fainter star. The planet belongs to that one. It's almost exactly the size of the Earth, is just a little closer to its sun than we are to ours, and the space telescope that your mommy helped build found oxygen in its atmosphere. That world has air that creatures like us could breathe." ``Who lives there?" asked the younger one. ``No one knows," the woman replied, ``but maybe they are looking at us, right now, wondering the same thing." }}
\end{quote}

\pagebreak

% ============================================================================
% Begin New PART - THE KNOWN EXTRASOLAR PLANETS
% ============================================================================

\chapter{The Known Extrasolar Planets}

%\setcounter{section}{0}

% - - - - - - - - - - - - - - - - - - - - - - - - - - - - - - - - - - - - - - - - - - - - - - - - - - - - - - - - - - - - - - - - - - -
\section{ Setting the Stage}

The familiar planets of our own planetary system have been revealed through 
hundreds of years of telescopic observation and by more than forty 
years of robotic space missions. We have movies of the swirling ammonia 
cloud storms on Jupiter, up-close views of a great dark spot on 
Neptune, images of the hellishly hot surface of Venus, 
and intrepid robots exploring the 
dusty panoramas and salmon skies of Mars. In the Solar System, the major
planets lie on dynamically stable, nearly circular orbits, with a clear
dichotomy between the rocky terrestrial planets in the inner regions and the
giant planets further out. This well-ordered configuration supports a
theory of planet formation in which the planets emerged from a largely quiescent
disk of gas and dust that orbited the newborn Sun.

In October 1995 at a conference in Italy, Michel Mayor and Didier Queloz 
announced their discovery of a planet circling the nearby sunlike star 51 Pegasi (Mayor and Queloz, 1995).
The planet, they claimed, was roughly 150 times more massive than Earth 
and traveled on an orbit that takes only 4.2 days to complete. The general
initial reaction was frank disbelief, as conventional wisdom in 1995
held that massive planets should be located much farther 
from their parent stars. Within days,  however,
the existence of 51 Peg b had been confirmed.
The discovery of this bizarre, completely unpredicted world spawned a 
new astronomical field: the study of alien planetary systems. 
Astronomers now know of over 260 planets, which populate 
systems of astonishing diversity. In addition to ``ultra-hot 
Jupiters" with orbital periods of just over a day, we know of frigid 
worlds only a few times more massive than Earth orbiting red dwarf stars, 
and bizarre multiple-planet systems that have likely experienced histories 
rife with planetary close encounters, collisions, and ejections. 
Extrasolar planets are enabling us to better understand how planetary systems 
form and evolve, and they are allowing us to place our own planetary 
system into the context of the galactic planetary census. Within the 
next ten years, we will know whether systems like our own are common or 
rare, and we will almost certainly have specific examples of alien 
Earths; terrestrial planets orbiting at distances from their parent 
stars where liquid water, and perhaps life, can exist. 

In this chapter we will briefly review some of the science highlights emerging 
from the amazing progress in extrasolar planet science achieved in the dozen 
years since the discovery of 51 Peg b.  We first give an overview of the 
characteristics of known exoplanets, including their orbital characteristics, 
masses, and sizes. We then consider the science of the objects themselves, 
including their atmospheres.  We also briefly discuss the role of disks in 
indirectly detecting extrasolar planets.   Theories of extrasolar giant planets 
are being tested almost daily by new discoveries, but our ability to detect and 
characterize extrasolar terrestrial planets hinges on our models of these 
objects.  We thus conclude this section with a look at the state of theory of 
extrasolar terrestrial planets. 

% - - - - - - - - - - - - - - - - - - - - - - - - - - - - - - - - - - - - - - - - - - - - - - - - - - - - - - - - - - - - - - - - - - -
\section{The Diversity of Exoplanets}

\begin{figure}
\begin{center}
\includegraphics[angle=0,scale=.9]{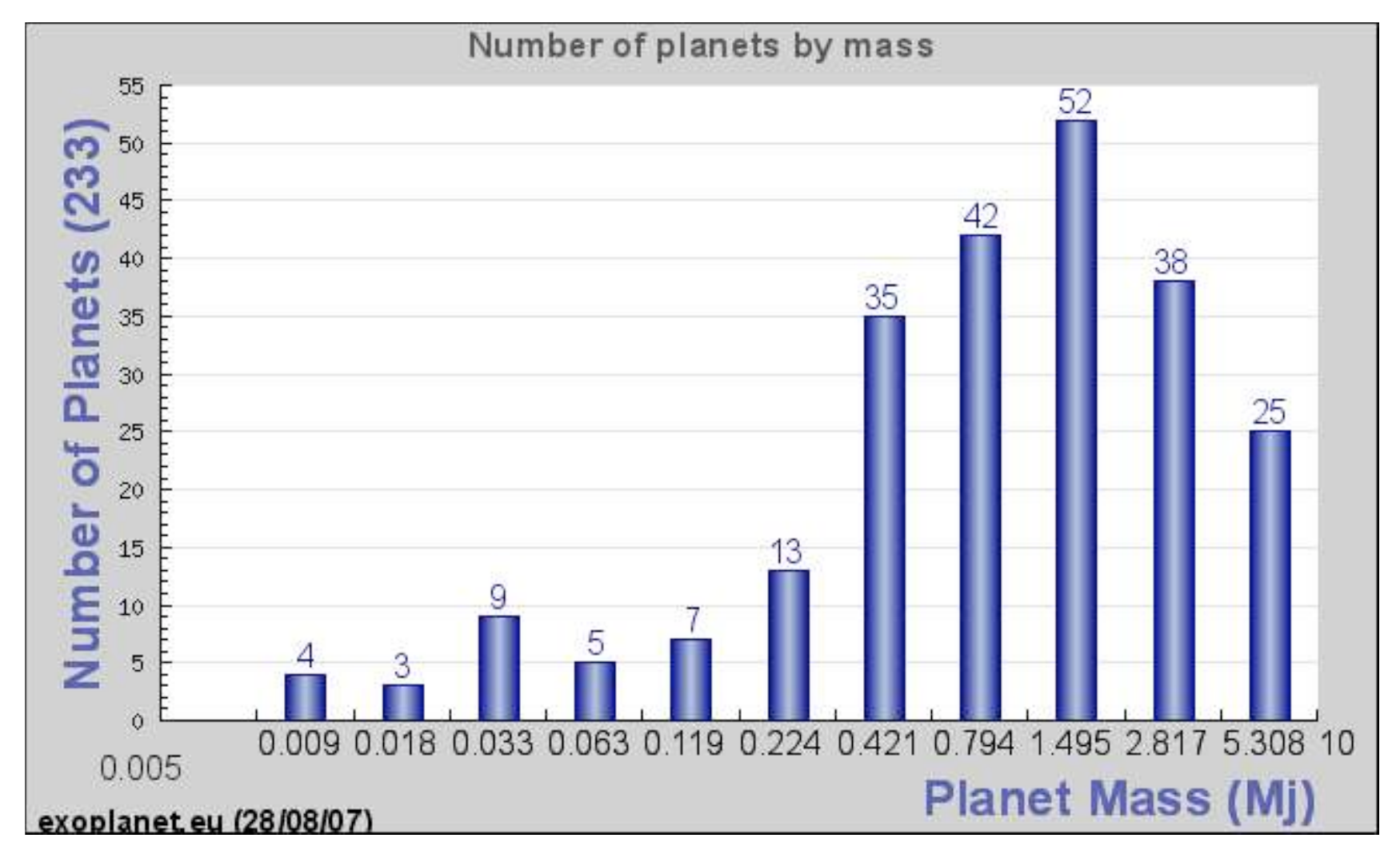}
%\plotone{allhist}
\caption{Masses of extrasolar planets.  }
\label{fig:allhist}
\end{center}
\end{figure}

The masses of all known extrasolar planets are shown in Figure 
~\ref{fig:allhist}.  The precise shape of this distribution, called the ``mass function," is still a matter 
of investigation and debate.  In particular
there are an insufficient number of smaller mass planets
detected to adequately constrain the complete mass function. As the number of known planets grows,
it is however becoming clear that those discovered so far
can be sensibly divided into a number of basic categories. 

\subsection{Ultra-Hot Jupiters}   There is increasing evidence that the planets with the very
shortest periods ($P<2\, {\rm days}$) can be meaningfully grouped into a category
of ``ultra-hot Jupiters"\footnote{Despite their nicknames, ``Ultra-Hot Jupiters'' and ``Hot Jupiters'' likely bear 
little resemblance to hotter versions of our own Jupiter.}.  These objects, of which six are now known, have 
average
masses more than 50\% greater than Jupiter, and are thus considerably
more massive than the population of conventional hot Jupiters (see below) found 
at slightly larger orbital distances.
It is possible that the extreme conditions under which ultra-hot Jupiters exist dictate that they contain more 
mass in order to survive evaporation for billions of years; alternatively, they could 
be forming through a somewhat different set of processes.

\subsection{Hot Jupiters}   These objects, characterized
by 2-7 day orbital periods and near-zero eccentricities, are estimated to
accompany roughly $0.5 \%$ of the main sequence dwarfs in the solar
neighborhood, and their average mass is approximately 1 $M_{\rm Jup}$ (Jupiter mass),
assuming a random distribution of viewing angles for the aggregate
of known members.  The prototypical ``hot Jupiter"
is the companion to 51 Pegasi.  To date, surveys
have found forty six planets in this category, with new ones currently
being added at a rate of several per month. 

Due to their proximity to
the parent star, hot Jupiters have a high {\it a priori} geometric probability of 
exhibiting transits (planets crossing in front of and behind their parent stars), 
and indeed, about 20 transiting examples are currently
known. The aggregate of transiting hot Jupiters displays a remarkable diversity
in planetary radii. Current theoretical models suggest that these planets 
should
all be similar in size to Jupiter, but the transit observations show a much wider size range than expected (for example, HD 149026 b has a
radius of 0.72 $R_{\rm Jup}$ (Jupiter radius), much less than expected, while TrES-4
has a radius of 1.74 $R_{\rm Jup}$, despite being only 85\% of
Jupiter's mass). It is believed that an object like HD 149026b, less massive than Jupiter and with an anomalously
small radius for a hydrogen-helium composition, must be dominated by heavy elements, whereas those
with larger than expected sizes are inflated by as-yet undetermined sources
of interior heat (Charbonneau et al., 2006). A full and definitive explanation for the diversity of observed
radii for the transiting extrasolar planets is one of the most exciting current
research areas in the exoplanetary field and is discussed further below.

\subsection{Eccentric Giants}   These
Jovian-mass planets travel on eccentric (non-circular) orbits in the region between 0.1
and several AU from their parent star; one Astronomical Unit is the distance from the Sun to the Earth. 
The Eccentric-Giant population 
constitutes the bulk of the current
planetary census. The average mass of the
known members of this group is $\sim 5$ $M_{\rm Jup}$, which is significantly
higher than the average mass of the Hot Jupiters (even when the observational
selection effects are taken into account).
Ongoing Doppler radial velocity surveys have excellent sensitivity to this 
population, and
so their statistical distribution is now quite well understood. Indeed, 
approximately 7\% of
the F, G, and K-type main sequence
stars in the solar neighborhood appear to be accompanied by an Eccentric
Giant, and thus it is important to gain a
definitive explanation of the processes that give rise to their odd orbital 
properties.
An important clue is almost certainly provided
by the fact that the parent stars of both the Eccentric Giants and
the Hot Jupiters tend to be significantly more metal rich than the Sun. Many 
Eccentric Giants are members of multiple-planet
systems, some of which show interesting planet-planet interactions on observable
timescales. The presence of more than one planet in a given system often allows for
inferences with respect to formation and evolution mechanisms that would 
otherwise
be difficult to make.

The standard paradigm of planet formation via
gas accretion from the protoplanetary disk onto protoplanetary cores, called the ``core-accretion" model,
is strongly supported by the observed planet-host star metallicity
connection. It
predicts that giant planets should begin
life in circular orbits. Therefore, the elongated, eccentric orbits
and observed multiplicity
common to many of the Eccentric Giants is likely telling us either that
violent orbital disruption is common,
or that strong eccentricity-pumping mechanisms are often at work in
protostellar disks. Alternately, it may turn out that
this subclass of
high-mass, high-eccentricity planets reflects formation via
gravitational instability
and direct fragmentation of the disk.

\subsection{Long-Period Giants}    Jupiter, Saturn, Uranus, and Neptune constitute the prototype objects for the
``Long-Period Giants" category. We define this group as being
characterized by long period ($\ge$
several years), nearly circular, orbits. Extensive, high-precision
Doppler radial velocity surveys have been operating for
a bit more than a decade, and therefore,
the extrasolar planets found to date have orbits
which are generally shorter than $P\sim5$ yr ({\it cf.} that of Jupiter at nearly 12 years. Thus, while our definition is somewhat arbitrary, it is convenient: extrasolar planetary systems with
multiple giant planets in long-period low-eccentricity orbits would mark the
long-sought true analogues of our solar system. 

As the time baseline for the radial velocity experiments has grown--and 
the experimental accuracy improved--giants at ever increasing orbital 
radii have been discovered.  The RV groups have indicated that many 
stars show slowly varying trends in their velocities, suggesting that 
announcements of true Jupiter- and Saturn-twins will eventually be 
forthcoming.  Such discoveries are important for two main reasons: they 
fill in our understanding of the arrangement and types of planets around stars ("planetary system architectures") and provide targets 
for direct detection and characterization.

First, more distant giants fill in our knowledge of planetary system 
architectures at the scale of our own solar system.  The masses and 
orbital elements of such planets provide insight into planetary system 
formation and later evolution by migration.  To understand the 
frequency of terrestrial planets, we need to understand how often  
giant planets migrate through the habitable zone, potentially 
disrupting terrestrial planet formation.  We do not yet know if 
planetary systems with (in some cases multiple) massive planets 
interior to and within the habitable zones are the exception or the 
rule.  Only by understanding the frequency of `Jupiters' on circular 
orbits beyond the habitable zone can this question be addressed.  
Radial velocity, with its already established long baselines and 
ability to constrain orbital properties, is uniquely well suited to 
answering this question.  The orbital dynamics of these systems can 
also be studied to understand the stability of orbits and the 
possibility of delivery of volatiles to their primary star's habitable 
zones.

The second great value of giant planets on wide orbits is that they are 
potentially directly detectable by the next generation of large 
ground-based telescopes (in the infrared) or by small, space-based 
coronagraphs (in the optical).  For example, the maximum angular 
separation of the giant planet 55 Cnc d from its primary is a 
tantalizing 0.39 arcsec.  Six other planets have angular separations 
greater than 0.2 arcsec; for 0.1 arcsec the number now stands at 
twenty.  Knowledge of the best time 
to attempt direct detection of such planets, in terms of their orbital separation, is invaluable.

But if orbits and masses are already known, why is direct detection 
important?  The simple answer is that planets are far more than masses 
on springs.  To give one example, the atmospheric composition of every 
solar system giant planet differs--by at least a factor of three--from 
that of the sun.  The diverse atmospheric composition of Jupiter and 
Saturn may point to the enrichment of these atmosphere by late-arriving 
planetesimals from the outer solar system.  The fingerprints--even if 
we do not yet fully know how to interpret them--of the planet formation 
process are imprinted on these planets.  Exploring how atmospheric 
composition varies between planets lying inside and outside of the `ice 
line'\footnote{The boundary in a circumstellar nebula exterior to which 
water is found as condensed ice.} or with mass or stellar properties, 
will dramatically expand our understanding of planetary formation 
processes.  Furthermore, we expect that there will be even greater 
diversity among extrasolar planets than we find in our solar system.  
Spectroscopy on directly detected planets is well suited to discovering 
the unexpected.

Reading the spectral fingerprints will be challenging, but not 
impossible.  New high-contrast adaptive optics coronagraphs
on current 8-m telescopes will be able to detect and spectrally 
characterize young ($< 1\,\rm Gyr$) giant planets through their 
infrared
emission. Similar instruments on the next generation of large
telescopes will be able to image mature planets either through
thermal emission or reflected light.  A 30-m telescope and an `extreme 
adaptive optics' coronagraph with near-infrared spectroscopy at a 
resolution of $R=100$ will be straightforward for many giant planet 
targets; higher resolution will be possible for some.  Small 
space-based optical coronagraphs should be able to target many giants at 
somewhat lower spectral resolution.
Spectra, even at low resolution, of extrasolar giants are sensitive to 
atmospheric temperature, metallicity, cloud structure, and mass.  The 
campaign to detect and characterize the warmer and more massive brown 
dwarfs (which share much in common with warm Jupiters) has already 
demonstrated this capability to interpret field objects.  As recounted 
above, even the limited spectroscopy possible on the hot Jupiters has 
already yielded new constraints their composition.

While our Task Force fully appreciates the great scientific potential 
for characterizing extrasolar giant planets from space, we recognize 
that this capability may not lie on the critical path to directly 
detecting and characterizing extrasolar terrestrial planets.  Any 
mission that will accomplish our primary goal will (if properly 
defined) also characterize numerous giant planets.  We thus do not call 
specifically for a cool giant planet characterization mission.  However 
if technological innovation enables an inexpensive new approach, or if a mission specifically designed to detect such planets turns out to lie on the critical path to characterization of Earths, we
would enthusiastically await its discoveries.  Given the potential very 
large scientific return, we do suggest that  an extreme adaptive optics 
planet characterization instrument be included as part of the early 
instrumentation complement on any 30-m class telescope.

\subsection{Hot Neptunes}     In the past three years, a new category of planet, the ``Hot Neptunes" 
has
emerged. These objects have masses in the 5-30 Earth-mass range, and have
been found primarily around low-mass stars. Notable members of this class 
include
55 Cancri ``e", which has an orbital period of 2.8 days, and is the innermost 
planet in 
a system containing at least five planets on low-eccentricity orbits. Another is 
Gliese 436 b, which has a 2.6-day orbit, is similar in mass to Neptune, and 
which has 
recently been 
found to transit its red-dwarf parent star. Structural models indicate that
Gliese 436 b is likely similar in composition to Neptune (Gillion et al., 2007). Although Neptune-mass objects are time-consuming
to detect with the Doppler method, the rapid pace of discovery indicates that 
they
are likely very common--an inference confirmed by the microlensing detection of two sub-Neptune mass planets. This abundance is in good concordance with the so-called
core accretion theory of planet formation, which posits that ``failed" giant-
planet
cores (those with little or no gas) should be common around a wide variety of main-sequence stars.

\subsection{Terrestrial Planets}      A final category, terrestrial planets, begins with our 
Solar System's four inner planets, Mercury, Venus, Earth, and Mars.  
Other Terrestrial Planets include three pulsar
planets (Wolszcan \& Frail 1992), and possibly objects such as
Gliese 581 c and Gliese 876 d. The
detection of terrestrial planets around external main sequence stars is one of 
the major
challenges for the next several decades and a focus of this report.

% - - - - - - - - - - - - - - - - - - - - - - - - - - - - - - - - - - - - - - - - - - - - - - - - - - - - - - - - - - - - - - - - - - -
\section{ Statistics of Known Planets }

As of this writing, 215 planetary systems with 263 planets are known. Twenty five of the systems have multiple planets. 
 Distributions
of the planets over mass, semimajor axis\footnote{The major axis of an an elliptical orbit is its longest 
diameter, and the semi-major axis is one half of the major axis; for the special case of a circular orbit, 
the semi-major axis is the radius of the orbit.} or projected separation, and
eccentricity are displayed below, although these distributions should
be interpreted with caution because of observational bias (discussed further below).       

\begin{figure}
\centering
\subfigure[] % caption for subfigure a
{
%    \label{fig:sub:a}
    \includegraphics[width=12cm]{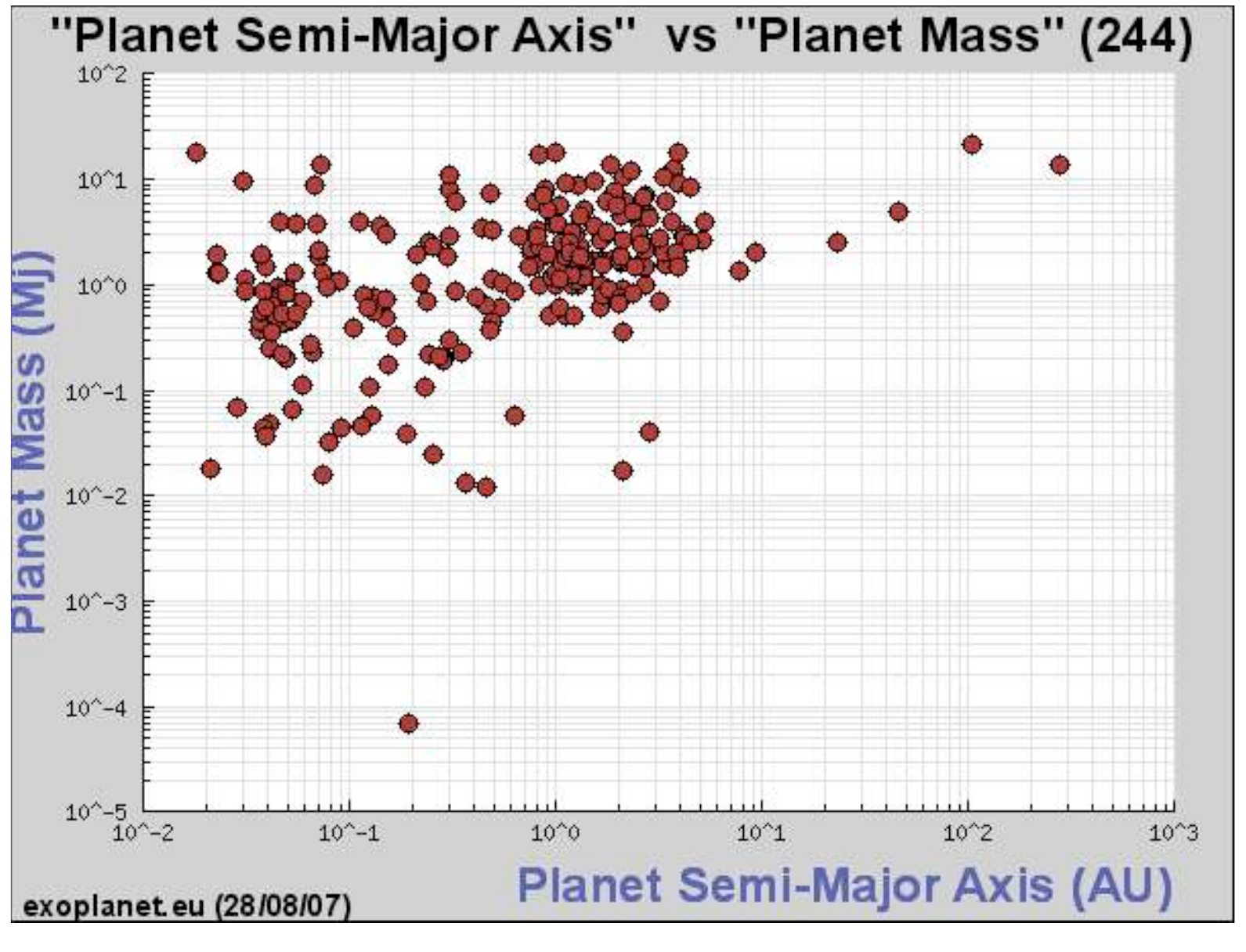}}

\subfigure[] % caption for subfigure a
{
%    \label{fig:sub:b}
    \includegraphics[width=12cm]{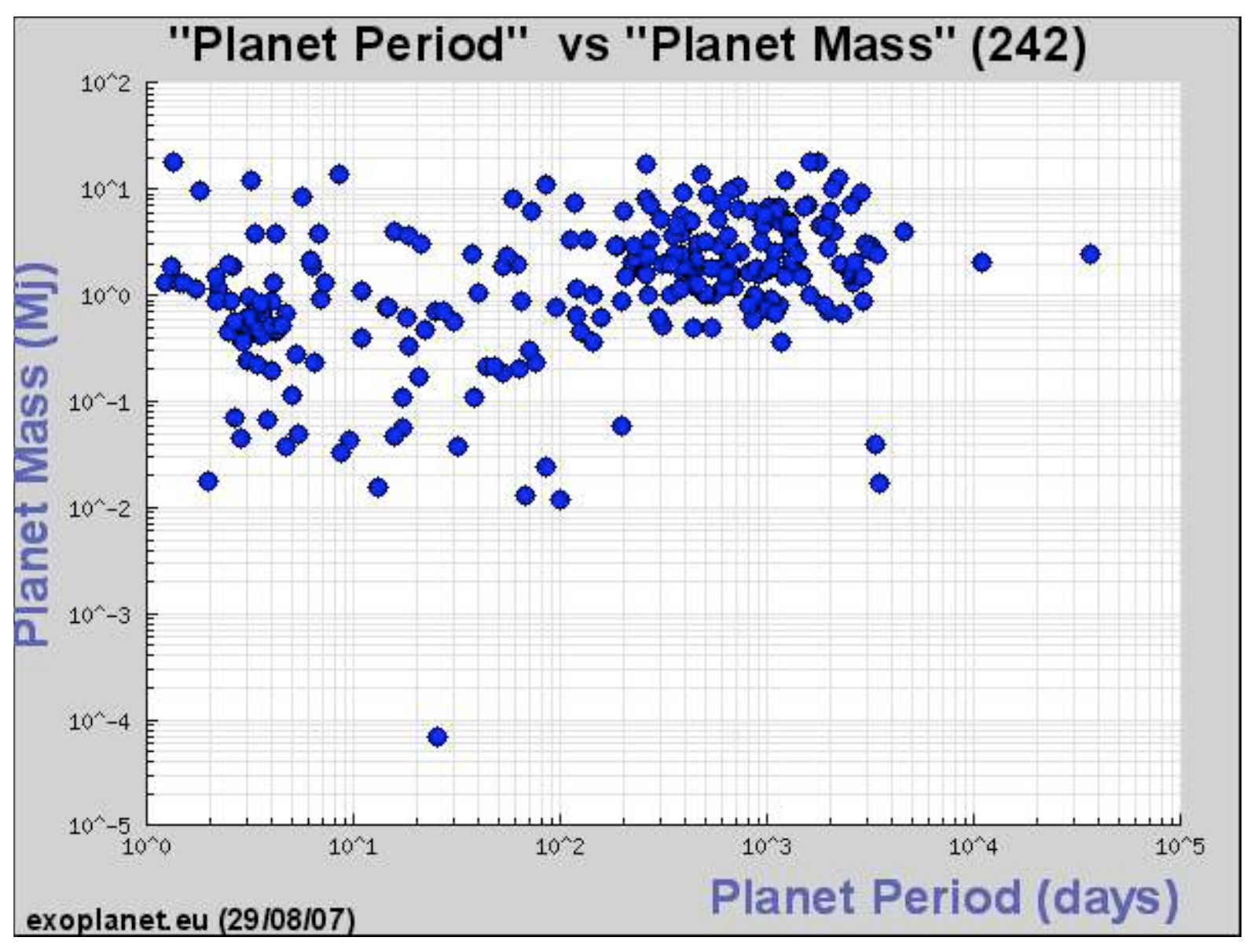}}
\caption{(a) Distribution of extrasolar planet masses as a function of
orbital semimajor axes. (b) Distribution of extrasolar planet masses as a function of
orbital period. }
\label{fig:allab} % caption for the whole figure
\end{figure}

\begin{figure}
\centering
\subfigure[] % caption for subfigure b
{
%    \label{fig:sub:c}
    \includegraphics[width=12cm]{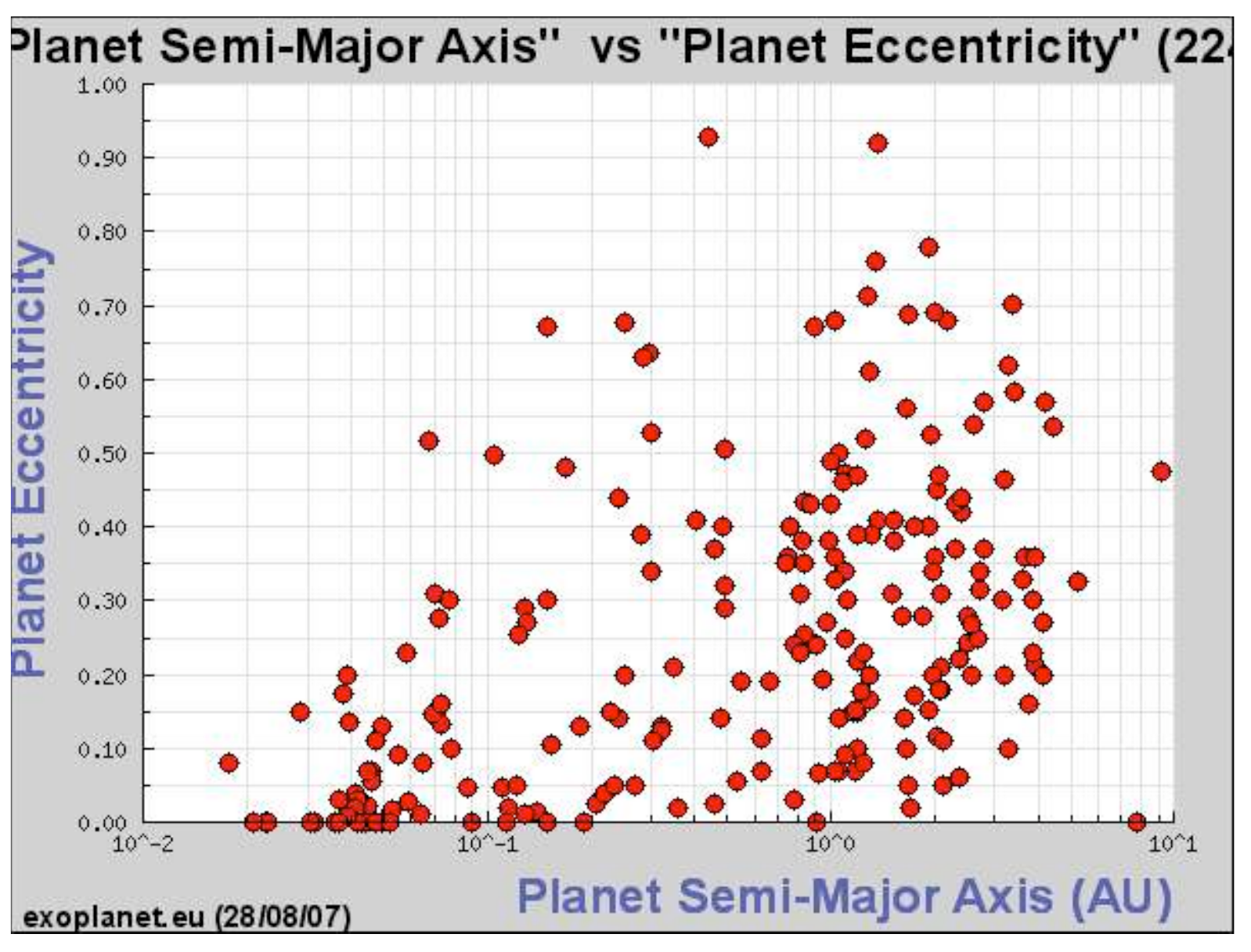}}
\subfigure[] % caption for subfigure b
{
%    \labelfig{fig:sub:d}
    \includegraphics[width=12cm]{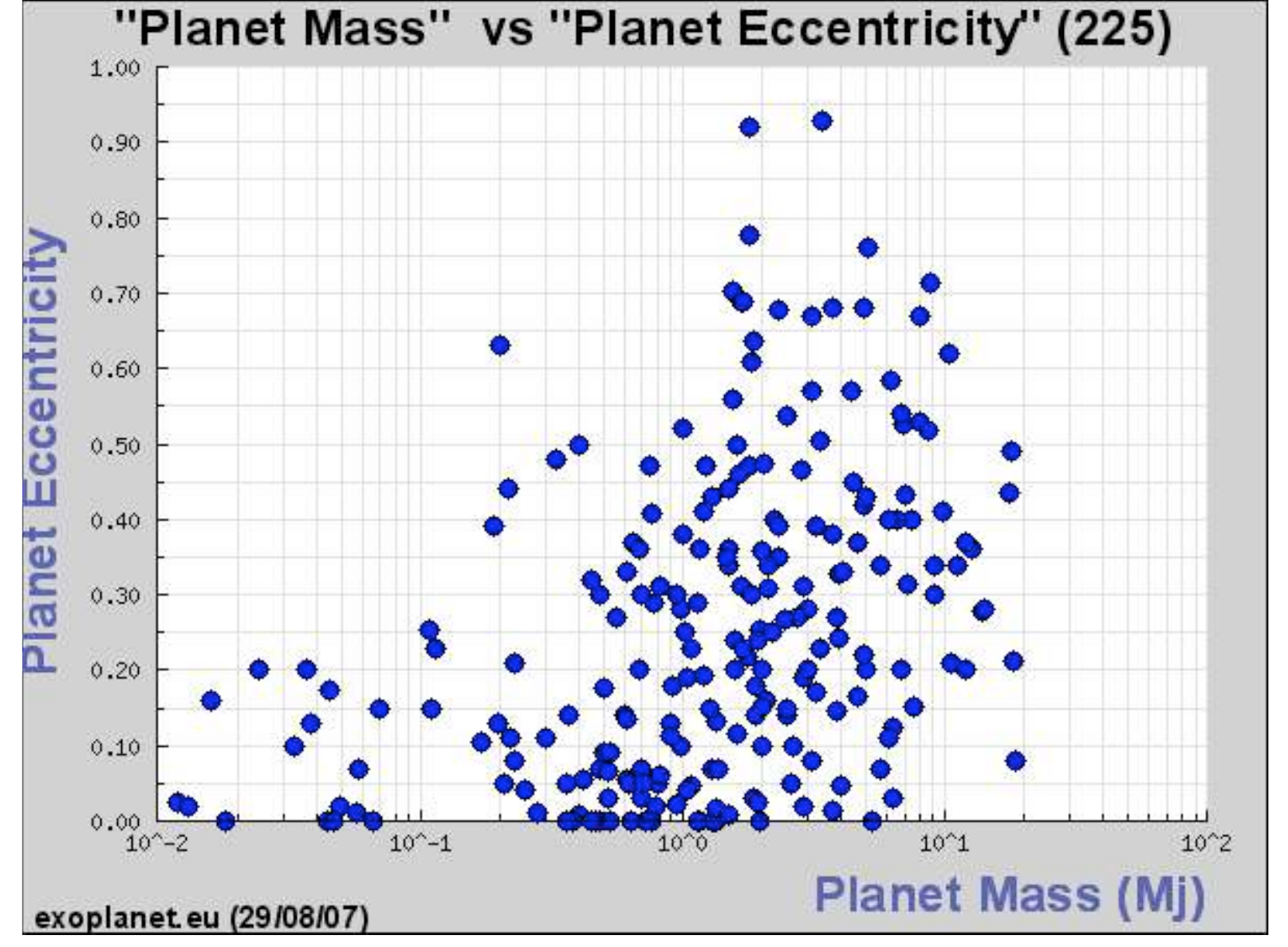}}
\caption{ (a) Distribution of extrasolar planet orbital eccentricities as a
function of orbital semimajor axis. (b) Distribution of extrasolar planet eccentricities as a function of
planet mass.}
\label{fig:allcd} % caption for the whole figure
\end{figure}

Fig.~\ref{fig:allab} panel (a) shows the distribution of orbital semimajor axes of known planets, and panel (b) the distribution of orbital periods.
Fig.~\ref{fig:allcd} shows in panel (a) the distribution of orbital
eccentricities as a function of semimajor axis for all the detected
planets for which eccentricities are known. (Eccentricities are unknown for planets 
detected via microlensing, intrinsically, and those detected so far from direct imaging.)  The median
eccentricity is 0.25, which is larger than the eccentricity of any Solar
System planet (excluding Pluto), even though it is biased low by planets at small semimajor axes whose orbits are likely circularized by tidal dissipation. It is 
still a puzzle why the solar system is comparatively so benign, although we are 
only now gaining the sensitivity to detect true solar system analogs, which 
could still be relatively common.

The largest known eccentricity is 0.93, for HD 80606b.  Some close planets have
significant eccentricities that should have been damped in times much less than the age of the host star; these are thought to be dynamically
excited by additional planets in the system.  The distribution of eccentricities
as a function of mass, shown in panel (d) of Fig.~\ref{fig:allcd}, does not display
any marked trend; this is likely a signature of the planet formation process .

\begin{figure}
\centering
\subfigure[] % caption for subfigure a
{\includegraphics[width=11.5 cm]{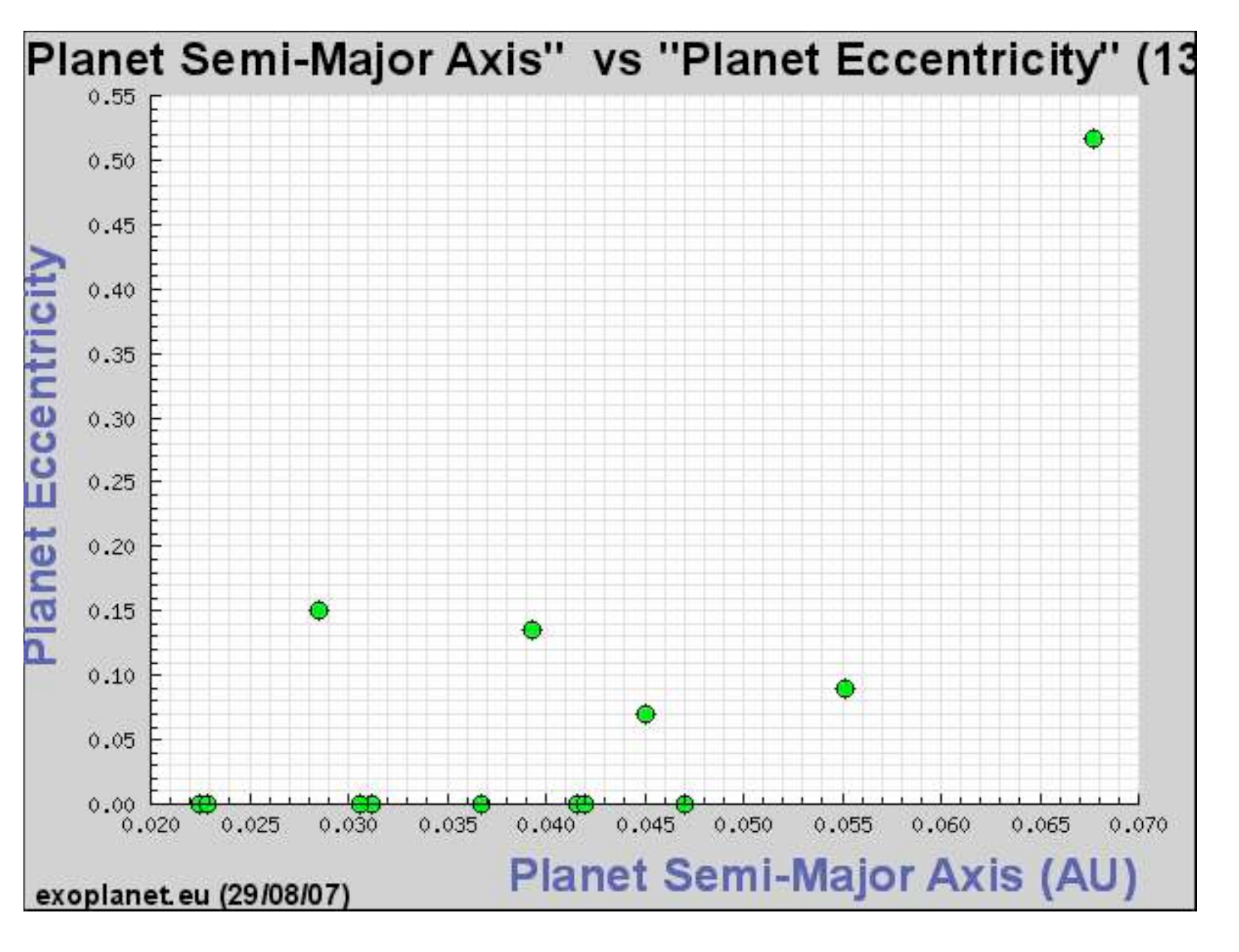}}
\subfigure[] % caption for subfigure a
{ \includegraphics[width=9.5 cm]{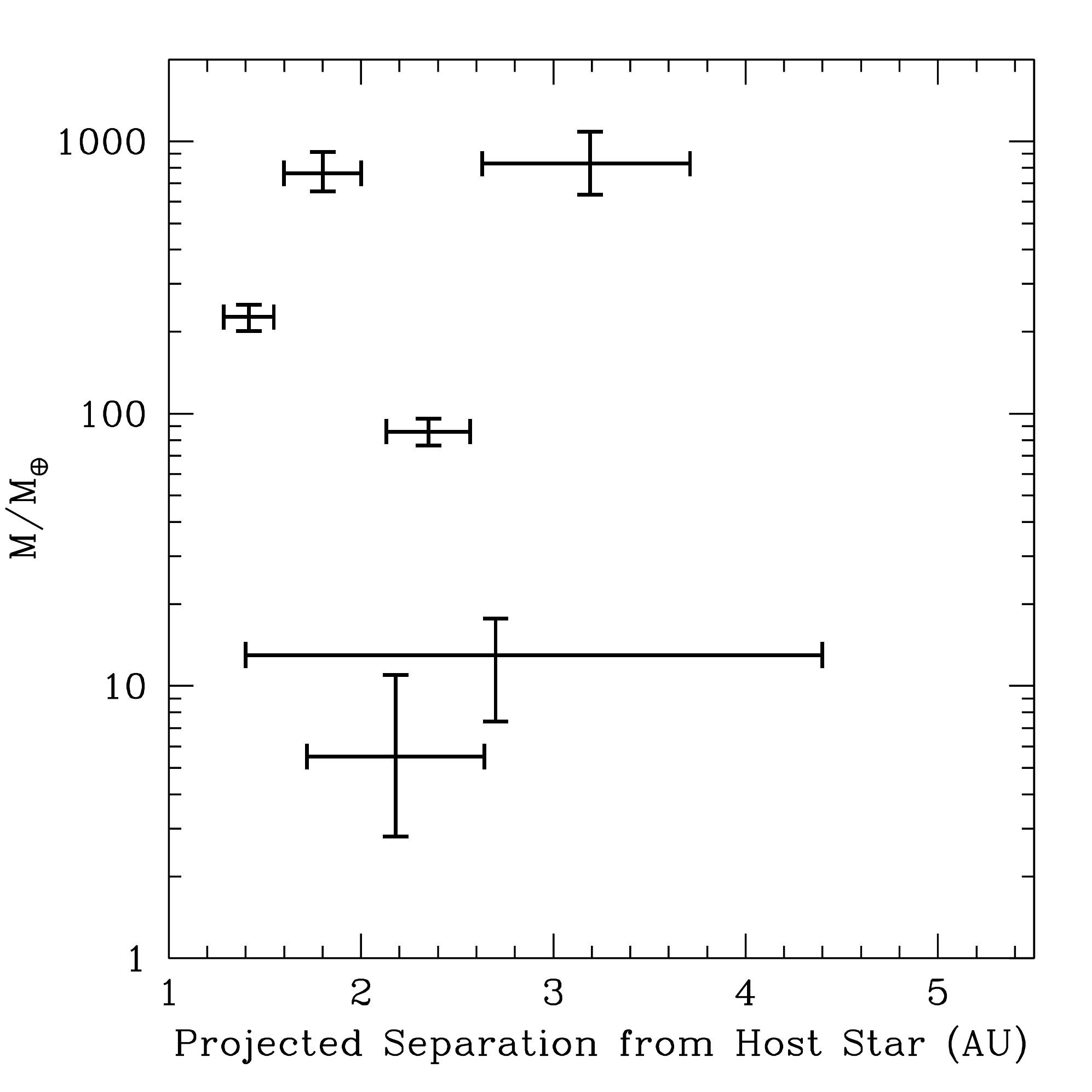}}
\caption{(a) Distribution of transiting planet eccentricities versus
semimajor axis.  (b) Distribution of microlensing-detected planet masses versus projected star planet separation.}
\label{fig:tranmiclensab} % caption for the whole figure
\end{figure}

There are technique-dependent biases.   Most of the detections  of extrasolar planets
have been via the radial velocity technique, so scatter plots limited to RV detections look much
like those for the complete set.  In contrast, an interesting trend emerges for 
the distribution of orbital eccentricity vs semimajor
axis for transiting planets (Fig. ~\ref{fig:tranmiclensab}) panel (a) shows that several have surprisingly large eccentricities that
imply that a second planet in the system is forcing the eccentricity
against the dissipative damping. Likewise for planets discovered by microlensing,  
panel (b) shows the distribution of planetary masses
as a function of projected separation of the planet from the host
star. The latter separation is a proxy measure 
of the semimajor axis, since
the projected separation has somewhat higher probability of being
close to the semimajor axis. The detection of the two
cold Neptune mass planets, despite lower probability of detection than
Jupiter mass planets, suggests that planets near this mass are common, although 
the number of detected planets by microlensing is still too small to draw firm 
conclusions.

%\begin{figure}
%\begin{center}
%\includegraphics[angle=0,scale=.6]{transitevsa}
%\plotone{transitevsa}
%\caption{Distribution of transiting planet eccentricities as a function of
%semimajor axis. \label{fig:transitevsa}}  
%\end{center}
%\end{figure}

%\begin{figure}
%\begin{center}
%\includegraphics[angle=0,scale=.6]{miclensmvsa}
%\caption{Distribution of microlensing-detected planet masses as a
%function of projected star planet separation.\label{fig:miclensmvsa}}  
%\end{center}
%\end{figure}

% - - - - - - - - - - - - - - - - - - - - - - - - - - - - - - - - - - - - - - - - - - - - - - - - - - - - - - - - - - - - - - - - - - -
\section{ Mass - Radius Relationships for Exoplanets}

\begin{figure}
\begin{center}
\includegraphics[angle=0,scale=.6]{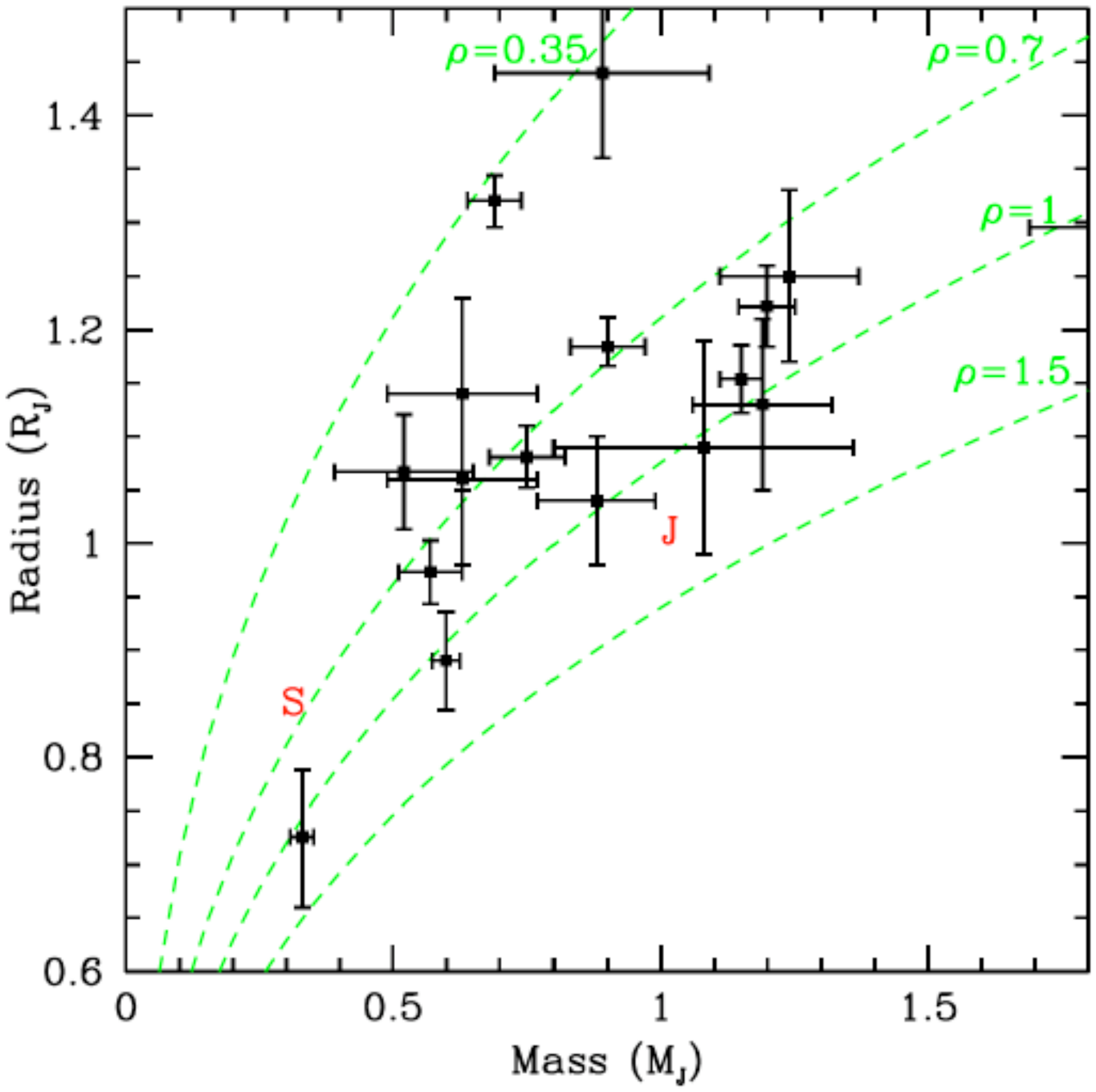}
\caption{Mass-radius diagram for giant exoplanets. The dashed lines are curves 
of constant density. Note the wide distribution of exoplanet masses and radii.  
S and J represent the positions of Saturn and Jupiter.
\label{fig:seager1sm}}  
\end{center}
\end{figure}

Over twenty exoplanets are known to transit their parent stars. Transiting 
exoplanets are the only ones whose physical characteristics can be 
measured with current technology.  Their masses and radii are the most basic physical 
parameters that can measured and have galvanized the field of exoplanet 
characterization (Figure ~\ref{fig:seager1sm}). Combined, they give us the 
planet's mean density. From the mass, radius, and/or density we may infer the 
planet's bulk composition. Most of the known transiting planets have low 
densities, so they must be composed almost entirely of hydrogen and helium - 
just like Jupiter. This is an important diagnostic for the 
exoplanets so extremely close to their parent stars, demonstrating that they are indeed Jovian-type planets whose structure is modified by their proximity to the parent star. On the other hand, several 
of the known transiting exoplanets have decidedly abnormal densities when compared with Jupiter. These anomalies have sparked voluminous research and 
raised interesting questions about the evolution of gas giant planets...

\subsection{Low-Density Gas Giants}  The first discovered transiting 
exoplanet, HD 209458b, has an unusually low density (0.4 $g/cm^3$) relative to the density of Jupiter, which is 1.33 g/cm$^3$. In other 
words, this planet has a very large radius for its mass. Planets are formed big 
and hot, contracting and cooling as they age. HD 209458b should have contracted and 
cooled to a size near that of Jupiter given the age of the star and hence, presumably, that of the planet; but instead it has a radius 30\% larger  than that of Jupiter. At least two other transiting giant exoplanets have 
an unusually low density relative to what simple models of energy balance with incoming stellar radiation would predict, suggesting that the interaction between gas giants and their parent stars is complex. Whatever mechanisms might be responsible for inflating these objects -- transport of energy by stellar winds into the interior, super-efficient convection driven by density and thermal differences, or enhanced tidal heating -- they do not seem to operate on all gas giant planets on close-in orbits. 

\subsection{High-Density Gas Giants} Several transiting gas giant planets have significantly higher densities than
would be expected for a pure H/He composition. One of these, HD 149026b, is so 
extraordinarily dense for its size, 1.4 g/cm$^3$, that up to 2/3 of it must 
consist of heavy elements. As such, HD 140926b is drastically different from all 
other observed giant planets, which are mostly hydrogen and helium. It is not understood how HD 149026b could have formed with so much heavy material - 
comparable to all of that contained in the solar system's planets combined.  The 
parent star is metal-rich, so perhaps its protoplanetary disk had more rocky 
material than the Sun's. Nevertheless, the existence of a giant planet with more 
than half heavy elements is a serious challenge to any planet-formation theory. A second anomalously dense exoplanet is HD 147506b. This exoplanet has a 
higher density even than than Earth's value of 5.5 g/cm$^3$, yet its large mass ($9 M_J$) and radius ($0.982 R_J 
\simeq 11 R_{\oplus}$) indicate it is composed mostly of H and He. The particular dependence of the compressibility of hydrogen and helium mixtures at ultrahigh pressures may be responsible, though this remains a speculative explanation. 

\subsection{Terrestrial Planets} The major uncertainties in the mass-radius relationship of terrestrial planets are the abundance of iron and the possibility that the planet is not mostly rocky but is a mixed rock-water or rock-ice world. The former problem is a familiar one in our own solar system: Mercury is denser than the much larger Venus, where compression of the rock plays an important role, so that the former must have a much larger proportion of iron than the latter (and than the Earth). Rock-ice or rock-water worlds have their analog in the large moons of the outer solar system--Ganymede, Callisto, and Titan--but no solar system analogs approaching the size of the Earth. Nonetheless, the equations of state are known with reasonable accuracy, so that a determination of the mass and radius of a transiting terrestrial-sized planet could immediately provide information on the bulk composition of the body, and hence fundamentally important constraints on its nature and evolution. 

\section{Characterization of the Atmospheres of Hot Jupiters}

Atmospheres of planets serve as gatekeepers, controlling the fate of 
incident radiation and regulating
the loss of thermal energy from a planet.  They are also archives, 
preserving gasses that reflect
the formation and the evolution of a planet.  Atmospheres thus lie at 
the heart of any characterization beyond mass of directly detected 
planets (even planetary radius measured from transits depends upon the 
atmospheric structure of the planet).  Since the hot Jupiters have been 
the first planets to be directly detected beyond the solar system, 
their characterization has nurtured a remarkable blossoming of research 
into the science of extrasolar planetary atmospheres.  Ultimately the 
characterization of of cooler Jupiters and terrestrial planets will be 
informed by the lessons learned from our experience with these 
unexpected planets.

As extrasolar planets transit their parent star, a small fraction of 
the stellar flux passes tangentially
through the planetÕs limb and upper atmosphere. The absorbing 
properties of the planetary atmosphere
(along a slant geometric path through the planetÕs limb) are added to 
the transmitted stellar
absorption spectrum. The stability and sensitivity required to detect 
such small variations favor
space-based observations and the first data were collected by the STIS 
spectrograph on the Hubble
Space Telescope and the IRAC photometer on the Spitzer Space Telescope. 
Such observations
detected first sodium and then water in the atmosphere of the hot 
Jupiter HD 209458b. The subsequent failure of STIS has limited the 
reach of transit spectroscopy, but future observations by
JWST, now under development, and perhaps a repaired STIS should enable the detection of 
important atmospheric species in
a much greater variety of transiting planets.  The abundance of 
particular species detected during transit reveal atmospheric 
composition and, through comparisons with the predictions of 
equilibrium chemistry, information on atmospheric dynamics and mixing.

Measurements of the reflected light from hot Jupiters as a function of 
orbital phase by MOST, COROT, and Kepler will also constrain the 
scattering properties of these planets.  This provides a window into 
the properties of any atmospheric condensates.  Clouds or hazes have 
been proposed as possibly accounting for the less than expected depth 
of the sodium line detected by STIS in HD 209458b, although other 
explanations are possible.   In any event, cloud layers are 
exceptionally challenging to model in all types of planetary 
atmospheres (c.f. \S V.2.2) and the hot Jupiters are thus already 
providing important insights.

The direct detection of planetary thermal emission in the near- and 
mid-infrared from about a dozen giant extrasolar planets
by {\em Spitzer Space Telescope} has provided an even clearer window 
into the atmospheric physics of extrasolar
planets.  These observations are today probing atmospheric composition, 
thermal structure, and atmospheric dynamics.  This exceptionally rich 
science yield is too extensive to review in detail here, but will 
certainly be considered one of the great legacies of this telescope.  
In our recommendations we support the continued operation of {\em Spitzer} after cryogen depletion to extend this bountiful harvest.

{\em Spitzer} has identified at least two classes of atmospheres found in hot Jupiters.  One class boasts  blazingly hot ($\gtrsim 2000 K$) atmospheric 
temperatures and large day to night temperature variations.  Another 
class exhibits cooler temperatures and smaller temperature variations, 
as measured by observations of thermal emission as a function of 
orbital phase.  Such phase measurements have even allowed for the 
construction of a crude temperature map of the surface of HD 189733b 
and more such observations are currently in process.  It has been 
proposed that this dichotomy results from the
exceptionally large range of incident flux (Figure ~\ref{fig:classes}) 
experienced by these planets.  Some atmospheres are so hot that 
titanium oxide (TiO), is present in the gas phase.  Like ozone in Earth's atmosphere, TiO absorbs 
incident radiation so strongly that it produces a very hot 
stratosphere.  In other somewhat cooler atmospheres where the TiO is condensed out, the planets lack the resulting hot stratosphere.  Such 
chemical and radiative processes must further be understood in terms of 
atmospheric dynamical models, which are also making great strides.  As 
these examples demonstrate, extrasolar planetary science is reaching 
surprisingly high levels of sophistication and complexity, only a few 
years after our first direct measurements of these atmospheres.  JWST 
and perhaps other space-borne infrared telescopes will certainly 
further extend these observations to include greater measurements of 
orbital phase variation and atmospheric composition and thermal 
structure.
As 
with solar system planets, we find that the interpretation of observations of 
exoplanets hinges on the interplay of atmospheric chemistry, physics, and 
radiative transport.  Understanding and appreciating processes that 
occur in many different planetary settings provides key insights into how planets work.  

\begin{figure}[h]
\begin{center}
\includegraphics[angle=0,scale=.4]{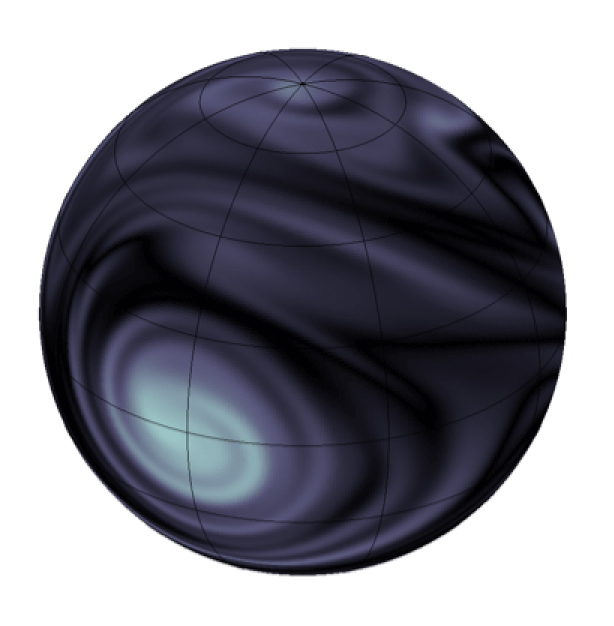}
\caption{Map of model atmospheric vorticity of the hot Jupiter HD 185269 b.  
Models of atmospheric dynamics of hot Jupiters such as this can be tested by 
careful measurement of the phase variation in emitted thermal flux over the 
orbital period of a hot planet by a space-based infrared telescope, such as 
Spitzer or JWST.  Figure from Langton and Laughlin (2008).\label{fig:vorticity} }
\end{center}
\end{figure}

\begin{figure}[h]
\hspace{2.0 cm}
\begin{center}
\includegraphics[angle=0,scale=0.8]{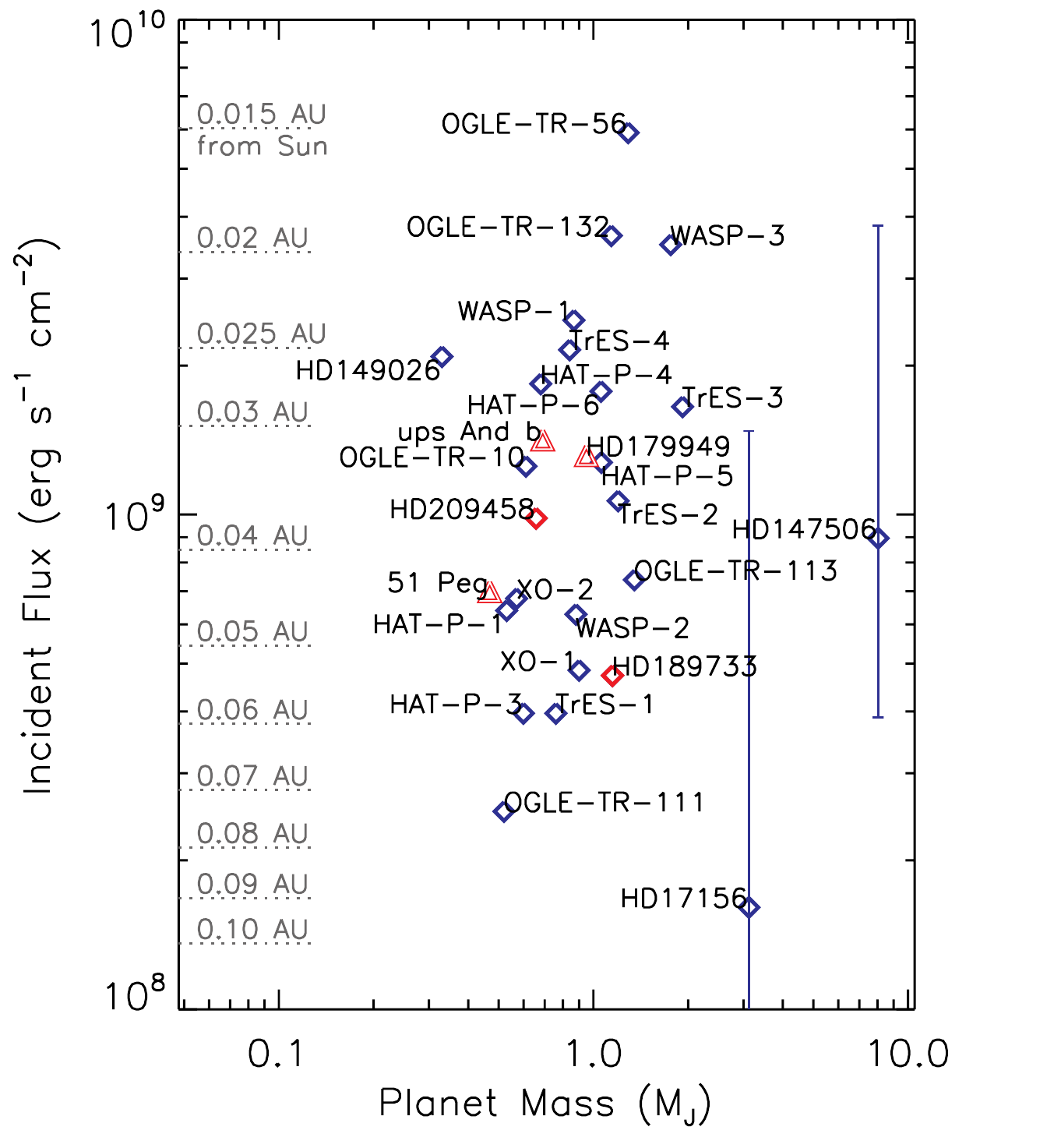}
\caption{Incident flux and planet mass for a collection of hot Jupiter 
planets.  The labeled dotted lines indicate the distance from the Sun 
that a planet would have to be to intercept this same flux. Diamonds 
indicate the
transiting planets while triangles indicate non-transiting systems 
(with minimum masses plotted). Red color indicates that Spitzer has 
observed the variation in thermal emission over an orbital period.   
 Figure modified from Fortney et al. (2007).}
\label{fig:classes} 
\end{center}
\end{figure}

Upcoming multi-wavelength time-resolved photometry for many hot Jupiters by the Spitzer Space Telescope, after its cryogen is depleted (the ``warm" Spitzer mission) will shed further light not only on these processes, but on weather 
patterns and atmospheric dynamics as well (Figure ~\ref{fig:vorticity}).  Long 
time baseline observations during a warm Spitzer mission have the potential of 
mapping the diversity of atmospheric dynamics for many of the hot Jupiters.  

\section{Evidence for Planets from Circumstellar Disk Structure}

Primordial disks are capable in general of forming planetary systems
similar in size and architecture to our own solar system. Thus, circumstellar 
dust and gas is a valuable indirect tracer of planets,
both during their formation and early orbital migration, as well as
during their multi-billion year lifetimes.    Spitzer and 2MASS have contributed enormously to our understanding of the depletion of
circumstellar dust.  On time scales
of just a few million years the observational diagnostics of
proto-planetary material disappear as small particles are incorporated
into larger bodies.  These then grow to terrestrial-sized planets and
giant-planet cores, which then eventually accrete much of the disk gas.It is becoming technically feasible to image
such recently formed giant planets either directly
through their thermal or accretion luminosity (via AO techniques), or indirectly
by imaging the gaps they are predicted
to produce in their disks (via millimeter interferometry). Long-baseline interferometry in the infrared has mapped disk geometries at spatial resolutions better than 1 AU, tracing the stellar infall of hydrogen gas and possibly the evaporation of water from small bodies. 

Debris disks surround 10-20\% of all solar type stars.
Because the survival time for small particles in the radiation and wind
environment of the star is small -- only a few hundred to a few thousand years
-- any such circumstellar dust observed in stars older than a few tens
of millions of years is recently formed and therfore must be continuously
generated in order for the disks to be detected. 
The observed debris particles are small,  a few to a few tens of microns
in size, a result of in situ collisions among unseen populations
of planetesimals.  The debris may result from the steady grinding
produced when planetesimal orbits are perturbed either by
the largest embryos in the planetesimal population or
by planetary mass bodies. Or in some cases it may
result from a single catastrophic collision that artifically raises
the mass in small dust particles over steady state evolution values.

For radiation-dominated disks, an inner cleared disk geometry is often used to
suggest the presence of a planet.  A sufficiently massive planet 
not only stirs the exterior planetesimals, increasing their velocity
dispersion such that dust-producing collisions occur, but also 
efficiently ejects or traps into resonances those dust particles that 
cross its orbit. Such scenarios prevent or at least impede material from reaching radii
much smaller than those where the dust is in fact detected.
In contrast, for collision-dominated disks, the location of
the planetesimal-stirring planets is less obvious, and requires detailed
modeling of individual systems once they can be spatially resolved
and thus further observationally constrained.

The dust inferred from mid-infrared continuum excess
is dominated by a cold component located exterior to $\sim$10 AU, typically.
In a few cases, debris disks are detected around mature stars
known to harbor planets.  For these systems, one can move
beyond the uncertainties inherent in modelling
spectral energy distributions by using
dynamical simulations that take into account the role of mean motion
and secular resonances of known planetary companions.

% ============================================================================
% Begin New PART - CRAFTING THE SEARCH
% ============================================================================

%\setcounter{section}{0}

\pagebreak
\clearpage

\chapter{Crafting the Search for A Habitable Planet}

% - - - - - - - - - - - - - - - - - - - - - - - - - - - - - - - - - - - - - - - - - - - - - - - - - - - - - - - - - - - - - - - - - - -
\section{Habitability}

The most exciting prospect for the near- to mid-term future is to search for evidence of Earth-sized planets around nearby stars and then ultimately to determine through spectroscopy whether any of them might be habitable. Any such search must begin with a definition of what is meant by the term ``habitable." 

\subsection{Defining Habitability and a Habitable Zone}

We assume that life is carbon-based and that it requires at least the transient presence of liquid water. We are, of course, cognizant of other possibilities, but since we must focus our initial search efforts, it makes sense to begin with this assumption.  The region around a star in which surface liquid water is stable is termed the habitable zone (HZ).  We can also define a continuously habitable zone (CHZ) as the region around a star that remains habitable for a specific length of time as the star evolves.

\subsection {HZ boundaries around Sun-like stars}
	The boundaries of the HZ can be defined either from theory, using climate models, or from observations. The theoretical problem has so far only been addressed using 1-D, globally averaged climate models, e.g., Kasting et al. (1993). Near the inner edge of the HZ, a planet is expected to develop a dense, H$_2$O-rich atmosphere. Photodissociation of H$_2$O in the stratosphere, followed by escape of hydrogen to space, leads eventually to depletion of surface water and loss of habitability. A conservative estimate for the inner edge of the HZ is 0.95 AU (Kasting et al., 1993). But a planet may remain habitable well within this distance if clouds, which are treated only parametrically within such a 1-D model, help to keep the planetÕs climate cool.

	Planets near the outer edge of the HZ are expected to build up dense, CO$_2$-rich atmospheres because loss of CO$_2$, which is determined by dissolution into water raindrops and chemical sequestration as carbonates, should slow as surface temperature decreases and rainfall consequently becomes less important.  The additional greenhouse warming by the CO$_2$ provides a negative feedback on climate that suggests that the outer edge of the HZ should be relatively far out, probably beyond 1.7 AU. Beyond this distance, CO$_2$ itself begins to condense, and so the negative feedback is lost. Only planets with internal heat sources and volcanism comparable to the Earth's can take advantage of this feedback. Small, volcanically inactive planets, such as Mars, may lose their CO$_2$ and become uninhabitable, despite being located within the HZ.

	An alternative way of estimating habitable zone limits is by looking at Venus and Mars. The semi-major axis of Venus is 0.72 AU, and this was evidently too close for habitability, as Venus has lost whatever water it once had. Conversely, although Mars is not habitable today (at its surface, at least), early Mars appears to have been habitable, based on the observation of fluvial channels in the heavily cratered southern highlands. Solar luminosity was lower at that time, so the solar flux at Mars' orbit was equivalent to the present flux at ~1.8 AU. Hence, empiricism suggests that the actual outer edge of the HZ is at or beyond this distance.
	
\begin{figure}
\includegraphics[angle=0,scale=0.8]{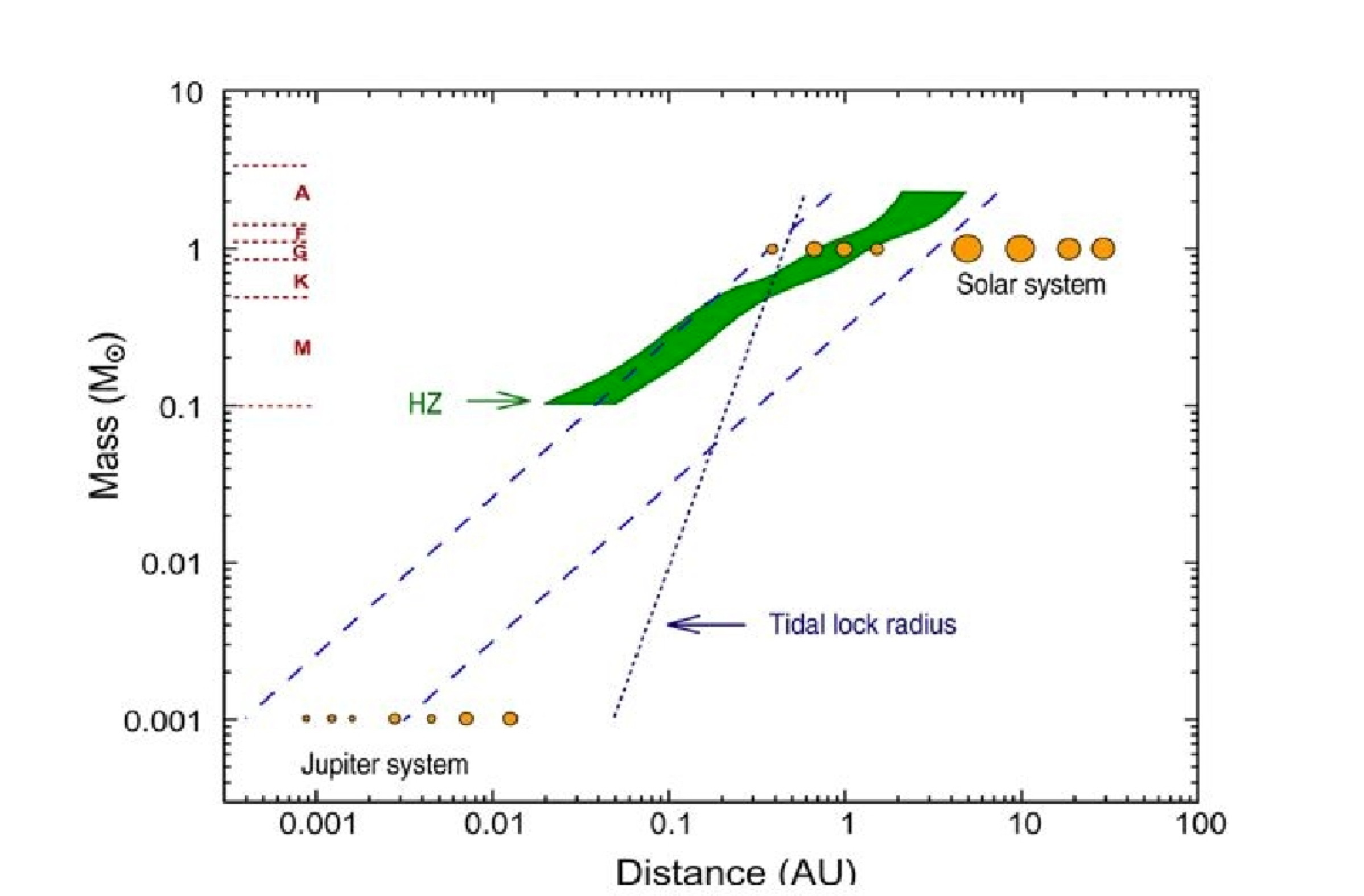}
\caption{The zero-age-main-sequence habitable zone around different types of stars. The vertical scale is the mass of the star divided by the mass of our Sun. The planets of our own Solar System, and the moons of the Jovian system,  are indicated by orange circles. Lettered boundaries on the upper left delineate the main sequence spectral types for stars of given mass. The HZ is plotted as a green-shaded region at the time when a star enters the main sequence, and so with time it will drift outward. The dotted curve shows the distance within which a planet's rotation would become tidally captured within 4.6 billion years. The dashed boundaries plotted are the theoretical ``runaway greenhouse" and ``moist greenhouse" limits defined in the text.\label{fig:hab-zone} }
\end{figure}

\subsection{ HZ boundaries for other stars}
	The same types of climate calculations performed for our Solar System can be done for other types of stars (Kasting et al., 1993). The results are as one might expect:  the HZ moves outward for more luminous stars,  inward for less luminous. While to first order the boundaries of the habitable zone scale as the square root of the stellar luminosity,  the spectral distribution of the stellar radiation plays an important role as well. Blue light is reflected more effectively by Rayleigh scattering, especially in the dense H$_2$O- or CO$_2$-rich atmospheres near the HZ boundaries, shifting the HZ closer to the star. Conversely, red light is scattered less effectively and is absorbed more effectively in the near-IR by H$_2$O and CO$_2$; hence, the HZ boundaries shift outwards.
	
	The boundaries of the HZ are plotted as a function of stellar mass in Figure ~\ref{fig:hab-zone}.  M-dwarfs pose particular problems for planetary habitability. The HZ around an M star lies within the tidal locking radius, within which planetary spins are likely to be captured (although some planets, like Mercury, may avoid this fate by becoming trapped in spin-orbit resonances in which the diurnal and annual periods are not equal to each other). Freeze-out of volatiles on the dark side of an M-dwarf planet may be a problem, at least for thin atmospheres. M-dwarfs are also chromospherically active, and their intense stellar winds, combined with the expected absence of magnetic fields on slowly rotating M-dwarf planets, may cause a planetÕs atmosphere to be stripped off over time. Nonetheless, M-dwarf planets the size or mass of the Earth remain interesting, partly because they are expected to be the easiest to find over the next 6--10 years.

\section{Detecting Exoplanets}

The process of detecting true earthlike planets involves three stages:
(1) {\it Detection}. The results of a single observing technology
will suggest that an earthlike exoplanet is present and provide
a few astrophysical parameters for the system.  (2) {\it Verification}.
A second observing technology confirms the presence of the planet and
through the combination of derived parameters, offers increased
understanding of the system. (3) {\it Characterization}. More
sophisticated (more complex, more expensive) observing technologies
are devoted to specific systems and the combination of all observations
delivers specific details such as radius, density, temperature, chemical
composition, etc., about the planetary system needed for astrophysical
interpretation.

We discuss here the six principal approaches for detecting, verifying, and characterizing extrasolar
planets,  providing an overview of the concepts and the projected capabilities over the timescale of the strategy in this section. A separate chapter compares the depths of search possible with each technique, while the Appendix to the report provides technical details.  The approaches are often conceptually divided into so-called ``direct" and ``indirect" techniques. Indirect detection encompasses observations of the host star for often-subtle effects revealing the presence of the planet. With the exception of transits, there is no detection and isolation of light from the planet itself; thus atmospheric properties and temperatures can only be inferred by modeling. With transits, some knowledge of the planetÕs size, atmospheric composition, and effective temperature is obtained and (through secondary eclipses) light from the planet is detected.

\subsection{Spectroscopic radial velocity method}

A planet and star orbit their common center of mass (or system center of mass), causing a periodic reflex motion of the star against the sky. For a system oriented nearly edge-on to the Earth, the component of the star's motion toward and away from the observer causes an oscillating \emph{Doppler shift}, so its atomic and molecular spectral lines are shifted slightly toward the blue and then the red. Telescopes outfitted with precise spectrometers can measure this small ``radial velocity" (RV) shift. 

\subsubsection{Current State of the Art: Optical Doppler Spectroscopy} 

In 1995, precise measurements of stellar radial velocities revealed 
the first exoplanet orbiting a main sequence star (Mayor and Queloz, 1995). 
Radial velocity (or Doppler) surveys have since discovered more than 90\% of the 
$\sim 250$ known exoplanets.  This technique is naturally most sensitive 
to planets that induce large reflex velocities in their host stars. 
Thus, detection efficiency is a trade off between several factors, including 
planet mass, stellar mass, the planetary orbital radius (or period), and the planet's orbital inclination to the obsever's line of site. 
An additional requirement for Doppler detection is that the star must be 
observed for at least one full orbital period, further favoring the discovery 
of short-period planets.

The first detected exoplanets were commonly several hundred times the 
mass of the Earth, akin to gas giant planets, and they orbited relatively 
close to their host stars.  Doppler surveys now collectively 
monitor about 3000 bright sunlike stars.  Among the surveyed stars,
the most easily detected planets (i.e., massive and with orbital periods 
shorter than 1 year) have already been detected.  The statistics for stars 
in Doppler surveys show that: 

\begin{enumerate}
\item About 1\% of sunlike stars have very close gas giant planets (or hot Jupiters). 
\item Gas giant planet formation is more efficient around high metallicity stars 
\item At least 15\% of stars have gas giant planets with periods shorter than 10 years
\item About half of the stars with one known planet have additional velocity variations 
indicating the presence of additional planets.
\item The number of planets increases with decreasing planet mass
\item Exoplanets exhibit a wide range of orbital eccentricity
\item Low mass M-dwarfs have fewer gas giant planets than solar type stars
\end{enumerate}

From the ensemble of known planets, it would appear that nature has not 
been entirely kind to Doppler planet hunters; the largest population 
of exoplanets are also the most technically challenging to detect. 
It is now clear that the number of planets rises steeply with decreasing 
planet mass. Among the massive gas giants, most are found at large orbital 
radii.  Both of these characteristics (low planet mass and large orbital 
radius) contribute to weaker radial velocity signals and additional 
complexity is introduced by the high occurrence of multiple planet 
systems. However, over the past decade, both the duration and precision 
of Doppler measurements have steadily increased revealing a record-breaking 
low mass planet of only 5 \mearth (Lovis \etal,  2007).
Several Neptune-mass planets have also been discovered in relatively 
short-period orbits and several new multi-planet systems have been 
discovered, including the 5-planet system orbiting 55 Cancri, a remarkable
example of how radial velocity measurements can unravel exoplanet architecture. 

RV measurements have also detected planets orbiting low mass Mdwarf stars. 
Statistically, M-dwarfs have fewer gas giant planets than solar type stars. 
GJ 876 stands out as an exception, with two gas giant planets and a third planet
(7.5 times the mass of the Earth) whose composition is difficult to guess. 
Neptune-mass planets orbit two other M-dwarf stars, so current data suggest that
M-dwarfs either have fewer planets or that planets orbiting M-dwarfs are 
substantially lower in mass than those orbiting solar-type stars. 

The observed characteristics of exoplanet populations provides an important
positive indicator of success for future exoplanet searches. There is clearly 
a significant reservoir of low mass planets that may extend into the terrestrial 
regime, as well as a substantial population of widely-separated gas giant planets 
that will be favorable targets for direct imaging techniques. 

\subsubsection{Infrared doppler spectroscopy} 

M-dwarfs are the most common star in the solar neighborhood and comprise 
about 70\% of stars in the Galaxy.  In addition to their sheer abundance, there are two key reasons that M-dwarfs are optimal
targets for Doppler planet searches. First, the stars have 
have low stellar masses, which result in larger reflex velocities 
for a given mass planet. Second, their low luminosity moves the habitable zone 
closer to the star, again resulting larger reflex velocity, coupled with shorter 
and more quickly detected orbits.  Therefore, M-dwarfs may offer the best 
detectability for Earthlike planets at habitable zone distances.  In addition, 
the ice-line in the protoplanetary disk of M-dwarfs is much closer to the star, 
alowing us to probe planet formation. 

We note that the width of the habitable zone scales with stellar luminosity, making
the HZ for late-type M-dwarfs comparatively narrow in linear (physical) dimension.  And while the proximity 
of the HZ increases the stellar velocity and improves detectability, this 
proximity also tends to synchronize planet orbits so that the planets keep the 
same face toward the stars.  However, it is not clear how either of these factors would 
affect extant biology, a topic which is beyond the charge to the ExoPTF.

A technical concern that supersedes the ability of IR spectrometers to demonstrate 
precisions approaching 1 \ms in the laboratory is the time variability 
of telluric lines in real observations.  Furthermore, stellar rotation begins 
to increase again for stars later than M4V (a relatively warm or ``early" M-dwarf). Broader stellar lines have reduced 
information content and could set an intrinsic stellar limit to Doppler precision 
and bear upon planet detectability for IR surveys. 
 
There are currently about 150 M-dwarfs on optical Doppler planet surveys. 
Because the stellar flux for M-dwarfs peaks in the near-IR between 1 - 2 micron, IR spectroscopy 
has been proposed as a new technique to search for planets orbiting the lowest mass 
stars.  The current precision with IR spectrometers ranges from about 80 \ms to 200 \mse and 
might be adequate to detect gas giant planets around these low mass stars. 
However, proposals for new IR spectrometers predict velocity precisions approaching 1 \mse, 
which if achieved would bring the detection of Earth mass planets around late-type main sequence stars within reach. 

Recent surveys (Lepine, 2005) have identified more than 
1000 M-dwarfs brighter than $V = 12$; the vast majority of these are earlier in spectral 
type than M4V and optical spectroscopy gives demonstrated precision that competes with 
the predicted precision of IR spectrometers with similar exposure times. For the $\sim$ 2000 known nearby M-dwarfs that are later in spectral type than M4V, (Lepine, 2005) IR spectroscopy holds 
tremendous promise to surpass optical spectroscopy, even on 10-m class telescopes 
because this population of stars have typical visual magnitudes of $V > 16$, but 
these stars are relatively bright at IR wavelengths. 
There are currently design studies for IR spectrometers underway now; if precisions better than 
10 \ms (approaching 1 m/s optimally) can be demonstrated we expect that this will be an important planet-hunting technique for 
relatively low cost. 

\subsubsection{Timing Radial Velocity Method}

Pulsars emit signals that are extremely regular, with a typical rate of change of only a second per ten million years. Because of this, small anomalies in the timing of pulsars can reveal the existence of planets orbiting the now-dead star. RV variations can also be detected by modulation of pulse arrival times from other pulsating parent stars.  The pulsar planets have only the pulse timing
for their detection.  Like the standard RV method, pulsar RV techniques are highly sensitive to masses (with the same caveats), as well as to orbits.

\subsubsection{Radial velocity: 0 -- 5 year timeframe}

Ground-based Doppler programs will continue to detect planets with 
masses down to a few \mearthe.  Optical spectroscopy will also identify 
multi-planet systems and seems unique in its sensitivity to 4 orders 
of magnitude in orbital periods (from 1 - 10000 days).  These detections 
lay the groundwork for target selection for space-based missions. 

A Doppler strategy employing extreme-cadence (thousands of observations) 
would help us to understand the floor of Doppler precision, the character 
of stellar noise at low frequencies and is poised to detect the lowest
mass exoplanets around nearby stars.  A particularly aggressive strategy 
has the potential to detect the first Earth mass planets at habitable 
zone separations for a few of the brightest stars. 

Design studies of IR spectroscopy should continue and if sufficient 
precision (i.e., better than 50 \mse) is achieved, this will be an important 
technique to search for Earthlike planets in the habitable zones of 
some of the nearest stars. Such searches should reveal the dependence of 
planet formation and evolution on stellar mass.

\subsubsection{Radial velocity: 6 -- 10 year timeframe}

We expect optical Doppler surveys to continue, but identify a 
fork in the road - if extreme cadence observations achieve 
precisions of a few \cms and low frequency stellar noise is 
essentially Gaussian, then this opens the option of dedicated 
radial velocity telescopes. 

If radial velocity noise has power at low frequencies, then 
RV observations will continue as support for other techniques. Although verification 
of Earth mass planets from the Kepler detections will probably not 
be possible with optical spectroscopy (because of the faintness of 
the stars combined with the low velocity amplitudes) limits can still 
be placed on some of these systems and verification of Neptune mass 
or higher mass planets will be possible.  If IR spectroscopy is able 
to achieve precisions aproaching 1 \mse, then this technique offers 
promise for Kepler follow-up.

\subsubsection{Radial velocity: 11 -- 15 year timeframe}

Continued improvement in the time-baseline and precision 
(particularly for the near-IR technique) will allow for probing 
to planets with smaller masses and larger orbits. 

\subsection{Astrometric Measurements}

The same stellar reflex motion that gives radial velocity signatures also gives rise to astrometric signatures -- looping motions across the sky.   Astrometric data do not have the $\sin(i)$ uncertainty associated with radial velocity
solutions.  Combining astrometry with radial velocity
yields unambiguous and precise orbital elements: together permitting confirmation
of the presence of Earth-mass
companions as well as constraining the planet's dynamical history based
on orbital elements such as the eccentricity. The first (still tentative) detection of an exoplanet by astrometry was announced as this report went to press (Docobo et al., 2008). 

The astrometric signature of Jupiter as seen from 10 pc away is 500 microarcseconds (``$\muas$"), easily seen by an astrometric mission such as Gaia. Because the amplitude of this signature increases with orbital radius, outer planets are more easily found by astrometry than by other indirect methods, but the limited duration of a space mission will sharply cap this improvement. We expect many giant planets to be detected this way, many of which will have been discovered earlier by spectroscopic RV surveys.

Finding earth-mass planets requires sub-microarcsecond sensitivity. As discussed in Section 11.5, a mission able to detect an Earth-size planet in an Earth-like orbit around each of at least 60 stars would need the capability for confident detection of astrometric signatures as small as 0.22 $\muas$ (zero-to-peak), using hundreds of observations per star over the mission life. Experimental results suggest this might be feasible, and preliminary studies presented to the committee by the NASA Jet Propulsion Laboratory suggested such a mission might be built for around \$1billion. If true, such a mission could make an invaluable contribution to planet-finding:

\begin{enumerate}
\item It would find and confirm specific planets and their orbits, so that a later direct-detection mission could characterize them more efficiently and cost-effectively, and
\item It would supply the planet mass, which is fundamental to understanding its nature.
\item If earth-like planets are rare, it would nevertheless provide a deep data set for understanding planetary system architectures, allowing a theoretical understanding of their formation processes.
\end{enumerate}

\begin{figure}
\centering
\subfigure[] % caption for subfigure a
{\includegraphics[width=10 cm]{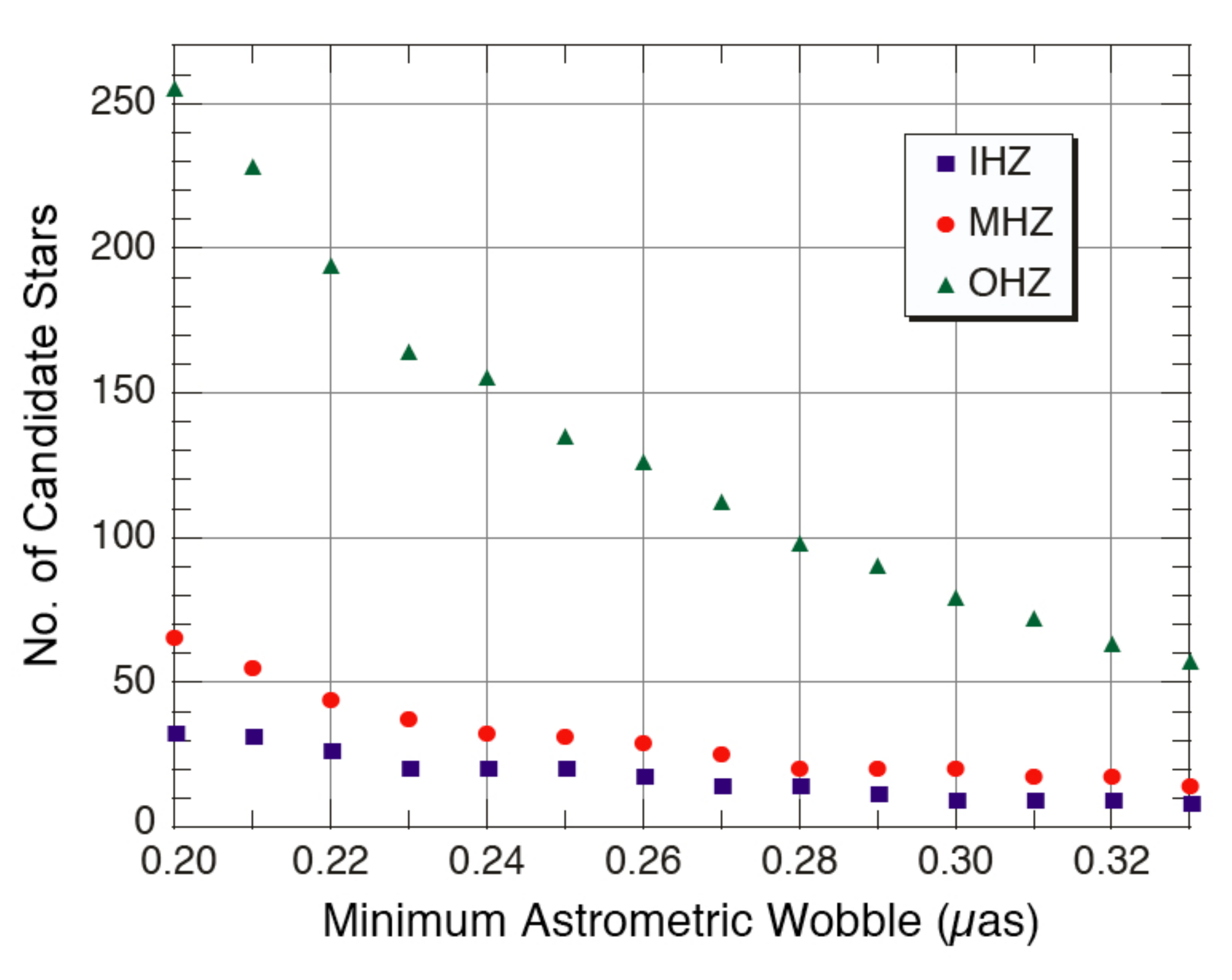}}
\subfigure[] % caption for subfigure a
{ \includegraphics[width=10 cm]{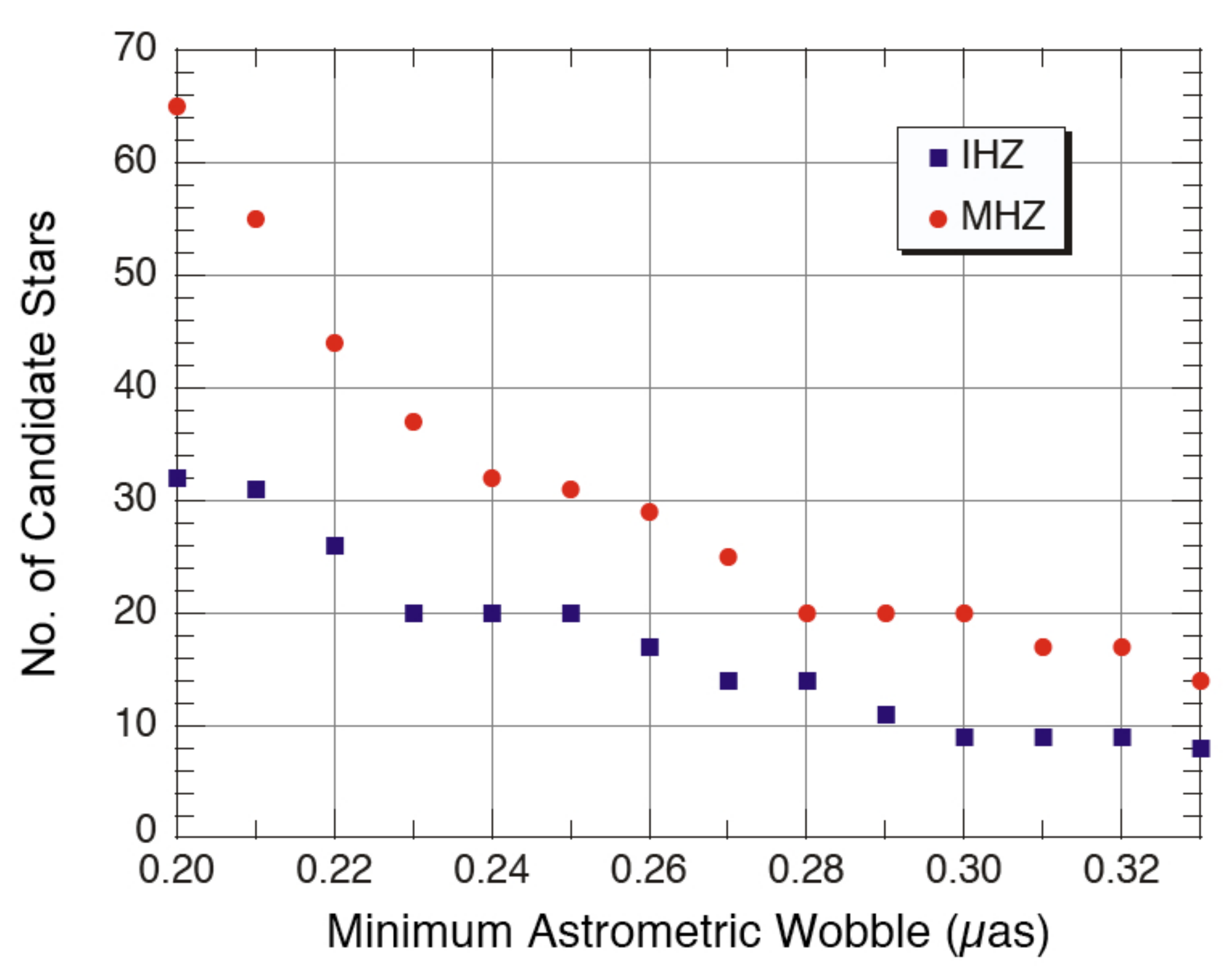}}
\caption{(a) The number of nearby F, G, and K stars, scaled by the square root of the solar luminosity, for which an Earth mass planet produces a wobble greater than the value shown on the x-axis for three selected orbital distances: 0.82 AU (IHZ or ``inner habitable zone"), 1 AU (MHZ or ``middle habitable zone"), and  1.6 AU (OHZ or ``outer habitable zone"). See Appendix 11.5 for details of the calculation. (b) Blowup for the IHZ and MHZ.}
\label{fig:Numstars} % caption for the whole figure
\end{figure}

\subsubsection{Astrometry: 0--5 Year timeframe}

Astrometric 
accuracy from ground-based telescopes is expected to improve to better than 100 $\muas$ on large
telescopes as the various sources of wave front error are understood. At this level of precision ground-based astrometry will principally concern itself with detection of giant planets, determining masses and orbital properties.  The best ground-based astrometric work will be done at several facilities with 8-meter or larger telescopes. The two 10-meter Keck telescopes (Keck-I) in Hawaii have a goal of 50-100 $\muas$ by 2009; a similar accuracy is predicted for the Large Binocular Telescope (twin 8.4 meter mirrors) in Arizona. 

Perhaps the most ambitious effort in ground-based astrometric interferometry
is the PRIMA instrument for ESO's Very Large Telescope Interferometer (VLTI).  This uses phase referencing in a dual beam configuration to measure relative position of a target star with
respect to a reference star within about 1 arcminute.  The projected
accuracy of 10 $\muas$ is nothing sort of amazing for a ground-based
facility, but still falls short of detecting exo-Earths around most of
the proposed target stars.  The Task Force found no other facilities that
would match or exceed the performance of VLTI within the 15-year time
frame that it examined.

Currently in an advanced stage of development is 
ESA's Gaia mission which is currently scheduled for a launch
late in 2011.  This mission uses two 1.45-m primary mirrors and a large
number of CCDs in its focal plane.  As exciting as its scientific returns
will be, it is projected to be sensitive to perturbations at the level
of 10 $\muas$ over the mission. A more ambitious mission being developed by JPL is the Space Interferometry Mission, or SIM, with a goal of making individual measurements with 1 $\muas$ single-measurement precision around a limited number of stars and a broader survewy at lower accuracy for a variety of astrophysical applications. At the time of publication of this report its programmatic status was uncertain, but some of the technology developments associated with SIM were presented to the Task Force by JPL in the form of a conceptual narrow-field astrometry capability in the 6-10 year timeframe that would match the 0.22  $\muas$ requirement for the Earth-mass search outlined above. 

\subsubsection{Astrometry: 6--10 Year timeframe}

In this time frame the roadmap envisions launch of a space-based astrometric system capable of observing Earth-mass planets around 60-100 of the nearest solar-type stars, requiring ability to detect astrometric signatures as small as 0.2 $\muas$ zero-to-peak over hundreds of observations of a given star. This number is determined by the Task Force in a calculation detailed in Appendix 11.5. Fig. \ref{fig:Numstars} provides a summary of the results of such a calculation in the form of the number of nearby F, G, and K stars that produce a wobble greater than a particular value, for a one-Earth-mass planet located at several distances of interest from these stars. 

The figure shows how our choice of a given required accuracy plays into the number of potential detectable planets. An ability to detect
only astrometric wobbles greater than 0.3 $\muas$, for example, results
in a drop by a factor of three in the number of potentially habitable planets detectable by astrometry relative to 0.22 $\muas$ . Depending on the actual width of the habitable zone in nature (that is, are Earth-mass planets at 1.6 AU habitable or is the zone narrower?), and the value of $\eta_\oplus$, such a drop might or might be a critical factor in planning an astrometric mission. So, If Kepler finds that $\eta_\oplus$ is much greater than
0.1,  a slightly relaxed specification on the astrometry mission may still
find enough candidate exo-Earths to justify the expense of a direct
detection mission.  This illustrates the importance of a paced approached to the
space missions. 

\subsubsection{Astrometry: 11--15 Year timeframe}

Assuming the space-borne astrometric mission described above is fielded in the second time epoch, no additional major space-based astrometric effort is envisioned in this time frame. Possible extension of the mission from the 6--10 year time frame to monitor longer-period orbits, and unexpected technical innovations leading to even more sensitive new missions are possible but not presupposed for the strategy outlined here.

\subsection{The Transit Technique}

A transit is the passage of a planet in front of the disk of its
parent star. During a transit, a small fraction of the starlight is
blocked. One may discover a transiting planet by monitoring a large
sample of stars, seeking periodic and short-lived dimming events.  One
may also seek transits of planets that have been discovered through
the Doppler technique but whose orbital inclinations are unknown.  In
both cases, the combination of transit photometry and Doppler
velocimetry reveals the planetary mass, radius, and hence its mean
density.  High-precision follow-up observations (some of which are
described below) can then reveal a wealth of information about the
planet and star that is not available for a non-transiting planet. The
unambiguous planetary mass and radius, along with the richness of
follow-up opportunities, are why transits are so highly prized.

Three important aspects of transits are:
\begin{enumerate}

\item The transit probability. Transits are rare, because the
  planetary orbital plane must be nearly edge-on as viewed from
  Earth. For randomly oriented orbits this circumstance occurs with a
  probability of approximately $R_\star/r$, where $R_\star$ is the
  stellar radius and $r$ is the planet-star distance at the time of
  conjunction. For a Sun-like star, the probability is $\sim$0.5\% for
  $r=1$~AU and $\sim$10\% for $r=0.05$~AU.

\item The transit depth. During a transit, a fraction
  $(R_p/R_\star)^2$ of the starlight is blocked. For a Sun-like star,
  this fraction is 1\% for a Jovian planet and $8\times 10^{-5}$ for
  an Earth-sized planet.

\item The transit duration. The transit lasts for a duration $\sim
  R_\star/v$ where $v$ is the orbital velocity of the planet.  The
  exact duration also depends on the impact parameter of the planet's
  trajectory across the stellar disk. For a Sun-like star, the transit
  of a close-in planet with $r=0.05$~AU lasts $\sim$2 hr, while the
  transit of a more distant planet at 1~AU lasts $\sim$12~hr.

\end{enumerate}

\subsubsection{Current State of the Art}

Approximately 30 transiting planets are known. Five of these planets
were originally discovered through the Doppler technique and were
subsequently found to transit their parent stars. The remaining 25
were discovered by virtue of their transits, by five different teams
employing small ground-based telescopes. All but one have orbital
periods smaller than 5 days; the exception is HD~17156 with a 21-day
period.

The main steps in the transit discovery process are: (1) monitor tens
of thousands of dwarf stars as continuously as possible for several
weeks or more; (2) apply signal-processing algorithms to search for
periodic transit-like signals; (3) rule out astrophysical false
positives (generally, variants on eclipsing stellar systems) through
low-resolution spectroscopy and higher-precision photometry; (4)
measure the planetary mass with high-precision Doppler measurements.
Step 3 proved much more difficult in practice than was originally
appreciated--the false positives can be ten times as numerous as the
true planets--but the various survey teams have gradually become much
more efficient in this process.  One lesson that has been learned by
the transit community, obvious in retrospect, is that the brightness
of the host star is a critical factor. Steps 3 and 4, as well as all
of the follow-up observations, require enormous and often prohibitive
telescope resources for host stars fainter than $\sim$13th magnitude.

The ground-based wide-field photometric surveys have detected planets
with transit depths ranging from 0.5\% to 3\%, using differential
image analysis and a variety of techniques to remove systematic
effects due to the atmosphere and instrument. The stars are almost all
FGK dwarfs, and the planets are almost all gas giants with masses
comparable to Jupiter and radii that are generally larger by 5-20\%.
Part of this radius excess is attributable to the intense radiation of
the nearby star, but this explanation by itself is insufficient to
explain the largest of the transiting planets, and many other theories
have been proposed.  In addition, there is one known case of a
transiting Saturn-sized object, and one known case of a transiting
Neptune-sized object whose parent star is an M-dwarf.

Follow-up transit science has been a very active subfield in the last
few years. Precise follow-up photometry has resulted in measurements
of the planetary radius with a typical precision of 3-5\%. Spectra
with very high signal-to-noise ratio have been taken in and out of
transit, and the slight differences between them have been used to
measure absorption by sodium atoms in the planetary atmosphere.  In
several cases the planetary orbit has been shown to be prograde with
respect to the stellar rotation, utilizing the ``Rossiter-McLaughlin
effect,'' a spectral distortion that is observed during
transits (see the Appendix on the transit technique for a more detailed explanation). Furthermore, a transiting planet is likely to undergo
occultations, also called secondary eclipses, when the planet is
hidden behind the star. Observations of secondary eclipses with {\it
Spitzer} have allowed planetary temperatures and even infrared spectra
to be measured.

\subsubsection{Transits: 0--5 Year Timeframe}

Several ground-based surveys are planning to continue finding giant
planets around main-sequence field dwarfs. The discovery rate is
currently $\sim$10~yr$^{-1}$, limited by the telescope resources that
are needed for follow-up spectroscopy. There does not seem to be any
obstacle to continuing this discovery rate for the next few years, as
long as giant planets remain of sufficient interest. Given the
difficulties of working with faint host stars, the narrow-deep transit
surveys (those that survey a narrow field of faint stars) are likely
to cease operation in favor of the wide-shallow surveys, a transition
that is already underway. Wide-field ground-based surveys with
somewhat larger telescopes and more sophisticated detectors may reach
down to Saturn-sized objects or even smaller (0.2--0.5\% transit
depths), although the needed precision for such observations has not
yet been demonstrated.

At least one ground-based program will specifically target small stars
(M-dwarfs), thereby enabling the discovery of smaller planets for the
same photometric precision. The difficulty here is that M-dwarfs are
faint and have a low surface density on the sky. The budding
``MEarth'' project will attempt to overcome these difficulties by
employing $\sim$10 0.3m telescopes to observe M-dwarfs one-at-a-time.
Several transiting rocky planets may be found in this manner, if their
occurrence rate is $\sim$10\% or higher (Nutzman \& Charbonneau 2007).

The European spaceborne photometer, COROT, will monitor $\sim$10
fields of $10^4$ stars, each for several months, and is likely to
discover scores of giant planets down to Neptune-sized planets.  If
the photometric performance conforms to theoretical expectations then
COROT may also discover transiting super-Earths.  The continuous
coverage and uniformly high quality photometry of COROT should allow a
good statistical estimate of planet frequencies, at least for
relatively short-period planets.

\subsubsection{Transits: 6--10 Year Timeframe}

The NASA Kepler satellite will complete its
mission of monitoring $\sim$10$^5$ stars with micromagnitude precision
for 3.5~yr. It is expected to find hundreds of transiting giant
planets, and will reach all the way down to Earth-sized planets or
even somewhat smaller. Confirmation of the smallest planets will
require time-intensive observations with the largest ground-based
telescopes equipped with the most stable spectrographs. A key goal of
Kepler will be a good measurement of $\eta_\oplus$.

Several teams are envisioning a wide-shallow equivalent of  
Kepler\, using a small satellite with a very large instantaneous
field of view. Such a mission could be launched during this time
period. The entire sky (or a large portion of it) would be searched
for transiting planets around the brightest stars, for which planet
confirmation is much easier and follow-up observations are more
rewarding. These surveys may be restricted to relatively short-period
planets because of the difficulty of monitoring the entire sky
simultaneously.

\subsubsection{Transits: 11--15 Year Timeframe}

By this time we expect that the brightest possible host stars, down to
at least 10th magnitude, will have been searched for transits.  The
{\it JWST}\, once launched, will be used to measure transmission and emission spectra
for the most favorable systems.

\subsection{Microlensing Method}

Microlensing occurs when an intervening star (lens) passes almost
directly between a more distant star (source) and the observer, and
the light from the source is gravitationally focused by the lens into
two images on either side of the lens. The separation of the images is
$\sim 1$ mas, and the images are generally not resolved. But the source is
magnified by the lensing, and it will appear to brighten and dim in a
bell shaped light curve as the source and lens approach and recede
from each other. Any planet a lens may have can perturb the
light curve by a large amount if it comes close to one of the images
and thereby reveals its presence. Observations following and sometimes during a planetary microlensing
event can often be used to characterize the host star sufficiently to
determine the mass of the planet and its projected separation from the
star in physical units.

The theory of microlensing is rigorous, and the technique for finding the
signatures of planets in ongoing microlensing events is robust.
With a space based survey, 
it is ultimately possible to determine the occurrence of planets as a
function of planet mass (down to $0.1M_{\oplus}$), type of host star
(for all F and G lens stars, most of the K lens stars, and some M lens
stars), and separation of the planet from its star from $a=0.5$ AU to
15 AU, where planets at larger separations may be difficult to
distinguish from free floating planets
(Bennett and Rhie, 2002).  

The statistics of planetary masses and separations  available
from a microlensing survey are
vital for constraining the theory of planet formation, and the
fraction of stars with terrestrial mass planets $\eta_{\oplus}$
available in the statistics will guide procedures for direct
detection and characterization of possibly habitable Earths. The
Kepler transit mission (http://kepler.nasa.gov) is expected to provide
$\eta_{\oplus}$ for semimajor axes $a<1$ AU, which complements the
microlensing sensitivity for the detection of more distant habitable
planets.  

So far 10 planets have been detected by microlensing with six, four
Jovian and two sub-Neptune mass planets, having been published
(Bond {\it et al.} 2004, Udalski {\it et al.} 2005, Beaulieu {\it et
al.} 2006, Gould {\it et al.} 2006, Gaudi et al. 2007b). The
last two Jovian planets discovered are members of the same system with 
mass ratios and, if the projected separations are comparable to the
semimajor axes of the the orbits, separation ratios comparable to the
those of the Jupiter/Saturn system (Gaudi et al. 2007b; Bennett et
al. 2007b). The remaining four planets have just been discovered in
2007 (D Bennett, private communication, 2007).

Current practice in the microlensing detection of planets (alerts
plus follow-ups) has thus demonstrated the unique prowess of the technique
in detecting low mass, cold planets below the sensitivities of radial
velocity, transit, and astrometric techniques.  In addition, it may have
indicated that more embryonic cores fail than succeed in becoming gas
giants. However, these successes have come only after a tremendous
amount of work by large groups of people. Monitoring one event at a
time is too inefficient, and the tremendous effort expended to detect
and characterize only six planets means that the
current practice 
cannot even approach the potential of microlensing to complete a
statistical census of planetary systems in the Galaxy. For this reason
a ``next generation'' ground based program and a space based program
that would greatly increase the efficiency of planet detection and
characterization have been proposed. The next generation ground based
program falls into the 0-5 year time frame, if construction of a
single additional telescope in S. Africa is begun soon.  

Lensing theory for a point source single lens is straight
forward and simple, and this theory and its generalization to a binary
lens, where one member of the binary is a planet, is outlined in
the Appendix. Here we point out the capabilities of microlensing surveys
for detecting planets in the 0-5 year, 6--10 year and 11--15 year periods. 

\subsubsection{Microlensing capabilities in the 0-5 year time frame.}

A next generation ground based program could replace the current
survey-alert-follow-up procedure with an array of 2 meter class, large
FOV telescopes in the southern hemisphere. A high frequency of high
precision photometric measurements of $>10^8$ stars in the Galactic
bulge field would simultaneously detect and monitor the events for
short term planetary perturbations.  For the detection and
characterization of Earth-mass planets, the intervals between samples
would have to be $\sim 10$ minutes at a photometric accuracy $\sim
1\%$ at $I\sim 21$. Bennett (2007b) has provided a
conservative estimate of planetary detections from a next
generation ground based survey, filtered to include only those
detected planets with sufficient lightcurve coverage at a sufficiently
high signal-to-noise ratio to enable an unambiguous determination of
the planet-star mass ratio $m/M$ and the separation of the planet from
the star in units of the Einstein ring radius $R_{E}$ (defined in
the Appendix).  

\begin{figure}[h]
\begin{center}
\includegraphics*[scale=0.5]{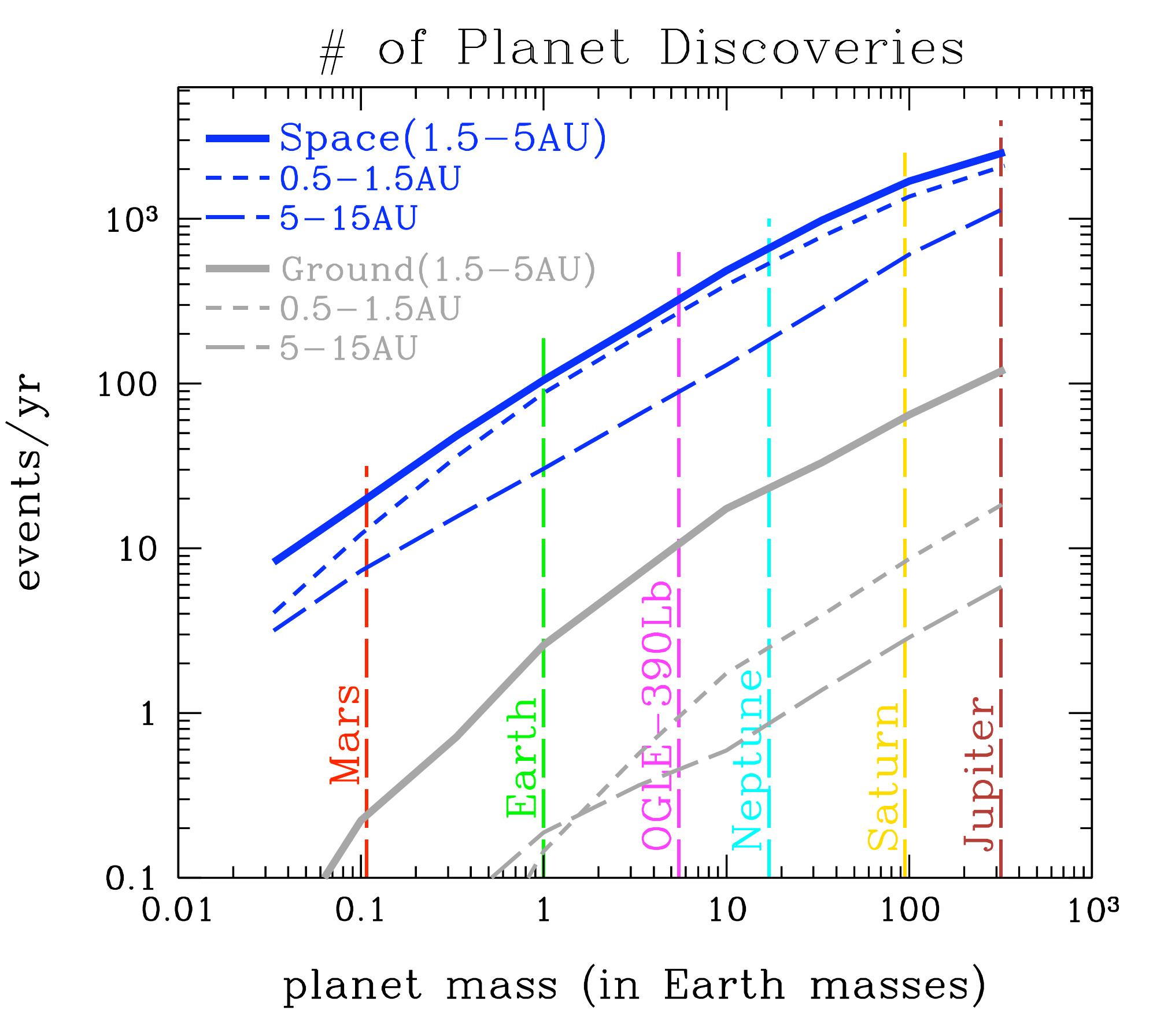}
\caption{Simulated rates of planet detection for a conservative next
generation ground based program (Bennett, private communication, 2007)
and for a space based program (Bennett, Anderson and Gaudi,
2007). (Figure from Bennett, 2007b)
\label{fig:bennett1}} 
\end{center}
\end{figure}

The gray curves in Fig. \ref{fig:bennett1} show the rate of planet
``discoveries'' as a function of planet mass for this realistic scenario
for a next generation ground based program--telescopes in 
New Zealand, S. Africa and S. America with characteristics similar to
MOA II or OGLE IV. The observing strategy (integration time {\it vs}
number of covered fields) is optimized for $m/M=
10^{-4}$.  The 
number of events/year in Fig. \ref{fig:bennett1} is determined under
the assumption that every lens star has an average of one planet of the given
mass and in the given range of semimajor axes. A significant number of Earth's might be discovered
in the lensing zone in this ground based program, but essentially none
could be found in the habitable zone of 0.5 to 1.5 AU. More massive
planets have a much higher discovery rate, where more than 400
Jupiter's would be found in a 4 year survey if all stars had a Jupiter
in the lensing zone, where our Jupiter happens to fall if the Sun were
at 4 kpc and the source at 8 kpc.   

The sensitivity of this  conservative next generation ground based
program discussed here is shown by the red curve with hash marks in
Fig. {\ref{fig:bennett2}, where the range of semimajor
axes for the detection of planets of given mass is indicated. The
sensitivity criterion is such that more than 10 planets are discovered
in a four year program if their mass and semimajor axis falls in the region
above the curve. (Note from Fig. \ref{fig:bennett2} that if Earths are
common near the lensing zone, the next generation ground based program
could detect 1 or 2 Earths per year.) 
The pink curve gives that sensitivity for the
current survey-alert-followup procedure with six of the ten planets
discovered by microlensing indicated in red. For comparison, the
range of sensitivity of the radial velocity (RV) search is indicated
in orange with the detected planets marked as lower bounds on the
masses. The range of sensitivity of the Kepler mission and a notional spaceborne astrometry mission are
indicated in blue and green respectively.  The blue squares are
transiting planets whose masses are known. The purple curve with hash marks is the limit of sensitivity for a proposed space based search (Bennett et al. 2004), comparable to that of the next-generation ground-based search. The space based search will be discussed below.

\subsubsection{Microlensing capabilities in the 6--10 year time frame.}
If the next generation ground based program discussed above is
continued, it will improve the statistics of $m/M$ and 
$r_p/R_{E}$ for planets in the lensing zone and beyond.  
The necessity of
using HST or adaptive optics on large telescopes, where there is fierce
competition for time, to obtain followup observations that enable
lens and planet masses and separations in physical units to be
determined means relatively few of the planets detected from the
ground will be so characterized.  However, a dedicated space based
microlensing 
search for planets can determine the type of star, the mass and
separation of most of the detected planets in physical units for $\sim
0.5 {\rm AU} < r_p < 15 AU$, where planets at larger separations are
detectable as free floating planets.  It would be at least 5 years
after funding begins for a space based mission to be launched, so this
capability naturally falls into the 6--10 year time frame.   

The blue curves in Fig. \ref{fig:bennett1} show the rates of planet
detection for the simulation of the space based microlensing survey,
where the procedure of the simulation is described in the Appendix. The
rate of discovery  
of Earth mass planets in the lensing zone is about 40 times that from the next
generation ground based survey, but the rate of discovery of Earths in
the habitable zone is higher by a factor of more than 600. This quantifies
a major advantage of the space based survey over the ground based
mentioned above. Low
mass planets like the Earth can be detected from the ground only if
the planet is in the small range of semimajor axes spanning the
average Einstein ring radius, whereas the MPF curve in
Fig. \ref{fig:bennett2} in the Appendix implies a sensitivity for
Earths down to a 0.25 AU separation. The reason for 
this is that the ground based survey generally requires moderately
high magnification $A >10$ to resolve the source star well enough to
get the moderately precise photometry that is needed to detect
planets.  A space based microlensing survey will mostly resolve the
source stars, so planets further from the Einstein radius can be
detected by their light curve perturbations at relative low
magnification from the lensing effect of the planetary host star. 

\begin{figure}
\begin{center}
\includegraphics*[scale=0.3]{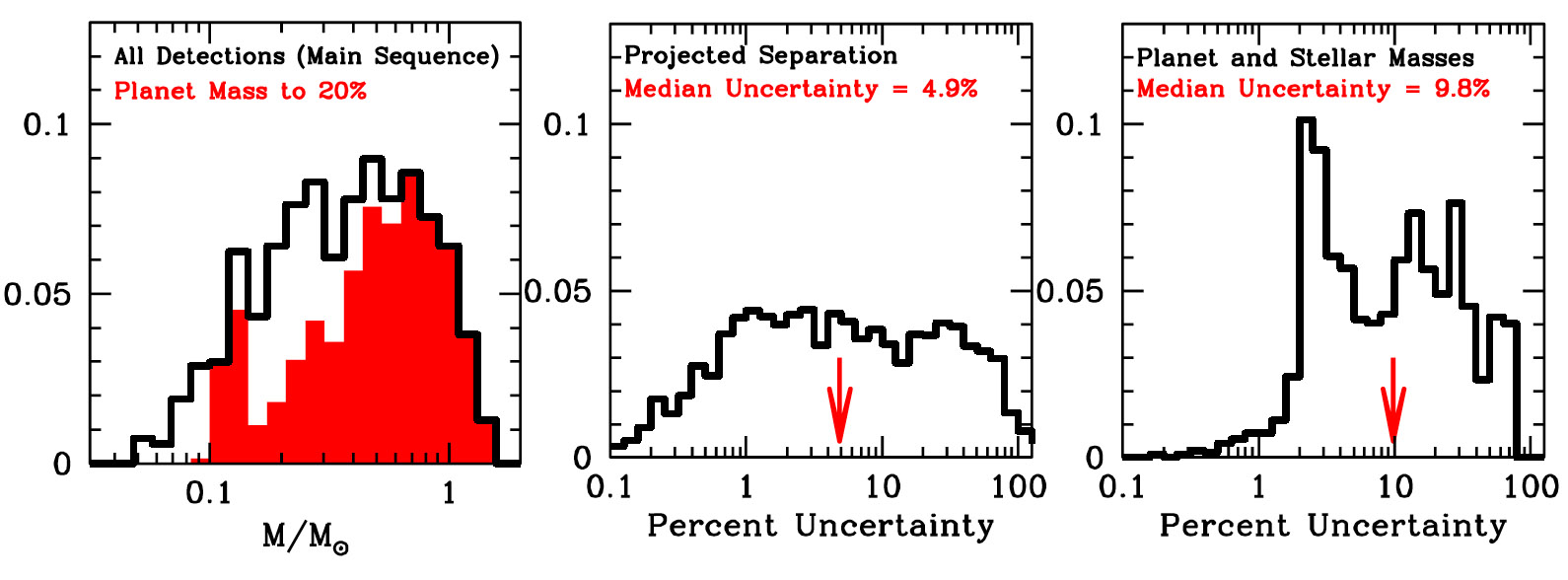}
\vspace{0.1in}
\caption{(a) Simulated distribution of stellar masses for stars with
detected terrestrial planets. The red histogram indicates the subset
of this distribution for which the masses can be determined to better
than 20\%. (b) and (c) Distribution of uncertainties in the projected
star-planet separations and planet and stellar masses respectively.
(Bennett, Anderson and Gaudi, 2007). \label{fig:bennett3}} 
\end{center}
\end{figure}

The space based program can collect enough additional data for most of
the planetary detections to find a complete gravitational lens
solution and determine the masses of the planets and their host stars
and the displacement of the planet in AU (Bennett, Anderson and Gaudi,
2007). This information depends entirely on the detection of the lens
star through the stability of the PSF in a space-borne telescope,
where the focal plane is dithered between images to reduce the pixel
undersampling of the field. The methods by which the complete
solutions of the systems are found are outlined in the Appendix. As an
example, the first panel of Fig. \ref{fig:bennett3} shows that
approximately 70\% of the masses for terrestrial planets in a
space based microlensing survey can be known to $\pm 20$\% (Bennett
{\it et al.} 2007). These complete solutions are the basis of
the above claim that a space based mission can collect complete
statistics on planetary masses and separations, including free floating
planets, as a function of stellar type and location in the Galaxy.
As pointed out above, this information, which is vital for 
theories of planet formation and, through $\eta_{\oplus}$, for
developing the strategy for the direct detection and characterization
of terrestrial planets orbiting the nearest stars, can be obtained by
no other technique. The distribution of planetary properties in
physical units can usually only be approximated in a statistical sense for
planets more massive than Neptune in a ground based survey, since most
of the lenses will not be detected because of limited HST time.

\subsubsection{Microlensing capabilities in the 11--15 year time frame}

If funding for a space based mission is delayed, and/or if the
fraction of stars having terrestrial planets is so small that an
extended mission is necessary to secure the statistics on the low mass
end of the planetary mass function, the space based capability could
spill over into the 11--15 year time frame. If a space based
microlensing search is successful, a continued ground based program
would not appear to be cost effective and would be less scientifically
rewarding. On the other hand, it is difficult to predict the advances
in observational techniques and data analysis that could materialize
in the next decade that could give microlensing a continuing important
role beyond the success of a space based mission.

\subsection{Optical Coronagraphy}

\vspace{-2.5mm}

Ultimately, the ability to {\em directly image} an extrasolar planet 
-- i.e., separate the light emitted or reflected by the surface or atmosphere of 
a planet from that of the star it orbits -- offers the greatest prospect for
characterizing these worlds.  In particular,
direct imaging offers the possibility of determining the colors and spectra of a
large number of planets, independent of their orbital inclination.  Such information
provides the means to distinguish between gas giants, ice giants, and terrestrial
planets under a variety of circumstances such as distance from the star, age, etc.
Moreover, direct imaging offers the possibility to study not only the atmospheres,
clouds, and surfaces of these planets, but also their variability.  Finally, {\em direct
imaging may be the only means for establishing the habitability or signs of
life on any exoplanet}.

To directly image extrasolar planets, a telescope must possess two key attributes:
(1) the angular resolution sufficient to spatially separate the planet
from its central star; and, (2) the means to adequately suppress the diffracted light
from the central star such that the planet
brightness is comparable to or greater than the residual diffracted star light.
Practical limits to the size of space-based mirrors force approaches to direct imaging
that divide according to wavelength.  At optical wavelengths, values of $\lambda/D$ 
(where $\lambda$ is the wavelength
of light and $D$ is the mirror diameter) are
sufficient to spatially resolve a planet in the habitable zone around nearby stars
using single monolithic mirrors that fit in current launch vehicles.
In the near-infrared, maintaining the same values of $\lambda/D$
requires a prohibitively large primary mirror, thus forcing
consideration of interferometric techniques, which is discussed in a subsequent
section.  Here we discuss both internal-occulter and external-occulter
coronagraphs.  An internal-occulter coronagraph blocks the
starlight using optical elements within a telescope, while an external-occulter
coronagraph blocks the starlight with a separate large starshade
positioned in front of the telescope, usually many thousands
of kilometers away.

The challenges of using an optical coronagraph for exoplanet imaging are four-fold.
First, for an Earth-sized planet orbiting another star, the star-planet brightness
contrast ratio is about 10$^{10}$:1 in the visible.  Any successful starlight
suppression technique must reduce this contrast ratio to approximately 1:1
and must do so at an Inner Working Angle (IWA) -- i.e., the smallest
star-planet angular separation at which this suppression level is achieved --
of $\approx\,$3-5$\,\lambda/D$ 
to avoid the need for a large diameter primary mirror ($\gtrsim$ 8-meter).  
Second, the throughput of the optical system for the off-axis planet-light must be as high as 
possible to limit integration times.  High throughput is especially important 
for exoplanet spectroscopy, where the number of photons per spectral resolution
element is further reduced.  Third, the optical bandwidth of the light
suppression technique -- i.e., the range of wavelengths over which the required
light suppression can be maintained -- should be as large as possible to enable
either broadband photometry ($\Delta\lambda/\lambda\:\gtrsim$10\%)
or the spectroscopic study of possible exoplanet atmospheric features
at various different wavelengths.  Finally, even with high throughput, 
collecting enough photons to detect the weak signals from extrasolar
planets often requires substantial integration times (many minutes to many hours).
During such integration times, it is essential that the residual stellar leakage
be kept extremely stable.
Atmospheric as well as thermal fluctuations experienced on the ground
prevent the required wavefront accuracy from being maintained except over very
short intervals, thus the need to conduct observations of Earth-sized
exoplanets from space. Even then, drift and vibration in the telescope cause varying 
stellar ``speckles" that can limit the ultimate sensitivity for planet detection. 

Wavefront accuracy and stability is a problem common to all internal optical
coronagraphs.  Wavefront errors produced by imperfections in the telescope
mirrors and coatings cause speckles in the image after the coronagraph masks have 
suppressed the starlight.  The intensity and variability of these speckles 
can easily be high enough to obscure
the faint exoplanet image. It is possible to suppress these speckles using aggressive
wavefront sensing and control with deformable mirrors, limited
only by optical disturbances that cause the speckles to move before the control
system can sense and remove them.  This race between disturbance and 
compensation sets a practical floor to the planet sensitivity of a coronagraphic
telescope.

The chief advantage of internal coronagraphs is their packaging simplicity -- all
of the hardware needed to detect an exoplanet is contained within a single
telescope assembly.  In principle, source targeting is
most straightforward and efficient in this configuration.  The challenge of this 
approach remains
achieving the stellar light suppression with the throughput and stability
necessary to detect an Earth-like exoplanet.  Moreover, since increased cost and 
complexity lessen the probability that a mission will be selected for flight, there
are clear advantages to achieving these requirements with a 4-meter class aperture
rather than an 8-meter class aperture.  Doing so will require light suppression
techniques with IWAs of $\sim\,$2$\:\lambda/D$ -- which have yet to be
proven in laboratory experiments -- rather than the current approaches which
come close to meeting the light suppression requirements at
$\sim\,$4$\:\lambda/D$.  
In recent years, at least a dozen new types of internal coronagraphs 
have been invented that might enable us to image an Earth close to a vastly
brighter star.   In addition, wavefront sensing and control techniques have 
steadily improved.
However, further development work will be necessary before it is clear 
whether any internal coronagraph incorporating these technologies can
demonstrate the required performance and be practically implemented.

External-occulter coronagraphs have been studied for years, beginning 
with Lyman Spitzer in the early 1960s.  The appeal of external coronagraphs 
lies with their potential to circumvent many of the light suppression problems
faced by internal coronagraphs by instead blocking the stellar light with a
free-flying occulter located between 20,000 and 70,000~km (set, respectively, by Fresnel vs. Fraunhofer diffraction) from the telescope. This would allow the use of a generic diffraction-limited visible-light telescope.
Early laboratory experiments have verified that this technique can produce
stellar light suppression down to about 10$^{-7}$ in air.  Experiments in vacuum
are beginning, and results are expected later in 2008.  The main drawback of the
the external occulter approach lies in its operational complexity relative to a single
spacecraft -- two vehicles must perform properly for this technique to work and
source targeting requires aligning the two spacecraft.  When many targets must be searched in a survey mode to find planets, travel time will limit the planet observing cadence; but this is offset by high telescope throughput and alternate observing strategies. Also, during travel time, the telescope conducts other astronomical observations. If there is a second occulter spacecraft, this time can be used for more planet observations.

A hybrid approach of an external occulter shadowing an internal occulter telescope might provide the best performance for the cost, but this must be studied. In any event, direct detection using coronagraphy remains a significant technical challenge.
No technique has yet demonstrated an end-to-end performance sufficient to detect
Earth-like exoplanets.  However, progress is being made and investment in several
promising techniques (involving both laboratory and possibly in-space development) will best ensure that at least one practical and affordable 
approach will emerge during the next 5 to 10 years.

\clearpage

\subsection{Mid-infrared nulling intererometry}

This approach is particularly useful at infrared wavelengths where coronagraphs would become huge and unwieldy. An infrared interferometer consists in its simplest form of two telescopes joined on a structure, or mounted on separate satellites that maintain a controlled distance by precision formation flying. More than two telescopes can be used, and indeed would likely be required for the application of detecting Earths around other stars. A nulling interferometer uses two telescopes to suppress the light from the parent star so that a planet orbiting that star may be detected. The starlight is suppressed by introducing a shift in the phase of the incoming light in one telescope of the interferometer, so that there is destructive interference for light that arrives aligned with the axis between the two telescopes. Off-axis light, as from a planet, experiences a different phase shift due to the optical path delay, which frustrates this destructive interference and allows a high throughput for the planet.

This concept was conceived by the Stanford radio engineer Ronald Bracewell in 1978 (Bracewell, 1978). The most important characteristic of such a nulling
interferometer is the quality of the destructive output, called the null depth
$N$. The observed starlight is reduced by a factor $N$ compared
to the starlight that would have been received by the two telescopes together operated without nulling. Very precise control of wavefront defects and precise matching of the interferometer arms are required to achieve a null depth of $\sim 10^{-5}$, the minimum performance for the spectroscopy of Earth-sized planets. Sensitivity at this level also relies on sophisticated chopping methods to subtract instrument and astronomical backgrounds that can obscure the small planet signal.

\subsection {Blind spots}

Astrometric and radial velocity observations have a blind spot centered on one year planetary orbits, because this corresponds to the parallax motion of the Earth's one-year orbit about the Sun.   The width of the blind spot for current state-of-the-art radial velocity observations at optical wavelengths is about 10 days out of  365 day period. For a star of one solar mass this corresponds to a blind spot in semimajor axis between 0.98 AU and 1.02 AU. Astrometric planet detection techniques have a blind spot for periods close to one year, for which the parallax signature due to the Earth's motion around the sun (typically 0.1 arcseconds) can swamp the signature due to a planetary companion (typically less than a $\muas$.) Although this prevents detection of a perfect Earth twin orbiting a 1.0 solar mass star, it has very little effect on the detection of planets in the habitable zone of stars even slightly more or less massive than the sun. We have simulated in Figure ~\ref{fig:astrom_blind} the effects of such blind spots. In this simple simulation, any planet with a period in the blind spot was completely undetectable.

\begin{figure}[h]
\begin{center}
\includegraphics[angle=0,scale=1.0]{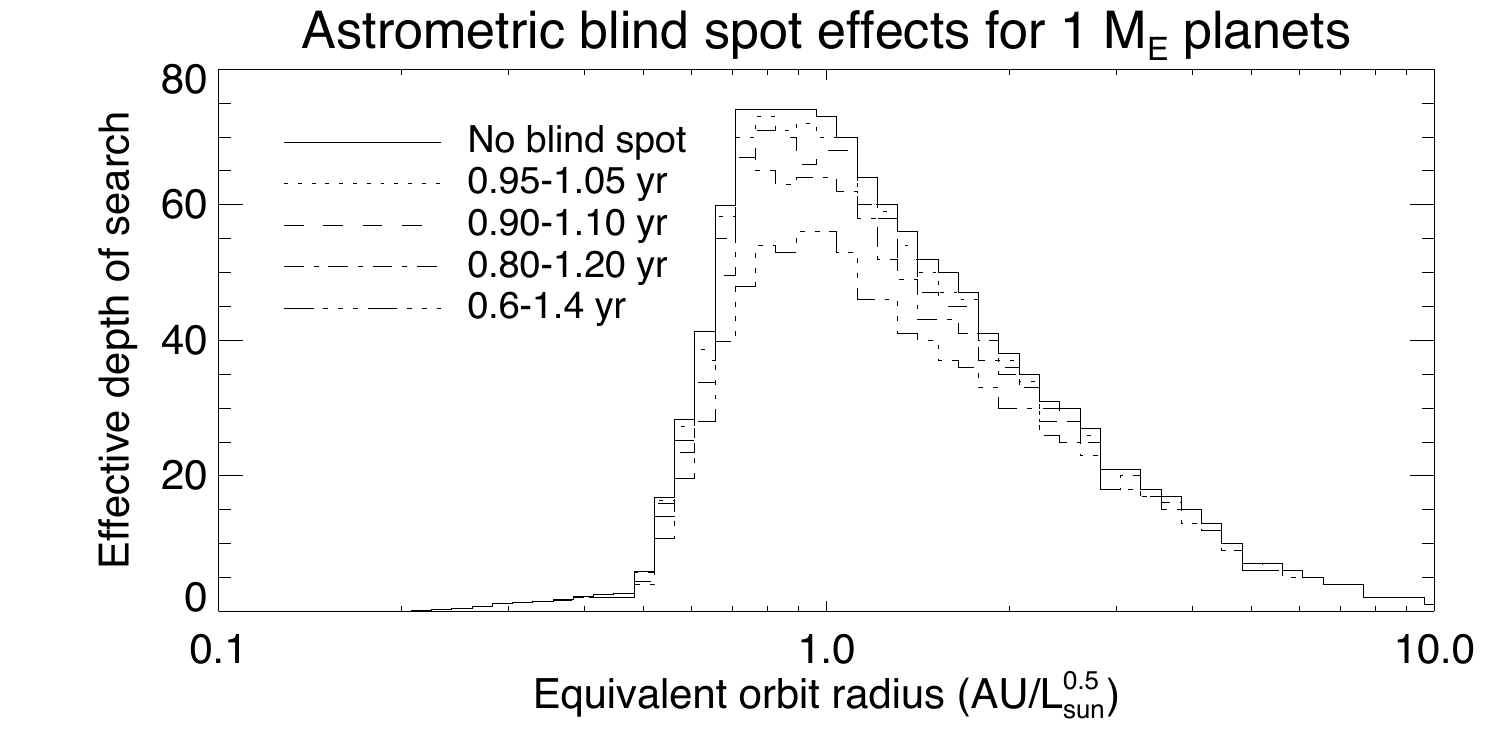}
\caption{Simulation of a sub-micro-arcsecond space-based astrometric mission searching for planets orbiting
nearby stars as discussed in Chapter 4. The figure plots the depth of search (effectively the number of stars around which a planet of a given mass and orbital separation could be detected) for 1 Earth mass planets as a function of equivalent orbital radius scaled by stellar luminosity (see Chapter 9). As can be seen, period blind spots of 0.95-1.05  years or 0.9 to 1.1 years have a relatively small effect on detection of planets within
the habitable zone.
\label{fig:astrom_blind}}  
\end{center}
\end{figure}

Direct detection approaches have a blind spot associated with the planet going inside the inner working angle or simply becoming too faint. The width of this type of blind spot is highly dependent on the particulars of any given observatory, and thus it is difficult to give a general magnitude of the effect for use in the depth of search comparisons described in Chapter 9. 

% - - - - - - - - - - - - - - - - - - - - - - - - - - - - - - - - - - - - - - - - - - - - - - - - - - - - - - - - - - - - - - - - - - -
\section{Characterizing Exoplanets Beyond Orbits, Masses, and Radii}

\subsection{Using Mass-Radius Relationships to  Constrain Compositions}

The combination of radial velocity mass and transit radii for over a dozen extrasolar planets is beginning to provide a picture of the range of densities of extrasolar planets.   (Figure ~\ref{fig:seager2sm}). While most extrasolar planets with mass and radii detected to date appear to be mostly hydrogen and helium objects, in some cases thermally expanded, at least one Saturn-mass body (HD~149026b0 is overdense and therefore probably possesses a significant fraction of heavy elements. Gl 436 b, discussed above, has a mass and radius very close to that of Neptune. 

\begin{figure}
\begin{center}
\includegraphics[angle=0,scale=.6]{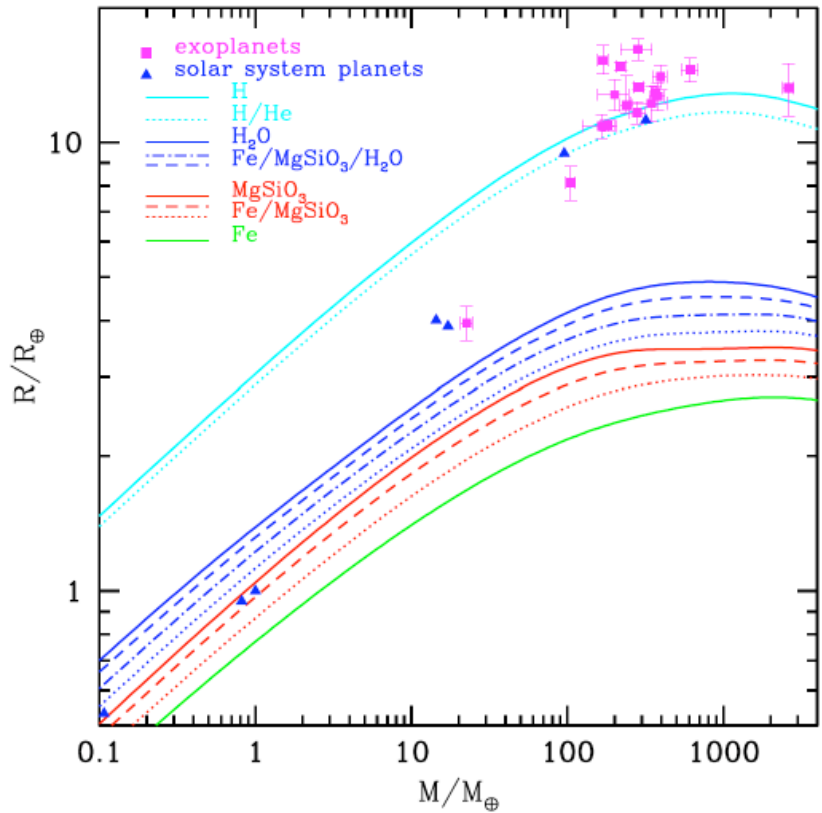}
\caption{Mass-radius relationships for planets.  The solid lines
are homogeneous planets. From top to bottom the homogeneous planets
are: hydrogen (cyan solid line); a hydrogen-helium mixture with 25\%
helium by mass (cyan dotted line); water ice (blue solid line);
silicate; and iron (green solid line). The non-solid lines represent various 
model
differentiated planets.  The blue triangles are solar system planets: from left 
to right
Mars, Venus, Earth, Uranus, Neptune, Saturn, and Jupiter. The magenta
squares denote the transiting exoplanets, including HD~149026b at 8.14
$R_{\oplus}$ and GJ~436b at 3.95 $R_{\oplus}$.   From Seager et al. 2007.
\label{fig:seager2sm}}  
\end{center}
\end{figure}

Due to the wide variety of plausible internal compositions, there will likely 
always be some uncertainty in identifying the bulk composition of transiting 
planets. For solar system planets, important constraints come from
gravitational moments and, for planets with significant gas envelopes, precise 
effective temperatures (at 1 bar); but for exoplanets, these are not obtainable. Thus, for example, mass and radius measurements alone are insufficient to distinguish between a water planet (one composed of 50\% or more water by mass) and a rocky 
planet with a massive H/He gaseous envelope. This is an important distinction because one could test the details of planetary formation models through the detection of water planets that originally may have formed at distances from their parent stars sufficient to ensure the stability of large amounts of water ice.  So-called ternary diagrams that describe the relationship between mass and bulk composition (introduced to exoplanets by Valencia et al. 
2007) can be used to reduce the uncertainty  (Figure ~\ref{fig:seager3}), if the presence of water as a significant bulk component of the planet could be inferred somehow (perhaps by spectra). 

\begin{figure}
\begin{center}
\includegraphics[angle=0,scale=.6]{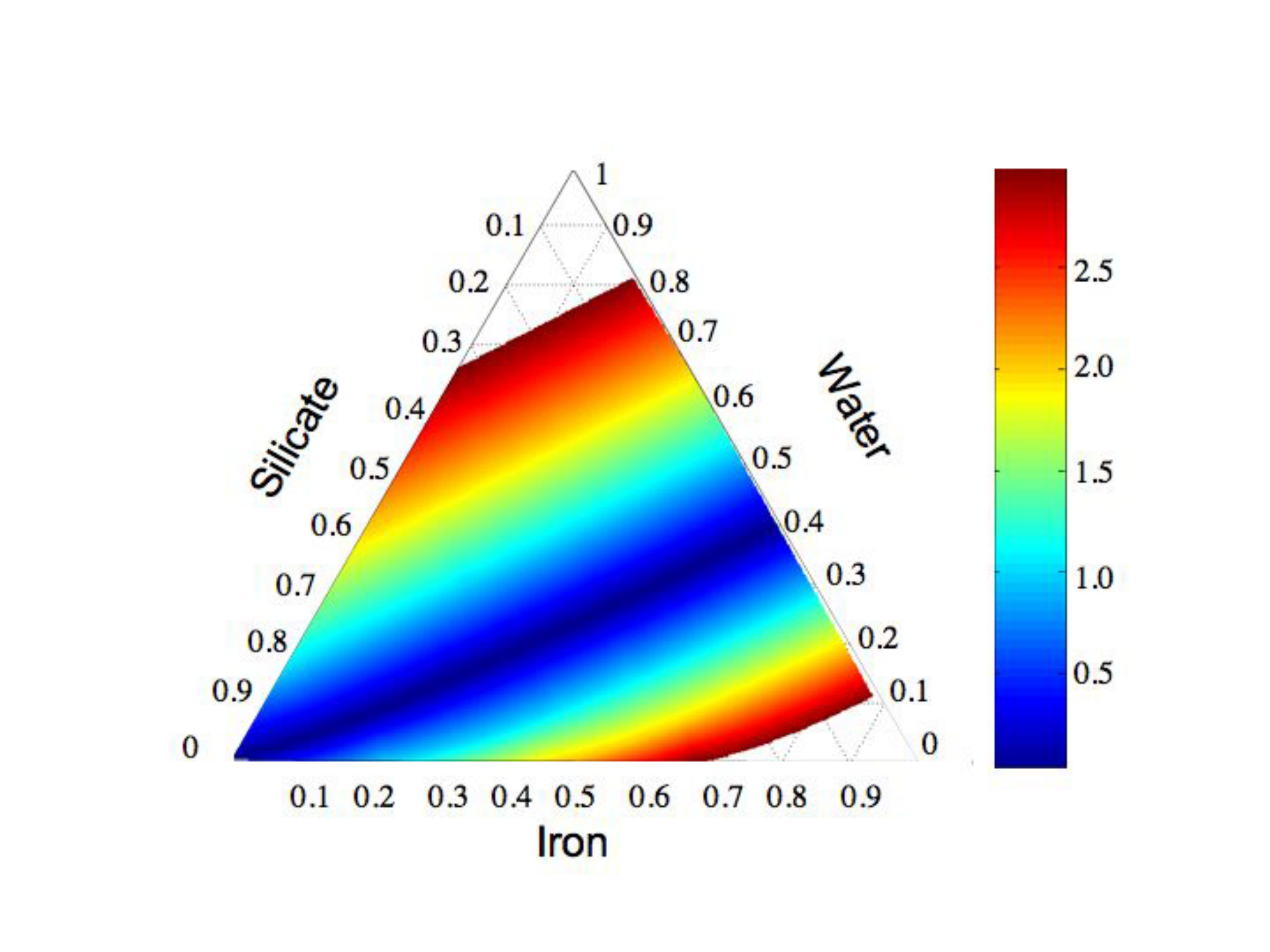}
\caption{A ternary diagram for a planet of 10 $M_{\oplus}$ and 2 $R_{\oplus}$ to 
illustrate the uncertainty in interior composition. The shaded contours are $1-
\sigma$ increments for a 5\% uncertainty in planet mass and radius. From Li and 
Seager in preparation.
\label{fig:seager3}}  
\end{center}
\end{figure}

\subsection{Spectroscopic Characterization of Habitable Planets}

Once planets have been identified within the HZs of nearby stars, interest will 
quickly shift to characterizing these planets spectroscopically. 
At visible and near-IR wavelengths, the initial goal would be to look for the 
presence of gases found in Earth's atmosphere, specifically, $\rm O_2$, $\rm 
O_3$, and $\rm H_2O$. Figure~\ref{fig:VEMfullres} shows what the Earth should 
look like at these wavelengths. Mars and Venus are shown for comparison. For 
Earth, a spectrometer with a resolution, $R = \lambda/\Delta\lambda$, of 70 
should be able to identify the strong $\rm O_2$ A band at $0.76\,\rm \mu m$, 
along with several different $\rm H_2O$ bands. The $\rm H_2O$ bands get stronger 
as one moves out into the near IR.

\begin{figure}
\centering
\subfigure[] % caption for subfigure a
{
    \label{fig:VEMfullres}
    \includegraphics[width=10.5cm]{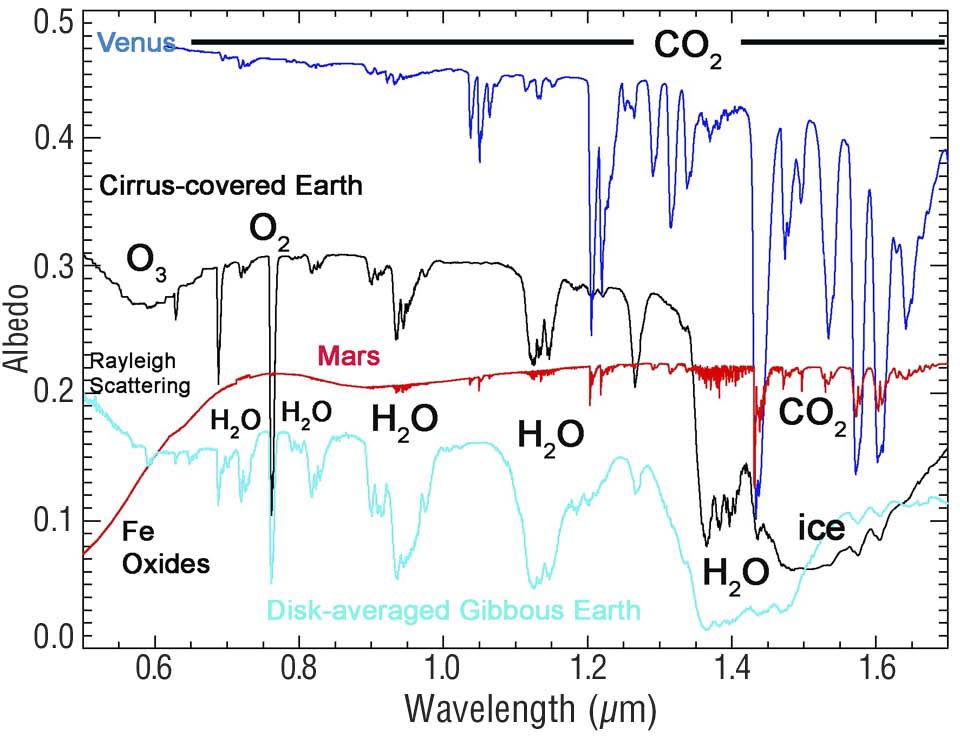}}
    \hspace{1 cm}
\subfigure[] % caption for subfigure c
{
    \label{fig:Thermal_IR_Venus_Earth_Mars}
    \includegraphics[width=4.5cm]{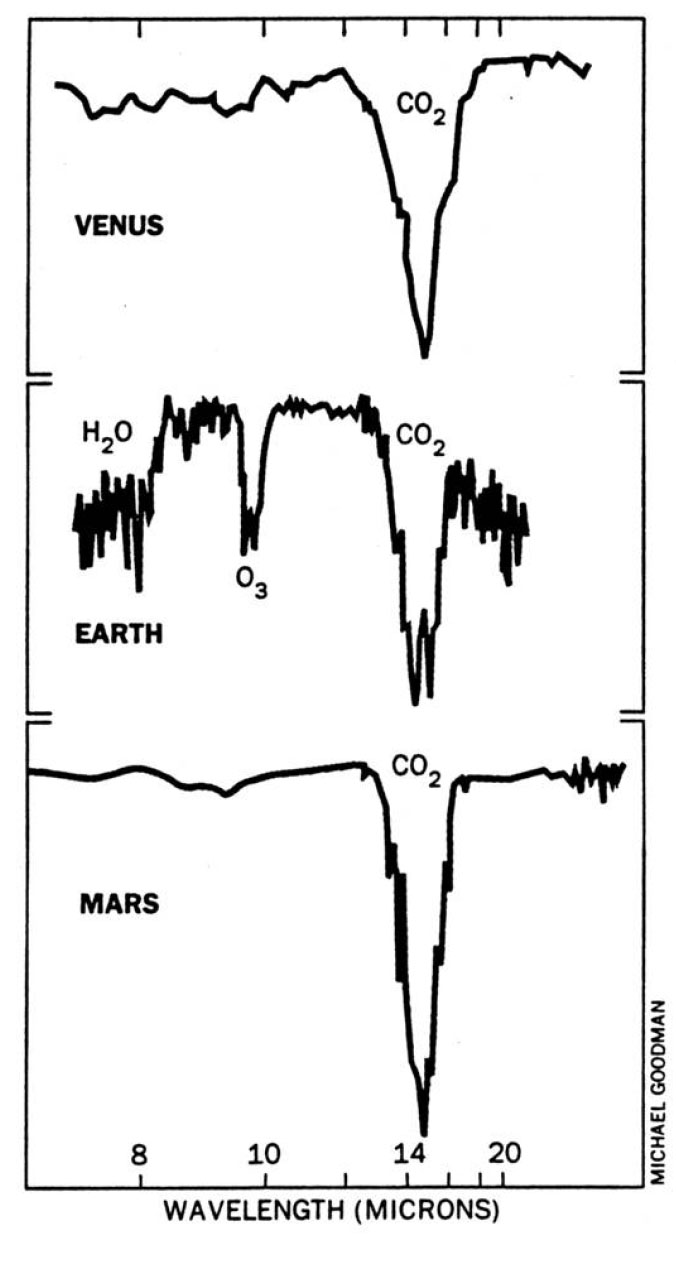}}    
\caption{(left)  Full-resolution synthetic disk-averaged albedo spectra of Venus, Earth, 
and Mars (from Meadows, 2006). Synthetic Earth spectra are shown for both 
uniform high cirrus cloud cover, and as a fit to Earthshine observations of the 
gibbous Earth. The Venus spectrum was approximated to a disk average and has 
been multiplied by 0.6 to fit the plot. The Mars and Earth spectra are disk-
averages of 3-D spatially- and spectrally-resolved Virtual Planetary Laboratory 
models of the Earth and Mars (Tinetti et al., 2005, 2006).  For the observed 
Earth, which was ocean-dominated with relatively little cloud cover, the 
Rayleigh scattering ($0.45-0.6\,\rm \mu m$) is pronounced, but the ozone is less 
apparent. The ozone absorption is much more pronounced for the Earth with cloud 
cover, increasing the difficulty of identifying the Rayleigh scattering 
component. (right) Thermal-infrared spectra of Venus, Earth, and Mars. Courtesy of R. 
Hanel, NASA GSFC.
}
\end{figure}

The information that is available at longer wavelengths is somewhat different. 
Figure ~\ref{fig:Thermal_IR_Venus_Earth_Mars} shows thermal-IR spectra of Venus, 
Earth, and Mars. For Venus and Mars, a low resolution spectrum shows only the 
strong 15-$\rm \mu m$ band of $\rm CO_2$. For Earth, one sees this same $\rm 
CO_2$ band, but one also sees evidence for $\rm H_2O$ at $6.3\,\rm\mu m$ and in 
the rotation band longward of $\sim 17\,\rm \mu m$.   $\rm O_3$ (ozone) has a strong 
absorption band centered at 
$9.6\,\rm \mu m$ that should be relatively easy to identify. $\rm O_3$ is formed 
photochemically from $\rm O_2$, and most of Earth's $\rm O_2$ comes from 
photosynthesis. Hence, the detection of $\rm O_3$ in the thermal infrared or of $\rm 
O_2$ itself in the near infrared could provide indirect evidence for life.
Of course, one would need to carefully consider all 
other possibilities before drawing any solid conclusions about the presence of life on a 
planet based on this type of evidence. The best spectroscopic evidence for life on 
another planet would be the simultaneous detection of $\rm O_2$ (or $\rm O_3$) 
along with a reduced gas such as $\rm CH_4$ or $\rm N_2O$. 
This profound disequilibrium condition is a powerful indicator of biological processes, but is harder to detect, and may require a follow-up, high resolution direct-detection spectroscopic 
mission.

Any discussion of the spectra of putative terrestrial-type exoplanets must conclude with a warning that, because we have so few examples of such atmospheres to study today, one must anticipate surprises when the first such spectra for an exoplanet are obtained. This was true for extrasolar giant planets, and it means that the spectral observations should not be so highly selective or tuned to particular features that significant discoveries would be missed. 

\clearpage
% - - - - - - - - - - - - - - - - - - - - - - - - - - - - - - - - - - - - - - - - - - - - - - - - - - - - - - - - - - - - - - - - - - 
% ============================================================================
% Begin New PART - ONGOING INVESTMENTS
% ============================================================================

\pagebreak

\chapter{Ongoing Investments in the Search and Characterization of Exoplanets}

To fully capitalize on the sizable investment that will be required to capture the first images and spectra of extrasolar planets,  ancillary investments are needed in theory, laboratory astrophysics, and fundamental astronomy. In this section we detail some of the required investments, but the list is not meant to be all-inclusive.  There may well be additional specific areas we have overlooked where small, focused research will improve the science yield of future planet hunting victories.  Nevertheless we believe that the areas highlighted here are of particular importance for interpretation of exoplanet observations and will impact a variety of areas of science, not just planet hunting.

% - - - - - - - - - - - - - - - - - - - - - - - - - - - - - - - - - - - - - - - - - - - - - - - - - - - - - - - - - - - - - - - - - - -
\section{Investments in Planetary System Architecture Theory}

The theory of planet formation is poised to make predictions about the occurrence and properties of
rocky planets, constrained by the properties of the observed giant exoplanets. To support planetary
detection and characterization, theoretical work on the formation of rocky planets, their dynamical
evolution, and the diversity of their interiors and atmospheres should be strongly supported.  In addition
since planet formation is governed by the characteristics of the protoplanetary disk, including other planets forming in the disk,
support for ongoing theoretical investigations into planet formation as a whole
will provide the necessary context for understanding the architectures of other planetary systems.

A short list of the interplay of observations of extrasolar planetary
systems and dynamical theories shows the importance of maintaining a
strong theory community for successful interpretation of the
distribution of planetary systems and our place among them.
Dynamical theory is used for the initial description of the properties
of extrasolar planetary systems found by the several observational techniques.  The distribution of orbital semimajor axes and
eccentricities in multiplanet systems usually requires a dynamical
fit, where initial conditions and the masses of the planets are solved
for in a least squares fit to the data.  Mutual
perturbations in near resonant configurations can sometimes be used to resolve the $\sin{i}$
ambiguity and determine the actual masses of the planets.  Dynamical
tests of stability can reject some best fit solutions. The
nature of perturbations in multiple planet systems can explain variations in transit signatures.

Already the wide variety of properties of extrasolar planets has
drastically revised our ideas about the formation and evolution of
planetary systems, where much current work has concentrated on
understanding how observed configurations came to be. For example, unexpected properties such as the close orbits of the hot Jupiters, or the high eccentricities of some exoplanets led
respectively to the realization of ubiquitous migration of planets
while the nebula was present, and dynamical means of eccentricity
excitation. The resonant configuration of one system, that around GJ 876, most probably was established by differential migration of
the two planets through their interaction with the nebula in which
they formed, which scenario may imply a constraint on the magnitude of
orbital eccentricity growth during migration, or a prompt dispersal
of the nebula after the resonance was established.  Further
modeling can predict the most common features such as the typical
numbers of terrestrial planets and giant planets. The relative number
of ice giants like Neptune compared to the number of gaseous giants
like Jupiter, can constrain the probability that a giant planet core
fails to accrete the nebular gas to become a gas giant.
These examples
show that dynamical theory is a necessary input to understanding what
the observations of extrasolar planetary systems indicate for the
formation and evolutionary processes of the systems in general, and
these observations in turn test theoretical predictions.

Embedded in this formation-evolution scenario is the modeling of
specific physical processes involved in going from disks, with
properties constrained by observation, to predictions of the
properties of planetary systems being observed. Observations of disk
structure can indicate the presence of planets inferred by the
dynamical theory of disk-planet interactions, where the planets
themselves give no other signature. There is the possibility that the
interplay of theory and space-based IR photometric observations
can contribute to a fundamental understanding of the nature of tidal
dissipation in gaseous planets.  Planets that are relatively close to
their host stars, but in highly eccentric orbits with very small
periastron distance, such as HD80606b and HAT-P-2b, are likely to be tidally
evolved to an asymptotic spin state with a rotational angular velocity
between the orbital mean motion and the orbital angular velocity at
periastron.  The value of this angular
velocity is strongly dependent on the tidal model assumed, but not on the disspation rate. Such models
of dynamic tides are highly uncertain, and eventual measurement
of the rotational angular velocity of such planets through IR
photometry could provide an important constraint on such models.

In the future, dynamical theory can predict the best observational
opportunities for direct imaging of the close planetary systems.
As observational technologies develop, details of some of the nearest
planetary systems may be sufficient to determine if the dynamical
constraints thought necessary to maintain a climate stability for the
development of life, and especially intelligent life on the Earth,
obtain for these other systems. These include the protection of the
Earth from frequent giant impacts late in its history by the Jupiter
shield, the presence of the Moon in stabilizing the Earth's obliquity,
and limiting the orbital and rotational variations for ice
ages.

The growth in the understanding of extrasolar planetary systems and a
determination of the probability of occurrence of systems like our
own, where intelligent life has evolved, will be best served by
a healthy balance of new observational constraints and theoretical
investigations that interpret and explain the new discoveries.

% - - - - - - - - - - - - - - - - - - - - - - - - - - - - - - - - - - - - - - - - - - - - - - - - - - - - - - - - - - - - - - - - - - -
\section{Investments in Understanding Planetary Atmospheres}

\subsection{Molecular Line Data and Laboratory Astrophysics}

Identifying and interpreting planetary spectra requires reference to lists of molecular and atomic spectral lines.  While band positions and relative strengths are generally well known for species of importance in the solar system, there are still important gaps in our understanding.   These gaps will limit our ability to model atmospheres and interpret observed spectra.

Most spectroscopic databases are built upon measurements taken at or
near room temperature, and theoretical calculations supply the missing
transitions that can become important at high temperatures. If only
room temperature databases are used, the transitions from highly
excited energy levels (``hot bands") would be missing and the true opacity at elevated
temperatures perhaps substantially underestimated.  Examples where such hot databases would be of importance
would be the interpretation of a Venus-like $10\,\rm M_\oplus$ planet with a deep carbon dioxide ($\rm CO_2$) atmosphere
or a hot Titan or giant planet with a methane-rich atmosphere.   Figure ~\ref{fig:methane}, for example, illustrates the differences between a standard molecular line list, an enhanced computational line list, and a low resolution laboratory spectrum of methane ($\rm CH_4$).  Even the modern computational list misses important bands in the near-infrared and the optical (not shown), which limits our ability to correctly interpret spectra of methane-rich atmospheres.

\begin{figure}
\begin{center}
\includegraphics*[scale=0.6]{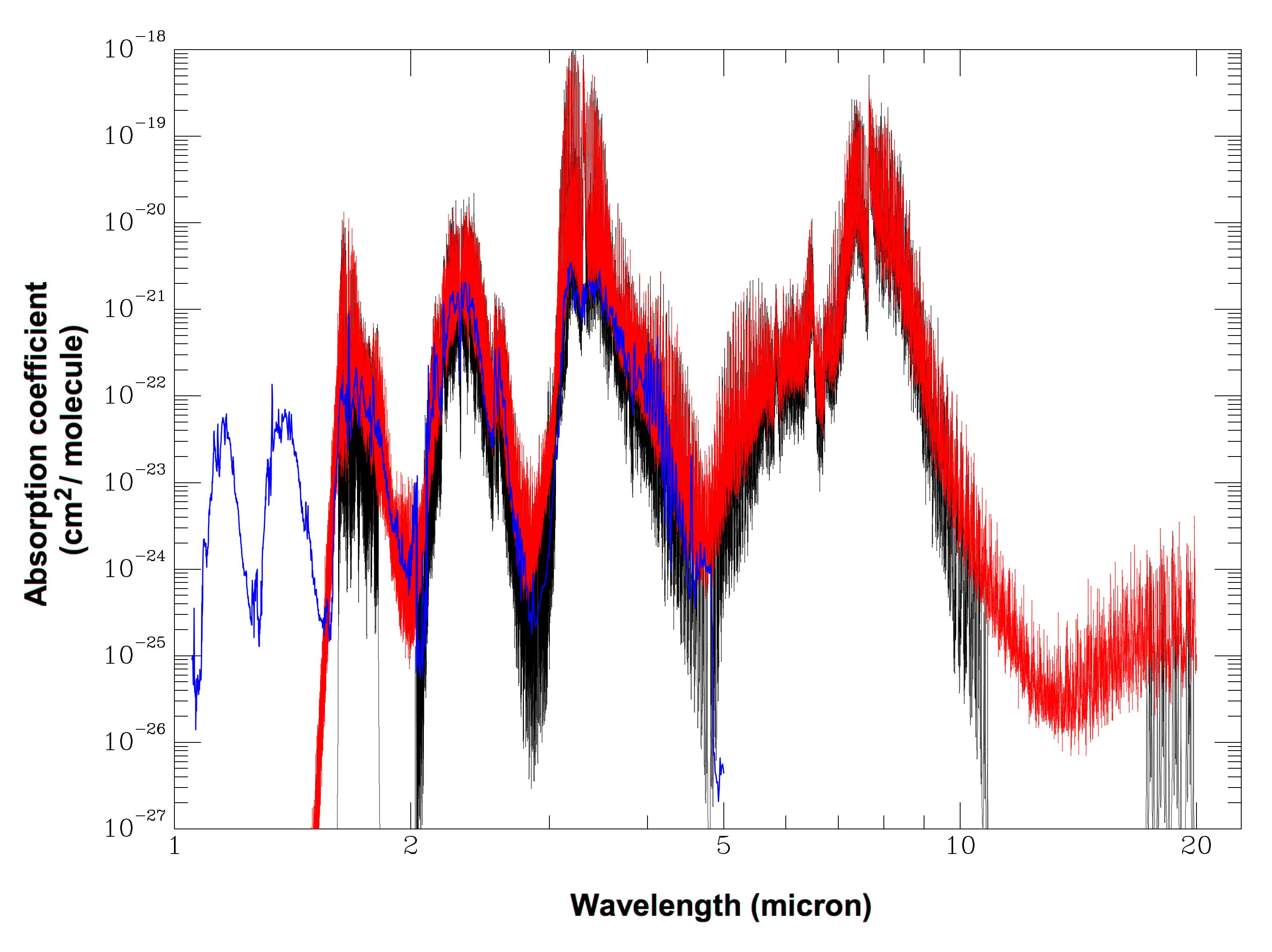}
\caption{Comparison of different sources of data on molecular opacity of methane.  The black line is derived from the standard ``HITRAN'' database.  The red line is a computational database that has not been verified by laboratory data, but points to many shortcomings in the standard line list (note the gaps and differing band strengths in HITRAN).  Both databases are missing important molecular bands of methane that have been observed in optical and near-infrared laboratory spectra (blue line).  Figure courtesy Richard Freedman.  \label{fig:methane}}
\end{center}
\end{figure} 

Perhaps the most important species for understanding atmospheric structure of extrasolar terrestrial planets is $\rm CO_2$.  Although the room temperature line list is well characterized, substantial uncertainties remain in the partition function and line strengths at higher temperatures that are relevant to the atmosphere of Venus and perhaps many extrasolar planets.  In addition the pressure-induced opacity of $\rm CO_2$ at high pressure is important, but has not been well characterized throughout the range of conditions likely to be encountered in extrasolar terrestrial atmospheres.  Other molecules of importance for the interpretation and modeling of extrasolar terrestrial and giant planet atmospheres which are lacking adequate opacity databases include $\rm CH_4$, $\rm NH_3$, and even the pressure-induced opacity of molecular hydrogen.  Today there is no methane line list in the optical available, only a pseudo-continuum absorption spectra and there are substantial uncertainties in ammonia opacity in the near-infrared.  
The line list for water is comparatively well in hand, even at very high temperatures, but substantial uncertainties remain in the line broadening.  Finally the standard spectroscopic databases for potential biomarker species (e.g.,  $\rm CH_3Cl$) are  known to be missing entire bands.  We therefore support renewed laboratory and computational studies of $\rm CH_4$, $\rm NH_3$, $\rm CO_2$ (particularly at tens to hundreds of bars pressure) and other species to better characterize the molecular line lists that will be required to interpret spectra of extrasolar planets.  Further theoretical work on the pressure-induced opacity of hydrogen is also sorely needed to interpret the spectra of giant planets.

\subsection{Cloud Modeling}

Clouds play a vital role in every solar system body with an appreciable atmosphere.  From Venus to Earth to Mars to Titan, clouds affect the planetary albedo, and hence surface temperature and observed brightness, and play major roles in controlling atmospheric thermal structure and dynamics. Modeling such effects requires knowledge of first order cloud characteristics such as particle size, vertical distribution, and areal coverage. The campaign to understand brown dwarf atmospheres--in some sense the first planetary science done on real objects outside the solar system--has clearly shown that a proper treatment of clouds (in this case composed of iron and silicates) is a requirement. 

Despite the crucial role clouds play in controlling the observed spectra of planetary atmospheres there is no generally accepted procedure for modeling cloud physics on global scales or incorporating clouds into radiative equilibrium models of planetary atmospheres.  Because of the history of the development of planetary science, cloud physics in the solar system has tended to focus on specific problems relating to individual atmospheres. In Earth's atmosphere, of course, highly detailed studies of cloud physics aim to address key questions for understanding climate change. Nevertheless clouds remain the leading source of uncertainty in global models of Earth's climate. Studies of other planetary atmospheres within the solar system tend to be less detailed and focus on localized concerns. This targeted approach to cloud modeling has left a surprising gap in our first-principles understanding of clouds in other planetary atmospheres that is required to understand the global properties of a given hypothetical planet.  As a result, efforts to model extrasolar atmospheres by necessity generally rely on fairly simplistic approaches.  It is well understood that high, thick cloud decks produce a brighter planet and smaller depths for most absorption bands.  Lower or thinner clouds have the opposite effect.  While adequate for the exploratory nature of the calculations to date, clearly more sophisticated approaches are needed to enable realistic models of atmospheres of extrasolar atmospheres. 
 
Since habitable planets, by conventional thinking, must have abundant supplies of water, the importance of water clouds is unmistakable. In addition as we move towards an era of studying perhaps dozens of extrasolar terrestrial planets the need for more generally applicable tools becomes clear.  
The expertise of the terrestrial atmospheric science community is a resource that must be tapped by interdisciplinary collaborations in order to address questions such as these. The challenge will come in identifying pathways for maximizing the value of existing tools in the new regimes.  Encouraging interdisciplinary interactions between planetary scientists, atmospheric scientists, and astrophysicists now will sow the seeds to maximize the scientific return from our first encounters with extrasolar terrestrial planets.

\subsection{Photochemistry}

Perhaps an even more challenging problem than cloud physics is atmospheric photochemical hazes.  All of the solar system giant planets
 are strongly influenced by hazes produced by the ultraviolet photolysis of methane.  Such photolysis may have been important earlier in Earth's history and might also play an important role in methane-rich atmospheres of extrasolar terrestrial planets.  As such, photochemical processes in extrasolar planetary atmospheres are deserving of additional study.
 
At Jupiter methane photolysis is the main driver of photochemistry since water, ammonia, and $\rm H_2S$ are trapped in clouds far below the upper atmosphere.  In the atmospheres of extrasolar terrestrial or giant planets, this may not always be the case.  In some atmospheres these molecules will be rapidly photolyzed, perhaps providing important sources for photochemical haze production. Ultimately haze optical depths depend upon production rates, condensation temperatures, microphysical processes, and 
mixing rates in the nominally stable stratosphere that in turn depend upon the atmospheric structure and poorly understood dynamical processes.  To date photochemical hazes have been neglected in most studies of extrasolar terrestrial and giant planet atmospheres. Since hazes can play very important roles in shaping atmospheric spectra--for example Titan--they are well deserving of additional laboratory and theoretical studies.  

% - - - - - - - - - - - - - - - - - - - - - - - - - - - - - - - - - - - - - - - - - - - - - - - - - - - - - - - - - - - - - - - - - - -
\section{Investments in Stellar Characterization and Astrophysics}

For almost every technique, the correct interpretation of exo-planetary observations hinges on knowledge of the properties of the parent star.  A transit-derived planetary radius, for example, is known only as accurately as the stellar radius of the primary.  Similarly, a radial-velocity-derived (minimum) planet mass is only as accurate as the primary's mass.  Understanding the evolution of a planet requires knowledge of the age and evolutionary history of the radiation field, in particular, of its primary star.  Stellar metallicity likewise affects planetary composition and evolution and the availability of volatiles.  There are many more examples.

The TPF project has compiled a list of some 2,300 nearby stars that are potential direct detection targets.  Fundamental astrophysical data are not available for many (roughly half) of these targets.  To optimize the choice of targets and maximize the eventual scientific interpretation of exoplanet observations, it is important to have an ongoing program to measure stellar parameters, particularly mass, radius, effective temperature, gravity, metallicity, age, activity level, and (where Hipparacos data are unavailable) distances to those nearby stars that may eventually be targets for direct or indirect detection of planetary companions.

% - - - - - - - - - - - - - - - - - - - - - - - - - - - - - - - - - - - - - - - - - - - - - - - - - - - - - - - - - - - - - - - - - - -
\section{Other Ongoing Investments}

Significant progress in the past decade has invigorated the field of exoplanet research, providing a context for exoplanet studies grounded in solid observational fact and rigorous theoretical modeling.   This report focuses on new initiatives, but we emphasize that many of the current activities should continue.    This is the case for so-called ``Origins" research, including the formation of planets, planetary evolution, and the chemistry of planets and their atmospheres.  The physics and chemistry of circumstellar disks is also extremely important in that these proto-planetary systems are often signposts to current or future planetary development.

Planetary astronomy is also a vigorous field that must be maintained.  This work, both within our Solar System and beyond, is critical for understanding the physics, chemistry, and evolution of planetary atmospheres.   Researchers are now beginning to map out temperature fields on exoplanets, laying the groundwork for atmospheric dynamics studies.  Yet the source of zonal wind patterns on our local giants is still being hotly debated. Only now, after decades of study, has local planetary astronomy reached an age and maturity that we can begin to understand the interactions of our primary star with the giant planet atmospheres of the giant planets, e.g., the seasonal variations of gas and ice giants.  The implications of such work will be significant for exoplanetary system architecture and evolution of true Solar System analogues (i.e., those with distant, long-period giant planets).  Even basic ``local'' issues such as the size and composition of Saturn's and Jupiter's cores is still in active discussion.

The field of exoplanets has captured the imaginations of students, becoming one of the fields most requested by entering graduate students.  At the same time, many established researchers are venturing into this opening research frontier.  Both forces together conspire to create very limited funding opportunities, particularly for the as-yet unestablished young scientists.  Fellowships that target exoplanetary work can help ensure the steady flow of brilliant young researchers into the field, while at the same time not cut into the funding streams desperately needed by the established science community.

% ============================================================================
% Begin New PART - FINDINGS
% ============================================================================
 
%\setcounter{section}{0}

\pagebreak

\chapter{Findings of the Exoplanet Task Force}

What follows are the major findings of the ExoPlanet Task Force.
The first five sections are organized according to each major
planet detection and characterization technique. The sixth section
discusses exoplanet topics not included in detection
and characterization techniques and the final section presents
a discussion on $\eta_\oplus$. All findings are in light of the 
motivating questions we list again here.

\bigbreak

\noindent {\bf Compelling Questions}

The fundamental questions to be addressed are listed here in priority order, based upon the Task Force's assessment of centrality to the charge, and the potential for public interest and excitement generated by discoveries. All three questions reflect scientific issues crucial to an understanding of the nature of planets, their origins, and their ubiquity in the cosmos. 

\begin{enumerate}

{\bf \item What are the physical characteristics of planets in the
habitable zones around bright, nearby stars?}

{\bf \item What is the architecture of planetary systems?}

{\bf \item When, how and in what environments are planets formed?}

\end{enumerate}

% - - - - - - - - - - - - - - - - - - - - - - - - - - - - - - - - - - - - - - - - - - - - - - - - - - - - - - - - - - - - - - - - - - -
\section{Radial Velocity Findings}

Out of all planet discovery techniques, radial velocity is the most
advanced on the path to finding low-mass planets orbiting nearby
stars. A 1 m/s  precision is the new
benchmark, with 0.1 m/s a goal. Precision as high as 0.6 m/s has enabled, at the time of writing,  the discovery of a planet with M $\sin i$ of 5.1 M$\oplus$.
orbiting a nearby M4V star. However, access to sufficient time with premier radial velocity instruments is currently the key bottleneck for the follow-up of transiting planets discovered by ground-based surveys and COROT, and the situation will be worse when Kepler begins producing candidates.
 
 \bigbreak
 
{\bf \emph{Finding}} {Radial-velocity can reach sub 1~m/s precision and
is the nearest-term path to finding terrestrial-mass planets around
low-mass stars. A significant investment in ground-based telescopes, equivalent to a dedicated 4-meter facility as well as additional NASA time on the Keck telescopes, could
enable radial-velocity surveys to find statistically-significant numbers of exoplanets with M $\sin i$ down to 10 M$\oplus$ around solar-type stars in the habitable zone. A program focussed on a selected set of late-type stars, down to early M-dwarfs, could find one or two dozen planets with M $\sin i$ approaching the mass of the Earth.}

\bigbreak

Main sequence M-dwarf stars are of keen interest because a given
radial velocity precision permits detection of a lower-mass planet
around an M star than around a G star. Although M-dwarfs are faint at
visible wavelengths (typically V=10 or fainter) the forest of titanium oxide (TiO)
lines provides a means for obtaining a radial velocity precision comparable to that
for G stars. Visible-wavelength radial velocity surveys are able to
monitor M-dwarfs brighter than M4; later (that is, less massive and less luminous) M-dwarfs require moving the
Doppler spectroscopy into the near-infrared. Near-IR, high precision
Doppler spectroscopy is currently under development.

\bigbreak

{\bf \emph{Finding}} {The potential of the near-IR radial velocity
technique for late M-dwarfs seems quite high but as yet unproven. 
Potentially this technique could survey the lowest mass nearby stars and 
discover dozens of planets with M $\sin i$ less than 10 M$\oplus$ in the 
habitable zone.  Additional
resources will need to be expended in order to make this technique
operational and understand its ultimate capability.}

\bigbreak

There are long-term prospects that radial velocity techniques will be
able to discover candidate Earth-twins (Planets with M$\sin i \sim 1 M\oplus$ in $\sim 1 AU$ orbits
around Sun-like stars), provided radial velocity precision continues to
improve and sufficient telescopic resources are available to collect the required data.  
However, the $\sin i$ ambiguity will always be present and the discovery 
of such objects will spur on the use of other techniques to determine the mass unambiguously. 

Since the sensitivity declines with increasing semi-major axis,
radial velocity will likely not be the technique of choice for finding
sub-Saturn-mass objects beyond 5 AU or Earths in the habitable zone around solar-type stars, even with
the predicted improvements over the 15-year time horizon considered
here. At the same time, however, the long time-baseline established
for radial velocity surveys will allow sufficiently massive planets at
large semi-major axes to be detected, providing key information on
planetary system architectures and targets for future direct detection
missions. We expect that in the third period (11--15 years) of the strategy's time horizon, RV will be superseded in this part of discovery 
space by other approaches such as extreme AO and advanced coronagraphs on ELT's. 

Past the ten-year time frame, radial velocity surveys will learn more
about the intrinsic noise limits set by the star.  In the event that
astrophysical noise impedes the detection of terrestrial planets, then
radial velocity observations will transition from a survey mode to a
support mode for photometric and astrometric space-based surveys. They
will be of high value in providing mass and hence density for transit
detections, in resolving ambiguities in multiple planet astrometric
signatures (see below), and in finding massive planets in orbits of
long (in excess of ten-year) periods.

\bigbreak

{\bf \emph{Finding}}. {The radial velocity technique will find giant planets at larger and 
larger semi-major axes as the time baseline increases over the 15 years considered here. 
However, the technique of choice for finding large period giant planets will 
likely become direct detection as, for example,  ELT's with extreme AO come on 
line.  Nevertheless, the value of the radial velocity technique
will remain high in the latter part of the next 15 years, as 
improvements in accuracy allow for smaller short-period 
planets to be detected, and as a support
mode for photometric and astrometric space-based surveys.  }

Exoplanet science and technology is a rapidly changing field. There is potential for new, transformational technologies that may significantly impact planet discovery. One such exampleÑthat of a new radial velocity techniqueÑarose just as this report was being completed. The new radial velocity technique enables precise wavelength calibration of high-resolution spectrographs using laser frequency combs. Demonstrated in the lab, the frequency comb spectrograph still needs to be tested on a telescope. If the significant challenges of long-term instrument stability and intrinsic stellar variability can be met, the precise calibration could enable discovery of Earth analogs around hundreds of sun-like stars. This development serves as a 
reminder that transformational techniques will arise, and may call for reevaluation of the relevant decision points in the strategy. 

% - - - - - - - - - - - - - - - - - - - - - - - - - - - - - - - - - - - - - - - - - - - - - - - - - - - - - - - - - - - - - - - - - - -
\section{Transits Findings}

The discovery of some two-dozen transiting exoplanets has
revolutionized the field, because they are
currently the only ones that can be physically characterized with
existing technology by measurements of the physical mass, radius,
temperature, and atmospheric composition. Transits that are followed-up or preceded by radial velocity or astrometric studies to obtain the mass of the planet allow the bulk density to be determined, which can be used to constrain the
planet's overall composition. The combined masses and radii of some 30
transiting hot Jupiter exoplanets have already challenged theories of
planet formation and evolution. The recent transit of a Neptune-mass
planet, revealing it to have the size and hence density of Neptune as
well, extends the quantitative study of extrasolar planetary properties significantly downwards in
mass.

In addition to the smaller mass ratio of planet to star around
M-dwarfs, the M star habitable zone is much closer to the star than
around G-dwarfs. Therefore, both the transit probability and the radial
velocity amplitude increase closer to the star, making the search for and
characterization of planets with properties akin to our home world, in the
habitable zone a realistic near-term goal. 

\bigbreak

{\bf \emph{Finding}} {Both ground-based and proposed, small space-based systems are capable of conducting a transit survey of a large number of nearby M-dwarfs ($\sim$ 1000). Detecting a population of low-mass transiting
planets orbiting nearby M-dwarfs, with
radial-velocity-measured masses, would provide an opportunity to find a
potentially habitable planet within the next decade.}

\bigbreak

Temperature and atmospheric composition are critical for identifying
habitable planets. Spitzer has made large advances in exoplanet
science, particularly in the field of comparative
exoplanetology, through transit observations. Temperatures of several Jupiter-mass exoplanets,
thermal orbital phase curves, and identification of water vapor are
among the highlights. Most relevant to the ExoPTF compelling questions
is that Spitzer, even absent its cryogen, can detect and measure the radius of a
transiting low- mass exoplanet, down to the equivalent of a few Earth radii. The warm-Spitzer radius and a
radial-velocity mass can be used to constrain the interior bulk
composition of the exoplanet. Spitzer is ideal for measuring planet
radii because limb darkening of the star is virtually absent at
near-IR wavelengths where M-dwarfs are bright.

\bigbreak

{\bf \emph{Finding}} {Spitzer, even after its cryogenic mission is over,
is the best telescope for studying transiting planets via
temperature and spectral absorption or emission features until the
launch of JWST. Of primary relevance to the ExoPTFs compelling
questions, Spitzer is able to detect transits and measure precise
planet radii for planets a few times larger than the Earth orbiting low-mass M-dwarfs.}

\bigbreak

Beyond radii, Spitzer has shown that extrasolar giant planet
atmospheric composition and temperature measurements are
straightforward for many exoplanets by exploiting the secondary
transit and primary transit transmission spectra. A larger space-based
telescope with high spectral resolution at visible through IR
wavelengths is required to enable the same
kinds of measurements on planets that are the size of 
the Earth, that is, Earth analogs, orbiting M-dwarfs. This need is satisfied with the James Webb Space Telescope currently under development for launch in the decade beginning 2010. Preliminary calculations based on the community white papers suggests that detection of CO$_2$ and O$_3$ would be possible with the near-infrared spectrometer (NIRSpec) over 30-50 hours per Earth-sized planet, and with the mid-infrared instrument (MIRI) over 200 hours. The latter is equivalent to a Hubble deep field observation, not unreasonable in view of the small number of targets and the significance of the observations.

\bigbreak

{\bf \emph{Finding}}. {The James Webb Space Telescope will have the capability 
for doing transit studies of Earth-sized planets orbiting in the habitable zone of 
M-dwarfs. These studies will include the size of the planet and, depending 
upon the characteristics of the planet and the ultimate performance of the 
telescope, detection of some atmospheric spectral features at low spectral 
resolution. Potentially such JWST observations, combined with RV studies 
to determine the mass, would represent the first characterization of a true 
Earth analog orbiting an M-dwarf. Such studies could be done prior to the 
start of the final, 11--15 year, time frame of the strategy.}

\bigbreak

The number of transiting planets available for discovery around nearby
bright stars is ultimately limited by statistics. The probability of a
planet to transit diminishes as $R_*/a$, where $R_*$ is the stellar
radius and $a$ is the planet-star separation (for a circular
orbit $a$ is the semi-major axis). A more distant--and hence
numerous--population of stars, while not amenable to followup
characterization, is the best way we have right now for a first
determination of the frequency of Earth-sized planets orbiting
different star types. The frequency of Earth-sized planets established 
on a statistically significant basis will be used to make key decisions on the 
size, approach, and the timing of development, of spaceborne direct detection missions.  

\bigbreak

{\bf \emph{Finding}}  {Kepler is the most reliable way to get a census
of Earth-sized planets around solar type and other main sequence
stars. This census will determine whether Earth-sized planets are a
frequent outcome of star formation, and hence the likelihood that
stars within the solar neighborhood harbor planets like our own. 
In addition, and prior, to Kepler, ground-based surveys of M-dwarfs may 
be able to obtain statistics on short-period transiting planets approaching the size of the Earth}

% - - - - - - - - - - - - - - - - - - - - - - - - - - - - - - - - - - - - - - - - - - - - - - - - - - - - - - - - - - - - - - - - - - -
\section{Microlensing Finding}

Microlensing sensitivity to planets peaks at the so-called ``Einstein
ring" which typically lies beyond the snow line at temperatures
comparable to Saturn, although it does retain significant
sensitivity a factor of a few closer and also out to infinite separations
(free-floating planets).  Even low-mass planets can give rise to a
pronounced signal, but the probability of this occurring declines at
low mass, so overall sensitivity basically scales as the planet mass.
Of the six published microlensing planets, two are cold Neptunes and
four are cold Jovian planets.  Two of the latter are in a single
Jupiter/Saturn analog system, with the similar mass ratios, separation
ratios and temperatures to the solar-system gas giants.  Therefore,
despite
the low number of detections, microlensing has already told us that
cold Neptunes are common and that solar-system analogs may be common.
Microlensing is also the only way to detect old, free-floating planets.

Present microlensing searches work primarily by identifying the
most promising 3\% of the roughly 700 events that are found in
wide-angle surveys each year,
and monitoring these intensively to look for planetary deviations.
These are primarily the ``high-magnification events", which are quite
rare.  The microlensing yield could be dramatically improved by
continuous monitoring of the wide-angle fields by wide-field cameras
on 3 southern continents.  Then more events would be monitored
well enough to find the, overall much more common, planetary deviations
in ``normal" microlensing events with sufficient light curve coverage
and signal to noise ratio to determine the planet-star mass ratio
$m/M$ and projected separation in units of
$R_E$. Fig. \ref{fig:bennett2} in the Appendix compares the sensitivity
of a microlensing survey using the triad of telescopes with that of current
practice.  Continuous coverage of all events by the three telescopes
is only possible near the middle of the bulge season, but 
additional small telescopes distributed in intermediate longitudes
could fill in gaps in the higher magnification light curves and yield
more events with the above sufficient coverage.     

Upgrades to existing survey telescopes are already establishing
2 of the desired triad of 3 wide-field telescopes. The addition of a
third telescope will provide the improved sensitivity shown in
Fig. \ref{fig:bennett2}.  

The next generation ground based microlensing survey will considerably
improve the rate of aquisition of statistics on the properties
planetary systems in the Galaxy. But the necessity of using the stable
point spread function of HST, with limited access to this telescope,
to constrain the planet mass and star-planet projected separation in
physical units means relatively few of the planets detected from the
ground will be completely characterized.  The promise of microlensing
to obtain the statistics of the properties of planetary system as a
function of stellar type and location in the galaxy in a reasonable
time frame will therefore require a dedicated space based survey
similar to that described elsewhere in this report. 

\bigbreak

{\bf Finding:} {\it A modest investment in ground-based microlensing,
equivalent to the commissioning of one dedicated 2-meter class
telescope, will increase the efficiency of planet detection and
enable a high return on the last decade's investment in microlensing.
Space-based microlensing is the optimal approach to providing a true
statistical census of planetary systems in the Galaxy, over a 
range of likely semi-major axes, and can likely be conducted with a
Discovery-class mission.}

% - - - - - - - - - - - - - - - - - - - - - - - - - - - - - - - - - - - - - - - - - - - - - - - - - - - - - - - - - - - - - - - - - - -
\section{Astrometry Findings}

The mass of an exoplanet is critical to determining whether it is
terrestrial in nature. It is essential for the preferred pathway of this strategy that a technique be available to identify 1 Earth mass planets for later spectroscopic study. Space-based astrometry is the only known method able to determine
unambiguously the mass of a planet, down to the mass of the Earth, for a
reasonably large sample size (of order a hundred stars). The number of
inner terrestrial planets is limited by dynamical stability--thus
making the planet mass and multiplicity determination by astrometric
observations tractable.
 
\bigbreak

{\bf \emph{Finding}} {A space-based astrometric search for planets, if sufficiently sensitive and with a sufficiently narrow parallax-caused blind spot,  is the
best all around technique to find planets the mass of the Earth around nearby
bright stars that are candidates for a direct detection follow-up; 
to measure the planet mass and orbit; and to determine $\eta_\oplus$ for low $\eta_\oplus$.}

\bigbreak

The habitable zone
around A, F, G, and K-type main sequence stars is accessible to 
to the astrometric technique, excepting the blind spot corresponding to one-year orbital periods.
Immediate preparation for an
astrometric space-based mission will enable a launch in the intermediate
time frame. The sensitivity for such a mission should allow confident detection of an astrometric signature as small as 0.2 $\muas$ for 60--100 stars.
The astrometric technique is insensitive to the effect of zodiacal 
dust emission from any given system, and the technique will yield a 
census of exoplanets in an as-yet unexplored range of parameter space for any value of $\eta_\oplus$.

\bigbreak

{\bf \emph{Finding}} {The European ``Gaia"
space mission will be a useful demonstrator of the ability to do spaceborne astrometry to find giant planets, and will contribute to the census of Jovian-mass planets around Sun-like stars.}

\bigbreak

Current work being done
at the European Very Large Telescope, Keck and LBT may enable Neptune-mass planets
to be detected from the ground.  Furthermore, current astrometric
accuracies are such that the technique is very useful in determing
stellar distances, and hence understanding the characterisitics of
host stars (distances, proper motions, luminosity). However, success in these efforts, and their extension to Extremely Large Telescopes, will require sustained funding efforts that must not fall through the cracks.

\bigbreak

{\bf \emph{Finding}} {Current work on precision astrometry at the largest ground-based facilities in the world holds promise for eventual astrometric detection of Neptune-mass planets, and important ancillary work on host stars. This work is an important component of the overall strategy for detection and characterization of extrasolar planets.}

% - - - - - - - - - - - - - - - - - - - - - - - - - - - - - - - - - - - - - - - - - - - - - - - - - - - - - - - - - - - - - - - - - - -
\section{Direct Imaging Findings}

Direct imaging means the direct detection of the radiation from the
planet, picked out from under the glare of the parent star, and is the 
technique of choice for studying the atmospheres of Earth-sized planets 
around solar-type stars. The state of the atmosphere is
critical to determine the nature of the planet and its potential for
habitability. Although the near-term goal of transiting Earths
orbiting M-dwarfs is our fastest path to detecting potentially habitable
exoplanets, \textit{ the Copernican revolution truly will be completed
when Earth-sized planets orbiting of order 1 AU from Sun-like stars are detected and characterized.} M-dwarf Earths are not the final word on the search for habitable planets because of potential habitability problems. These include: tidal locking and the related lack of a magnetic field for slow rotators; atmospheric loss from stellar wind sputtering; flares from active M-dwarfs; and volatile deficiency caused by rapid planetary accretion and high-velocity impactors in a planet--forming environment close to the star. Under the rubric of direct 
imaging we include adaptive optics (for ground-based observations), 
coronagraphy/occulters, and interferometry (including nulling). 

\bigbreak

{\bf \emph{Finding}} {Direct imaging of terrestrial planets is essential to addressing the fundamental question of whether habitable Earth-sized planets exist around Sun-like stars.}

\bigbreak
 
A space-based facility is essential for directly imaging Earth-sized
planets orbiting sun-sized stars because of the limitations imposed by our own atmosphere
atmosphere in detecting a potentially habitable Earth-sized planet orbiting a star that
is 10 billion times brighter than the planet.  Direct imaging is the
most costly and complex of all the discovery techniques.

\bigbreak

{\bf \emph{Finding}} {The most promising way to mitigate the cost of
space-based direct imaging is: 1) to identify targets before the direct
imaging mission is flown and  2) to pursue continuous technology
development and planning for such a space mission}

\bigbreak
 
The limitation of direct imaging is that it does not determine the planet's mass, and provides an ambiguous constraint on the planet's radius. While using the planet's color and absolute brightness
has been suggested to constrain the planet radius and adopting density
assumptions to constrain the planet mass, the ExoPTF finds this is not
sufficient. The currently known exoplanets have a wide variety of
densities, making any density assumptions confusing to inferring the
planet masses. 

\bigbreak

{\bf \emph{Finding}} {The ExoPTF finds that a capability to determine 
mass, specifically
space-based astrometry described above, is necessary in conjunction
with the direct imaging platform.}

\bigbreak
 
Whether the space-based direct imaging experiment should be coronagraphy (with or without an external occulter) or interferometry needs to be decided early in the strategy but not immediately, and should be based primarily on (a) technological readiness) and (b) determination of $\eta_\oplus$ by COROT  or Kepler or both. The distribution of exo-zodiacal emission may be determinative as well, though both approaches are affected by large amounts of dusty emission.  While the strategy laid out here simplifies the direct detection task and hence makes more attractive an external occulter system it is premature to recommend one approach over the other at this juncture.  However, it is important for the affordability of the strategy that the system employed be sized to the required task based on the considerations given above. 

\bigbreak

{\bf \emph{Finding}} {It is premature to recommend coronography or interferometry as the follow-on to the astrometric mission at this juncture, however a decision should be made on one or the other approach by the second epoch of the strategy that is, the 6--10 year timeframe.}

\bigbreak

Direct imaging from the ground is limited by rapidly changing
atmospheric turbulence; even with advanced adaptive optics (AO) and 30 to
40-m diameter telescopes, Earth-mass planets will remain out of reach
for any significant sample of target stars.  Nonetheless, ground-based
adaptive optics will play an important role on the path to Earth
twins, and in the study of giant planets.

High-contrast AO systems planned for the US-led Gemini
Observatory and the European Very Large Telescope will be able to
detect young (less than 1 billion years age) Jovian planets (1 to 10
times the mass of Jupiter) through their infrared radiation.  Future
Extremely Large Telescopes (ELTs) 25-40m in diameter would be able to
detect and spectroscopically characterize mature giant planets (down
to the mass of Saturn or even Neptune) in the inner parts of nearby
solar systems, study the ongoing process of planet formation around
the youngest stars. 

\bigbreak

{\bf \emph{Finding}} {The ExoPTF supports the efforts of 
ground-based observatories to advance the field of exoplanet science
through the development of advanced adaptive optics techniques.}

\bigbreak

The James Webb Space Telescope (JWST) is planned to incorporate several
coronagraphic modes, but neither the telescope nor the instruments were
optimized for this capability. Its design tradeoffs yield modest coronagraphic performance when compared to next-generation ground-based coronagraphs. However, with its low thermal backgrounds, JWST will have access to
wavelength ranges almost completely inaccessible from the Earth, in
particular the 4-5 micron range in which models predict that even mature
(1 Gyr) extrasolar planets will have strong thermal emission. For a 1
Gyr M0V star, its estimated sensitivity is 1 Jupiter mass at $\sim$ 2
arcseconds and 2 Jupiter masses at 0.5 arcseconds. For a survey of 100
low-mass (below 0.5 solar mass) stars younger than a billion years, the median target
distance would be 20 pc. Hence JWST would be able to detect a
Jupiter-mass companion at 40 AU and 2 Jupiter masses at 10 AU around a
median star in this sample. The long-wavelength channel and the ability
to observe faint low-mass stars provides a useful complement to
ground-based ExAO systems. In addition, the experience of HST shows that
the PSF stability of a space-based coronagraph is extremely powerful for
imaging of diffuse circumstellar debris disks. 

\bigbreak

{\bf \emph{Finding}} {JWST, under development for launch in the next decade, will provide direct detection capabilities for giant planets complementary to those of envisioned ExAO systems.} 
%------------------------------------------------------------------------------------------

\section{Circumstellar disk findings}

Direct imaging of dust belts within the habitable zone region around 
other stars--called ``exo-zodis"-- is important because excessive dust 
emission could limit future space direct detection missions.   
Figure ~\ref{fig:ExoPTFzodi} shows a twin of our solar system, 
observed obliquely (60 degrees), with a zodiacal emission at two levels: 
that of our own system, and ten times larger (10 zodis, where one zodi equals emission levels from the zodiacal dust in our own 
solar system). With the order of 
magnitude increase in the amount of dust emission, the higher noise and expected spatial variation are a severe challenge for observing an Earth 
with direct detection approaches. 
Therefore, it is imperative that a statistically significant survey of exo-zodis around
solar-type stars be made, 
prior to significant programmatic commitment to a space-based direct detection system. 

\begin{figure}
\begin{center}
\includegraphics*[scale=0.6]{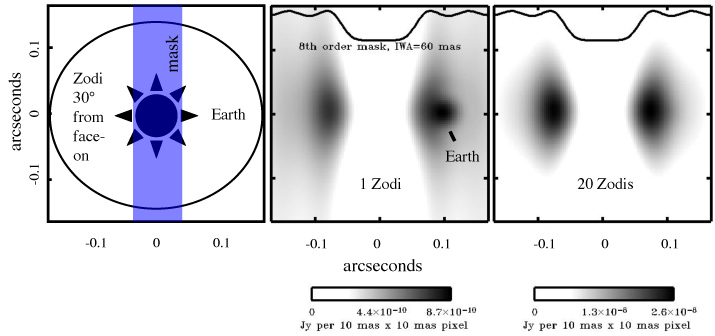}
\caption{Simulation of the detectability of an Earth at 1 AU around a solar twin, with zodiacal
emission (center) of 1 zodi and (right) 10 zodiis. The system is observed at an inclination angle of 60
degrees (left). Figure constructed by Mark Kuchner.  \label{fig:ExoPTFzodi}}
\end{center}
\end{figure} 

Our present knowledge suggests that at least 10-20\% of FGK stars have cold dust with fractional luminosities ( $L_{dust}/L_{star}$) more than $10{^-5}$ or roughly 100 times the solar system level, with higher percentages having lower, currently undetectable dust levels.  However, this dust is located at many tens of AU, rather than true ``zodi"  near the location of Earth-analog planets.  The latter is constrained to be present in at least 1-2\% of FGK stars at $L_{dust}/L_{star}$ levels more than $10^{-3}$ with, we can assume, higher percentages at lower, currently undetectable dust levels if the trends for the cooler Kuiper-Belt analog dust are followed for the warmer Asteroid-Belt analog dust.  Figure ~\ref{fig:limits2} illustrates present/projected dust sensitivity in the parameter space of $L_{dust}/L_{star}$ vs $T_{dust}$. 

\begin{figure}[h]
\begin{center}
\includegraphics*[scale=0.6]{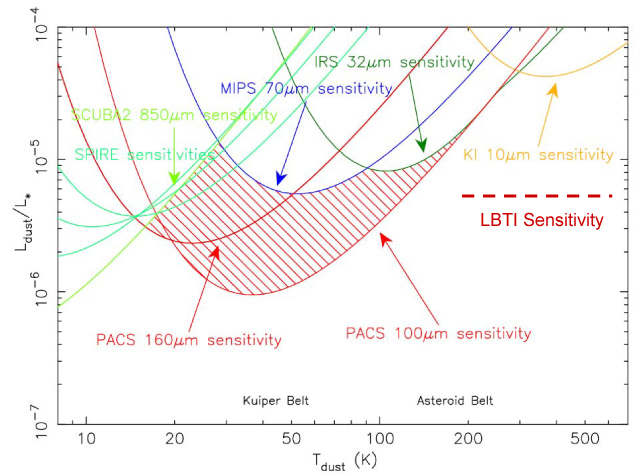}
\caption{Sensitivity limit to dust around other stars achieved with current state-of-the-art techniques.   MIPS and IRS are instruments on Spitzer while PACS and SPIRE are instruments on Herschel;  SCUBA is a ground-based sub-millimeter instrument.  These are all sensitive to dust levels only several hundred to many thousand times that of the cool Kuiper Belt.  LBTI and KI are ground-based interferometers and are sensitive to only tens to hundreds of times the warm Asteroid Belt or ``zodi" dust.   \label{fig:limits2}}
\end{center}
\end{figure} 

The nulling interferometer mode of the twin Keck telescopes allows
sensitivity to dust disks at a level of tens of zodis. 
Current plans for the US Large Binocular Telescope Interferometer LBTI are
to enable exo-zodi
detection down to a sensitivity of 3-10, though with an uncertain start date. Experience with the Keck telescopes shows
that the sensitivity of such instruments can be reduced when
they move from laboratory to observatory, so the ultimate LBTI sensitivity must await telescope testing.
Funding to complete the LBTI and continue to operate the Keck
interferometer is uncertain at the time of this
writing. Further in the future, ELT's could measure and map the
structure 
of exo-zodis at least down to 10 zodi sensitivity. However, suitable
instruments are not currently prioritized as first-light capabilities
for
either US ELT project. 

\bigbreak

{\bf \emph{Finding}} {Early knowledge of exo-zodis down to or below 10 zodis sensitivity level, around nearby, bright sun-like stars, is crucial
in sizing spaceborne direct imaging systems and determining the
likelihood that any given system with an astrometrically detected
Earth-mass planet will have dust emission sufficiently small to allow
examination of the planet by direct imaging.} 

The study of circumstellar disks is informative of the processes of planet formation 
and of the presence of young planets in disk systems.  Such studies will be pursued through a variety 
of existing and planned, space- and ground-based facilities. 

\bigbreak

{\bf \emph{Finding}}  {The execution of a robust program of 
disk science is an essential complement to the goals here of direct detection of 
Earth analogs and the mapping of planetary system architectures in mature systems.}

% - - - - - - - - - - - - - - - - - - - - - - - - - - - - - - - - - - - - - - - - - - - - - - - - - - - - - - - - - - - - - - - - - - -
\section{Other Exoplanet Science findings}

Whether or not a planet is habitable depends on the
atmospheric conditions which generally require interpretation via
model atmospheres, due to the non-linear nature of radiative
transfer. Likewise, using observed planetary system architectures
to understand planet formation requires detailed theoretical models.

\bigbreak

{\bf \emph{Finding}} {Theory is vital to interpretation of observational
findings and ultimately answering the compelling questions. Theory
programs provide a high return for a modest investment.}

\bigbreak

Training the next generation of researchers is
essential to carrying out the 15 year strategy outlined here, and its logical follow-ons. 
While doctoral programs are a core part of the research infrastructure in the United States and Europe, attracting  the most talented future workers in the field requires postdoctoral programs that are prestigious and tied directly to key elements of the program. These exist today and need to be maintained or even expanded. 

\bigbreak

{\bf \emph{Finding}} {The ExoPTF finds
that the prestigious fellowships such as the
Michelson and Spitzer fellows are a critical support for the most
promising young researchers in key areas of astronomy today, including exoplanet research.}

\bigbreak

% - - - - - - - - - - - - - - - - - - - - - - - - - - - - - - - - - - - - - - - - - - - - - - - - - - - - - - - - - - - - - - - - - - -
\section{Findings on $\eta_{\oplus}$ and Planning}

The ExoPTF extensively discussed the significance of
$\eta_{\oplus}$, the fraction of sun-like (i.e., F, G., K) stars that have at least one planet in the habitable zone.We find that the frequency of terrestrial planets in
habitable zones of stars is important in understanding our place in
the cosmos. $\eta_{\oplus}$ is even more relevant as the gateway to
answering our most compelling question--about potentially habitable
terrestrial planets around nearby bright stars. The frequency of
Earths is critical to knowing how and when to proceed to find and
characterize habitable planets orbiting the nearest stars. We find
Theory cannot adequately constrain $\eta_{\oplus}$; observations
are ultimately needed.  The prevailing view that $\eta_{\oplus}$ is not low is
bolstered by radial-velocity and microlensing discoveries--evidence
that points toward a high frequency of planets less massive than 
Neptune, that might be terrestrial-type planets.

Space-based astrometry should be conducted both for searching for Earths around nearby bright stars and
for studying planetary system architectures. If $\eta_{\oplus}$ is not low the
astrometric mission will very likely find Earths around the bright stars for later
direct imaging followup. If $\eta_{\oplus}$ is low then an astrometric
mission with the sensitivity to detect Earths at 1 AU will find planets of larger mass (gas giants, Neptunes) in orbits inaccessible to other techniques, and hence contribute unique information on
planetary system architectures of solar neighborhood stars. 
\bigbreak

{\bf \emph{Finding}} {A space-based astrometric mission should be
developed and flown regardless of prior results constraining $\eta_{\oplus}$, conditioned on a rigorous determination that it is feasible at 1 Earth-mass sensitivity for 60--100 stars (confident detection of a 0.2 $\muas$ signature over hundreds of observations of a given star during the mission lifetime). }

\bigbreak 

The ExoPTF recommends that a space-based direct detection mission
be finalized and flown when either $\eta_{\oplus}$ or targets are
known. Because the observational evidence for a high $\eta_{\oplus}$
is accumulating, and due to the long lead time for developing a direct
imaging mission, the ExoPTF finds that technology development
should be supported even before $\eta_{\oplus}$ is known. 

If astrometry at this level proves infeasible within this time period, consideration should be given to proceeding instead with a direct detection mission. Each type of mission -- internal occulter coronagraph, external occulter coronagraph, and nulling interferometer -- can also search for planets before characterizing them. However, the complexity, duration, cost and thus risk of the direct detection mission becomes higher. In addition, some means of determining the masses of the directly detected planets would have to be eventually implemented.

\bigbreak

{\bf \emph{Finding}} {Technology development for a space-based direct imaging mission should proceed early and aggressively. The scope of the direct detection mission depends on the findings of the 
astrometry mission.  If, for some reason, an astrometry mission is not
executed, then an assessment of whether a direct detection mission should
proceed will depend upon measured values of $\eta_{\oplus}$ as well as the
exo-zodi environment around target stars, among other factors}

% ============================================================================
% Begin New PART - STRATEGY
% ============================================================================

\pagebreak

\chapter{Strategy Recommended by the ExoPTF}

Based upon the findings assembled above, derived
from 85 community white papers, briefings from experts in
the field, and deliberations during five meetings, the Exoplanet Task
Force has developed a strategy that will achieve three important goals
by the end of the fifteen year planning horizon:

\begin{enumerate}

\item To a high degree of statistical significance, $\eta_{\oplus}$ will be determined around a broad range of main sequence stellar types.

\item The architectures of planetary systems, both in the nearby solar neighborhood of order $10^6$ cubic parsecs, and in the larger galactic bulge, will be constrained down to sub-Earth masses for semi-major axes out to (for G-dwarfs) several astronomical units. 

\item Provided $\eta_{\oplus}$ exceeds 0.1, at least one Earth-sized planet will have its mass and basic atmospheric (or oceanic) composition characterized. 

\end{enumerate}

The strategy pursues two
parallel tracks as described below: one focused on low mass stars
(M-dwarfs) and the second on Sun-like (F, G, and K  dwarf) stars (Fig. ~\ref{fig:two-pronged}).  It is crafted such that
significant discoveries for exoplanets are made whether $\eta_{\oplus}$ is
large (at least 0.1) or small ($\leq{0.1}$) (Fig. ~\ref{fig:two-pronged1}).   Finally, if $\eta_{\oplus}$ is large for both 
M-dwarfs and Sun-like stars, then the first
characterization of an Earth-sized planet around an M-dwarf occurs in
the middle of the planning horizon, and the first characterization of an
Earth around a Sun-like star occurs at the end of, but still within, the
planning horizon.  Thus, the strategy has a high likelihood to enable completion of the Copernican Revolution, in the sense of detection and study of an Earth-sized planet, within 15 years.

\begin{sidewaysfigure}
\centering
\subfigure[] % caption for subfigure a
{\includegraphics[width=10 cm]{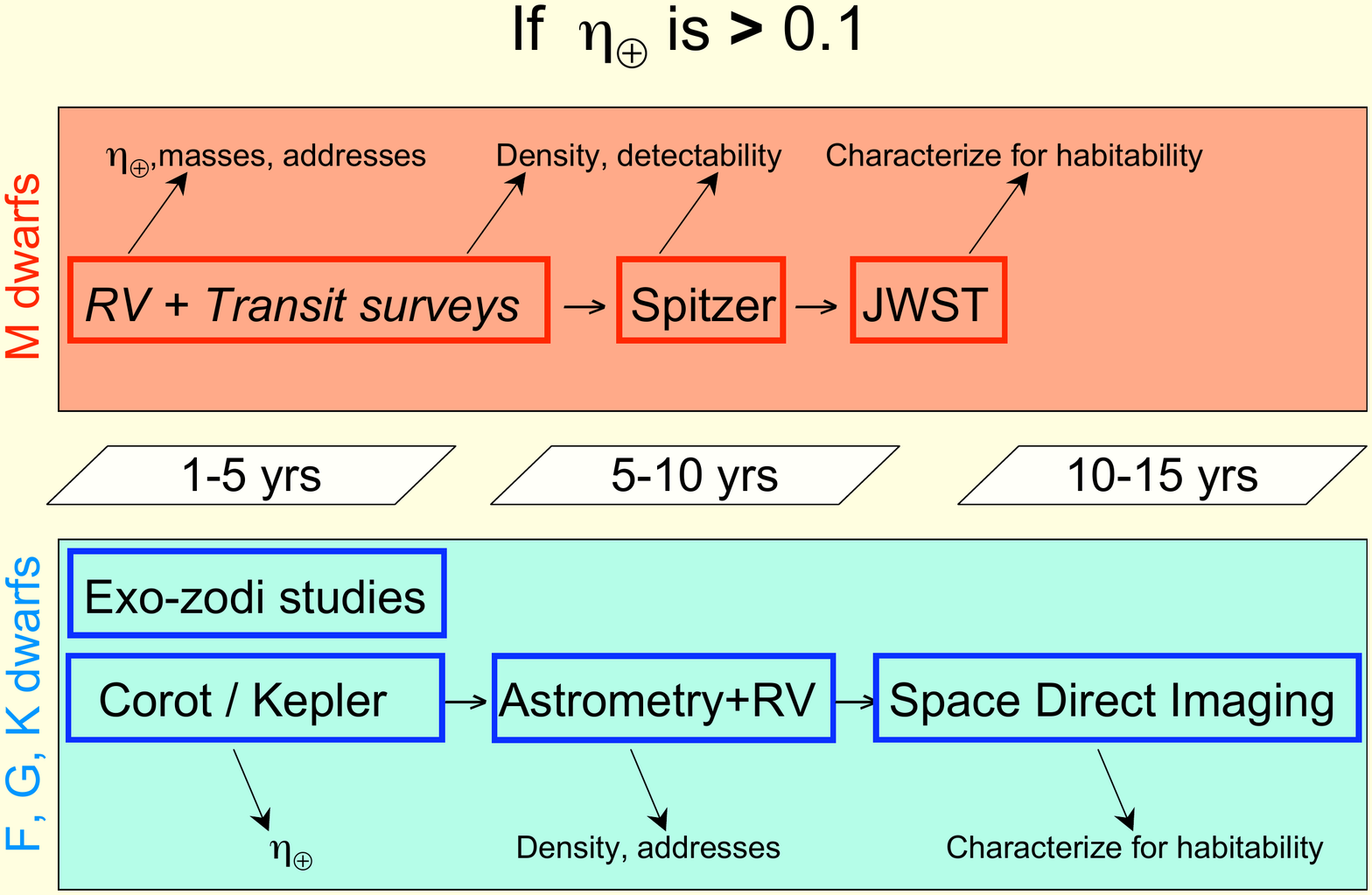}}
\subfigure[] % caption for subfigure a
{ \includegraphics[width=10 cm]{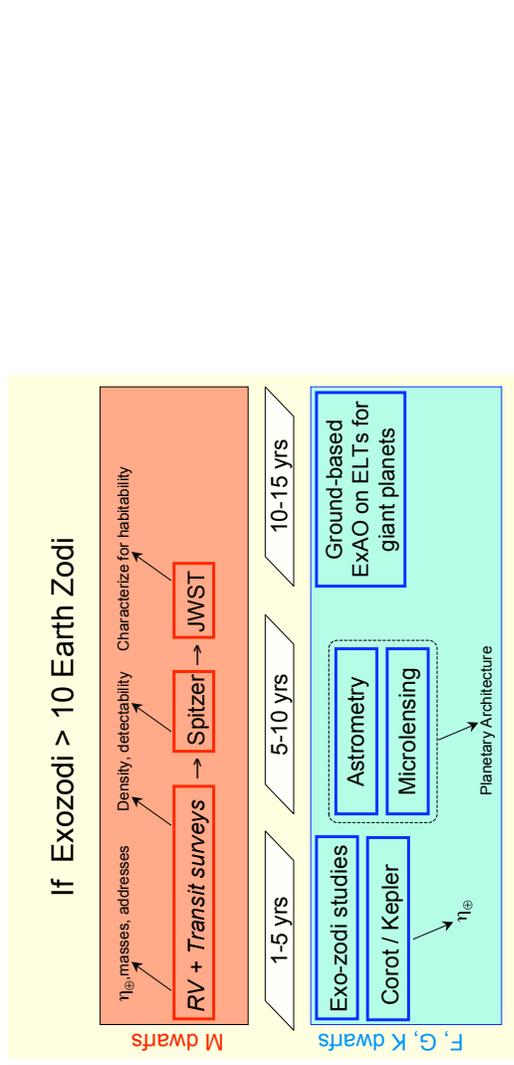}}
\subfigure[] % caption for subfigure a
{ \includegraphics[width=10 cm]{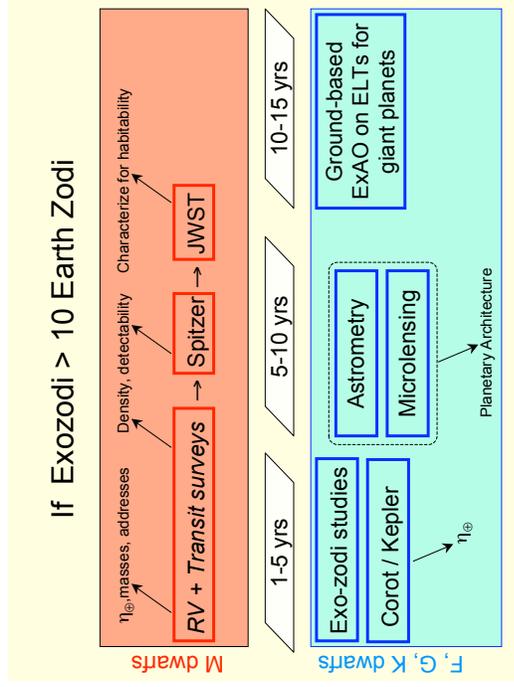}}
\caption{Details of the strategy under the assumptions  (a) $\eta_{\oplus}$ is large; (b) $\eta_{\oplus}$ is small, and (c) exo-zodiacal emission is significantly larger than solar system zodiacal emission. Not all elements of the strategy are shown in each case; the narrative description should be carefully read to understand the full scope of the proposed effort. }
\label{fig:two-pronged1} % caption for the whole figure
\end{sidewaysfigure}

The strategy is outlined in narrative form below under the assumption that all the recommendations detailed in the following section are followed. 

% - - - - - - - - - - - - - - - - - - - - - - - - - - - - - - - - - - - - - - - - - - - - - - - - - - - - - - - - - - - - - - - - - - -
\section{Strategy in the 1--5 year time frame}

In this time horizon, radial velocity (RV) surveys from the ground with projected technological improvements should allow for the detection of objects a few times the mass of the Earth around stars in the spectral classes F,G, and K.
Earth-mass objects may be found around M-dwarfs, either through extension of optical RV to early M-dwarfs or development of near-infrared doppler spectroscopy for late M-dwarfs.  

At the same time,  transit surveys conducted with automated systems in space or on the ground that allow for broad sky coverage and quick-response to a potential event, should target nearby M-dwarfs and in particular those with with RV signatures. Their goal is detection of transits for planets one to several times the size of the Earth that could then be candidates for characterization with Warm Spitzer or JWST.  

In order to proceed with a mission in the second epoch, NASA must begin its in-depth review of astrometric instrument performance requirements and capability to assess the feasibility of a  $\mu$as class astrometric mission with the first year of the strategy. If needed, additional technology demonstrations should be defined and completed in the first epoch. If possible, an abbreviated Formulation Phase should be completed, so that the Implementation Phase can begin leading to operations in the 6--10 year time frame.

It is expected that extreme AO systems with coronagraphs will begin to come on line during this time on 8-10 meter class telescopes, and will permit
direct detection of warm Jupiter-sized or larger objects for planetary
architecture studies.  Since these will be combined with coronagraph and wavefront
control technology research, they provide a ground-based precursor to a potential space-borne direct detection
system.
 
Likewise, essential measurements of the exo-zodiacal dust levels around Sun-like stars, preferably down to 10 zodis sensitivity, must begin in the 1--5 year time frame to determine the likelihood that systems containing planets will be clean enough for coronagraphic direct detection. 

Ongoing development of JWST and commissioning of cryogen-depleted (``warm")
Spitzer as a transit-characterization system for exo-planets should be
pursued by NASA during this time frame as well. If suitable targets of
interest have been identified by the transit survey for Spitzer--"super-Earths" several
times our own planet's diameter orbiting M-dwarfs--then transit
observations of these can be attempted as soon as possible as a preliminary step toward
JWST transit studes of Earth-sized planets around M-dwarfs.

The launch and prime mission results from Kepler's transit survey during this epoch will provide the first determination of $\eta_{\oplus}$ for Sun-like stars in the solar neighborhood, with high statistical significance. If $\eta_{\oplus}$ is high, technology development of a relatively modest space-based direct detection system to detect and make spectra of Earth-sized planets can proceed in earnest for deployment in the third time period. This will be either coronagraphic or interferometric depending on the degree of technological readiness and suitability given the COROT/Kepler results.  If $\eta_{\oplus}$ is low enough that the astrometric mission of the next time period is unlikely to detect more than a few (or zero) Earth-mass targets for direct detection, consideration should be given to doing a more ambitious search and detection mission, perhaps in concert with the Europeans, but beyond the planning horizon of this strategy. If for some reason Kepler and/or COROT have not yielded the information required to size the direct detection mission, technology work for that mission continues but its scoping is deferred to the results of the spaceborne astrometric mission in the second time epoch. 

A ground-based microlensing network should be completed prior to the end of the first five years, through the addition of a 2-meter telescope, allowing for coverage sufficient to allow probing of $\eta_{\oplus}$ on a galactic scale, as a precursor to a much more statistically complete survey obtainable with a space-based mission. 

A survey of the exo-zodiacal emission of Sun-like stars must also begin in this time frame to assist in determining the feasibility of direct detection approaches for Earth-like planets. If typical exo-zodi emissions are found to be less than 10 zodis, this would be no impediment to the deployment of such systems in the 11--15 year time frame. If typical emissions are larger, consideration should again be given to a larger scale direct detection mission, perhaps with international cooperation, if such a mission can overcome the exo-zodiacal problem, sometime beyond the timeframe of the present strategy. 

By the end of the first time period, a determination should be available from multiple approaches as to whether $\eta_{\oplus}$ is sufficiently high (in excess of 0.1) that a reasonable expectation would exist that in the remainder of the planning horizon, Earth-sized planets could be detected around nearby Sun-like stars and then characterized. Depending on the results of the M-dwarf survey, study of super-Earths around M-dwarfs could be completed by warm Spitzer, and targets for JWST identified.

% - - - - - - - - - - - - - - - - - - - - - - - - - - - - - - - - - - - - - - - - - - - - - - - - - - - - - - - - - - - - - - - - - - -
\section{Strategy in the 6--10 year time frame}

In this time period humankind could, under the recommended strategy, obtain its first look at the properties of an Earth-sized planet around a neighboring star--in this case an M-dwarf--provided that surveys in the first five years of the strategy have obtained one or more suitable targets. JWST should be used to detect spectral features during transits of one or more Earth-sized planets around M-dwarfs if these have been identified by transit surveys. The strategy moves beyond M-dwarf Earths, on which life may be limited due to potential habitability issues (refer to Appendix or section on habitability), to find true Earth analogs.

The strategic centerpiece for F,G,and K dwarf stars in this time period is the launch of a spaceborne astrometric system capable of confident detection of a 0.2 $\muas$ astrometric planet signature for 60--100 Sun-like stars. We envision this to be a mission that heavily emphasizes sub-microarcsecond
planet-finding science, thereby limiting its scope and hence
cost when compared with concepts such as the Space Interferometry Mission (SIM) that has broader goals across astrophysics.  The astrometric system must detect planets down to or below an Earth mass, which can then be observed directly with a spaceborne direct detection mission in the third time period. It will provide the target list for the direct detection mission, already being developed based on COROT and Kepler results, or will provide the necessary scoping and technology decision point (coronagraphy/occulter versus interferometer) should COROT and Kepler not have provided the necessary information. 

If the astrometric mission finds any earth-like planets accessible to the direct detection mission being developed at the time, the main thrust of the strategy to move that mission to launch in the third epoch can be pursued. If $\eta_{\oplus}$ exceeds 0.1, one would expect to find multiple systems with potentially-habitable Earths so that a direct detection mission would be kept busy studying at least several such planets. By providing mass and orbital information on these planets, the astrometric system also will provide essential boundary conditions for understanding, in concert with the direct detection spectral observations, their potential habitability. The astrometric measurements should be supported by ground-based RV observations of the same targets, which will help resolve potential ambiguities in the astrometric signature of multiple planet systems. 

If $\eta_{\oplus}$ is low, say much less than 0.1, and no Earth-mass planets are found by the astrometric mission, the latter should focus instead on obtaining planetary system architectures for nearby stars, since results obtained to date from radial velocity and transit studies virtually ensure that planets from Neptune-mass upwards will be available for discovery and study. Development of direct detection from space might then shift to larger systems that could look deeper but be deployed later than a system scaled for the more optimistic scenario sketched above; alternatively, more resources might be invested in ground-based or simple space-based systems capable of direct detection and study of giant planets. Likewise, for the M-dwarfs, a low value of $\eta_{\oplus}$ might imply that no transiting Earth-sized planets have been found within the reach of JWST, or super-Earths for study with warm Spitzer. Since both facilities will have a number of other science areas to address, the contingency plan is simply that more time be available to these science areas--inlcuding the study of giant planets-- in the absence of Earth-sized targets around M-dwarfs. 

If many or most Sun-like stars possess significant (greater than 10 zodi) zodiacal emission, the strategy proceeds as for the low $\eta_{\oplus}$ case, with direct detection technology development refocussed on overcoming the zodiacal emission problem through larger systems deployed beyond the strategy's planning horizon.

Parallel to the activities described above, a space-based microlensing mission selected in the first epoch and launched in this second epoch could determine $\eta_{\oplus}$ in the galactic bulge with much higher statistical confidence than could be done from the ground. This would provide information as well on planetary system architectures--that is, mass and separation of planets
from their host stars as a function of stellar type and location in
the galaxy--on galactic scales. The scope of the mission is such that it should be selected under a line such as Discovery and not the larger, proposed astrophysics probes. By this time ground-based extreme-AO and coronagraphic systems should move into an advanced state of development, ready to be deployed on 20-30 meter class telescopes nearing completion. 

% - - - - - - - - - - - - - - - - - - - - - - - - - - - - - - - - - - - - - - - - - - - - - - - - - - - - - - - - - - - - - - - - - - -
\section{Strategy in the 11--15 year time frame}

If $\eta_{\oplus}$  is much less than 0.1, then this time period would see no major flight programs, with ground-based extreme AO and coronagraphic systems on 20-30 meter-class telescopes imaging and studying Jovian- and Neptunian-class planets, while RV and transit surveys continue and microlensing maps the large-scale statistics of planets and planetary systems. Technology work and collaborations with international partners such as ESA on large-scale direct detection and characterization missions, possibly interferometric, continue toward a launch beyond the report's planning horizon. 

If $\eta_{\oplus}$  exceeds 0.1, then in all likelihood the astrometric mission would have found Earth-sized planets around Sun-like stars suitable for advanced characterization.  A space-based coronagraphic or interferometric system should be launched to take spectra of these targets to see if they are similar to Earth, that is, are within the size/mass range defined earlier and possess an atmosphere at least with water and carbon dioxide. Because the astrometric mission would have  already determined the locations (``addresses") and orbits of the candidate planets, the spaceborne direct detection system is relieved of the burden of searching; its mission is to point at those stars and study the planets. This greatly simplifies a direct detection mission in several ways: 

\begin{itemize}
\item{Coronagraph options, if selected as the technique for direct detection,  become much more attractive and less expensive because the observatory IWA need not be overspecified (to capture planets in a wider range of orbital phase), and external coronagraphs are not quite so pressed for agility and sky coverage to revisit stars often.}
\item{It is unnecessary to make multiple visits to a single target to check whether it is a background object or a lump in an exo-zodiacal cloud}
\item{The target list can be ordered and prioritized ahead of time to maximize efficiency in moving from one target to another and in observing those targets whose orbital phases make them most observable early in the mission (thereby minimizing mission risk) .}
\end{itemize}

The determination of true mass via astrometry also provides baseline information required in understanding the nature of the planet, since geological recycling of volatiles, atmospheric evolution and escape, and atmospheric energy balance are all affected by the planetary mass. 

% - - - - - - - - - - - - - - - - - - - - - - - - - - - - - - - - - - - - - - - - - - - - - - - - - - - - - - - - - - - - - - - - - - -
\section{Accomplishments of the recommended strategy}

The recommended strategy ( Fig.~~\ref{fig:recommended_program}) provides discoveries throughout the planning horizon. The spectrum of an Earth-sized planet may be obtained as early as the middle epoch of the planning horizon, and again for solar-type stars toward the end of the third epoch; but even in the event that $\eta_{\oplus}$   is small, the strategy provides for discoveries that dramatically extend our understanding of the architecture of planetary systems and Earth's place therein. 

\begin{figure}[h]
\centering
\includegraphics[scale=0.4]{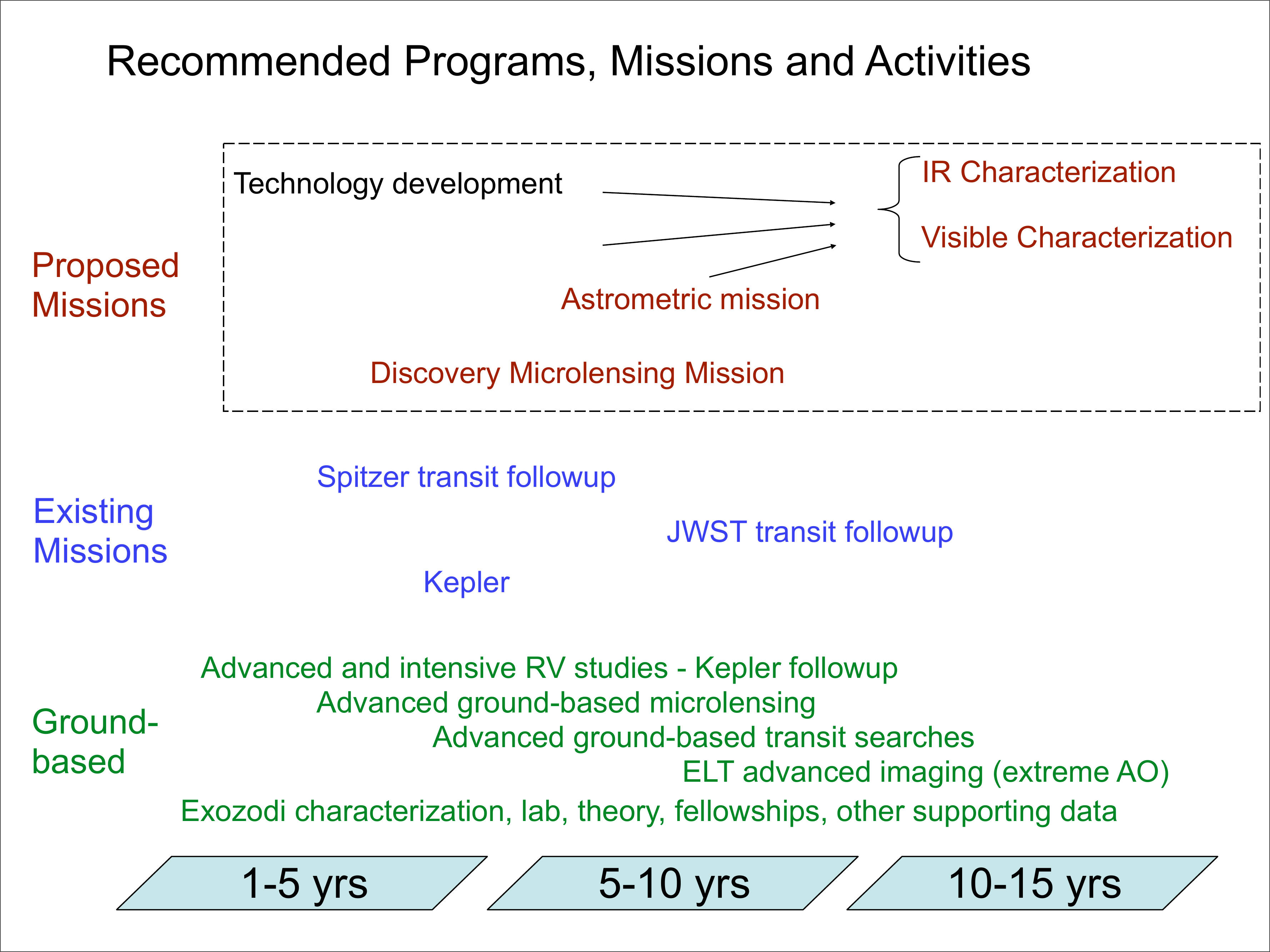}
\vspace{1mm}
\caption{Graphical depiction of the recommended strategy parsed in time periods and according to ground-based, space-based or existing assets.}
\label{fig:recommended_program}
\end{figure}

Key to the continuity of discoveries in a complex milieu of ground- and space-based programs, existing and future space assets, technology development and key programmatic decisions are two things: the parallel tracks of M-dwarf and F,G,K-dwarf efforts, and the placement of spaceborne astrometry prior to the direct detection mission in the latter. Two tracks, pursued in parallel in which one relies more on existing assets or those under development, while the second requires significant new capabilities, provides robustness to the strategy. The failure of one track to yield significant results, for whatever reason, is largely decoupled from the other track. And the M-dwarf track provides a high probability of mid-term results on Earth-sized planets that would fuel continued enthusiasm for the other, more demanding, track. 

The placement of astrometry prior to direct detection confers the key advantage that the strategy is less sensitive to the outcomes of the Kepler mission and an exo-zodi survey with this ordering. A census of planetary masses and orbits could be done by spaceborne astrometry whether Earth-like planets are common or rare, given that we know that the overall occurrence of planets is large enough that targets will be available. And the distribution of exo-zodiacal emission among sun-like stars, key to designing the direct detection mission, is not an issue for the astrometry. Because neither one of these parameters will be known well prior to the middle (at least) of the 1--5 year first time period of the strategy, reversing the order of the two types of missions would necessarily entail a delay in the entire program, since it is assumed here NASA cannot afford to fly both in the middle time period. 

Astrometry will provide a sensitive  means of developing a target list of planets for the direct detection mission, along with an unambiguous mass determination and orbital parameters for such planets. While one could argue that, all else being equal, direct detection missions could do likewise for astrometry, we believe the cost for a delayed astrometry mission would rise dramatically, as would the cost for a sufficiently capable early direct detection mission.

The strategy relies heavily, in particular in the early years but throughout the planning horizon, on ground-based observational programs. Table~1 summarizes the number of nights on and approximate status of key ground-based assets that would be called into play. The table lists very approximate survey times for representative ground-based programs simulated in Chapter 9 (``Depth of Search"). For each search,
we list the number of targets surveyed, number of nights requried for different telescope sizes,
and the status of the instrument needed - existing, under construction, or new. For reference, 1 night of 8-10m telescope time at TSIP rates is approximately $\$100,000$, one night of 3-5m is approximately $\$25,000$. 

The first program is a 
survey of moderate-mass stars searching for long-period Jovian planets (5 milliarcseconds/year, 1 m/s, 15 years), carried
out with either the Keck HIRES instrument or a new dedicated and optimized
Doppler spectrograph on a 4-m telescope. The second is a high-candence survey optimized
to detect low-mass planets orbiting early M-dwarfs (100 observations per target over 5 years) using either the 2-m Automated Planet Finder telescope or 
a new dedicated 4-m instrument. The third is a similar survey for lower-mass stars using an
IR spectrograph (based on Jones white paper: 60 minutes for 1m/s on J=12). The fourth is followup of transit events  
detected by the Kepler spacecraft (4 candidate 2 \mearth habitable zone planets, 40 2-10 \mearth). The fifth is similar followup of 1-10 Earth mass 
events detected by an all-sky transit search emphasizing the nearest M-dwarfs (All-sky transit doppler followup follows candidate events down to 1 \mearth, assuming
$\eta_{\oplus}$ = 0.1 and false-positive rate is 8\%, and typical primary star is 0.3 \msun). The sixth project is a Large Binocular Telescope interferometer survey of extrasolar zodiacal
dust surrounding nearby stars spanning a range of spectral types, sensitive to ~10 times
solar zodiacal levels. The final two projects are extreme adaptive optics coronagraphy surveys studying giant planets,
either young self-luminous planets (0-1000 million years age, 1-10 Jupiter masses) using an
8-m telescope and the Gemini Planet Imager instrument, or a search for mature planets above
0.03 Jovian masses using a similar instrument on a 30-m telesope. 

While the requirements are ambitious, the Task Force believes that the programs are doable and are consistent with planned enhancements of US ground-based capabilities. Where international partnerships would assist in such programs, the opportunity to form such collaborations should be encouraged.

\begin{table}[t]
TABLE~1.  Cost in nights of major ground-based programs associated with the strategy\\*[1.5mm]
\begin{tabular}{|clc|c|c|c|} \hline
%\rule{0mm}{6mm} & \multicolumn{5}{|c|}{} \\*[2mm] 
\rule{0mm}{6mm} Projects & Targets & Tel. size & Nights & Instr. status \\*[2mm] \hline
\rule{0mm}{6mm} Visible doppler & 2000 & 10--m & 50/year & Existing \\*[2mm] 
\rule{0mm}{6mm} Jupiter search & 2000 & 4--m & 25--50/year & new HARPS-type spectr. \\*[2mm] \hline
\rule{0mm}{6mm} Vis. doppler & 100 & 2-m & 800 total & in construc.  \\*[2mm] 
\rule{0mm}{6mm} Low-mass star + pl. search & 200 & 4-m & 200 total & new HARPS-type spectr. \\*[2mm] \hline
\rule{0mm}{6mm} IR doppler & 240 & 8-m & 100 total & New prec. IR-spectrogr.\\*[2mm] \hline
\rule{0mm}{6mm} Kepler doppl. follow-up & 100 & 4-m & 300 total & HARPS-type spectr. in constr. \\*[2mm] \hline
\rule{0mm}{6mm} All-sky transit doppl follw-up & 150 & 4-m & 110 total & new HARPS-type\\*[2mm] 
\rule{0mm}{6mm} All-sky transit doppl follw-up & 150 & 8-m & 15 total & new-precision IR spectr. \\*[2mm] \hline
\rule{0mm}{6mm} LBTI exo-zodi srch & 68 & 2 x 8-m & 60 total & LBTI nuller in constr.   \\*[2mm] \hline
\rule{0mm}{6mm} Gemini plan. imag. & 2000 & 8-m & 200 total & GPI in constr. \\*[2mm] \hline
\rule{0mm}{6mm} ExAO mature giant & 1000 & 30-m & 200 total & ExAO system for 30-m\\*[2mm] \hline
\end{tabular}
\end{table}

Several circumstances could stretch the timeline relative to what is recommended here. One is a reduced effort in the early years, in terms of delays in mapping exo-zodis or technology development necessary for the astrometric survey or the direct detection mission. Unforeseen technological roadblocks in the major elements of the strategy, for example in the spaceborne astrometry or direct detection, might also require extension of the schedule or even rethinking of the strategy itself. For example, should the astrometric precision required to detect an Earth-mass planet turn out not to be feasible, but Kepler finds a large value of $\eta_{\oplus}$, one option would be to fly a direct detection mission before or instead of an astrometric survey mission, even with the additional burden of searching for candidates. Such an approach carries higher risk than the strategy outlined here. An alternative that arises should the achievable astrometric precision be within a factor of a few worse than specified here is that one might accept a reduced number of target stars for which detection of Earth-mass planets would be feasible with the spaceborne astrometric mission, recognizing that interesting discoveries at higher planetary masses would still be in the offing for more distant target stars. Partial or complete technical failures in planned missions such as Kepler or JWST would also affect portions of the strategy, though overall it would remain robust. In general, the two-pronged and stepped approach considered here provides robustness and flexibility in responding to surprises along the way. 

The implementation of this strategy--resulting potentially in the detection and preliminary characterization of an Earth-like planet in the habitable zone of another star--will be of profound scientific and philosophical significance.  We assert here that the goals are attainable within decades, but its accomplishment will require a sustained commitment of fiscal resources. We expect, given the universal nature of the questions being addressed, that this strategy will be pursued as a collaboration among international partners. But the United States' pivotal role in large space science endeavors over the history of space exploration suggests that the nation must lead the effort if it is to be completed in a timely fashion. 

% ============================================================================
% Begin New PART - RECOMMENDATIONS
% ============================================================================

\pagebreak

\chapter{Recommendations}

Recommendations of the Task Force to implement the strategy are below.
The recommendations are divided into different time epochs, the 1--5
year, 6--10 year, and 11-15 year time frames.  While we do not
strictly prioritize the recommendations, we do order them---within
each time epoch---according to the priority order of the compelling
science questions introduced above.  Where appropriate, we further
organize the recommendations into the F, G, K (i.e., sun-like)
strategy and the M star strategy. 

\subsection{A. Recommendations for 1--5 Years}                                                                                                                                                                                                                                

\bigskip

{\bf  A. I. a.  What are the physical characteristics of planets in the
habitable zones around bright, nearby {\it F, G, K} stars?}

\bigskip

{\bf \emph{Recommendation A. I. a. 1}} {\it Sufficient
investment in ground-based telescope time for radial velocity
measurements to enable the discovery of low-mass exoplanets down to
Earth mass planets orbiting bright stars.  The required precision for the detection of Earth analogs is substantially
better than 1 m/s.  In the first time period, we recommend feasibility
studies for extreme Doppler precision (down to several cm/s) for bright
star targets.  It is also critical to continue surveys for planets of all detectable masses
with a target list well in excess of 1000 stars, with 3000 observations
per year total.}

\bigskip 

The investment could be NASA time on the Keck telescope, including
programs such as the NASA/Keck $\eta_\oplus$ program, as well as the
dedication of underutilized and existing 3-to-4 m class telescopes
with high-precision and high-throughput spectrographs.  A focus on
bright stars including the brightest M-dwarfs is the fast track to
finding a potentially habitable exoplanet.

\bigskip

In order to implement two key spaceborne capabilities---astrometry and
direct detection---within the 15 year timeframe of the strategy, technological
studies and mission development are required early on. NASA has
already invested substantially in technologies for both, and
particularly for spaceborne astrometry, so that there is a strong
foundation already for completing the additional technology
developments needed to conduct astrometric and direct detection
missions in space.

\bigskip

{\bf \emph{Recommendation \hspace{0.05in} A. I. a. 2}}
{\it Preparations should begin for a space-borne astrometric mission
capable of surveying between 60 and 100 nearby main sequence stars
with the goal of finding planets down to the mass of the Earth orbiting
their parent star within the habitable zone - i.e., approximately
0.8 to 1.6 AU, scaled appropriately for stellar luminosity.  Achieving
this goal will require the capability to measure convincingly wobble
semi-amplitudes down to 0.2 microarcseconds (``$\muas$") integrated over the mission lifetime.
Space-borne astrometry is currently the only technique that can distinguish
masses in the range of 1-10 earth mass. To this end, a rigorous 
technical feasibility
assessment should be undertaken immediately, and any additional
required technology development beyond what has been accomplished to
date should be completed promptly, leading to an implementation phase
start in the intermediate 6-10 year time period. The feasibility assessment should include a critical analysis of stellar systematic effects, such as starspots. }

\bigskip

{\bf \emph{Recommendation \hspace{0.05in} A. I. a. 3}} 
{\it Technological development of space-borne
direct detection capabilities to ultimately find and characterize Earth-sized
planets should be undertaken at the start of the strategy. This
includes visible wavelength internal and external coronagraphs and IR
nulling interferometers.  A key enabling technology for internal
coronagraphs is advanced wavefront sensing and control; support must
be sufficient to assess the viability of internal coronagraphs
operating at an inner working angle IWA $< 3.5\lambda$/D. (The IWA is
the minimum angular separation from the star at which the observatory
can detect a planet.) Additional technologies that need attention
include, but are not limited to, next-generation deformable mirrors,
low-noise detectors, coronagraphic masks, and ultraprecise optical
surfaces. A key enabling technology for external occulters is
validated diffraction modeling. Support must be sufficient to complete
demonstration scalable to flight dimensions. Additional technologies
that need attention include, but are not limited to, alignment
sensors, deployment methods, high-specific-impulse thrusters, and
studies of plume effects. Also needed are mission models assessing the
science harvest as it depends on system size. }

\bigskip

Because both the astrometry and direct detection technologies are so
crucial to the strategy, ongoing technological development should be
supplemented with in-depth reviews by experts in the various fields.

{\bf \emph{Recommendation \hspace{0.05in} A. I. a. 4 }}{\it NASA should
establish a blue-ribbon panel, consisting largely of physicists or
optical scientists with expertise in wave optics, to evaluate various
coronagraph and wavefront control concepts and ensure that no
fundamental physical effect has been overlooked in planning for an
optical wavelength, direct detection mission. An equivalent panel
should be established for direct detection by interferometry. }

\bigskip

Sizing of direct detection systems, indeed their feasibility, for
studying planets around sun-like stars depends on how typical is our
solar system's dust emission--that is, what is the distribution of  zodiacal emission around other stars.

{\bf \emph{Recommendation \hspace{0.05in} A. I. a. 5 }}
{ \it Invest in a census of exo-zodi systems
around solar-type stars that might be targets for exoplanet searches.}

\bigskip

\noindent {\bf  A. I. b.  What are the physical characteristics of planets in the
habitable zones around bright, nearby {\it M } stars?}

\bigskip
This parallels Recommendation A. I. a. 1, but for the smaller M-dwarfs.

\bigskip

{\bf \emph{Recommendation \hspace{0.05in} A. I. b. 1 }}{ \it In view of
the fact that M-dwarfs might harbor the most detectable Earth-sized
planets, search the nearest thousand M-dwarfs (J $\leq 10$) for
transiting low-mass exoplanets with radial-velocity measured masses. }

\bigskip

Near-IR spectrographs are potentially the best way
to find terrestrial mass planets in the habitable zones of main
sequence stars later than M4; optical spectroscopy is
competitive and already available for earlier spectral types.
Develop IR spectrographs with a target precision of
1 m/s for radial velocity surveys of late M-dwarfs.  A near-term
demonstration of 10 m/s is critical to validate this technique.

{\bf \emph{Recommendation \hspace{0.05in} A. I. b. 2 }} {\it Develop
near-IR spectrographs with 1 m/s precision for radial velocity planet
surveys of late M-dwarfs, once feasibility at 10 m/s precision has been
demonstrated.}

\bigskip

Given the possibility that some M-dwarfs might harbor planets whose
properties could be studied with warm-Spitzer and/or JWST, these space
assets form an important part of the strategy.

{\bf \emph{Recommendation \hspace{0.05in} A. I. b. 3}} {\it Continue to operate Spitzer as a warm observatory for
characterizing low-mass transiting planets around
main sequence stars, particularly around M-dwarfs.}

\bigskip
\noindent {\bf A. II. What is the architecture of planetary systems?}

\bigskip

In this near-term period, progress toward a complete understanding of
planetary systems requires further development of two ground-based
techniques: microlensing, which will provide a census of planetary
mass and orbital separation as a function of stellar type and Galactic
environment; and extreme AO systems, which will enable the direct
detection of young giant planets.  Ground-based extreme-AO
coronagraphy on 8-m class telescopes represents the next major
scientific and technical step beyond the current generation of HST and
ground instruments.  Similar instruments on future Extremely Large
Telescopes (ELTs) will further advance the state-of-the-art and
provide a proving ground for technology that may ultimately be part of
an optical wavelength, direct detection approach.  Maintenance of US involvement in facilities that utilize or advance these technologies is key to addressing all three key scientific questions and to obtain full
benefit from the data.

\bigskip

{\bf \emph{Recommendation \hspace{0.01in} A. II. 1}}{ \it Increase
dramatically the efficiency of a ground-based microlensing network by
adding a single 2 meter telescope.}

\bigskip

{\bf \emph{Recommendation \hspace{0.05in} A. II. 2}}{ \it Continue
technological developments and implementation of key ground-based
capabilities such as extreme AO in the laboratory and on 8-m class
telescopes. Future ELT's should be designed with the capability to do
extreme-AO coronagraphy in mind. In the very near-term, establish a blue-ribbon panel to determine in detail the requirements, costs and opportunities associated with ground-based work on exo-planets.}

\bigskip

\noindent {\bf A. III. When, how and in what environments are planets formed?}

\bigskip

{\bf \emph{Recommendation \hspace{0.1in} A. III. 1}}{ \it Maintain
U.S. involvement in key facilities including Herschel and ALMA for
studies of disks, and continue support for archival analysis of
relevant Spitzer, Chandra, Hubble, and ground-based data.  }

\bigskip

{\bf \emph{Recommendation \hspace{0.05in} A. III. 2}} {{\it Sustain a healthy
level of support for ground-based, space-based, and theoretical
(see Section D) investigations of star and planet formation. }\bigskip

\bigskip

\subsection{B. Recommendations for 6--10 Years}

The recommendations here follow from the accomplishments of the first
time period and the logic of the strategy summarized above.

\bigskip

\noindent {\bf B. I. a. What are the physical characteristics of
planets in the habitable zones around bright, nearby {\it F, G, K}
stars?}

\bigskip

{\bf \emph{Recommendation \hspace{0.05in} B. I. a. 1}} {\it Launch and
operate a space based astrometric mission capable of detecting planets down to the mass of the Earth around 60--100 nearby
stars, with due consideration to minimizing the width of any blind
spot associated with Earth's parallax motion. (This requires a mission precision, over many visits to a given star, as small as 0.2 microarcseconds.)  }

\bigskip

{\bf \emph{Recommendation \hspace{0.05in} B. I. a. 2}}{ \it Contingent on
the latest knowledge of $\eta_\oplus$ and exo-zodi brightness for
potential target stars, move spaceborne direct detection into the
advanced formulation phase to enable a mission launching in the 11--15
year time frame.}

\bigskip

Note: if Kepler suffered a sufficiently serious mission failure,
development of the direct detection mission should proceed forward
based on COROT and ground-based resultsif they indicate a likelihood of
high $\eta_\oplus$. Likewise, if the astrometric mission fails or
turns out to be infeasible, pursuit of space-based direct detection in
the final time period would require a sufficiently large
$\eta_\oplus$ based on COROT and Kepler to give a reasonable
probability of mission success.

\bigskip

\noindent {\bf B. I. b. What are the physical characteristics of
planets in the habitable zones around bright, nearby {\it M} stars?}

\bigskip
Given the possibility that some M-dwarfs might harbor planets whose
properties could be studied with warm-Spitzer and/or JWST, these space
assets form an important part of the strategy.

{\bf \emph{Recommendation \hspace{0.05in} B. I. b. 1}} 
{ \it Use JWST to characterize Earth-sized
transiting planets around M-dwarfs.}

\bigskip

\noindent {\bf B. II. What is the architecture of planetary systems?}

\bigskip

{\bf \emph {Recommendation \hspace{0.05in} B. II. 1}} {\it Move planetary
system architecture studies and multiple-planet statistics beyond the
3 to 5 AU ``ice-line" (also called the ``snow-line") boundary for G-type stars by continuing
long-time-baseline Doppler spectroscopic studies.}

\bigskip

Section 4.2.4 and its associated appendix show that a spacebased microlensing mission has significant advantages over a ground-based network in being able to collect complete statistics on planetary masses and separations, including free-floating planets, as a function of stellar type and location in the Galaxy. 

{\bf \emph{Recommendation \hspace{0.05in} B. II. 2}} {\it  Without
impacting the launch schedule of the astrometric mission cited above,
launch a Discovery-class space-based microlensing mission to determine
the statistics of planetary mass and the separation of planets from their
host stars as a function of stellar type and location in the galaxy,
and to derive $\eta_\oplus$ over a very large sample. }

\bigskip

{\bf \emph{Recommendation \hspace{0.05in} B. II. 3}} {\it Begin
construction of a 30 meter telescope to do optical direct detection of
giant planets to understand planetary system architecture and planet
formation, and invest in appropriate instrumentation for planet
detection, characterization, and disk studies.}

\bigskip

\noindent {\bf B. III. When, how and in what environments are planets formed?}

\bigskip

{\bf \emph{Recommendation \hspace{0.05in} B. III. 1}} {\it Implement
next-generation high spatial resolution imaging techniques on
ground-based telescopes (AO for direct detection of young low mass
companions and interferometry for disk science).}

\subsection{C. Recommendations for 11--15 Years}

\bigskip

\noindent {\bf  C. I. What are the physical characteristics of planets in the
habitable zones around bright, nearby {\it F, G, K} stars?}

\bigskip

{\bf \emph{Recommendation \hspace{0.05in} C. I. 1}} {\it Provided
$\eta_\oplus$ is high and typical exo-zodi emission sufficiently low,
conduct space-based direct detection and characterization for
Earth-mass planets found by astrometry, in the habitable zone of
nearby solar-type stars.}

\bigskip

Given the current state of development of direct detection approaches,
the coronagraph/occulter appears to be more mature and less costly
than interferometric techniques. However, before a choice of which
direct detection approach to pursue is made, additional technological
studies as well as better constraints on $\eta_\oplus$ and exo-zodi
emission should be obtained.  

\bigskip

Should any of the following occur, namely $\eta_\oplus$ low, exo-zodi
emission high, or no astrometric candidates are found which are
suitable for direct detection missions, then technology studies of facilities capable of direct detection of Earth-sized planets around more
distant or more difficult targets are warranted.

{\bf \emph{Recommendation \hspace{0.05in} C. I. 2}} {\it Begin development
of a more ambitious space-based direct detection system, with
international collaboration where appropriate, to be launched beyond
the report's time horizon. Such a mission would either follow up on
the successful direct detection work begun in this period of the
strategy, or be used to overcome technological or observational (e.g.,
low $\eta_\oplus$) impediments that prevented such detection and
characterization in the report's 15-year time horizon. }

%{\bf \emph{Recommendation}} {Begin development of a more ambitious
%space-based direct detection system, with international collaboration
%where appropriate, to be launched beyond the report's time
%horizon. Such a mission would either follow up on the successful
%direct detection work begun in this period of the strategy, or be used
%to overcome technological or observational (e.g., low $\eta_\oplus$)
%impediments that prevented such detection and characterization in the
%report's time horizon. }

\bigskip

\noindent {\bf C. II. What is the architecture of planetary systems?}

\bigskip

{\bf \emph{Recommendation \hspace{0.05in} C. II. 1}}} {\it Should any of the
following occur, namely $\eta_\oplus$ low, exo-zodi emission high, or
no astrometric candidates are found which are suitable for direct
detection missions, then pursue with stronger emphasis studies of larger
planets via ground-based direct detection, and studies of the architecture of planetary systems with
ground- and space-based tools including microlensing.}

\bigskip

\noindent {\bf C. III. When, how and in what environments are planets formed?}

\bigskip

{\bf \emph{Recommendation \hspace{0.05in} C. III. 1}}
 {\it  Invest in technology for the
next-generation observations of planet-forming disks and disks where
young planets reside. These capabilities should include, for example,  sensitive
(equivalent to Spitzer and Herschel) far-infrared interferometric
observations from space to achieve the resolution of ALMA closer to
the peak of the dust spectral energy distribution. A particular need for the far-infrared is technology investment to increase detector performance. }

\subsection{D. Ongoing All Years}

\bigskip

Theoretical work is the tool to interpret the findings resulting from
the observational and technological recommendations in sections
1.1 through 1.4 above and to put them into a broader context.

\bigskip

{\bf \emph{Recommendation \hspace{0.05in} D. 1}} {\it A strong theory
program is essential to address all three compelling questions. Theory
programs include, planet atmosphere and interior studies, laboratory
astrophysics, n--body and hydrodynamic codes with large computational
demands to study planet formation, evolution, and dynamical evolution,
and stellar astrophysics (e.g. nearby young star samples, stellar
ages).}

\bigskip

{\bf \emph{Recommendation \hspace{0.05in} D. 2}} {\it NASA and NSF should
provide support for activities that maximize the knowledge return from
the data and train new scientists in the field, including theoretical studies
(see D. 1.), stellar properties surveys, and competitive fellowships
for young researchers.}

% ============================================================================
% Begin New PART - FIGURES OF MERIT
% ============================================================================

\pagebreak
\clearpage

\chapter{Depth of search comparisons}

In a sense habitable planets the size of the Earth around Sun-like stars occupy an ``anti-sweet" spot in detection space. They are too far from their parent stars with respect to RV, but too close with respect to astrometry.  Likewise, they are too far from their parent stars with respect to transits, too close with respect to microlensing. And for direct detection these are, simply, very faint objects. Yet a quantitative comparison of the reach of various techniques in terms of projected capabilities over the lifetime of the strategy suggests that the problem, while difficult, is not hopeless. 

To allow direct comparison of the science reach of various techniques for detecting and characterizing exoplanets, we have developed a standardized depth of search figure as a function of planet mass $M$ and luminosity-scaled orbital semi-major axis $A_E$. Mission sensitivity is often plotted as a single line in $M, A$ space, but this two-dimensional representation has several limitations. It misses both the fact that some detection approaches can survey very different samples of stars (compare a typical ground-based Doppler survey of thousands of stars to proposed space-based astrometric missions probing perhaps a hundred stars) and that some techniques (e.g. transits) have low probabilities of detecting any given planet and significant variations in detection probability across their accessible phase space.
Third, plotting detectable mass as a function of raw semi-major axis $A$ makes comparison of different target star samples difficult, since the key regions of other solar systems (such as the habitable zone) of course vary with spectral type.

To overcome this third issue, we plot not as a function of $A$ but as a function of equivalent semi-major axis $A_E=A/\sqrt{L}$,
where L is the luminosity of each target star. Since the energy input of the planet scales as $L/A^2$, this causes planets of the same effective temperature to be at a single location on the diagram irrespective of stellar type. In particular, in
units of AU and solar luminosity (the standard unit set we use),  the habitable zone will range from $A/\sqrt{L}$=0.75 to $A/\sqrt{L}$=1.8 over nearly the entire range of main sequence stars considered here. 

For each star $j=1,\ldots,N$ in a survey of $N$ stars, we calculate the probability 
$P_{j}(M,A_E)$ that a planet of given mass and semi major axis would be detected. 
This will depend on the properties of the planet (either uniquely 
defined for a given $M$ and $A$, or with some assumed distribution of e.g. planet ages), on the appropriate sensitivity metric such as minimum detectable orbital velocity amplitude or maximum detectable planet contrast, and on the distribition of properties of the system, such as orbital inclination. The details of the distributions used are discussed below for each individual technique. The \textbf{depth of search} $S(M,A_E)$ is simply the 
sum $\sum_{j=1}^{N}P_{j}(M,A_E)$. 
%I am not latex-savvy enough to get it to place the sum bounds properly
This quantity, though it produces a 2-dimensional distribution 
rather than a single figure of merit such as ``number of Earths 
discovered'' has the advantage of requiring no assumptions to be made 
about the expected distribution of planet masses or orbits. Some 
assumptions may have to be made about planet and star properties, but 
these are typically well known (e.g. stellar age distributions) or have 
relatively little effect on the result (e.g. planet eccentricity.)
Examples of the calculation for different search techniques are 
discussed below. 

This depth of search can be 
interpreted directly in two ways. The least biased interpretation is to think of this as a plot not of planets but of number of stars - the equivalent number of stars surveyed for each $(M,A_E)$ if the search had a 100\% probability of detecting any planet.  This allows an easy comparison between a deep survey (such as current Doppler surveys, with high detection probability for thousands of stars) to a shallower survey or a survey with non-uniform detection probabilities.
Alternatively, the output can be thought of in terms of planets - if every target star had a planet of mass $M$ and semi major axis $A_E$, $S(M,A)$ is the number of planets that would typically 
be discovered by the survey. For example, the expected number of Earth 
analogs to be found by a TPF-like mission would be given by the average 
of $P$ over the habitable zone $A_E$ and $M$ range times the fraction of stars 
posessing such a planet, $\eta_{\oplus}$. To guide this interpretation, we have 
shown the rough boundaries of the habitable zone for the typical 
targets in each survey. This allows one to see how sensitivity to habitable planets varies with their mass and
orbital properties, 

More broadly, $S$ provides a measure of the statistical robustness of each 
survey technique. The gold standard for this is large-scale Doppler 
surveys, for which $S > 1000$ over a broad range of $M$ and $A$. 
Surveys with such reach can expect to
find hundreds of planets and in turn allow statistical studies of the 
properties of the planet population. By contrast, a survey with $S=10$, 
even over a very broad range of masses, would expect few if any 
detections. 

We will typically display this quantity on a logarithmic color map, with the top color set to
correspond to $S=10000$ for large-scale statistical surveys such as Doppler planet detection, where
the main goal is measuring the planet distribution with high significance. For surveys that allow characterization of individual planets, particularly in the habitable zone, the top color will be set to 
$S=100$.

To calculate $S$ for a exoplanet experiment oriented towards initial 
detection, we must correctly calculate the number of systems $N$ that 
the experiment could survey over its intended duration. Such a 
detection experiment will typically measure 2-3 numbers -- the 
planet's mass or radius, orbital separation and perhaps eccentricity. 
The next step is characterization, which we generally define as 
measuring more than one physical property of the planet -- for 
example, radius or magnitude as a function of wavelength. By this 
definition, only the transiting exoplanets have yet been characterized. 
For a characterization mission, the relevant $S$ is not the number of 
stars that could be surveyed in a given mission lifetime, since not 
every star will have a planet worth characterizing; instead, it is 
more interesting to plot the number of systems whose planets 
theoretically could ever be characterized (possibly with some 
reasonable assumptions about integration times.) 

\section{Doppler}

By far the most successful planet detection technique to date has been 
Doppler searches for the reflex motion of a planets parent star; these searches are now reaching 
single-measurement precisions of 1 m/s for 
large target samples. This technique is also one of the most straightforward to model. We present several current or possible Doppler searches here. For each,
we generate a target sample with a range of masses based on current ongoing Doppler survey target lists. We define a survey by its duration $D$, expected number of measurements per star $n_0$, with each star having a poisson 
distribution of actual number of measurements $n$ about $n_0$.  Given 
these, a single-measurement precision $\sigma$ and a desired false-alarm 
probability $F$ we calculate the maximum detectable doppler signature 
using equation 10 of Cumming et al (2004.) For each star this gives a 
minimum detectable $M \sin i=M_S(A_E)$ as a function of $A_E$. The quantity $Pj
(M,A_E)$ is then simply the probability that $\sin i > M_S(A_E)/M$. (If $M < M_S$, 
this
probability is of course zero.) Adding together the $P_j$ produces the 
survey size $S(M,A_E)$. 

We have made several simplifying assumptions, but these are reasonable 
for this overview. The Cumming formula is valid only for large and 
randomly-distributed sets of measurements. Although we set a maximum 
and minimum detectable period, we have ignored particular periods that 
may have low detectability due to scheduling constraints (e.g. 1-year 
periods.) 
We have assumed only circular orbits; Cumming 2004 shows that moderate 
eccentricity has only a weak effect on initial detectability. The 
formula used gives only the required signature for inital detection --
once a star shows evidence of a planet, more observations may be needed 
to fully constrain its orbit, but a well-designed survey will reserve 
time for such allocations.

Figure ~\ref{fig:figAAAabc}  shows the depth of search of several different Doppler projects. Panel (a) is a rough representation of current surveys; 2000 stars with masses around 1 solar mass, each sampled twice per year at 3 m/s precision over 10 years. Panel (b) shows an extrapolated large-scale survey, with either a dedicated instrument on a 4-m class telescope or a larger allocation of 8-10m telescope time; 2000 stars observed 5 times per year. 
The last two panels show Doppler searches optimized for low-mass planets with multiple repeated observations of low-mass stars. Panel (c) is a representation of the $\eta_{\oplus}$ program combining Keck and Automated Planet Finder observations, 
with 100 low-mass stars, each observed intensively for one year with 100 measurements. The stellar masses are estimated for the actual target list of
this project and range from 0.3-0.7 \msun. Again, a dedicated 4-m class telescope and instrument would scale the science reach by a factor of 2 or more by allowing more and fainter stars to be observed. 

Panel (d) shows a similar near-IR doppler survey, targeting stars from 0.3 to 0.1 \msun. Some 120 stars are sampled 5 times per year over 5 years with 1 m/s single-measurement precision. Such a survey would require 100-300 nights on an 8-m telescope, broadly consistent with the goals of the Gemini PRVS near-IR doppler instrument. It has the potential to discover true Earth-mass planets in the habitable zone of low-mass stars, with the probability that some will be characterizable with follow-on transit searches. However, to be competive with visible-light Doppler surveys such an IR program must achieve a single-measurement precision of $~1 \mse$. 

Finally, Figure ~\ref{fig:starmasshist } shows the different stellar populations probed the varying Doppler programs
discussed here. 

\begin{figure}
\centering
\subfigure[] % caption for subfigure a
{
    \label{fig:figAAAa}
    \includegraphics[width=6.8cm]{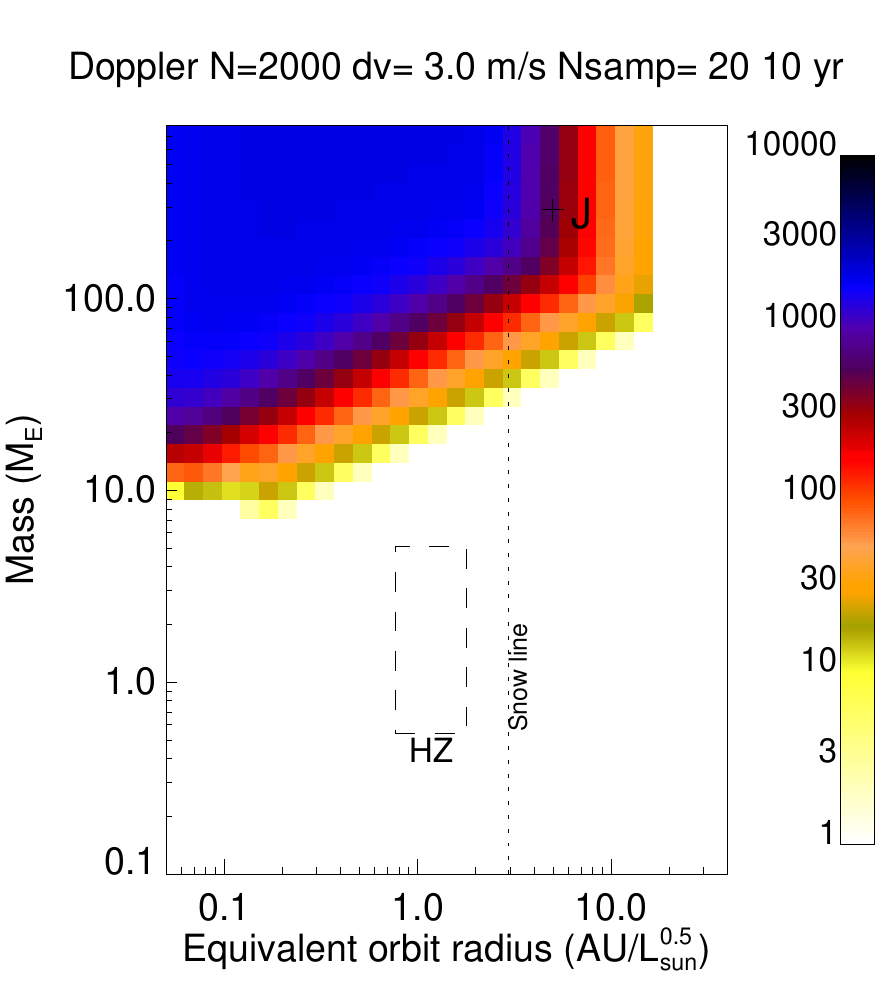}}
\hspace{1cm}
\subfigure[] % caption for subfigure b
{
    \label{fig:figAAAc}
    \includegraphics[width=6.8cm]{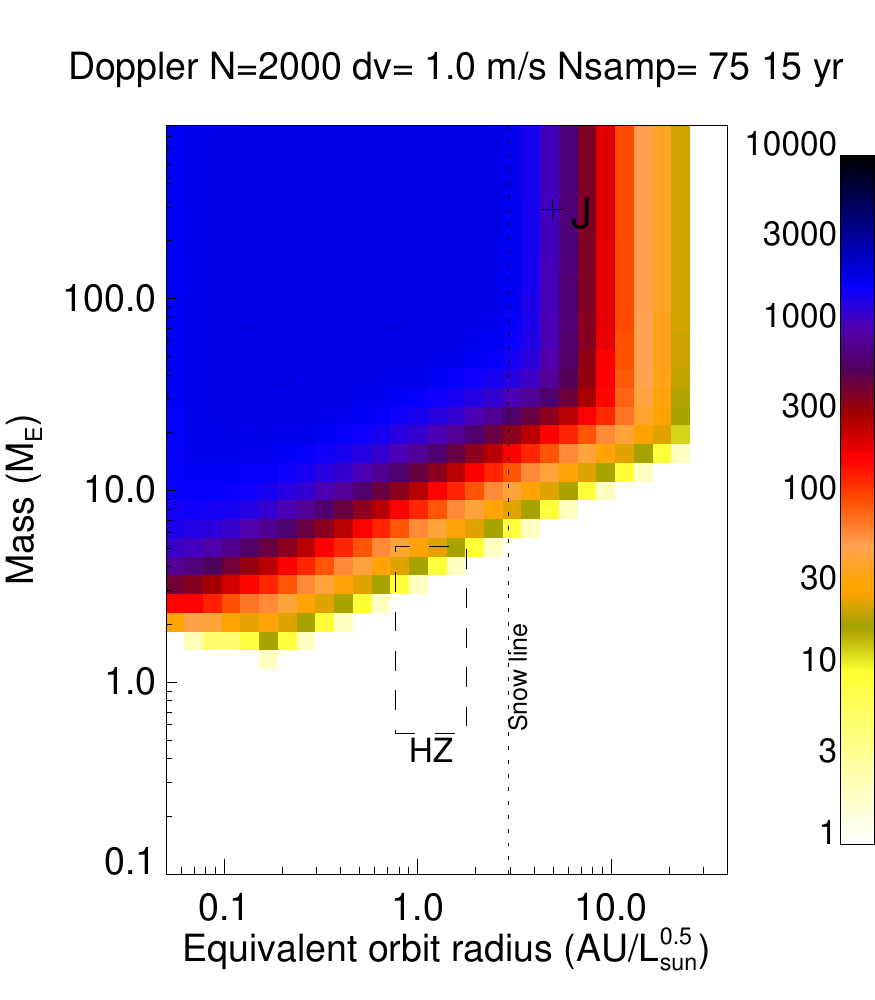}}
\hspace{1cm}
\subfigure[] % caption for subfigure c
{
    \label{fig:figAAAd}
    \includegraphics[width=6.8cm]{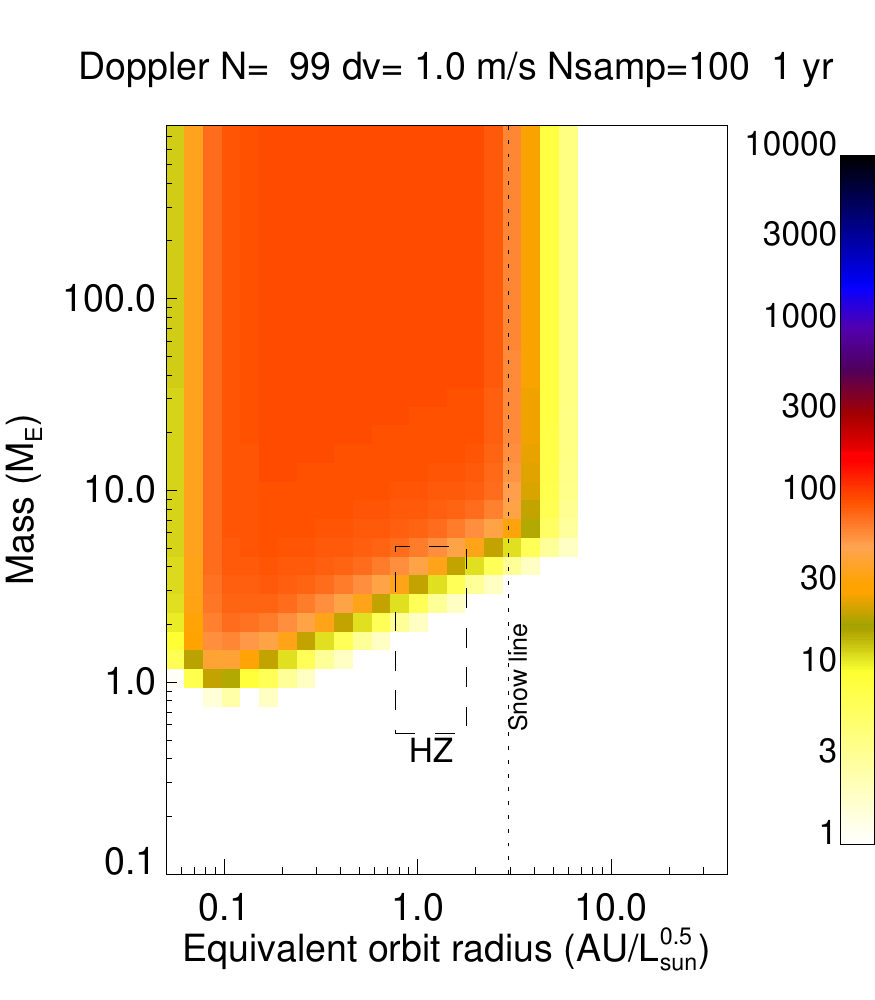}}
\hspace{1cm}
\subfigure[] % caption for subfigure c
{
    \label{fig:figAAAe}
    \includegraphics[width=6.8cm]{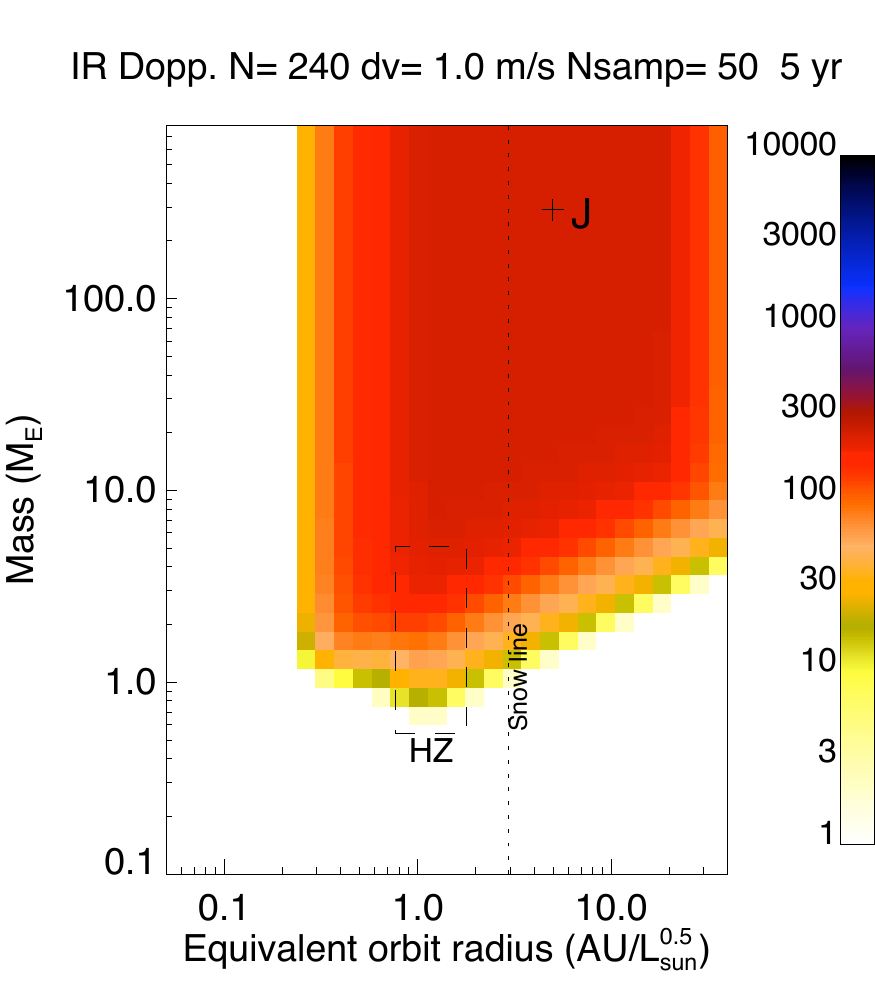}}
\caption{(a)  Depth of search for a representative current Doppler spectroscopic
survey with N=2000 solar-type stars observed twice per year with 3 \mse precision over ten years.. On these and the other comparable  figures, we plot the product of  survey completeness and number of
targets as a function of M and equivalent orbital semi-major axis scaled by stellar luminosity $A_E=A/\sqrt{L}$.  (b) Depth of search for optimistic future Doppler program, dedicated 4-m telescope with 2000 stars observed 5 times per year with 1 m/s precision over 15 years.
(c) Depth of search for an $\eta_{\oplus}$ program  observing 100  stars less than a solar mass for 1 year
each with 100 measurements (d) IR doppler $\eta_{\oplus}$ program measuring 120 0.3 to 0.1 \msun stars 5 times per year with 1 m/s precision.}
\label{fig:figAAAabc} % caption for the whole figure
\end{figure}

\begin{figure}[h]
\centering
\includegraphics[scale=1.0]{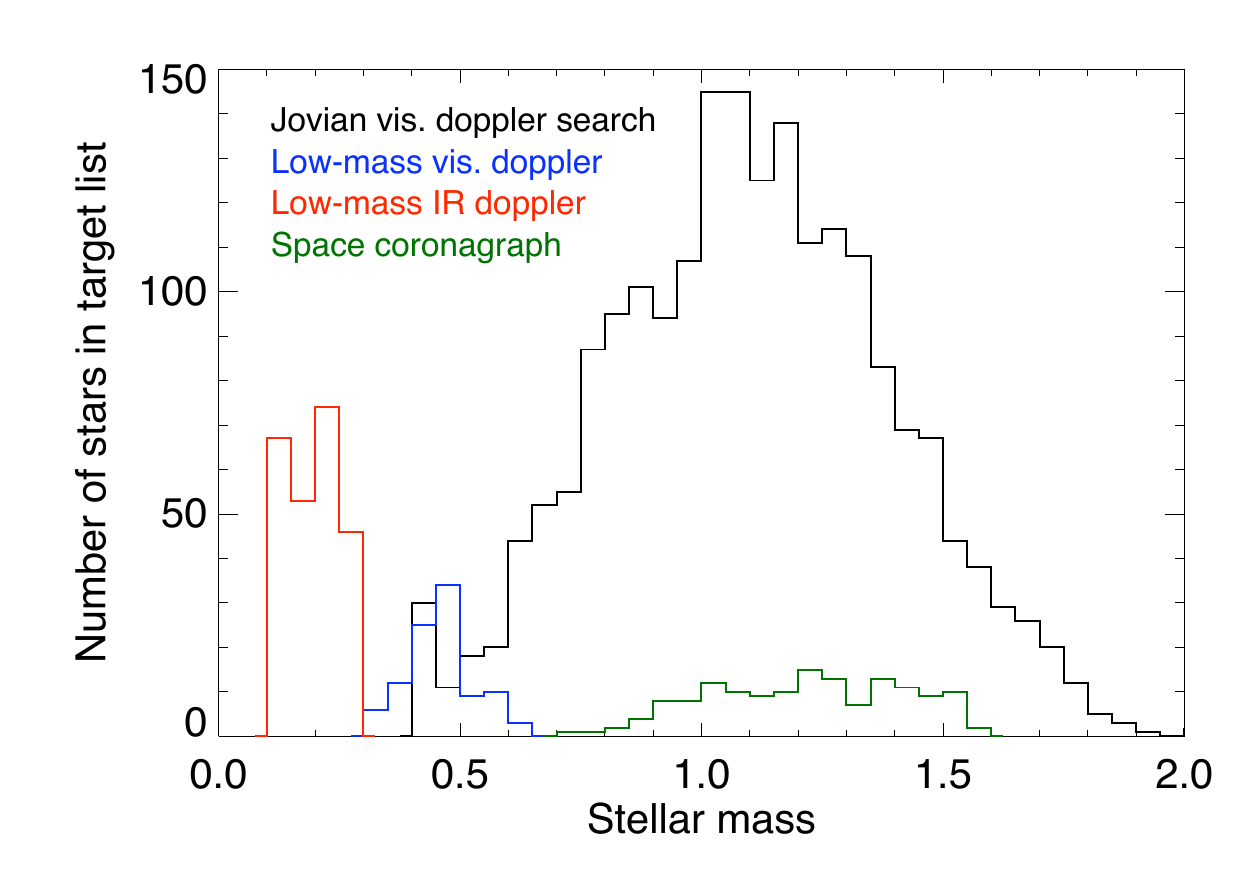}
\vspace{1mm}
\caption{The histogram of target star masses is plotted for a large-scale Jovian-planet
doppler search (black), a concentrated visible-light low-mass planet search (blue), and a
near-IR doppler search targeting the lowest-mass stars (red). Only the latter two have any
significant sensitivity to terrestrial planets. For reference, the target sample for a 4.0-m
space-based coronagraph is also shown (green).}
\label{fig:starmasshist }
\end{figure}

\section{Transit surveys and studies of individual planets}

In a transit survey, the relative flux of many stars are monitored as
continuously as possible, in order to seek the periodic,
low-amplitude, short-duration flux decrements that are characteristic
of transit events.  Stars that exhibit this type of photometric
behavior are candidate transiting systems. They must be distinguished
from grazing eclipsing binaries, high-mass ratio eclipsing binaries, and triple systems including an eclipsing pair,
through follow-up RV spectroscopy.

The most ambitious transit survey currently funded is of course the Kepler
mission. This will target 140,000 moderately-distant main-sequence stars with 
sufficient photometric precision to identify sub-earth-sized transits. Figure~\ref{fig:figBBB} (a) 
shows the predicted survey depth for an extended 6-year mission, illustrating  Kepler's
ability to search a significant portion of the habitable zone. By contrast to Figure~\ref{fig:figAAAabc} parts (c) and (d), 
most of the HZ planets probed by Kepler orbit mid-type FGK stars.
The Kepler results were produced for us by the mission team using their current mission simulator
code, with the results as a function of planet radius and true semi-major axis broken down by individual spectral type. We 
then rescaled according to luminosity and the mass-radius relationships discussed below.

\begin{figure}
\centering
\subfigure[] % caption for subfigure a
{
%    \label{fig:figBBBa}
    \includegraphics[width=6.8cm]{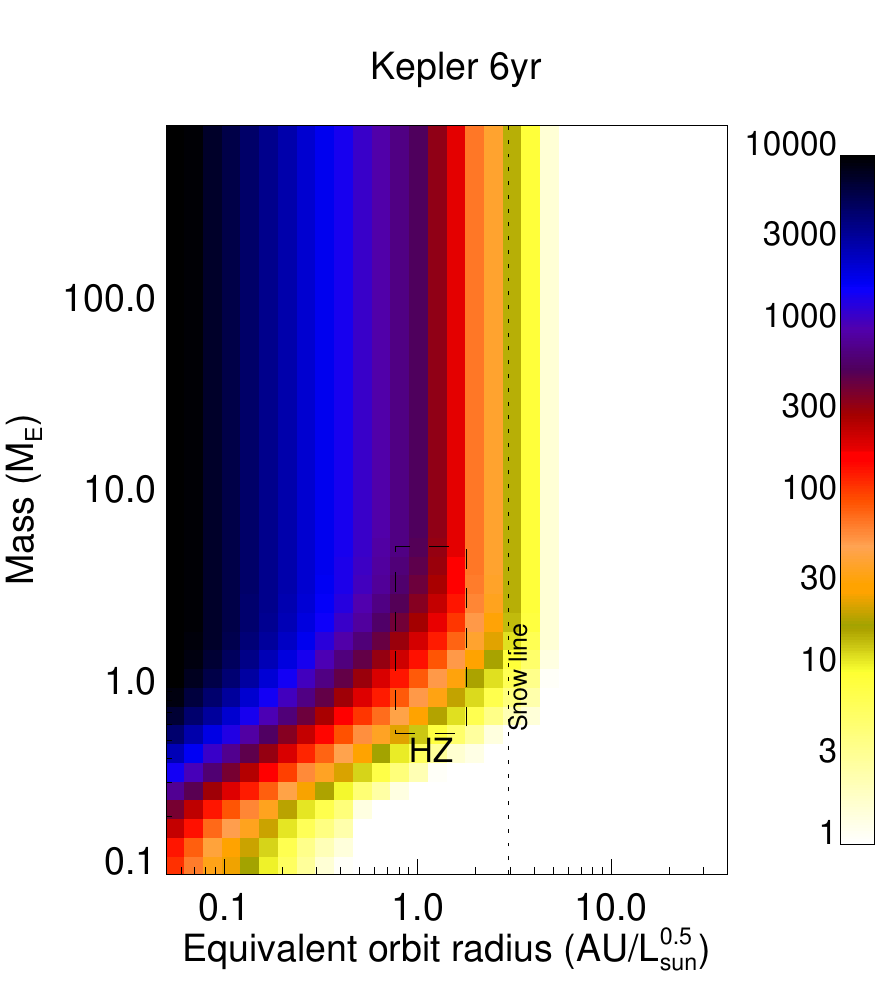}}
\hspace{1cm}
\subfigure[] % caption for subfigure b
{
%    \label{fig:figAAAc}
    \includegraphics[width=6.8cm]{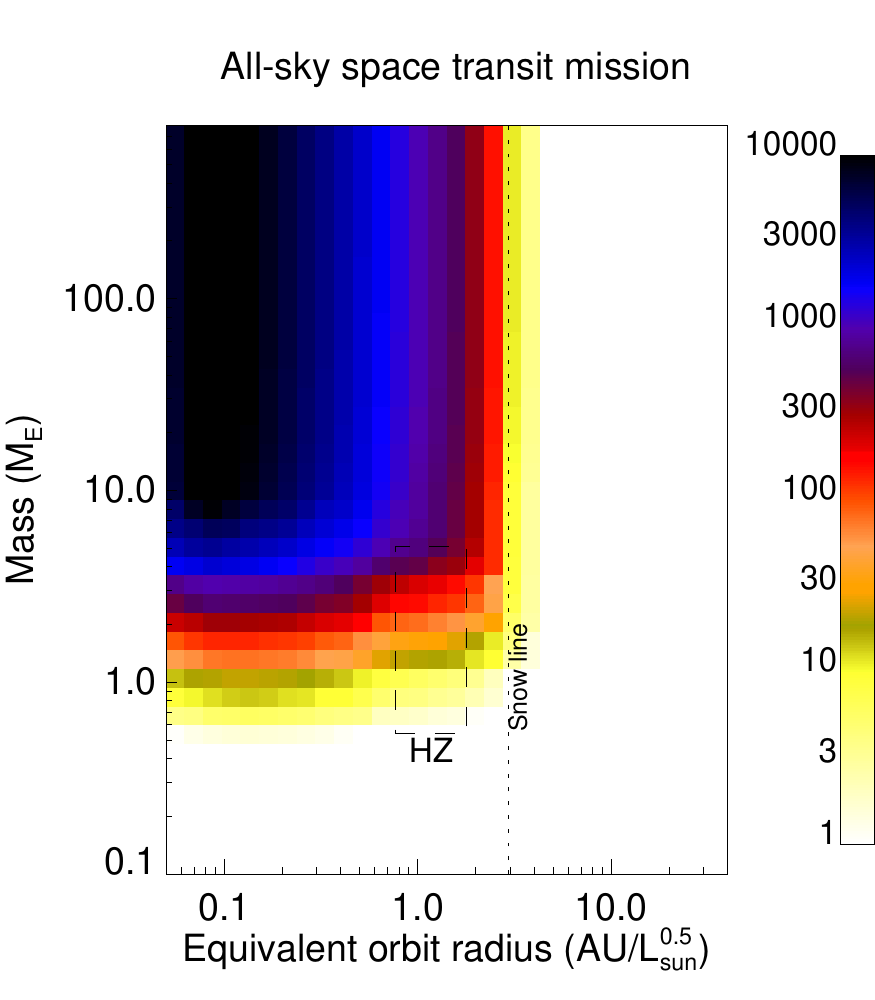}}
\hspace{1cm}
\subfigure[] % caption for subfigure c
{
%    \label{fig:figAAAd}
    \includegraphics[width=6.8cm]{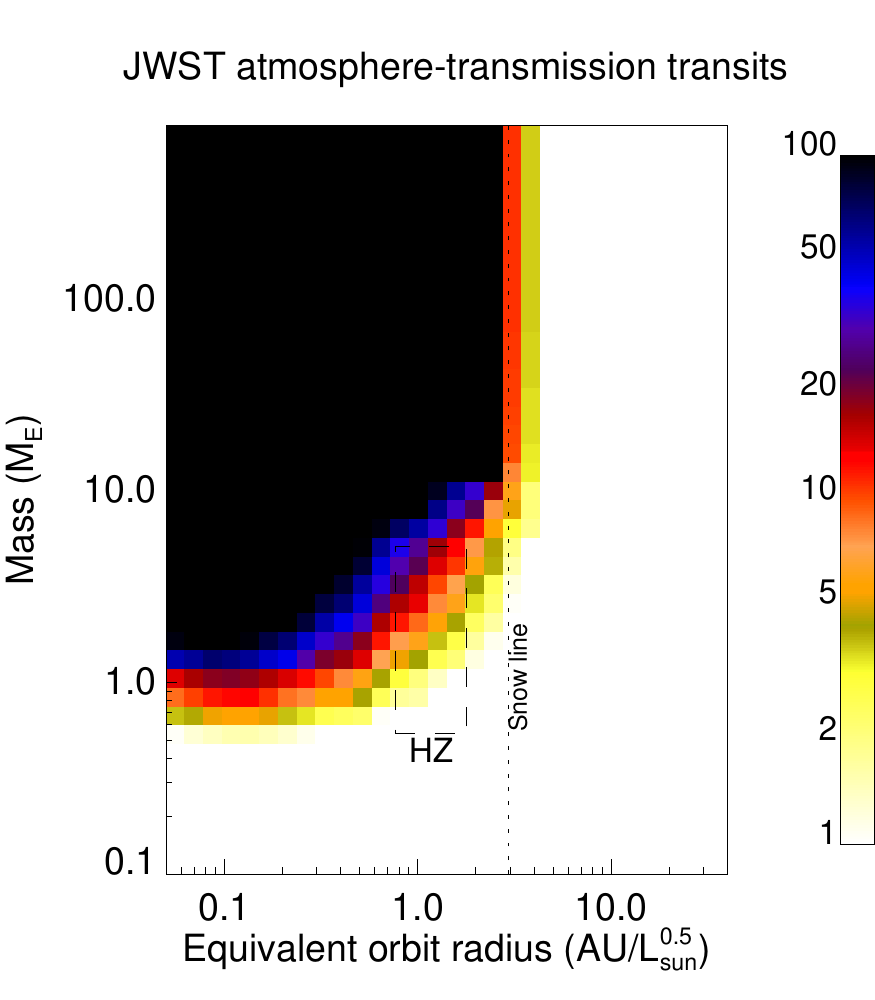}}
\hspace{1cm}
\subfigure[] % caption for subfigure c
{
%    \label{fig:figAAAe}
    \includegraphics[width=6.8cm]{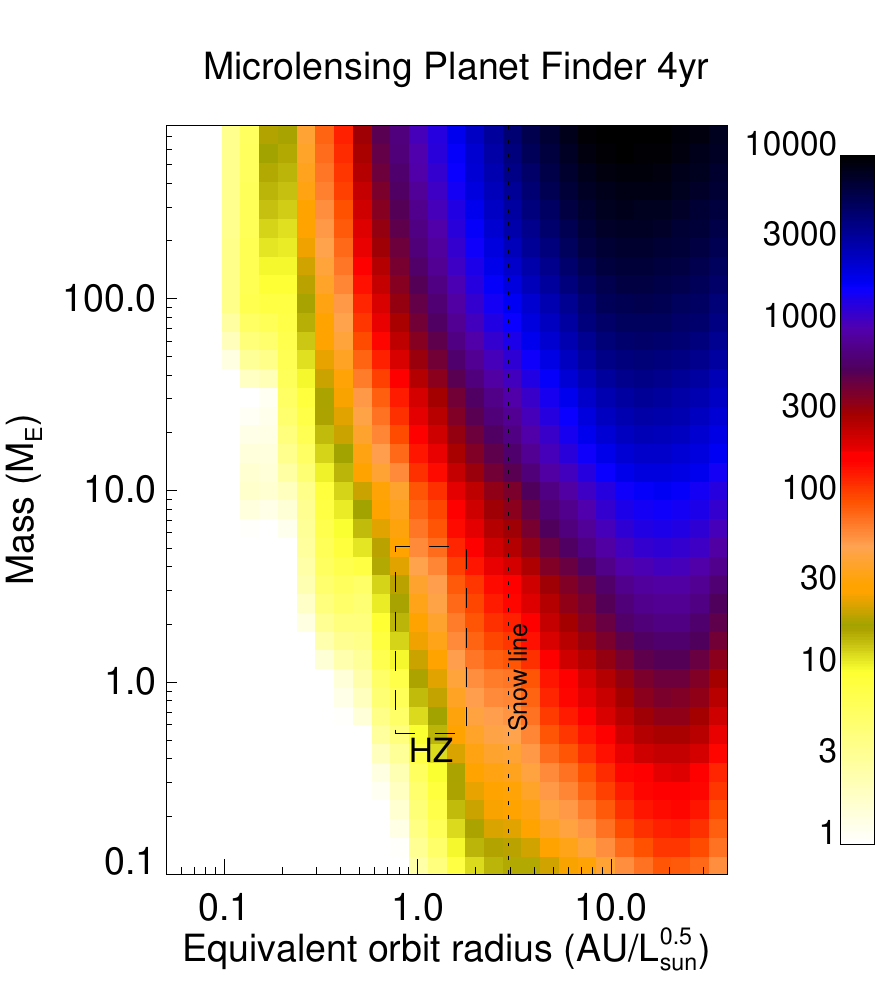}}
\caption{(a)  Depth of search for a 6-year Kepler mission. Original data courtesy W. Borucki.
(b) Depth of search for an all-sky survey mission searching 
for the nearest transits with effective collecting area of 300 $cm^2$
continuously
monitoring the entire sky. Target population is drawn from a
solar-neighborhood model.
To achieve a detection, each transit must be observed at SNR $> 7$ and the
target star must
 have V $< 15$ to allow spectroscopic confirmation. (c) Targets from  sample in panel (b) that would have a putative planet characterizable
in transmission during transit using JWST NIRSPEC instrument; 
R=100 and SNR=10 averaged over ten transits (Note change in plot range).
(d) Sensitivity of a four-year Discovery-class microlensing planet detection mission.}
\label{fig:figBBB} % caption for the whole figure
\end{figure}

The most scientifically exciting frontier
in the study of transiting planets is the detection of terrestrial planets
in the habitable zones of nearby M-dwarfs, where the transit depth and
star brightness allow detailed follow-up characterization, including possible
measurements of atmospheric transmission spectra using JWST. The first
step in the study of their systems is their discovery via an all-sky survey,
most likely space-based. 

We predict the effective sample size for such a mission using a statistical
rather than a Monte Carlo approach. 
In our idealized transit survey, all of the stars on the sky down to a
certain limiting magnitude ($V<15$) are monitored continuously for
transits. The magnitude limit is imposed to make RV spectroscopic
follow-up observations feasible; apart from imposing this limit, we
ignore the false positive problem, and assume that candidates are
followed up and that the transit events are successfully distinguished
from the imposters. We also impose a limiting distance from the Earth
($d<100$~pc) because the nearby stars will have measurable parallaxes
and therefore there will be much less uncertainty in the stellar
properties.  Uncertainty in the stellar properties translates directly
into uncertainties in the planetary properties, making more distant
systems of limited interest for detailed
characterization. \footnote{The distance limit also makes our
  calculations much simpler by obviating the need to take Galactic
  structure into account. In what follows we assume that the stellar
  density given by the luminosity function is uniform within the
  survey volume.}
Since we are concerned primarily with the relatively
deep transits around late-type stars, which will almost always be photon-
limited, we do not impose any additional systematic noise floor.

The remaining parameters of the idealized survey are the bandpass and
overall throughput of the survey instrument, and the maximum
detectable orbital period (which is a function of the survey
duration). We assume that the survey will be conducted in the $I$
band, reasoning that redder optical bands are preferred to increase
the signal-to-noise ratio for the interesting case of low-mass
stars. Our fiducial throughput is such that 220 photons per second are
recorded from an $I=15$ star; this translates into an effective area
of about 300~cm$^2$ over the $\Delta\lambda/\lambda=0.19$ bandpass.
We further assume that the maximum detectable period is 120~days; all
shorter periods are detected with 100\% efficiency and longer periods
are not detected. In reality, detectability will be a declining
function of period, depending on the specific observing cadence and
pattern of time coverage, details that we ignore for simplicity. One
might imagine a space-based survey or a longitudinally-distributed
network in which stars are monitored continuously for 120~days without
interruption.

To determine the number of transit detections as a function of the
planetary mass and semimajor axis, we need to know the properties of
the stars in the sample.  We adopt the luminosity function for stars
ranging from $M_V = 3.5$ to 18.5 (spectral type F5 to M7) from Reid \&
Hawley (2000, page 296), which is binned in increments of $\Delta M_V
= 1$. For each bin, we assign the stars a spectral type, radius, mass,
luminosity, $V-I$ color, and $V-K$ color (needed for determining the
prospects of infrared characterization; see below). The spectral types
were assigned using the empirical polynomial relation of Henry et
al.~(1994) between $M_V$ and spectral type. The radius, mass,
luminosity, and $V-I$ color were then assigned by interpolating Table
4.1 of Reid \& Hawley (2000). The $V-K$ color was assigned by
interpolating the color-magnitude relations of N.~Reid\footnote{ {
http://www.stsci.edu/\~inr/cmd.html} } and were checked against the
tables in Cox (1999, page 151).

For each stellar bin, we ask: at what distance can this type of star
be placed and still offer a great enough photon rate for transits to
be detectable?  We define ``detectable'' as a signal-to-noise ratio of
10 or greater in the measured transit depth, i.e., $\delta/\sigma >
10$, where $\sigma$ is the photometric uncertainty in the average flux
recorded throughout a single transit duration. Here and elsewhere in
these calculations we assume that the dominant error is Poisson noise
from the target star.  The transit depth is computed from the stellar
radius and planetary radius. Since the planetary mass is the input
variable, we must adopt a mass-radius relation. For planets with
M $< $10 M$_{\rm Earth}$ we use the planetary mass-radius relation for rock-iron
planets from Fortney et al.~(2007).  For M $>10$ M$_{\rm Earth}$ we assume a giant planet with a hydrogen
envelope. 
The transit duration is computed
from the stellar mass and the planetary semimajor axis; an average
impact parameter of 0.5 is assumed.

This maximum distance is compared to 100~pc and to the distance at
which the apparent $V$ magnitude is 15, and the minimum of those three
distances is taken to be the radius of the survey volume for stars of
that type. The volume of this region is multiplied by the luminosity
function (stars per pc$^{3}$) and then by the transit probability
[$(R_\star + R_p)/a$], to obtain the expected number of detected
transiting planets for that particular choice of stellar type,
planetary mass, and planetary semimajor axis, under the assumption
that all stars have such a planet.  Each pixel of the resulting completeness
diagram shown in panel (b) of Fig.~\ref{fig:figBBB} displays the sum of the results over all stellar types from
$M_V=3.5$ to 18.5 (spectral types F5 to M7).

The small stars are of special interest because they offer greater
contrast for detecting and characterizing small planets at infrared
wavelengths. For this reason, we have repeated the entire calculation
after restricting the stellar sample to M-dwarfs.  In addition, we have
constructed diagrams showing the subset of the collection of
detected transiting planets which can be characterized in detail
using JWST.

The subset we consider here is intended to be an order-of-magnitude
approximation of the targets for which JWST and NIRSpec can be
profitably used to characterize the absorption spectrum of the planet,
via ``transmission'' spectroscopy.  The basic idea is to seek a
wavelength-dependence in the transit depth that is caused by
absorption features in the planetary spectrum. As a rough estimate of
the size of this effect, we suppose that when observing at the central
wavelength of a strong absorption line, the effective planetary radius
grows by 10 atmospheric scale heights ($10 kT_p/\mu g_p$, where $g_p$
is the planetary surface gravity and $\mu$ is the molecular mass of
the species in question). For concreteness we use the scale height of
water vapor. The wavelength-dependent change in transit depth is
therefore $\delta = 2\pi R_p \times 10H / \pi R_\star^2$.  We treat
$\delta$ as the ``transit depth'' for the event. 
We imagine using NIRSpec
in its low-resolution mode ($R=100$) at $\sim$2~$\mu$m, which we
assume will record 1~e$^{-}$/s per resolution element when
observing a star with AB mag$=24.5$ (P.~Jakobsen, memo dated 10 Sep
2007). We estimate the AB magnitudes of the stars in our model using
AB$_K - K = 1.9$. We imagine observing 10 secondary eclipse events, with
equal time spent in and out of eclipse, and we require a SNR of 10 per
resolution element in the difference spectrum to be classified as
``characterizable.'' 

The result is shown in panel (c) of Fig.~\ref{fig:figBBB} Thus, both of these subsets are defined by a threshold count rate that
can be achieved, given the throughput of JWST and NIRSpec. In reality
there may be many other limiting factors besides the count rate, such
as the need to read out the detector quickly and avoid saturation. By
ignoring these, our subsets may include some systems that cannot be
observed in practice, at least not without employing a nonstandard
observing mode. Some of these issues have been taken into account in
simulations of NIRSpec transmission spectroscopy by Valenti et
al.~(White Paper). These authors investigated a few fiducial cases of
planet-star combinations in much greater detail than we have done here
for our completeness diagrams. However, our results are broadly consistent
with those of Valenti et al., particularly since statistically the majority of
characterizable targets are likely at the faint end of the distribution. 

\section{Gravitational microlensing}

Of all the proposed space-based survey techniques, gravitation microlensing has the potential to sample the largest number of stars over broad range of masses and semi-major axes. Modeling microlensing sensitivity is a complex exercise beyond the resources of this Task Force, so we have relied on models provided by David Bennett. These represent the detection capabilities of a 4-year next-generation ground-based microlensing survey (current capabilities augmented by an additional 2-m telescope and significant follow-up telescope time) and the proposed Microlensing Planet Finder space mission. As can be seen in Figure ~\ref{fig:figBBB} (d), though this technique provides only (non-repeatable) detections its effective survey size is large - one of few  techniques that could provide statistical information comparable to Doppler
surveys, particularly good for measuring the properties of the outer parts of planetary systems near the snow line.

\section {Astrometry} 

Precision astrometric detection of planets in principle scales in a similar fashion to Doppler detection, with multiple
measurements allowing detetion of astrometric signatures below the single-measurement precision $\sigma$. Monte Carlo 
simulations carried out by the JPL Navigator Program group give a minimum detectable astrometric amplitude in each axis of 
$5.3\sigma/n^{0.5}$, where$ \sigma=0.9$ $\muas$ is the single-measurement astrometric accuracy and $n$ is the number of independent measurements. This formula is in broad agreement with the Cumming (2004) Doppler result. We have applied this threshold 
to a list of 74 nearby stars, with the number of observations set to allow detection of low-mass terrestrial 
planets, ranging from 100 to 1300, and calculated the detectable $M$,$A$,$i$ range for each target star. We include a blind spot with a full width half maximum of 20\% in period space; it might be smaller.   Figure~\ref{fig:figDEEE} panel (a) shows the reach of such a precision astrometric survey. Such an 
astrometric mission has significantly greater detection capability than coronagraph missions of equivalent cost 
over a broad range of $A$ for low-mass planets.

\section{Coronagraphy and direct detection}

As with microlensing, a detailed simulation of a TPF-like mission is 
beyond the scope of this Task Force's studies, so we have chosen to
model a series of simplified space coronagraph missions. For a given 
coronagraph, we assume an absolute contrast floor of $0.6\times10^{-10}$, an 
inner working angle of $2.5 \lambda /D$ (for still-unproven agressive 
coronagraphs such as Phase-Induced Amplitude Apodization) or $3.5 \lambda /D$ 
(for more mature coronagraph options such as band-limited Lyot 
or shaped-pupils). Since spectral characterization of oxygen features 
is crucial to study of terrestrial planets, we evaluate at a wavelength 
of 800 nm.

For a planet of a given mass, its radius is calculated using relations in Fortney et al 2007, with planets below 10 M$_{\rm Earth}$ assumed to be rock/iron 
mixes and planets above 10 M$_{\rm Earth}$ treated as gas giants. Planet visible-light geometric albedo has been set at 0.3 for all planet types - subsequent refinements could 
include varying giant planet albedo with separation after Sudarsky et 
al. Planet contrast at phase angle alpha is calculated with a 
Lambertian phase function

$$C(\alpha)=p\,(r/a)^2(\sin(\alpha)+(\pi-\alpha)\cos(\alpha))/\pi$$

where $p$ is the geometric albedo and $\alpha$ is the phase angle as observed from Earth.

Target FGK stars are taken from the TPF nearby stars lists. 
Each star is assigned 20 uniformly-spaced visits over 1 year. (In 
practice, a TPF mission might use optimal visit scheduling to reduce 
this number.) We populate the target star with planets at random 
inclinations and starting orbital phases, then calculate $P_j(M,A_E)$ by 
determining for what fraction of these possible orbits any of the visits succeeds in detecting the planet. We retain any target for which the average completeness in the habitable zone M,A range is greater than 10\%. In addition, we have added a shallow giant-planet survey of a larger sample of targets with $10^{-9}$ contrast floor. 

Figure~\ref{fig:figDEEE} panel (c) shows the effective survey size for a 2.5-m telescope
with an agressive coronagraph (2.5 $\lambda / D$) -- the upper end of a Probe-class mission. (Performance of a
4.0-m telescope with a conservative coronagraph would be similar, as would a 2.5-meter telescope with an external occulter of average radius 16-17 meters at 27,000 km distance.) Averaging over the whole habitable zone, the depth of search is about ten. Such a mission could only expect to characterize an Earthlike planet if $\eta_{\oplus}$ is high.

\begin{figure}
\centering
\subfigure[] % caption for subfigure a
{
    \label{fig:figDDD}
    \includegraphics[width=7cm]{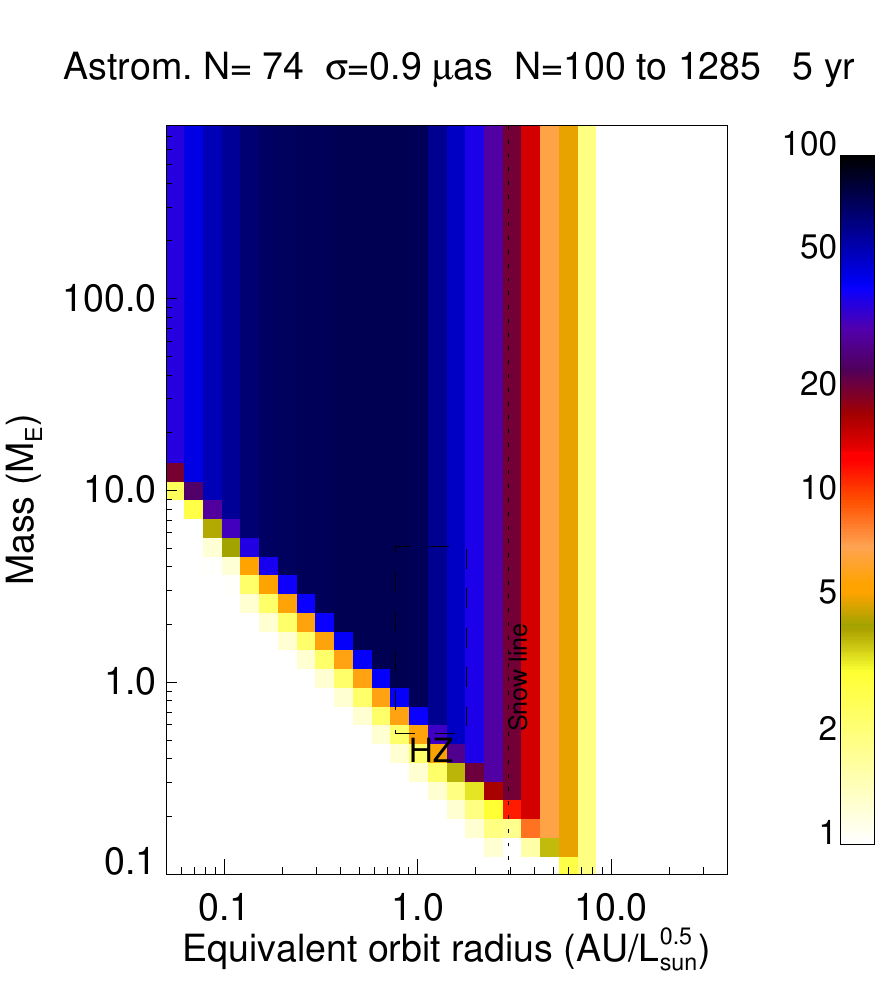}}
\hspace{1cm}
\subfigure[] % caption for subfigure c
{
    \label{fig:figEEEc}
    \includegraphics[width=7cm]{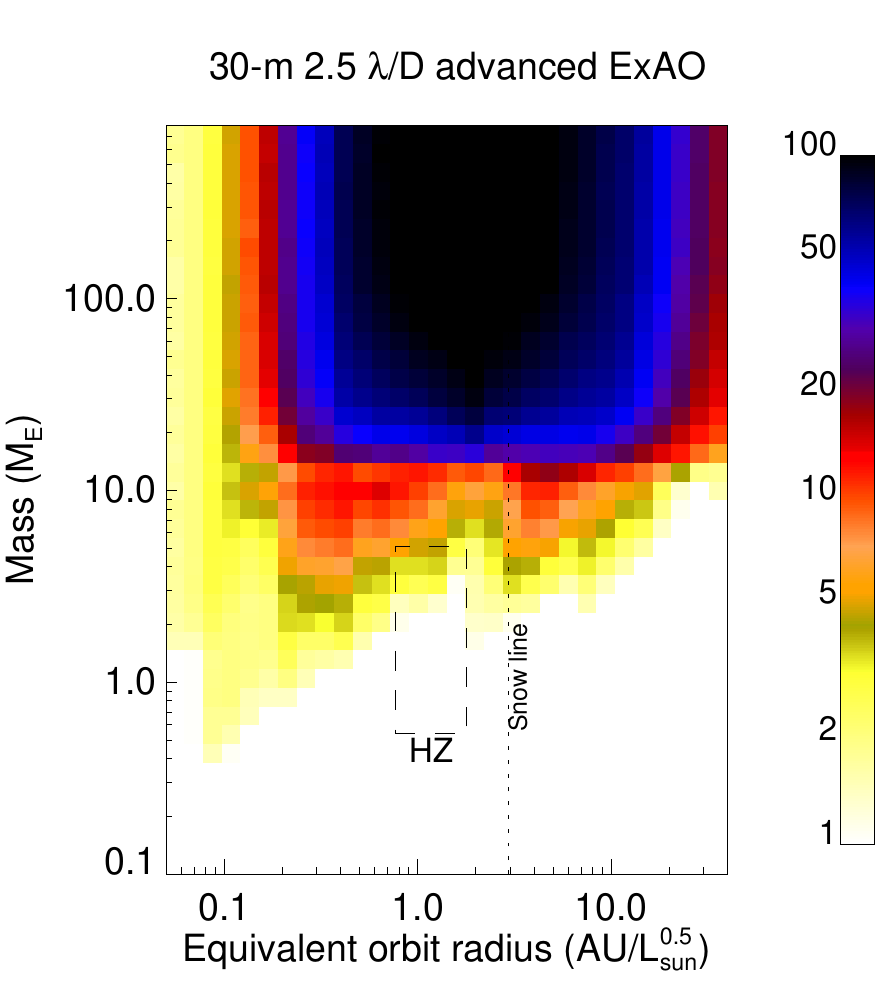}}    
\hspace{1cm}
\subfigure[] % caption for subfigure b
{
    \label{fig:figEEEa}
    \includegraphics[width=7cm]{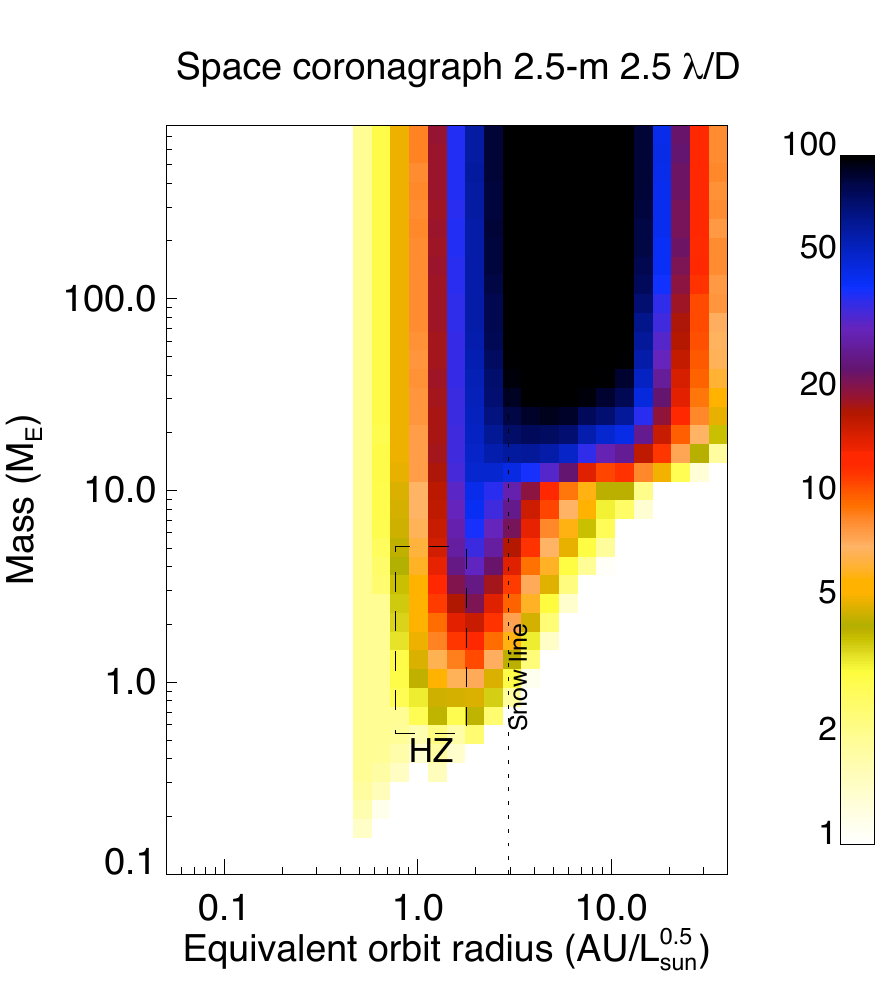}}
\hspace{1cm}
\subfigure[] % caption for subfigure c
{
    \label{fig:figEEEb}
    \includegraphics[width=7cm]{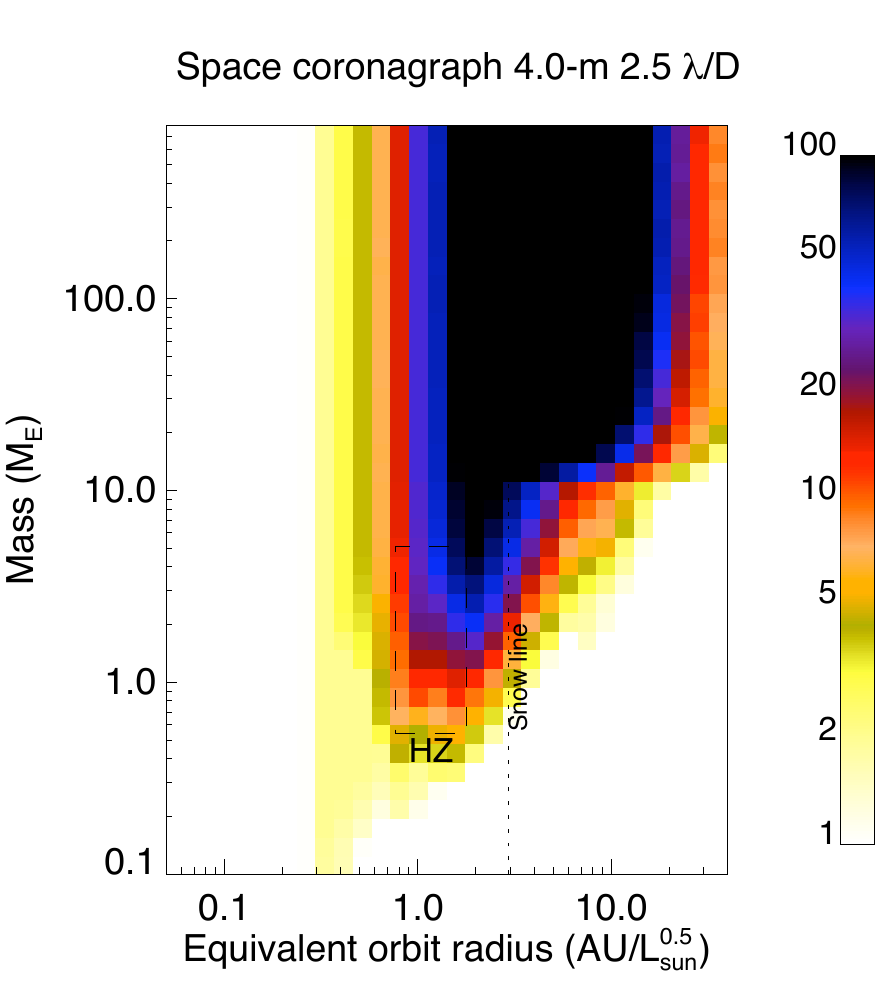}}
\caption{(a)  Depth of search of an astrometric planet search mission with 0.9 $\muas$ single-measurement precision and 100-1200 visits per target star.
(b) Sensitivity of a 30-m ground based telescope equipped with advanced adaptive optics and a $2.5 \lambda/D$ coronagraph. Although this has no sensitivity to earthlike planets, it can characterize (at high spectral resolution) giant planets over a broad range of separations.
(c) Depth of a space coronagraph mission with a 2.5-m primary mirror and an inner working angle of 2.5 $\lambda/D$. (A 4.0-m mission with IWA 3.5 $\lambda/D$ would be similar.)
(d) Sensitivity of a 4.0-m primary mirror with a $2.5 \lambda/D$ coronagraph or a 40 meter diameter external occulter at 40,000 km distance.
}
\label{fig:figDEEE} % caption for the whole figure
\end{figure}

Figure~\ref{fig:figDEEE} panel (d)  shows a more capable 4.0-m telescope with a similar 
coronagraph, representing a flagship-class mission. The corresponding external occulter system might have an average radius of about 20 meters at a distance of 40,000 km.

Our simulations are only intended as a easily-plotted guide for 
comparing different missions over a broad range of planet properties, 
though the results agree to within 50 $\%$ with more detailed JPL 
simulations. 
Since the visit schedule has not been optimized and since the exact 
throughput of instruments and coronagraphs is unknown, we do not 
calculate the integration times needed for each star - instead, our 
sample is limited by contrast floor and inner working angle. This is a 
reasonable simplification for two reasons. First, detailed simulations 
by JPL have shown that for blind surveys, most coronagraphs can easily 
allocate enough integration time to each target to nearly exhaust the 
accessible sample in 2-3 years. Second, as discussed in our strategic roadmap, we favor a TPF mission 
in which targets have been pre-selected by astrometric observations, in 
which case the relevant sample size is the number of stars around which 
a potential planet ever exceeds the contrast and IWA thresholds for 
spectral followup. Small coronagraphic missions will nonetheless often be 
photon-limited in their ability to characterize any planets discovered.  

We have also simulated the performance of a ground-based ``extreme'' adaptive
optics (ExAO) coronagraph system on a 30-m telescope. The averaged ExAO 
point spread function performance comes
from an analytic model similar to that in Guyon 2005 but with some scaling errors 
present in the Guyon paper corrected (O. Guyon and L. Poyneer, private communications).
The AO system simulated is a two-stage system with a 4 kHz visible-light pyramid wavefront
sensor a 1 kHz IR interferometer WFS, an optimal Fourier-modal controller, and a128 by 128
actuator deformable mirror, similar to the system
described for the Thirty Meter Telescope Planet Formation Imager in Macintosh et al. (2006).
Atmospheric speckle noise is assumed to be attenuated through multi-wavelength imaging
with an integral field spectrograph, so the dominant noise term is Poisson photon noise of the
residual halo. End-to-end throughput was 16\% and the atmospheric Fried parameter was $r_0=14$cm,
typical for Cerro Pachon. The resulting contrast for a I=6 mag star in a 1 hour exposure varies from $1\times 10^{-7}$ to 
$2\times 10^{-9}$ across the dark hole region. These analytic predictions are in broad agreement with
more detailed numerical simulations of ExAO performance given in Macintosh et al. 2006, though
the instrument simulated there was somewhat more conservative with a lower throughput and slower
first-stage WFS. It is also assumed that contrast (irrespective of magnitude) hits a systematic floor
of  $5\times 10^{-9}$ from quasi-static image artifacts and Fresnel effects within the instrument. 
The coronagraph IWA is assumed to be $2.5 \lambda/D$. The operating wavelength is \textit{Y} band, 1.04 microns.
Overall, these represent moderately optimistic assumptions about ExAO performance. Planet properties
are as described above, but with giant planet albedo set to 0.1 for $A_E<1$ and 0.2 for $A_E>1$. To allow
direct comparison with space coronagraphs, only mature planets seen in reflected light are considered.
The target sample consists of 500 bright ($I<9$ mag) southern-hemisphere stars selected from the TPF nearby-stars target list. Each star is re-observed four times, representing 200 nights of telescope time. It can be seen that ground-based ExAO is an effective approach to detecting and then spectrally characterizing a large sample of giant planets, though it has essentially no sensitivity to rocky planets.  An ExAO system operating in the near-IR would also detect a very large number of young ($<1$ Gyr) 
self-luminous planets of Jovian mass or above.
%could add a figure for this.  

Figure~\ref{fig:figDEEE} panel (b) shows the results. At these performance levels
rocky planets are only detectable at small angles around 1-3 of the very nearest and brightest stars. In spite of
these limitations, the extremely small inner working angle provided by the large telescope gives it the ability to
detect and spectroscopically characterize a large number of gas and ice giant planets, particularly at smaller semi-major
axis. With a 30-m light-gathering capability, detected planets could be characterized at spectral resolutions from
R=50 to 500.

% ============================================================================
% Begin New PART - DETAILED DISCUSSION OF HABITABILITY deleted
% ============================================================================
\pagebreak

\clearpage

% ============================================================================
% Begin New PART - DETAILED DISCUSSION OF OBSERVATIONAL TECHNIQUES  deleted
% ============================================================================

\pagebreak

\clearpage

% - - - - - - - - - - - - - - - - - - - - - - - - - - - - - - - - - - - - - - - - - - - - - - - - - - - - - - - - - - - - - - - - - - -
% - - - - - - - - - - - - - - - - - - - - - - - - - - - - - - - - - - - - - - - - - - - - - - - - - - - - - - - - - - - - - - - - - - -

% - - - - - - - - - - - - - - - - - - - - - - - - - - - - - - - - - - - - - - - - - - - - - - - - - - - - - - - - - - - - - - - - - - -

% ============================================================================
% Begin New PART - CIRCUMSTELLAR DISKS
% ============================================================================

\pagebreak

\chapter{Circumstellar Disks}

As is evident from the diversity of exo-solar planetary systems
(e.g. orbital parameters, planet sizes, and probably compositions)
the solar system planets populate only a small portion of available
physical parameter space.  To understand the {\it scope} of the observed diversity 
requires development of various planet detection techniques, as discussed
in other areas of this report.  To understand the {\it origin} of the observed
diversity requires investigation of the formation and early evolution 
of planetary systems.  Circumstellar dust and gas is a valuable indirect tracer 
of planets, both during their formation and early orbital migration, as well as
during their multi-billion year lifetimes.

Another way of phrasing it is that young circumstellar disks provide the material 
and the environment (i.e. the initial conditions) for planet formation and early 
planetary system evolution.  Older debris disks 
are proxies for the planets themselves since the dust that is observed 
should be removed from the systems on time scales shorter than the inferred
ages, unless replenished via collisions of planetesimals that 
are stirred from circular orbits by more massive planets. 

The processes associated with star and planet formation are amenable 
to study via observations at all available spatial scales,
at all wavelengths from x-ray to radio, and at spectral resolutions 
from broad band photometry to high dispersion.
Relevant investigations include those of 
molecular clouds, cores, and chemistry, the physics of star formation,
young stars/brown-dwarfs, binaries, and clusters, 
first light, feedback to the cloud, circumstellar disks,
disk accretion/outflow, disk turbulence and possible fragmentation, 
the evolution of protoplanetary gas and dust, the growth and migration 
of planets, and the formation and evolution of debris dust analagous 
to that in the Solar System's Asteroid Belt ($\sim$2-4 AU) 
and Kuiper Belt ($\sim$40-65 AU).

% - - - - - - - - - - - - - - - - - - - - - - - - - - - - - - - - - - - - - - - - - - - - - - - - - - - - - - - - - - - - - - - - - - -
\section{Proto-planetary ("Primordial")  Disks}

Primordial disks have total masses of order the so-called
``minimum mass solar nebula".
They are thus capable in general of forming planetary systems
similar in size and architecture to our own solar system.

The properties of primordial disks are inferred through study of spectral 
energy distributions e.g. Dullemond et al (2007)[in Reipurth et al., 2006; hereafter ``PPV" ],
direct images of the dust e.g. Watson et al (2007)[PPV,
as well as interferometric techniques, e.g.
Millan-Gabet et al (2007)[PPV], Malbet (2008)[arXiv:astro-ph/0708.3359],
and finally via gas diagnostics, which trace most of the mass at young ages
e.g. Dutrey et al (2007)[PPV], Najita et al (2007)[PPV].

Our general picture for planet formation is one of planetary core
accretion followed by gas capture.  This involves
brownian motion, sedimentation, and drift followed by
collision (inelastic) and coagulation, agglomeration and some fragmentation,
gravitational interaction and final oligarchic growth, then gas accretion
which provides most of the planetary mass. 
Key observations include those of the evolution of
the gas to dust ratio, the dust size distribution in the disk, and 
density, temperature, and composition all as a function of $r,\theta,\phi$
in the disk.  

Only the earliest stages of planet formation processes are
discerned from studies of primordial dust and gas. 2MASS and Spitzer
have contributed enormously to our understanding of the depletion of
circumstellar dust. Herschel and ground-based mid-infrared and radio
telescopes will do the same for gas dissipation. On time scales
of just a few million years the observational diagnostics of
proto-planetary material disappear as small particles are incorporated
into larger bodies which grow to terrestrial sized planets and
giant planet cores which then eventually accrete much of the disk gas.
Indeed, it is becoming technically feasible to image
such recently formed giant planets either directly (via AO techniques)
through their thermal or accretion luminosity, or indirectly
(via millimeter interferometry) by imaging the gaps they are predicted
to produce in their disks. Long-baseline interferometry in the infrared 
has mapped disk geometries at spatial resolutions better than an astronomical 
unit, tracing the stellar infall of hydrogen gas and possibly the evaporation of 
water from small bodies. 

Quite rapidly during the planetesimal agglomeration/accretion process,
however, a new regime is entered in which the large bodies
collide and re-generate the small particles which are once
again observable via their thermal re-radiation of stellar photons.
These are known as debris disks which appear as early as 3-10 million
years and persist for many billions of years.  Such debris disks surround
10-20\% of all solar type stars.
Because the survival time for small particles in the radiation and wind
environment of the star is small -- only a few hundred to a few thousand years
-- any such circumstellar dust observed in stars older than a few tens
of millions of years is recently formed and therfore must be continuously
generated in order for the disks to be detected.

The observed debris is a few to a few tens of microns
in size and results from in situ collisions among unseen populations
of planetesimals.  The debris may result from the steady grinding
produced when planetesimal orbits are perturbed either by
the largest embryos in the planetesimal population or
by planetary mass bodies. Or in some cases it may
result from a single catastrophic collision that artifically raises
the mass in small dust particles over steady state evolution values.

A long appreciated class of objects, so-called transitional disks,
have received substantial attention from current facilities; they are
thought to be in the process of clearing their disks, perhaps from
the inside out.  Future work will involve studies of the evolution of
the snow line and understanding of the condensation sequence with
metallic and silicate minerals in the inner disk, water ice and
hydrated minerals in the mid-disk, and carbon dioxide, methane, 
and other ices in the outer disk. 

Seeing the planets directly is difficult because of the contrast levels
and spatial scales (at the nearest star forming regions, 140 pc distant, 
the Sun-planet distance is 0".43 for Neptune, 0".074 for Jupiter, 
and 0".014 for Earth) involved.  Benefiting the case
is that young planets ($<$100 Myr) are 4 orders of magnitude brighter than
their counterparts around older field stars, likely even brighter
during a critical early accretion stage.
Wolf \& Klahr 2005 discuss possibilities.

Rather than observing the planets themselves, progress may be made by 
observing structures in the circumstellar disks.  Such perturbations 
in disks are caused by dynamical interactions between the young
planet/s and the disk, are larger than the planet, hence more readily observed.
Structure or gap inference can be achieved via scattered light imaging 
in the ultraviolet or near-infrared, thermal re-emission in 
the mid-infrared to millimeter, or via in high dispersion $R=10^5$ spectra 
sampling gas lines at 5-30 $\mu$m.
Difficulties in interpretation can arise due to
geometry dependencies, especially inclination, and extinction.
Existing and near-future facilities suffer sensitivity limitations 
for the detection of planet-induced distortions in young disks
(given typical target distances).  Possibilities are discussed by
Wolf et al. (2007) for example.

Another route for planet detection in young disks is interferometric 
detection of the stellar reflex motion.
Disk-planet tidal interactions are expected, which lead to 
orbital migration of young planets.
Once the disk dissipates, the processes associated with young solar system
evolution (planetary orbital evolution, ejection, planetesimal clearing, 
impacts, etc)  become invisible except to dynamical techniques capable
of better than $10 m/s$ precision velocities or $\mu$as astrometry.

% - - - - - - - - - - - - - - - - - - - - - - - - - - - - - - - - - - - - - - - - - - - - - - - - - - - - - - - - - - - - - - - - - - -
\section{Post-planetary (``Debris") Disks}

It is crucial to understand whether most circumstellar disks form planets.   Further  
critical issues following planet formation and the dispersal of 
primordial disks include: the evolution of planetary orbital parameters,
which affects disk structure, the survival of formed planets, and the
planetary system architecture established by ages of $\sim$100 Myr.
Evolution of the detritrus occurs on Gyr time scales with
collisional evolution of planetesimals responsible for the
formation/sculpting of remaining debris dust.

The most obvious signature of our mature
solar system is the large surface area of its dust disk, which emits at 
fractional levels relative to the star of $10^{-7}$ in $\sim$250K dust 
(Asteroid Belt) and $\sim$50K dust (Kuiper Belt). From studies 
of exo-solar debris disk systems, it is inferred that warm debris dust
with 6--10 AU evolves more quickly ($<$300 Myr) than colder more distant
debris dust ($>$Gyr). 
Our current state of knowledge of these processes is described in
Meyer et al (2007)[PPV] and Moro Martin et al (2007)[arXiv:astro-ph/0703383].

For disks in which energy and momentum transport are dominated by radiation, an inner cleared disk geometry is often used to
suggest the presence of a planet.  A sufficiently massive planets 
not only stirs the exterior planetesimals, increasing their velocity
dispersion such that dust-producing collisions occur, but also 
efficiently either ejects or traps into resonances those dust particles that 
cross its orbit. 
Such scenarios prevent or at least impede material from reaching radii
much smaller than those where the dust is in fact detected.
For disks in which collisions dominate, the location of
the planetesimal-stirring planets is less obvious, and requires detailed
modeling of individual systems once they can be spatially resolved
and thus further observationally constrained.

The dust as inferred from mid-infrared continuum excess,
is dominated by a cold component located exterior to $\sim$10 AU, typically.
In a few cases, debris disks are detected around mature stars
that are known to harbor planets.  For these systems, one can move
beyond the degeneracies inherent in modelling
spectral energy distributions of reprocessed photons by using
dynamical simulations that take into account the role of mean motion
and secular resonances of the known planetary companions. 

The detection of planet-induced distortions in debris disks, 
given typical target distances (a few hundred known objects within 50 pc),
is much more promising than for the younger proto-planetary disks.
As for the young disks, observations in both scattered stellar light and
thermal emission from dust are possible, and can indicate material 
in the habitable zone.

\section{Planets and Disks}

Circumstellar dust and gas is a valuable indirect tracer of planets,
both during their formation and early orbital migration, as well as
during their multi-billion year lifetimes.

Primordial disks have total masses of order the so-called
``minimum mass solar nebula".
They are thus capable in general of forming planetary systems
similar in size and architecture to our own solar system.
Only the earliest stages of planet formation processes are
discerned from studies of primordial dust and gas. 2MASS and Spitzer
have contributed enormously to our understanding of the depletion of
circumstellar dust. Herschel and ground-based mid-infrared and radio
telescopes will do the same for gas dissipation. On time scales
of just a few million years the observational diagnostics of
proto-planetary material disappear as small particles are incorporated
into larger bodies which grow to terrestrial sized planets and
giant planet cores which then eventually accrete much of the disk gas.
Indeed, it is becoming technically feasible to image
such recently formed giant planets either directly (via AO techniques)
through their thermal or accretion luminosity, or indirectly
(via millimeter interferometry) by imaging the gaps they are predicted
to produce in their disks. Long baseline interferometry in the infrared has mapped disk geometries at spatial resolutions better than an astronomical unit, tracing the stellar infall of hydrogen gas and possibly the evaporation of water from small bodies. 

Quite rapidly during the planetesimal agglomeration/accretion process,
however, a new regime is entered in which the large bodies
collide and re-generate the small particles which are once
again observable via their thermal re-radiation of stellar photons.
These are known as debris disks which appear as early as 3-10 million
years and persist for many billions of years.  Such debris disks surround
10-20\% of all solar type stars.
Because the survival time for small particles in the radiation and wind
environment of the star is small -- only a few hundred to a few thousand years
-- any such circumstellar dust observed in stars older than a few tens
of millions of years is recently formed and therfore must be continuously
generated in order for the disks to be detected.

The observed debris is a few to a few tens of microns
in size and results from in situ collisions among unseen populations
of planetesimals.  The debris may result from the steady grinding
produced when planetesimal orbits are perturbed either by
the largest embryos in the planetesimal population or
by planetary mass bodies. Or in some cases it may
result from a single catastrophic collision that artifically raises
the mass in small dust particles over steady state evolution values.

For radiation-dominated disks, an inner cleared disk geometry is often used to
suggest the presence of a planet.  A sufficiently massive planets 
not only stirs the exterior planetesimals, increasing their velocity
dispersion such that dust-producing collisions occur, but also 
efficiently either ejects or traps into resonances those dust particles that 
cross its orbit. 
Such scenarios prevent or at least impede material from reaching radii
much smaller than those where the dust is in fact detected.
For collision-dominated disks, the location of
the planetesimal-stirring planets is less obvious, and requires detailed
modeling of individual systems once they can be spatially resolved
and thus further observationally constrained.

The dust as inferred from mid-infrared continuum excess,
is dominated by a cold component located exterior to $\sim$10 AU, typically.
In a few cases, debris disks are detected around mature stars
that are known to harbor planets.  For these systems, one can move
beyond the degeneracies inherent in modelling
spectral energy distributions of reprocessed photons by using
dynamical simulations that take into account the role of mean motion
and secular resonances of the known planetary companions.

% - - - - - - - - - - - - - - - - - - - - - - - - - - - - - - - - - - - - - - - - - - - - - - - - - - - - - - - - - - - - - - - - - - -
\section{Recommendations for Circumstellar Disks}

Understanding of disk gas/dust evolution requires broad wavelength
coverage.  Current and near-term facilities such as 
Spitzer, SOFIA, Herschel, SMA, CARMA, ALMA can contribute.
The goal is to relate disk mass, size, and surface density to the masses,
semi-major axes, and composition of formed planets. Just as 2MASS and Spitzer
have contributed enormously to our understanding of the depletion of
circumstellar dust, so will Herschel, JWST, and ground-based mid-infrared and radio
telescopes do the same for gas dissipation. Millimeter-wave interferometry at new facilities such as ALMA will significantly advance the census of disks with gaps and other structures indicative of the presence of planets.

Imaging of disks, including warps, distortions, and other structures requires
high spatial resolution. This can be achieved via high contrast 
and high spatial optical to near-infrared scattered light observations,
or near-/mid-infrared to millimeter thermal emission 
(e.g. Chara, KI/VLTI, MMT/Keck nulling, ALMA).
Direct detection of young giant planets themselves 
may be possible in atmospheric opacity minima especially 1.6um, 4.5um, 11um
(e.g. via ``extreme" AO coronographic and/or polarimetric techniques at 
e.g. Palomar, Keck, Gemini, JWST in certain ranges of parameter space, and MMT).

%vlti/amber  nir   2-3 mas
%vlti/midi   mir  10-20 mas
%30m        opt/nir  lambda/D
%alma        mm     0.01" at shortest wavelength
%spirit/specs submm
%sma        submm  0.1-0.3"

{\subsection {Circumstellar Disks in the Next Five years}

Historically, this field of study is moderately well supported through major 
missions (e.g. HST, Chandra, Spitzer) and NASA/NSF grants. However, the scope
and the complexity of the research has increased, and the number
of productive researchers has grown significantly
over the past few decades -- especially since the discovery 
of exo-solar planets -- while the available funding has increased 
only moderately, not at all, or has been cut in some prominent programs.

Recommendation: Sustain a healthy level of support for ground-based,
space-based, and theoretical investigations of star and planet formation.
The relevance of these investments has never been higher.

Spitzer results on dust disks will be synthesized and Herschel follow-up 
conducted at higher spatial resolution and somewhat longer wavelengths.
ALMA will begin commissioning.
Gemini Planet Imager will become operational and achieve contrast
levels of $10^{-6}$ to $10^{-7}$ in the inner 1"
(compared to current levels of $10^{-3}$ to $10^{-4}$).

Recommendation: Maintain U.S. involvement in the near-term facilities
Herschel, ALMA, and funding for archival analysis of relevant 
Spitzer, Chandra, Hubble, and ground-based data.  

Recommendation: Support ancillary work in theory (e.g. n--body and hydrodynamic
codes requiring computational horsepower, young planet atmospheres), 
lab astrophysics, and stellar astrophysics (e.g. nearby young star samples,
stellar ages).

\begin{sidewaysfigure}[h]
\centering
\subfigure[] % caption for subfigure a
{\includegraphics[width=10cm]{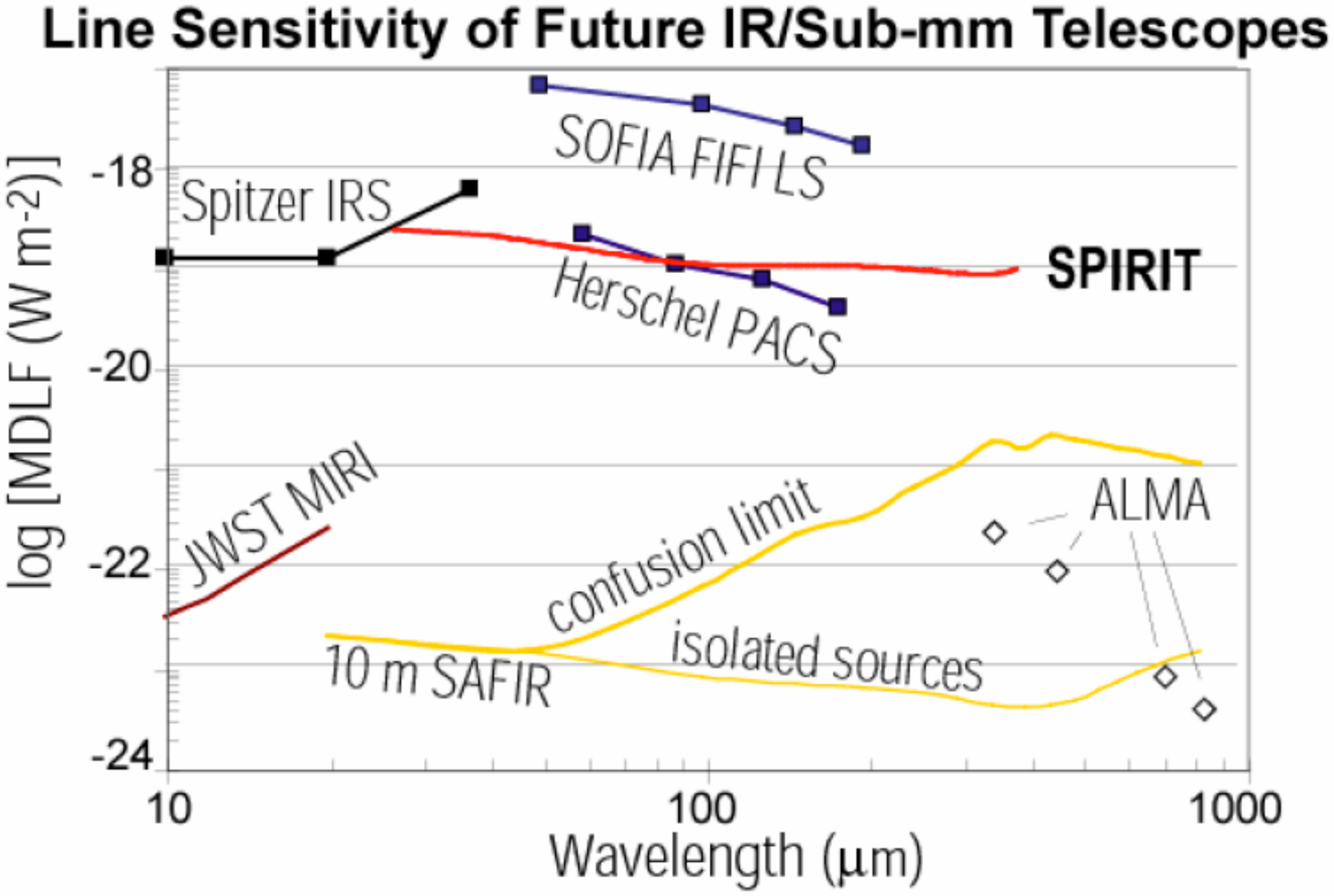}}
%\hspace{0.5cm}
\subfigure[] % caption for subfigure a
{ \includegraphics[width=10cm]{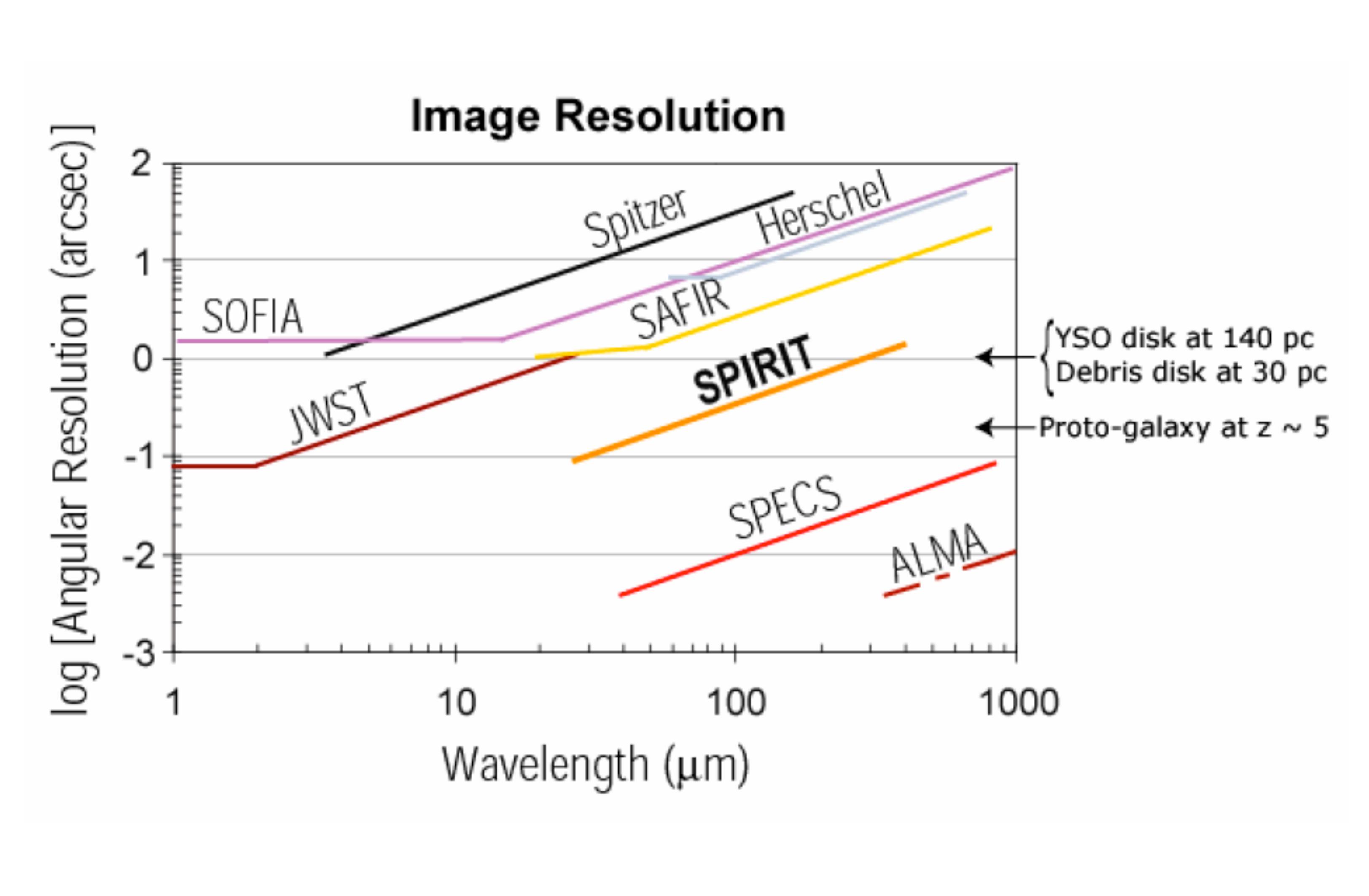}}
\caption{(a) Minimum detectable flux available with
current, next-generation, and
future infrared and submillimeter
facilities relevant to disks.  Quoted values are
for a $10^5$ sec exposure, except $10^4$
sec for SOFIA. (b) Angular
resolution available with
current, next-generation, and
future infrared and
submillimeter observatories.
SPIRIT is a proposal for an infrared space inteferometer; its presence on the figure does not constitute an endorsement by the Task Force. Figure from Leisawitz et al.}
\label{fig:disk-wave} % caption for the whole figure
\end{sidewaysfigure}

{\subsection {Circumstellar Disks in the 6- to 10-year Timeframe}

ALMA will become operational for young disk imaging.
SOFIA is nearing operational status and will be capable of
high dispersion gas spectrosocpy of the star formation process as well as
thermal imaging of disks at higher spatial resolution than Spitzer and
equivalent to Herschel. (Fig. ~\ref{fig:disk-wave})

Recommendation: Implement next-generation high spatial resolution
imaging techniques on ground-based telescopes (AO for direct detection
of young low mass companions and interferometry for disk science).

Recommendation: Invest in high resolution mid-infrared spectrographs
on next generation large aperture (20-30m) telescopes.

{\subsection {Circumstellar Disks in the 11- to 15-year Timeframe}

JWST thermal imaging capabilities will be exploited.  
Ground-based ELT's will be operational and with ``extreme AO" coronographic 
techniques can image young giant planets.  In the same way that Keck contrast 
and sensitivity to companions exceeds that of HST, a 30m telescope predicted
to have contrast levels of $10^{-7}$ to $10^{-8}$ in the inner 1" may be
operational and will rival JWST in certain regimes of parameter space,
especially inner working angle.

Recommendation: Invest in technology for a sensitive (equivalent to Spitzer
and Herschel) far-infrared interferometric space mission 
 to achieve the resolution 
of ALMA closer to the peak of the dust spectral energy distribution.

Recommendation: Invest in technology for a sensitive far-infrared
large aperture telescope to extend sensitivity
orders of magnitude below Herschel at comparable wavelengths, and 
near/below JWST and ALMA sensitivities at adjacent wavelengths.

% ============================================================================
% Begin New PART - APPENDICES
% ============================================================================

\pagebreak

\cleardoublepage

\bigskip

\bigskip

\chapter{Appendices}

\setcounter{section}{0}

% - - - - - - - - - - - - - - - - - - - - - - - - - - - - - - - - - - - - - - - - - - - - - - - - - - - - - - - - - - - - - - - - - - -
\section{Habitability}

The most exciting prospect for the near- to mid-term future is to search for evidence of habitable planets around nearby stars and then ultimately to determine through spectroscopy whether any of them are actually inhabited. Any such search must begin with a definition of what is meant by the term ``habitable." Here, we shall take the conservative approach and assume that life elsewhere is carbon-based, as it is on Earth, and that it requires at least the transient presence of liquid water. One should, of course, be cognizant of other possibilities (National Research Council, 2007), but if we hope to focus our initial search efforts at all, then it makes sense to begin with this assumption.

This assumption, by itself, is not very limiting. Both liquid water and life may exist in the subsurface on Mars, as well as on Jupiter's moon Europa. It is difficult or impossible to detect such life remotely, however, especially at the limited spectral resolution anticipated for future planet-finding telescopes. Hence, we further restrict our attention to planets on which liquid water is present at the planet's surface. Surface life can alter the composition of a planet's atmosphere in detectable ways, especially if it is powered by photosynthesis, as it is on Earth. The region around a star in which surface liquid water is stable is termed the habitable zone (HZ) or ecosphere. We shall use the former term here, as this also allows us to define the continuously habitable zone (CHZ) as the region around a star that remains habitable for a specific length of time as the star evolves (Hart, 1979; Kasting et al., 1993). Some authors, e.g., Franck et al. (2000) have gone one step further and attempted to define different habitable zones for different types of planets. However, for an initial search in which little is known about the planets at the start, it is probably best not to attempt this level of detail.

\subsection{Habitable zone boundaries within our Solar System}

\subsubsection{Theoretical Constraints}

Estimating the boundaries of the habitable zone for our Sun and for other stars is a tricky modeling task, and the calculations that have been done so far, e.g. Kasting et al. (1993), which were done with one-dimensional climate models, must be considered preliminary. Such 1-D calculations can account for the effects of clouds only parametrically, which limits their accuracy. Fortunately, information is also available from observations of other planets in our own Solar System, especially Mars and Venus, and these empirical constraints can help guide our thinking. Future 3-D climate modeling may eventually improve our estimates of habitable zone boundaries, although as all climate modelers are aware, clouds (which are often sub-grid-scale in size) still pose significant problems in 3-D models.

The physical processes that determine the boundaries of the HZ are reasonably well understood. The inner edge of the HZ is thought to be set by loss of water, either by way of a so-called ``runaway greenhouse" in which the entire ocean evaporates to form steam (Rasool and DeBergh, 1969; Ingersoll, 1969), or through a ``moist" greenhouse in which the ocean remains liquid but the stratosphere becomes wet (Kasting, 1988). The corresponding inner boundaries of the HZ are 0.84 AU and 0.95 AU, respectively (Kasting et al., 1993). In either case, photodissociation of stratospheric H$_2$O by stellar UV radiation, followed by escape of hydrogen to space, causes the ocean to be lost with a geologically short period of time (tens to hundreds of millions of years). The actual time scale for escape is uncertain because the details of the (hydrodynamic) hydrogen escape process are not that well understood (Tian et al., 2005; Catling, 2006). This is another area where future advances in modeling might lead to better understanding.

For an uninhabited planet, the outer edge of the HZ is thought to be determined by condensation of CO$_2$. Rocky planets towards the outer edge of the HZ are expected to build up dense CO$_2$ atmospheres, provided that they are endowed with substantial carbon inventories and with enough internal heat to sustain volcanism (Kasting et al., 1993). They are constrained to do this by a negative feedback loop:  Lower surface temperatures lead to slower evaporation and rainfall, causing silicate weathering to slow down. Weathering of silicate rocks on the continents, followed by deposition of carbonate sediments in the ocean, is the primary long-term loss process for CO$_2$. If silicate weathering becomes too slow, volcanic CO$_2$ should accumulate in a planet's atmosphere until an active hydrologic cycle is re-established. Hence, planets that are farther out in their star's HZ should tend to accumulate enough CO$_2$ in their atmospheres to keep liquid water stable at their surfaces and to keep a hydrological cycle running. This negative feedback loop also helps to keep a planet's climate stable as its parent star brightens during its main sequence lifetime. Without this feedback loop, HZs around stars would be extremely narrow (Hart, 1979). With this feedback loop included, HZs appear to be relatively wide (Kasting et al., 1993). However, if a planet is too far from its parent star, CO$_2$ itself begins to condense and the stabilizing feedback disappears. For our Sun, the corresponding ``maximum greenhouse" limit occurs at ~1.67 AU, somewhat beyond the orbit of Mars. The HZ might be further extended by the presence of other greenhouse gases, e.g., CH$_4$; however, no stabilizing abiotic feedback mechanism has been identified in this case. On Earth, most CH$_4$ is produced by methanogenic bacteria. If analogous organisms were present on an extrasolar planet, this might allow it to remain habitable beyond the boundaries of the conventional, CO$_2$/H$_2$O-based HZ.

\subsubsection{Constraints from Models}

Although climate models can be used to estimate the boundaries of the HZ, the theoretical estimates are not very reliable. The 1-D climate models that have been used to estimate these boundaries (Kasting, 1988; Kasting, 1991) are incapable of accurately simulating the effects of clouds on the planetary radiation budget. H$_2$O clouds are expected to play a major role in the radiation balance near the inner edge of the HZ, while CO$_2$ clouds may play a similar role near the outer edge. A CO$_2$ cloud of the right optical depth covering 100 percent of a planet's surface could extend the outer edge of the HZ to 2.0-2.4 AU (Forget and Pierrehumbert, 1997; Mischna et al., 2000). This does not seem likely, however, as CO$_2$ clouds (like H$_2$O clouds) form mainly by condensation in updrafts, and hence are unlikely to fill a planet's entire atmosphere. By contrast, the sulfuric acid clouds on Venus and the organic haze on Saturn's moon Titan are both photochemically induced and can therefore yield nearly complete cloud coverage.

\subsubsection{Empirical Boundaries}

An alternative way of estimating HZ limits is to base them on observations of Venus and Mars. Venus' semi-major axis is 0.72 AU. Radar maps of Venus' surface suggest that liquid water has not been present there for at least the last 1 billion years (Solomon and Head, 1991). The Sun was ~8 percent dimmer back at that time, according to standard solar evolution models (e.g., Gough, 1981). Thus, the solar flux at Venus' orbit then was equal to that at a distance of 0.72 AU(1/0.92)$^{0.5} \simeq$ 0.75 AU today. This, then, is a more optimistic estimate for the inner edge of the HZ, as compared to the 0.84-0.95 AU predicted by theory. But the empirical estimate for the inner edge boundary is almost certainly over-optimistic, as Venus is not, and was not then, habitable.

The empirical outer edge of the HZ is based on the observation that Mars, which orbits at 1.52 AU, looks as if it may well have been habitable at or before 3.8 billion years ago ( Pollack et al., 1987). The Sun is thought to have been ~75 percent as bright at that time. Hence, the solar flux hitting Mars back then was equivalent to that at a distance of 1.52 AU(1/0.75)$^{0.5} \simeq$ 1.8 AU. This distance may be taken as an empirical outer edge of the HZ. It should be borne in mind, though, that planets might still be habitable at even larger orbital distances if other greenhouse gases are present or if CO$_2$ clouds warm the surface, as discussed above. And early Mars looks as if it was indeed habitable, which also suggests that this limit is too conservative.

\subsection{Habitable zone boundaries for other stars}

The HZ boundaries quoted above apply to planets within our own Solar System. However, as the characteristics of the parent star changes, so do the possibilities for habitability.  

\subsubsection{General considerations}

Our Sun is a 4.5 billion-year-old G2V star with an effective temperature of ~5700 K. Planets orbiting earlier- or later-type stars will have their HZ boundaries shifted in a predictable way by the change in the star's spectrum. Blue stars emit more of their radiation at short wavelengths where absorption by atmospheric gases (particularly H$_2$O) is low and where Rayleigh scattering is high. Such planets should therefore have higher albedos than Earth, all other factors being equal. This moves their HZ boundaries inward by approximately 10 percent (Kasting et al., 1993, Table III). Planets orbiting red stars should have lower albedos than Earth for the same reason, shifting their HZ boundaries outwards by roughly this same amount. 

Most current HZ models 
are derivatives of the one published by Kasting et al. (1993). The conclusion of 
that study was that the HZ around our own Sun extends from at least 0.95 AU out 
to $\sim 1.4\,\rm AU$. Although the (conservatively estimated) inner edge of the 
HZ is not far from Earth's current orbital distance, the outer edge is 
relatively far out in this model because it is assumed that volcanic $\rm CO_2$ 
will accumulate in the atmosphere of a rocky planet as its surface 
temperature cools. Indeed, this outer limit is probably too close, as it assumes that the $\rm CO_2$ clouds that form near the outer 
edge of the HZ would cool the climate. If such clouds warm the surface, as 
predicted by Forget and Pierrehumbert (1997) (also Mischna et al., 2000), then 
the HZ outer edge may lie as far out as 2.0 AU or beyond. Again for concreteness the Task Force defines the HZ around our own Sun as extending from 0.75 AU to 1.8 AU.

\subsubsection{Special considerations for planets around M-dwarfs}

Planets orbiting M-dwarfs pose special problems for planetary habitability. The problem that has been discussed most frequently and for the longest time (e.g., Dole, 1964) is that of tidal locking. Planets whose semi-major axes lie within the dotted curve in Figure~\ref{fig:hab-zone} are expected to develop captured rotations within the lifetime of our own Solar System, 4.6 billion years (Kasting et al., 1993). An airless planet orbiting in this manner would have one side that was permanently hot and sunlit and one side that was cold and dark. However, transfer of heat by winds and/or ocean currents could lessen this temperature difference, particularly if the planet's atmosphere is dense and contains radiatively active gases such as CO$_2$ and H$_2$O (Joshi, 2003). Some planets may also rotate non-synchronously within this distance for various reasons. Mercury, for example, rotates 3 times for every 2 orbits because it is trapped in a spin-orbit resonance.

A second potential problem for M-dwarf planets is the loss of atmosphere caused by sputtering from stellar winds. M-dwarfs, young ones in particular, are much more chromospherically active than our Sun and have frequent large flares. During these times, the stellar wind can be very intense, and it may be able to strip off the atmosphere of planets orbiting within the star's HZ (Khodachenko et al., 2007). This is particularly true if the planet is spinning slowly, so that it lacks a magnetic dynamo and an accompanying intrinsic planetary magnetic field that might hold off the stellar wind.

Recently, Lissauer (2007) has pointed out two other potential habitability problems for M-dwarf planets. First, their future HZs are predicted to be hotter than those around G and K stars during the time when planets are forming because accretion times are faster and because low-mass stars remain brighter than their main sequence luminosities for longer periods of time after they have formed. This factor by itself may cause their innermost planets to be deficient in volatiles compared to the terrestrial planets in our Solar System. Second, impact velocities of planetesimals during accretion are predicted to be higher than around G and K stars, again because the HZ is so close to the star. High impact velocities may also result in less retention of volatiles.

Despite these acknowledged problems, rocky planets orbiting within the HZs of M-dwarfs could still be habitats for life. Only by observation will we learn whether any of them are actually inhabited. As discussed elsewhere in this report,such planets may be detected in the very near future either by ground-based radial velocity searches or from transits. And they may have another feature of great interest: the simultaneous presence of $O_2$ and CH$_4$. Lovelock (1965) suggested that the simultaneous detection of these molecules would be the strongest possible remote evidence for life. M-dwarf planets may exhibit this signature. As a consequence of the dearth of near-UV radiation from these stars, the tropospheric chemistry of an accompanying  planet would be very different from that of Earth (Segura et al., 2005). In particular, the hydroxyl radical, OH, is expected to be much less abundant. In an $O_2$-rich atmosphere like that of Earth, OH is the main photochemical sink for CH$_4$. Hence, given a similar biological source as on Earth, CH$_4$ on an M-dwarf planet could accumulate to very high concentrations - hundreds of ppmv, as compared to just 1.7 ppmv on the modern Earth. If such an atmosphere could be observed spectroscopically, and if the planet being observed was close to the size and reflectivity of the Earth, it may be possible to simultaneously observe both $O_2$ (or $O_3$) and CH$_4$. 

\section{Probability of Habitable Planets - definition of $\eta_\oplus$}

A considerable amount of work has been done over the past 15 years to estimate 
the chances of finding habitable planets.   The prospects for finding habitable planets 
are generally characterized by a parameter
 $\eta_\oplus$ (pronounced ``eta sub earth"),  the fraction of stars that have at least one potentially habitable planet. For concreteness the Task Force defines a potentially habitable planet as one that is close to the size of the Earth
 and that orbits within the stellar habitable zone. Close to Earth-sized means between 1/2 and twice the radius of the Earth or in terms of mass between 0.1-10 times the mass of the Earth. These two definitions are equivalent if a fixed density equal to that of the Earth is adopted. 
 
The circumstellar 
habitable zone itself is defined as the region around a star in which such a planet can maintain liquid water on its surface.  Although planets (or 
moons) outside this zone could conceivably support life in subsurface 
environments, it is unlikely that such life could be detected remotely. Hence, 
the conventional habitable zone is the region that is most relevant to future 
searches for extraterrestrial life. The width of the habitable zone 
has been studied with 1-D climate models, and by examining the nature of the two terrestrial planets just inward and just outside the orbit of the Earth. Venus orbits at 0.72 AU and thus receives 
$\sim 1.93$ times as much sunlight as Earth. It is almost entirely devoid of 
water and is thought to have lost its initial water by way of a runaway 
greenhouse effect. It provides empirical evidence that the HZ inner 
boundary lies outside of this distance, but how far is not known. Evidence that Mars had a warmer climate early in its history in spite of the fact that the Sun's luminosity was 30\% less at the time suggests that much of its evolution toward present day conditions was driven by its small size; an Earth-sized planet at the orbit of Mars might have remained habitable, indeed out to perhaps to 1.8 AU based on climate models. 

It therefore is reasonable to define the circumstellar habitable zone as extending from 0.7-1.8 AU for a star with the present luminosity of the Sun, scaling as the square root of the luminosity of the star. We do not consider the tighter bound associated wit the so-called continuously habitable zone, defined so that a planet within it is stably habitable for billions of years. Arguably this will be narrower than the habitable zone we use here but by what amount is very poorly understood.

 It is exciting that researchers have by now come very close to 
detecting Earth-sized planets within the HZs of their parent stars. The most 
interesting discovery to date is a 5.1 Earth-mass planet orbiting the star 
Gliese 581 (Udry et al., 2007). This planet, Gliese 581c, was discovered using 
the radial velocity technique; hence, its measured mass is a lower limit on its 
true mass. The star itself is an M3V star with a mass of $0.31\,\rm M_{\odot}$ 
and a luminosity of $0.0135\,\rm L_\odot$. The semi-major axis of the planet's 
orbit is 0.073 AU; hence, the solar flux incident upon it is $\sim 3470\,\rm 
W\,m^{-2}$, or 2.55 times that at Earth's orbit. This particular planet is 
almost certainly not habitable, despite early press releases that suggested that 
it might be. Nonetheless, even though Gliese 581c is 
not habitable, the fact that astronomers have identified such a planet suggests 
that many more such planets will be found by the radial velocity technique over 
the next 5--10 years. If so, some of these will very likely be within the HZs of 
their parent stars.

The value of  $\eta_\oplus$ is not presently known. It is considered to 
be a key parameter for the design of future terrestrial planet characterization 
missions, such as NASA's proposed Terrestrial Planet Finders. We consider a large value of $\eta_\oplus$ to be 0.1 or higher; suh a value as shown below allows for a high probability of success in detecting and characterizing Earth-sized or Earth-mass planets that might be habitable. It requires that the probability density for planets go up by an order of magnitude in logarithmic space from the known Doppler planets, but in view of the incompleteness of Doppler searches in terms of both mass and semimajor axis, it would not be implausible to find $\eta_\oplus$ to be this high or even higher arond Sun-like stars.

\subsection{Characterizing Habitable Planets Spectroscopically}

Once planets have been identified within the HZs of nearby stars, interest will 
quickly shift to characterizing these planets spectroscopically. In the optical and near-infrared part of the spectrum, coronagraphy with or without an external occulter is the techniquer of choice, whereas in the thermal infrared nulling interferometry would be used.  

In a visible/near-IR mission, the initial goal would be to look for the 
presence of gases found in Earth's atmosphere, specifically, $\rm O_2$, $\rm 
O_3$, and $\rm H_2O$. Figure ~\ref{fig:VEMfullres} shows what the Earth should 
look like at these wavelengths. Mars and Venus are shown for comparison. For 
Earth, a spectrometer with a resolution, $R = \lambda/\Delta\lambda$, of 70 
should be able to identify the strong $\rm O_2$ A band at $0.76\,\rm \mu m$, 
along with several different $\rm H_2O$ bands. The $\rm H_2O$ bands get stronger 
and stronger as one moves out into the near-IR. The broad ozone band in the mid-
visible is considered difficult to detect, as it is easily masked by clouds and it is difficult to 
discriminate from the Rayleigh scattering increase in albedo at short 
wavelengths.

The information that is available at longer wavelengths is somewhat different. 
Figure ~\ref{fig:Thermal_IR_Venus_Earth_Mars} shows thermal-IR spectra of Venus, 
Earth, and Mars. For Venus and Mars, a low resolution spectrum shows only the 
strong 15-$\rm \mu m$ band of $\rm CO_2$. For Earth, one sees this same $\rm 
CO_2$ band, but one also sees evidence for $\rm H_2O$ at $6.3\,\rm\mu m$ and in 
the rotation band longward of $\sim 17\,\rm \mu m$ . $\rm H_2O$ is considered 
difficult to identify on an extrasolar planet, however,  because of particular technical difficulties occurring at each end of the wavelength region covered by a TPF-I type mission. 
Importantly, though, $\rm O_3$ (ozone) has a strong absorption band centered at 
$9.6\,\rm \mu m$ that should be relatively easy to identify. $\rm O_3$ is formed 
photochemically from $\rm O_2$, and most of Earth's $\rm O_2$ comes from 
photosynthesis. Hence, the detection of $\rm O_3$ in the thermal-IR or of $\rm 
O_2$ itself in the near-IR could provide indirect evidence for life.

As soon as one makes a statement like the one above, one must be careful 
to qualify it. There may be ``false positive'' for life, that is, it may be 
possible under some circumstances to accumulate $\rm O_2$ abiotically in a 
planet's atmosphere without any biological production. An example is an early 
Venus-type planet that loses a lot of water quickly by way of a runaway 
greenhouse. In this scenario, water vapor is photolyzed by solar ultraviolet 
radiation. The hydrogen escapes to space, and most of the oxygen remains behind 
in the planet's atmosphere. Although we would hopefully be able to identify such 
an abiotic planet by its location near or inside the inner edge of the star's 
habitable zone, it is evident that one would need to carefully consider all 
possibilities before drawing any conclusions about the presence of life on a 
planet based on this type of evidence.

If we go one step further, the best spectroscopic evidence for life on 
another planet would be the simultaneous detection of $\rm O_2$ (or $\rm O_3$) 
along with a reduced gas such as $\rm CH_4$ or N$_2$O (Lovelock, 1965). Because of the difficulty of detecting CH$_4$ and N$_2$O, it is unlikely that a first-generation TPF-type mission  will be able to measure this combination of gases in an 
atmosphere. However, 
this might make a good goal for a follow-up, high resolution direct-detection spectroscopic 
mission that could be done at some later time.

\section{Doppler spectroscopy: Optical }

Radial velocity planet search programs employ high resolution spectrometers to measure 
Doppler-shifted light from M-dwarfs. This technique has been described in popular literature 
as the ``wobble'' method because the back-and-forth velocity of the star is used 
to infer the presence of an unseen planet. 

Fundamentally, this is a dynamical technique; the gravity of the planet tugs the host star  
about a common center of mass. Our ability to detect the planet with RV measurements depends 
on the amplitude of the stellar velocity, which is a function of the mass of the 
star, the mass of the planet, and the separation between the two objects. The velocity 
semi-amplitude, $K$, can be conveniently expressed in familiar units: the orbital period, $P$,
in years, the stellar mass, $M_{star}$, in solar mass units and the planet $M_{p}$, in 
Jupiter mass units: 

\begin{equation}
K = { 28.4329 \over \sqrt{(1-e^2)} } {\left( P \over 1 \ {\rm yr} \right)^{-1/3} } 
 { \left( M_{p} \sin i \over M_{Jup} \right) }  
 { \left( M_{star} \over M_{\odot} \right)^{-2/3}} 
\end{equation}

The velocity amplitude is attenuated by the constant factor $\sin i$, which describes the 
tilt of the orbital plane relative to our line of site.  Orbital inclination is defined
to be zero for face-on orbits, and $90^\circ$ for edge-on orientations. Orbital inclination 
is random, and orbits that are nearly face-on comprise a small 
fraction of random orbits inscribed on a sphere.  The probability that an orbit has an inclination 
in the range $\Delta i = i_1 - i_2$ is given by: ${\cal P}(\Delta i) = \cos i_1 - \cos i_2$. Thus, 
$\sim 90$\% of all orbits are expected to have an inclination greater than $30^\circ$; for these 
orbits, the measured velocity amplitude is within factor of two of the true stellar velocity.

For massive planets with large velocity amplitudes, orbital inclination has a small 
impact on the detectability of the planet. To illustrate, we adopt 5 \ms as the threshold for 
detectable velocity amplitude. For this case, 90\% of Saturn-mass planets closer than 1 AU (or 
jovian-mass planets out to a few AU) with randomly oriented inclinations still present 
Doppler amplitudes greater than 5 \mse and are detectable. 
However, as the reflex stellar velocities decline below 10 \ms (e.g., stars with sub-Neptune-mass planets 
in short period orbits or jovian mass planet beyond 5 AU), there is a steep drop in the detectability 
of planets, simply because orbital inclination 
attenuates the observed velocity amplitude to less than 5 \mse. Observational incompleteness 
arising from orbital inclination is statistically $\sim 50$\% for planets with associated 
velocity amplitudes below 10 \mse. This compounds the detection challenge for  
sub-Neptune mass planets in short-period orbits.  If planets with masses less than 
10 \mearth exist in nearby habitable zones,  only spaceborne astrometry or direct detection likely would be able to find them.  (but see Brown, 2008). 

Using optical wavelength spectrometers, the Doppler technique is most robust for 
chromospherically inactive stars with spectral types ranging from 
late F-type stars to mid M-type stars.  Chromospherically active stars (i.e., young stars) 
are problematic because the photospheric noise can contribute background radial velocity 
variations that far exceed dynamical signals represented by the current ensemble of 
detected exoplanets.  Only the most massive and closest planets (rare in the current 
exoplanet population) can produce detectable reflex stellar velocities. 

Radial velocity 
precision is seriously degraded for more massive main sequence stars 
for three reasons: the hotter stars have fewer absorption lines, the lines 
that are present are broadened by the more rapid stellar rotation
typical of massive stars, the massive stars are statistically younger and have more 
chromospheric activity and intrinsically higher astrophysical ``jitter.'' The best way 
for current Doppler techniques to detect planets orbiting more massive stars is as they 
cross the subgiant branch.  As subgiants, the stellar photosphere are cooler, rotation has 
slowed, and the intrinsic photospheric jitter is only slightly worse than for main sequence stars.
Surveys of stars on the giant branch have also been successful at detecting exoplanets,
despite the increase in background photospheric noise to about 30 \mse.

Stars that are later in spectral type than about M4V pose a challenge for optical 
spectroscopy because they are intrinsically faint at optical wavelengths.
These stars are best observed with IR spectroscopic techniques (discussed below).

Radial velocity detection efficiency is strongly enhanced by improving Doppler precision
and the current state-of-the-art precision is now about 1 \mse.  Empirically, 
velocity semi-amplitudes of detected planets are more than five times the single measurement precision. 
In addition, Doppler observations must also span at least one 
orbital period.  As a result, Doppler planet hunters work their way out to wider separations by extending the time 
baseline of the observations.  The Sun exhibits reflex velocities of 
0.09 \ms (9 \cmse) and 12.5 \ms
from Earth and Jupiter respectively. Of course, the periods of these velocity variations 
are quite different: one year for the Earth, and 11.86 years for Jupiter. This example 
illustrates the challenge for radial velocity techniques: even with 1 \ms Doppler 
precision and an edge-on orbit, the Earth would not be detectable; and while the signal 
from Jupiter could be detected, this detection would require twelve years of stable, 
high precision Doppler measurements to map out the planet orbit. Most Doppler programs 
are run on shared facilities that undergo regular hardware upgrades. While upgrades 
improve both the data quality and the Doppler precision, it is worth noting that long 
time series data tends to be uneven in quality (with older data having poorer precision).

The fact that Keplerian orbital periods increase with semi-major axis makes it possible 
to extract multi-planet signals as long as each planet induces a detectable reflex 
velocity in the host star. Of the known exoplanet systems, about one third are known 
to have at least 2 planets orbiting the host star.  This surely represents a lower 
limit; given the time baseline of most Doppler programs, it is likely that true 
Jupiter analogs would have eluded detection in all except a small fraction of observed 
stars.

\subsection{Current Status} 

There are a handful of high resolution optical spectrometers that have
demonstrated routine velocity precision approaching or better than 1 \ms
(HARPS, Keck-HIRES, AAT-UCLES, VLT-UVES) and additional telescopes
armed with high resolution spectrometers have demonstrated RV precision
in the 2-7 \ms range (e.g., HET, Lick-Hamilton, Magellan-MIKE, Subaru-HDS,
Thuringia State Observatory). The Doppler planet search surveys at all of
these telescopes are currently oversubscribed and demand for high resolution
spectroscopy is expected to increase, both because the detection of low mass
planets requires a substantially larger number of observations and
because of an increase in demand for verification from other techniques.
For example, it is not clear how high precision Doppler programs will be able to
follow up the flood of transit candidates expected from the Kepler mission.

Precision radial velocities should continue to serve as a front line detection 
technique and will provide support for space missions. Prescreening targets 
with radial velocity observations can identify the more massive components in planetary 
systems and can reduce the number of free parameters in multi-planet astrometric orbits 
or identify good targets for direct imaging missions.  This is one of the assessment 
tools that will help with the interpretation of transit candidates and astrometric 
detections.

\subsection{Current Limits} 

As noted, state-of-the-art Doppler precision is about 1 \mse, yet it is not clear what 
limits that precision. As an example, the triple Neptune system orbiting 
HD 69830 (Lovis \etal, 2007) is a K0V star with a triple Neptune system detected 
with 74 Doppler observations with single measurement precision of 0.6 \mse. 
The radial velocity amplitudes are 3.5, 2.7, and 2.2 \ms for planets with  
orbital periods of 8.7, 31.6, and 197 d respectively. The RMS 
of the residuals multi-Keplerian fit is an impressive 0.8 \mse. In another 
case, the detection of a 14 \mearth mass planet in a 9.55 d orbit ($K = 3$ \mse) was 
discovered as part of an asteroseismology study of $\mu Ara$. These low RV amplitude  
detections demonstrate the beginning of a new strategy in Doppler planet searches: 
the use of high cadence observations to beat down the single measurement 
precision by the square root of the number of observations. The ability of 
this strategy to push to amplitudes well below the single measurement precision
and to detect Earth analogs hinges on (1) the ability to control 
systematic noise on multi-year timescales and (2) the (poorly known) 
degree of low frequency ($< 10^{-4}$ mHz) photospheric noise.
 
Excellent data exist for the Sun regarding high frequency noise sources 
arising from p-mode oscillations with periods of minutes, but it has been 
intrinsically difficult to obtain the same quality data sets for the disk-integrated
Sun. We do not know if timescales associated with (for example) convective overturn will 
impede extremely high cadence observations of otherwise chromospherically quiet stars. 

UVES observations of $\alpha$ Cen A, a G0V main sequence star, similar in age to the Sun,  
show p-mode oscillations with amplitudes varying from 1 to 3 \ms (Butler \etal 2004). 
In contrast, observations of $\alpha$ Cen B show that the peak amplitude 
in this K0V star is much lower, reaching only 0.08 \ms (Kjeldsen \etal 2005). 
A total of 179 observations, with a median cadence of 58 seconds was obtained 
for $\alpha$ Cen B, yielding an average noise in the power spectrum of 0.0149 \mse, or 
1.49 \cmse.  As a model for exoplanet searches, this suggests that radial velocity 
planet searches could press to much lower mass planets, albeit requiring a large 
investment of ground-based telescope time. The known photospheric noise sources 
for $\alpha$ Cen A and B lie in a frequency range of 7.5 to 15 mHz, far below 
the frequency range associated with most planetary systems.

An observing strategy to obtain thousands of RV measurements has not yet 
been carried out because it would seriously impact the highest 
precison Doppler programs that each monitor of order 1000 stars. However, it 
is important to better understand the floor to Doppler precision and the 
power spectrum of noise from typical stars.  This will demonstrate the degree
to which Doppler observations can support space born astrometry and direct imaging 
missions.

\subsection{Synergy with Other Techniques} 

The Doppler technique is a powerful complement to other detection methods, such as photometric 
transit observations. The first transiting planet was detected with photometric 
follow-up of the Doppler-detected planet, HD 209458b.  The planet with the largest 
known core, 70 \mearthe in the Saturn-mass planet HD 149026b, was detected first in a radial 
velocity survey. The planet with the smallest known radius, GJ 436b, was discovered 
with Doppler observations. And most recently, a planet with high eccentricity in a 
22 day orbit, HD 17156b, was discovered in radial velocity surveys and observed to 
transit during the close periastron passage.  The small transit depths of 
HD 149026b, GJ 436b, and HD 17156b would have been challenging to detect first with  
photometric surveys. For the $\sim 20$ (a number that is steadily growing) 
transits detected first in photometric surveys, radial velocity measurements 
were critical in establishing orbital parameters and planet masses. 

It is worth noting that while transiting planets provide a wealth of 
information about exoplanet structure and atmospheres that cannot be derived 
any other way, statistically, only 2\% of the exoplanets detected with Doppler 
observations are expected to transit.  Currently, 20\% of Doppler-detected 
exoplanets have orbits shorter than 10 days where 10\% are expected 
to transit. In keeping with this statistical expectation, 5 of 250 Doppler-detected 
planets have been observed in transit; 98\% of planets detected in radial 
velocity surveys could not have been detected with any other current technique. 

One technique that could begin confirming and extending Doppler detected planets
over a similar range of semi-major axes is spaceborne astrometry. More than simply 
resolving the $\sin i$ ambiguity, spaceborne astrometry has the ability to detect 
planets with masses down to 1 \mearth at habitable zone distances from host stars. 
If planetary systems are complex (like the solar system, with 8 planets; 55 Cancri,
with 5 planets; Upsilon Andromedae, with 3 planets; $\mu$ Arae, with 4 planets;
and GJ 876 with 3 planets) then high precision Doppler observations will help 
constrain the number of free parameters in the astrometric solution.  Our own 
solar system is dynamically full of planets; there are few niches where even a low
mass planet could exist. It has been suggested that this may be the case with 
other planetary systems.  If true, this bears directly upon $\eta_{\oplus}$ and would 
transform our understanding of planet formation and evolution, providing important 
guidelines for searching for planets. 

\section{Doppler spectroscopy: Near-infrared (IR)} 

Radial velocity surveys of low mass stars have a strong appeal. Because the velocity 
semi-amplitude, $K$, is inversely proportional to ${\rm M_{star}^{2/3}}$, 
$K$ increases by a factor of $\sim 5$ when comparing identical planets orbiting 
late M-type stars to those orbiting the Sun. Thus, for a fixed
Doppler precision of $1 m s^{-1}$, lower mass planets (and planets in wider 
orbits) should be most easily detected around the lowest mass stars. 

Surveys of low mass stars are also appealing because these stars constitute 
the bulk population of our galaxy.  Among the $\sim$150 star within 8 parsecs, 
$\sim$120 are M-dwarfs, while only 15 are G dwarfs.  

The key challenge for traditional Doppler surveys is that M-dwarfs are 
intrinsically faint at visible wavelengths. Approximately 200 M-dwarfs are on 
collective optical Doppler surveys; these stars require relatively longer exposure 
times than typical FGK type stars, making them more expensive to observe. 
However, the flux of M-dwarfs peaks in the $1 - 2 \mu$m wavelength range. As a 
result, M-dwarfs are bright targets for IR spectroscopy, making this
an attractive new technique to survey several hundred M-dwarfs. An M4V star 
at 20 parsecs has a V-band magnitude of 14, which is beyond the reach of 
optical surveys, but a J-band magnitude of 9, making it accessible with 
IR spectrscopy on a large telescope. This is still a Doppler technique, 
measuring changes in radial velocities, but now using spectral lines at 
infrared wavelengths. 

Precision radial velocity surveys at IR wavelengths also have important
synergy with Kepler. Low mass stars also have smaller stellar radii, so 
transit event from Earth-radius planets will be more easily detected by Kepler
around late type stars. IR spectroscopy would enable followup and 
characterization of these critical Kepler candidates.   

Simulations using model atmospheres (McLean \etal 2007) show that photon-limited 
precision in YJH bands should be achieveable down to 1 m s$^{-1}$.  In practice, 
IR spectroscopy has yet to demonstrate this precision.  Some of the challenges are:

\begin{enumerate}
\item {\it Telluric Absorption.} 
Unlike optical windows, the infrared regions are densely populated with 
telluric absorption lines: molecular lines from the Earth's 
atmosphere. These telluric lines fluctuate in depth, particularly with changing 
column densities of water vapor, on timescales of hours. Independent studies 
suggest that selective line masking and monitoring and modeling of telluric lines 
through the night will an effective approach and that $1 m s^{-1}~$ precision 
can still be reached in M-dwarfs.  

\item {\it Rotational Velocity.}
Stellar rotation rates are low for older G, K and early M-dwarfs. However, by 
mid-M type, the typical $v \sin i$ is $5 km s^{-1}$. The average M8V (a cool or ``late" M-dwarf) star has
a rotation rate of $10 km s^{-1}~$ (e.g. Basri \etal 1996, Basri \& Marcy 1995, 
Delfosse \etal 1998). And at the very end of the main sequence, the early L dwarfs
have rotation rates that generally exceed $20 km s^{-1}$.  Higher rotation 
rates result in broader spectral lines that degrade velocity precision. 
Modeling the achieveable Doppler precision shows that it will decrease 
to somewhere between $2 - 6 m s^{-1}$ as $v \sin i$ approaches $15 km s^{-1}$. 
While it is possible to select slowly rotating M-dwarfs, the probability for 
selecting face-on orbits becomes greater and low $M \sin i$ values may 
not correspond to low mass planets. 

\item {\it Stellar Variability.} 
Cool star spots on the surfaces of rotating stars can result in 
shifting spectral line centroids that masquerade as planets in any
(optical or IR) Doppler analysis. However, late type M-dwarfs tend to be 
more photometrically variable than FGK type stars, with cool spots 
and hot flares. Thus, it will be important to carry out additional
tests to confirm the interpretation of a dynamical origin for 
IR Doppler signals. Such tests include photometric monitoring 
and line bisector analysis. One benefit of working in the IR is
that the contrast in flux between cooler spot regions and the rest
of the photosphere is suppressed at IR wavelengths.  Thus, it may
be less likely that spots will produce significant shifts in 
the line centers (i.e. there may be less confusion in the Doppler
measurements) and it should be straightforward to identify spurious 
photospheric variability.

\end{enumerate}

\subsection{Current Status} 

The current design studies for IR Doppler techniques involve
either high resolution IR spectrometers or modest resolution 
spectrometers equipped with an Externally Dispersed Interferometer (EDI).

\section{ Astrometry}

\subsection{Introduction to Astrometry}

The current status of stellar astrometry is that there are about $10^5$ 
star-position determinations with an accuracy in the range of milliarcseconds, but fewer
than $10^2$ with an accuracy in the range of a few hundred microarcseconds.
Of the former, most were determined by ESA's Hipparcos mission, and of
the latter most were determined by a very few ground-based observatories
and some by the Hubble Space Telescope.  Whereas these are useful for
understanding the astrophysics of host stars, they are of little interest
in the discovery of Earth-mass planets in the habitable zones of nearby
stars.  The astrometric signal from an Earth-mass object in a one year orbit around
a one solar mass star at ten parsecs distance is 0.3 $\muas$.

Although only indirectly relevant to the search for Earth mass planets,
the future for ground-based astrometry is promising.  Several
projects are schedule to collect photometric and astrometric data
over large portions of the sky, down to reasonably faint limiting
magnitudes, and with a cadence that enables solutions for both proper
motion and parallax.  These projects include Pan-STARRS (let by the
Institute for Astronomy at the University of Hawaii), SkyMapper (led
by the Mount Stromlo and Siding Spring Observatory), and LSST
(led by the LSST Corporation).  These milliarcsecond-class data will assist
in characterizing host stars as well as expedite the selection of
candidates for planetary searches, but will be of little use in the search for
Earth-mass companions.

The Task Force determined two key factors in identifying the importance
of astrometry as a tool for searching and characterizing Earth-mass
planets.  First, the mass was deemed to be the single most important
parameter for the physical characterization of a planet.  Astrometric data are
critical for removing the $\sin(i)$ uncertainty from radial velocity
solutions.  Second, astrometry can produce unambiguous and precise orbital elements; this is crucial to constraining the astronomical factors that determine whether a planet is habitable. (For multiple planet systems, the combination of astrometry and radial velocitywill be useful in untangling potential ambiguities associated with the various orbits, and it is expected that radial velocity will be used where possible to follow up on the astrometric discoveries). 

Another factor in the importance of astrometric searches for planets
is the need to thoroughly examine stars that would support extremely
sensitive follow-up characterization missions should planets be found.
Nearby stars are important because direct imaging missions would have
the highest probability for success for these.  Bright stars are
important because the high flux enables characterization based on
reflected light, secondary transits, or other subtle effects.  Since
many of the astrometric missions studied by this Task Force support
only a relatively small number of targets (perhaps 100 or so), the
need for an astrometric survey is not overly dependent on the value
of $\eta_\oplus$.

The ExoPTF cannot overemphasize the difficult nature of 1 $\muas$
class astrometry.  In the optical, the diffraction limit of a modest
1-meter aperture is about 100,000 $\muas$, and that for a
flagship class mission of a 10-meter aperture (probably asymmetric)
is 10,000 $\muas$.  An astrometric interferometer would measure angles by optical phase delays; a 1 $\muas$ uncertainty means roughly 0.0006 radians of optical phase, or 0.05 nm. Clearly, high signal-to-noise ratio
operation, many visits, and extreme instrumental stability are needed
to deliver an astrometric stability that is such a small fraction of
the size of the point spread function.  It appears that this level
of performance is not obtainable from the ground using the current
types of deployed or proposed technologies, and even a space-borne platform is pushing the state of the art to achieve this goal.  The Task Force recognizes the various astrometric development programs, and in particular to the NASA Navigator Program, which continue to work down to this ambitious goal. 

\subsection {Ground-based Astrometry}

As noted above, no ground-based astrometric programs have claimed an
accuracy better than 100 $\muas$ for the mean position
of a stellar image with respect to one or more reference stars. Atmospheric turbulence (i.e., seeing)
limits both the size and stability of the point spread function,
which can be mitigated through AO correction, and astrometric 
accuracy is expected to improve to better than 100 $\muas$ on large
telescopes as the various sources of wave front error are understood. At this level of precision ground-based astrometry will principally concern itself with detection of giant planets, determining masses and orbital properties.  The best ground-based astrometric work will be done at several facilities with 8-meter or larger telescopes. The two 10-meter Keck telescopes (Keck-I) in Hawaii have a goal of 50-100 $\muas$ by 2009; a similar accuracy is predicted for the Large Binocular Telescope (twin 8.4 meter mirrors) in Arizona. 

Perhaps the most ambitious effort in ground-based astrometric interferometry
is the PRIMA instrument for ESO's Very Large Telescope Interferometer (VLTI).  This uses phase referencing in a
dual beam configuration to measure relative position of a target star with
respect to a reference star within about 1 arcminute.  The projected
accuracy of 10 $\muas$ is nothing sort of amazing for a ground-based
facility, but still falls short of detecting exo-Earths around most of
the proposed target stars.  The Task Force found no other facilities that
would match or exceed the performance of VLTI within the 15-year time
frame that it examined.

\subsection {Space-based Astrometry}

Various space-based optical astrometry missions have been proposed. USNO's
FAME satellite was accepted as a NASA MIDEX mission, but was terminated
after Preliminary Design Review.  Currently in an advanced stage of development is 
ESA's Gaia mission which is currently scheduled for a launch
late in 2011.  This mission uses two 1.45-m primary mirrors and a large
number of CCDs in its focal plane.  As exciting as its scientific returns
will be, it is projected to be sensitive to perturbations at the level
of 10 $\muas$ over the mission.  Hence, the ExoPTF cannot recommend this
as the type of mission needed to find Earth-mass planets around even
the nearby stars.  Despite this limitation, the committee emphasizes that the Gaia
mission is extremely important, and is expected to make new discoveries across a range of astronomical disciplines. 

An astrometric planet detection mission proposed by JPL and based on its SIM project is one example on paper of a mission that could provide the required precision and sensitivity. (The relationship of the planet detection mission proposed by JPL  to SIM is described further in Chapter 4). 
The planet detection mission features a design optimized for accurate observations of a relatively
small number of stars (of order 60) which incorporates the cost
savings associated with a restricted astrometric mission.  The targets
for such a mission would include the nearest stars, the brightest stars,
and selected other stars where the probability of low mass planets
is high.  These targets are the ones where the
detection probability is high, and the possibility of following up
detections with other observing technologies such as radial velocity
and high-accuracy photometry is highest.

\subsection {Astrometric Search for Nearby Earth-Mass Planets: Requirements}

\vspace{-1.5mm}

Because stars orbit about the center-of-mass
of the star plus planetary system, the angular reflex motion of a star
(as measured against more distant background stars) is a response to
the number, mass, radial distribution, and orbital parameters of the
members of the planetary system.  As such, astrometry holds the potential
to decipher the architecture of extra-solar planetary systems.
However, unlike the radial velocity
technique, which is best suited to detecting massive planets in short-period
orbits around lower mass main-sequence stars, the amplitude of the 
stellar astrometric
perturbation induced by a planet increases with the planet's orbital radius
and period.  Thus, astrometric searches are more sensitive to long-period
planets than radial velocity surveys if their time baseline is long enough.

Astrometry, like radial velocity measurements, is of use to the field of 
exoplanet
research regardless of $\eta_\oplus$.  If $\eta_\oplus$.
turns out to be small, i.e., $\ll\,$0.1, then interest will shift to 
understanding
why earths are rare and knowledge of
planetary system architectures will be crucial to
this understanding.  However, if $\eta_\oplus$$.\:\simgt\;$0.1, then 
attention will focus
on which nearby stars harbor Earth-mass planets and the planetary orbital
parameters.  If such planets are relatively common, then the locations and
times of maximum star-planet angular separations determined by an 
astrometric
mission will make the follow on direct detection planet characterization 
missions
more cost-effective and observationally efficient.  However, for an 
astrometric mission
to serve as a useful precursor to a direct detection coronagraphic or 
interferometric
mission, it must be sufficiently capable to obtain the locations and 
orbital elements
of any nearby Earth-mass planets to a distance at least as far as that 
envisioned
for a direct detection mission that might be realized within the next 
15-20 years.

To estimate the required sensitivity of such an astrometric mission,
for each F, G, or K main sequence star in the solar neighborhood, it is 
possible
to calculate the angular size of the orbit (for direct detection) and 
the astrometric
wobble (for astrometric detection) of a one-earth-mass planet around 
that star.
Three cases are considered for the planet's orbital radius: the inner 
habitable zone
(IHZ), mid-habitable zone (MHZ), and the outer habitable zone (OHZ).
These radii correspond to 0.82 AU, 1 AU, and 1.6 AU (see Kasting \etal\ 
1993),
respectively, around our sun and scaled to radii around other stars 
according to
formulation provided in Kasting \etal.
%$\sqrt{L_*/L_{\odot}}$, where $L_*$ is the luminosity of the star and
%$L_{\odot}$ is the luminosity of the sun.
The semiamplitude of the star's astrometric wobble is given by

\begin{equation}
\frac{\alpha}{\rm arcsec} \equiv \frac{m}{M_*}\:\frac{a_p}{\rm AU}\:
\frac{\rm pc}{D}~,
\end{equation}

\noindent where $m$ is the mass of the planet; $M_*$ is the mass of the parent
star, assumed to scale with luminosity as ($L_*/L_{\odot}$)$^{1/3.5\,}$;
$a_p$ is the semimajor axis of the planet; and, $D$ is the distance
to the system from the observer (e.g. Sozzetti \etal\ 2002).   An 
examination
of 472 F, G, and K single (i.e., not binary) main sequence stars in the
solar neighborhood
yield the results shown in Fig.~\ref{fig:astromet} and summarized in 
Table~2.  Table~2 shows three things: (1) the total number of F, G, and
K stars around
which a 1 Earth-mass planet at scaled orbital radii of 0.82, 1.0, and 1.6~AU
would produce an astrometric wobble of 
$\geq\,$0.22, $\geq\,$0.40, and $\geq\,$0.80 $\mu$as; (2) the total
number of F, G, and K stars around which a planet could be spatially
separated with a 5-meter diameter aperture
and an Inner Working Angle (IWA) of 3$\times\,\lambda/D$ at
5,000, 8,000, and 10,000$\:$\AA; and, (3) the total number of such stars
around which a 1 Earth-mass planet could be detected using both techniques.
Of particular interest are the numbers of Earth-sized planets for which it
might be possible to study O$_2$ at $\sim\,$7,800 \AA, and H$_2$O near
10,000 \AA.

These sample cases illustrate strategies in which the projected 
capabilities of a direct detection mission -- in this case a coronagraphic
mission -- and the sensitivity requirements of the preceding astrometric
mission are tuned to match.
For example, an astrometric mission able to measure a wobble
$\geq\,$0.22$\:\mu$as could find 30\ti\etae\ stars with an earth at a scaled
distance of 1$\:$AU, for which a direct detection mission
mission, with an IWA of 99 milliarcsec (``mas" in the table), could 
search for O$_2$.  These results also suggest that a total-mission
astrometric sensitivity of approximately 0.22$\:\mu$as or better will be required
in order to either detect or rule out the presence of an Earth-mass planet
in the inner- or middle-habitable zone around a statistically significant
number of F, G, and K stars.

\begin{figure}
\vspace{-21mm}
\centering
\includegraphics[scale=0.5]{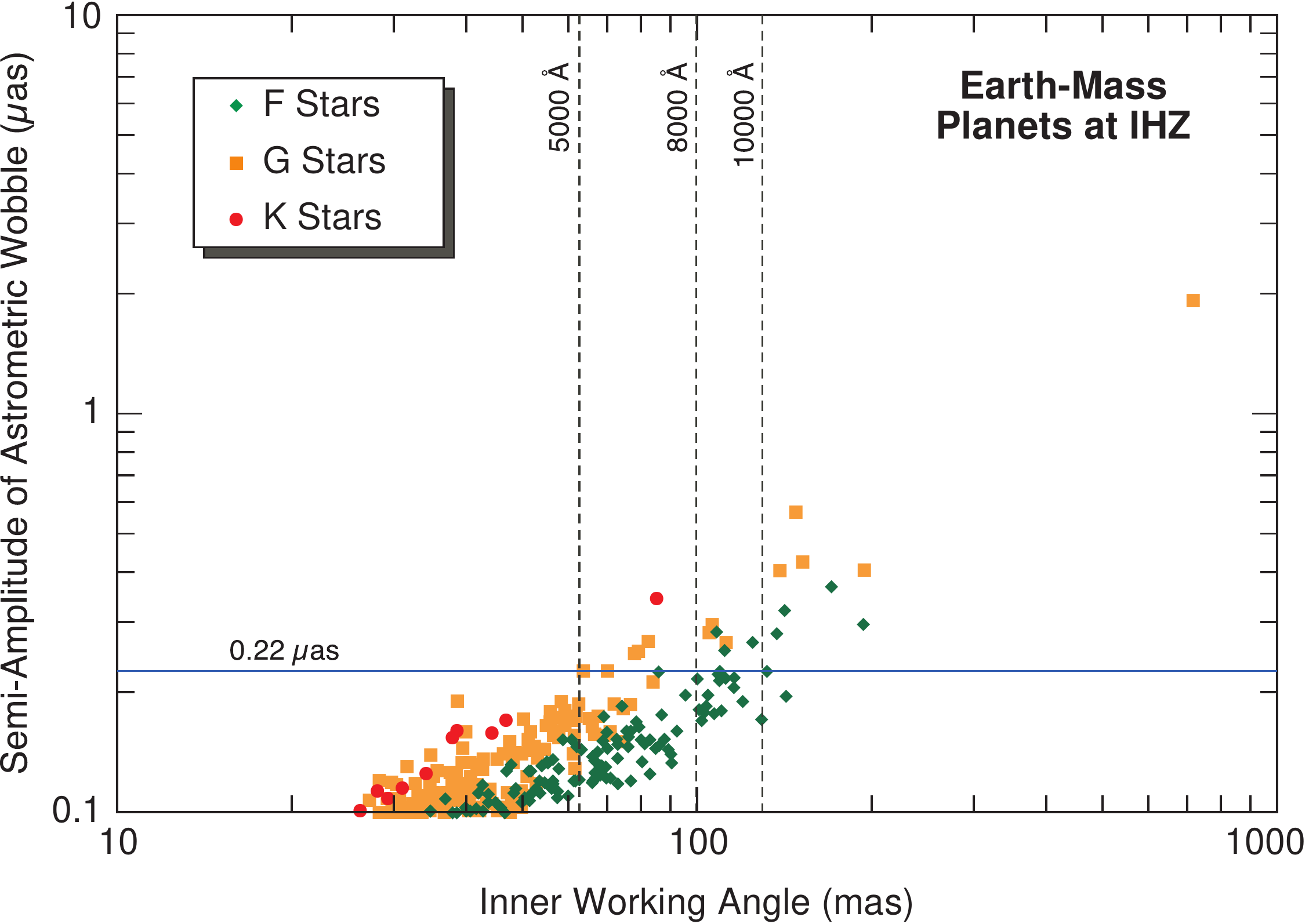}
\vspace{0.3mm}
\caption{Inner working angle (IWA) and astrometric signature for an earth-mass planet at a scaled IHZ orbit of 0.82 AU around each of the 472 main sequence 
F, G, and K Hipparcos-catalog stars examined.
The X-axis plots the IWA required for direct detection of a planet at the 
scaled IHZ edge at maximum elongation; the Y-axis plots the required 
full-mission sensitivity to planetary astrometric signatures for detection of an 
earth-mass planet at the scaled IHZ. Vertical lines are shown at 62, 99, and 
124 mas, appropriate for a 5-meter diameter coronagraphic telescope
with an IWA$\,=\,$3$\lambda/D$ at $\lambda\,=\,$5000, 8000, and 
10,000 $\rm \AA$, respectively. For stars plotted to the right of each vertical line, 
a planet at the IHZ could be directly detected at this wavelength. Similarly, for stars plotted above the horizontal line, an earth-mass planet at the scaled IHZ would produce an astrometric signature of 0.22 $\mu$as or more. There are 
26\ti\etae\ stars above this line; if orbits for these planets could be determined also, then direct detection observations could be scheduled at maximum elongation. 
Such a mission could see the signature of  O$_2$ in 18\ti\etae\ of these stars 
and H$_2$O in 10\ti\etae\ of them with little wasted observing time.  The 
vertical line at 99 mas line and the horizontal line 0.22 $\mu$as line cross in the 
middle of the F-G star distribution, indicating these mission parameters are well
matched.}
\label{fig:astromet}
\end{figure}

\begin{table}[t]
\begin{center}
TABLE~2.  Potential Number of Earth-Mass Planets Detectable\\*[1.5mm]
\begin{tabular}{|l|ccc|ccc|ccc|} \hline
\rule{0mm}{6mm} & \multicolumn{9}{|c|}{Number of Candidate Stars} \\*[2mm] 
\cline{2-10}
\rule{0mm}{6mm}   & \multicolumn{3}{|c|}{~~~IHZ~~~} & \multicolumn{3}{|c|}{~~~MHZ~~~}
    & \multicolumn{3}{|c|}{~~OHZ~~} \\*[0.2mm]
 &  \multicolumn{3}{|c|}{(0.82 AU)} & \multicolumn{3}{|c|}{(1.0 AU)} & 
    \multicolumn{3}{|c|}{(1.6 AU)} \\*[2mm] \hline
    
\rule{0mm}{6mm} Astrometric Detections: & & & & & & & & &  \\*[0.4mm]

\rule{0mm}{6mm}\phantom{0}\hspace{8mm}@$\geq\,$0.22$\mu$as &
 \multicolumn{3}{|c}{26} & \multicolumn{3}{|c}{~~44} & 
  \multicolumn{3}{|c|}{194} \\*[0.9mm]

\rule{0mm}{6mm}\phantom{0}\hspace{8mm}@$\geq\,$0.40$\mu$as &
  \multicolumn{3}{|c}{~~6} & \multicolumn{3}{|c}{~~~~8} & 
  \multicolumn{3}{|c|}{~~31} \\*[0.9mm]

\rule{0mm}{6mm}\phantom{0}\hspace{8mm}@$\geq\,$0.80$\mu$as &
  \multicolumn{3}{|c}{~~2} & \multicolumn{3}{|c}{~~~~2} & 
  \multicolumn{3}{|c|}{~~~4} \\*[0.9mm] \hline
 
\rule{0mm}{6mm} Direct Detections: & & & & & & & & &  \\*[0.4mm]

\rule{0mm}{6mm}\phantom{0}\hspace{8mm}@ ~~5,000$\rm\,\AA$ 
(IWA $\geq$ 62 mas) & \multicolumn{3}{|c|}{99}  & \multicolumn{3}{|c|}{144}  &  
   \multicolumn{3}{|c|}{312} \\*[0.9mm]

\rule{0mm}{6mm}\phantom{0}\hspace{8mm}@ ~~8,000$\rm\,\AA$
(IWA $\geq$ 99 mas) & \multicolumn{3}{|c|}{31} & \multicolumn{3}{|c|}{~~44} & 
   \multicolumn{3}{|c|}{140} \\*[0.9mm]

\rule{0mm}{6mm}\phantom{0}\hspace{8mm}@ 10,000$\rm\,\AA$

(IWA $\geq$ 124 mas)~ & \multicolumn{3}{|c|}{13} & \multicolumn{3}{|c|}{~~28} & 
   \multicolumn{3}{|c|}{~~88} \\*[2mm] \hline

 &  \multicolumn{9}{|c|}
{\rule{0mm}{6mm}When the Astrometric Wobble is Greater Than or Equal to:} \\*[1.6mm]
\cline{2-10}
\rule{0mm}{6mm} Detectable {\em both} Astrometrically  & 0.22 &  0.40 & 0.80 & 
0.22 &  0.40 & 0.80 & 0.22 & 0.40 & 0.80 \\*[-2.9mm]
~~~~and Directly:\rule{0mm}{6mm}  & $\mu$as & $\mu$as & $\mu$as & $\mu$as & 
  $\mu$as & $\mu$as & $\mu$as & $\mu$as & $\mu$as \\*[0.6mm]
\cline{2-10}

\rule{0mm}{6mm}\phantom{0}\hspace{8mm}@ ~~5,000$\rm\,\AA$
(IWA $\geq$ 62 mas) & 24 & 5  &  1 & ~~42 & 7 & 1  & 187 & 30 & 3 \\*[0.9mm]

\rule{0mm}{6mm}\phantom{0}\hspace{8mm}@ ~~8,000$\rm\,\AA$
(IWA $\geq$ 99 mas) & 18 & 5 & 1 & ~~30 & 7 & 1 & 134 & 30 &  3 \\*[0.9mm]

\rule{0mm}{6mm}\phantom{0}\hspace{8mm}@  10,000$\rm\,\AA$
(IWA $\geq$ 124 mas) & 10 & 5  &  1 & ~~24 & 6 & 1 &  ~~86 & 29 & 3 \\*[2mm] \hline

\end{tabular}
\end{center}
\end{table}

\subsection {Astrometry at Other Wavelengths}

A White Paper for doing astrometry at radio wavelengths was received; at present this approach appears to be limited to accuracies of 10 $\muas$ suitable for the detection of giant planets.  The planned Square Kilometer Array may reach values of a microarcsecond, as cited in the White Paper. The potential for microarcsecond accuracy at radio wavelengths from the ground would be of keen interest in combination with optical astrometry from space, and efforts toward achieving this are therefore to be encouraged.

\section{Transits}

When extrasolar planets are observed to transit their parent stars,
we are granted unprecedented access to their physical properties. It is only for
transiting planets that we are permitted direct and highly accurate estimates of
the planetary masses and radii, which provide the fundamental constraints on
on models of their physical structure. In particular, precise determination 
of the planetary radius may indicate or at least constrain the planet's overall composition, 
that is, whether it is a solid rocky body, a Neptune-like object with a mixed rock-water-gas
composition, or a full-fledged gas giant like Jupiter. Determination
of the radius, in turn, can speak to the validity of the canonical model of 
planetary formation, which involves gas accretion onto a core of ice and rock embedded in
a protoplanetary disk. Furthermore, transits provide
one of the potentially most sensitive methods
for planetary detection, and are indeed likely, within a time frame of roughly six
years, to lead to the first detections of Earth-sized planets in habitable
orbits around solar-type stars.
For a general review of the scientific results and prospects arising from transits, see
Charbonneau et al. (2007).

Most of the presently known transiting planets have been detected by Doppler spectroscopy as well.  Radial velocity detection
has the advantage that the observed signature of the target object is 
generally unambiguous. Its disadvantage is that it requires substantial
observing time on large telescopes to identify each planetary system,
and only then can the relatively inexpensive process of searching
for photometric transits begin. Direct photometric transit 
searches simultaneously monitor large numbers of stars in a given 
field of view, but suffer from a very high rate of astrophysical false
positives.

The transit method has been the subject of a great deal of excitement for
over a decade, and is currently one of the most mature techniques for planetary
detection and characterization. The major scientific benefits of the transit
technique are accruing right now, and we expect that the next decade will continue
to be an extremely fruitful time. The time for transits is {\it now}, however. Over 
a time frame of a decade or longer, we expect that other methods (microlensing,
astrometry, direct imaging, spectroscopy) will take the lead in observational exoplanet research.

As of this writing, 24 transiting extrasolar planets are known. Five
of these--HD 209458 b (Charbonneau et al. 2000), HD 149026 (Sato et al. 2005),
HD 189733 (Bouchy et al. 2005) Gl 436b (Gillon et al. 2007), and HD 17156 
(Barbieri et al. 2007)-- were initially
detected using the radial velocity method, and were then discovered to transit
after the parent stars were monitored photometrically during potential transit
windows stemming from the RV-determined orbits. All five of these planets
lie north of the celestial equator, suggesting that photometric monitoring of
deep-southern RV-detected planets may be significantly incomplete. RV-detected
transiting planets are extraordinarily valuable because they provide the best
opportunity for follow-up and for synergistic studies with transit spectroscopy from the ground or space.

To date, six research groups have been successful in discovering planetary
transits with ground-based photometric surveys, 
and about twenty systems are currently
in operation. It is interesting to note that all of the transit discoveries from these surveys arising from parent stars brighter than V=14 are north of the celestial equator, indicating that effort in the Southern Hemisphere is lagging behind coverage in the North.

The OGLE group has been the most successful survey to date, having discovered
five planets, OGLE-TR-10,56,111,113, and 132 (Konacki et al. 2003, 2005,
Pont et al. 2004, Bouchy et al. 2004a, 2004b). The OGLE III experiment monitors
120 million stars in the direction of the galactic bulge for all types
of photometric variability, and the survey detects microlensing events as
well as transiting planets. The small field of view (35'$\times$35') and the
large number of stars surveyed leads to planet-bearing stars that are
dim (V~15-17), distant, and expensive to study with other methods. Nevertheless, the
group has had both the earliest and most productive successes. It seems
possible, however, that as wider-field surveys come up to speed, the 
expense of following up on faint candidates may make narrow-field
surveys less competitive.

The TrES (Trans-Atlantic Exoplanet Survey) is a network of three small-aperture
telescopes at locations in Arizona, California, and Canary Islands. This
survey is sensitive to photometric planet detections 
around V=10-12 magnitude stars, and
has had good success with their multi-tiered strategy for screening
out false-positives. The survey
monitors roughly 10,000 stars at a time in a 6-degree square field-of-view
and has to date discovered four planets
(Alonso et al. 2004, O'Donovan et al. 2006, 2007, Mandushev et al. 2007).
All of these planets have radii equal to or exceeding Jupiter's radius, 
including TrES-2 which lies in the Kepler field of view, and TrES-4, which
has a radius of ~1.7 Jupiter radii that is large enough to present 
serious challenges to current structural models for strongly irradiated
Jovian planets.

The WASP and HATnet surveys are similar in basic design to TrES, and have
been similarly successful. WASP has detected two planets (Collier-Cameron
et al 2006), and HAT has now detected six (Bakos et al. 2007 a,b,c Torres et
et al. 2007, Noyes et al. 2007). 
HAT-P-2b is remarkable for its bright ($V\sim7.5$) parent
star and the eccentric orbit of the planet itself. Both properties lend
themselves to extremely interesting follow-up opportunities with the
Spitzer telescope.

Another successful transit project (The XO survey, McCollough et al.
2006, 2007) also surveys wide fields on the sky, and aims for target
stars in the V=9-12 range. It is slightly different from the other programs
in that it employs a drift-scan approach. It has published two planets
XO-1 and XO-2, and has announced a third (XO-3, announced at the 2007 AAS
meeting, but with orbital parameters as-yet unbpublished.)

Finally, the Transitsearch.org collaboration has also successfully identified
a transiting planet. Transitsearch's strategy is to
enlist ad-hoc groups of amateur astronomers in global campaigns to detect
transits by observing known planet-bearing stars at the dates and
times when the radial velocity solutions indicate that transits may
possibly occur. This technique works well for planets that have orbital
periods in the 10 to 100 day range, and  has been used to determine that the
$P=21.22$-day planet orbiting
HD 17156b is transiting (Barbieri et al. 2007). 

Large-scale photometric transit surveys produce lots of candidate planets,
the majority of which are inevitably false positives (for example,
grazing eclipsing binary stars and background eclipsing binary stars).
Brown et al. (2003) provide a good overview of realistic
detection statistics and false-positive rates for wide-field surveys.
As more surveys work through their pipelines and produce candidates, there
becomes an increasing need for confirmational follow-up with large
telescopes equipped with high-precision spectrographs (e.g. Keck, HARPS,
VLA, Subaru, etc.) RV confirmation of candidate transits will
likely be both the most expensive part and the pipeline bottleneck for 
these surveys. Indeed, it is important to stress that the Doppler velocity
and transit methods are inextricably linked. Progress in transit 
research will proceed only as quickly as radial velocity observing time is made
available.

\subsection{Prospects for transit surveys}

During the next five years, it is expected that a large number of additional
transiting planets will emerge from ongoing ground-based wide-field surveys. Indeed,
over the past year, the inventory of known transiting planets has more than
doubled, as projects such as HATnet, TrES and Exo start to reach the full 
production end
of their observational piplelines. As the number of planets reaches the 
threshold for
statistical comparisons, interesting trends (or possible trends) have started to 
emerge. For example, there appears to be an inverse correlation between planetary
mass and orbital distance among transiting planets, and it 
also appears
that the metallicity of the parent star correlates inversely with planetary size
at given planetary mass. 
Further ground based observations will be able confirm these
trends and place them on far firmer statistical footing.

%% start JW

Since there are already 20 known examples of transiting planets,
significant progress in this field will require either an
order-of-magnitude increase in the number of discoveries (thereby
elucidating the trends mentioned above), or the discovery of
fundamentally different types of transiting planetary systems.  About
16 of the known systems can be safely classified as apparently
isolated, close-in gas giant planets, with masses and radii comparable
to Jupiter and orbital periods shorter than 5 days.  Therefore, some
obvious ways to make significant progress are to discover (i) hundreds
or more of transiting gas giants; or to devise efficient methods for discovering
(ii) longer period planets, or (iii) smaller planets. We discuss the
prospects for each of these possibilities in turn.

Obtaining a sample of hundreds of transiting gas giants within about 5
years would be difficult but is within the realm of possibility for
the current large-scale photometric surveys, if they receive
continuing support and also funding to expand by a factor of a
few. The discovery rate is already accelerating: four such planets
were discovered in 2006, and six in the {\it first half} of 2007.  If
current surveys continue operating, it seems safe to predict a
discovery rate of 10 per year for 5 years, for a total sample size of
about 70.  While it is true that ground-based observations impose
severe limits on photometric precision and consequent biases on the
sample of discovered planets (see, e.g., Gaudi 2005), the 0.5-2\%
transit depths of gas giants around FGK stars are readily detectable.
The limitations imposed by the day-night cycle and by the vagaries of
the weather can be largely overcome by operating a longitudinally
distributed network (such as TrES, HAT or Los Cumbres Observatory Global Telescope Network).  Furthermore, as mentioned
above, the bottleneck in the transiting planet discovery process is
currently the need for RV follow-up observations to
eliminate false positives and measure planetary masses. This is the
area that would seem to require additional funding, through the
construction or acquisition of automated 1-2m RV
telescopes.  The follow-up problem would not be eliminated by
performing the photometry from space, although it might be alleviated
to some degree because a lower false-positive rate may result from the
higher-quality initial photometric database.

One reason why finding longer-period transiting planets would be
interesting is because their mass-radius relation could be compared
with that of the close-in planets. At a long enough period, the
structure of the planet would be totally unaffected by stellar
insolation, in contrast to the current sample, in which insolation is
a key factor in the planetary energy balance. In addition, one could
seek transiting planets in the habitable zone, which generally occurs
at significantly longer periods than the 2-5 day periods of the
current sample. Furthermore, for planets at large orbital distances, 
large satellites
can exist in stable, potentially even habitable orbits. 
If such moons are Mars-sized or larger, they
can potentially be detected using space-based, or even ultra-high-cadence
ground-based photometry.

However, finding long-period
transiting planets is a game with rapidly diminishing returns because
the geometric transit probability declines inversely with the orbital
distance (and therefore as $P^{-2/3}$ for a fixed stellar mass). One
must observe many more stars, and therefore generally fainter stars,
than the current surveys. Obviously, the frequency of transits also
decreases as $P^{-1}$. Greater patience is required to observe
multiple transits and a reasonable fraction of the doppler spectroscopic
orbit, and there are fewer opportunities for high-precision
characterization observations. This is essentially the reason why
transit science is expected to become relatively less important in
comparison to astrometry and direct detection over the next 11--15
years. Surely transiting planets will emerge as rare and wonderful
special cases, but transit science owes its fantastic and
disproportionate impact in recent years to the unanticipated existence
of close-in giant planets.

Over the next 6--10 years the frontier will be ever-smaller planets.
Here there seem to be three basic options: (i) invest in more precise
radial velocimetry to increase the discovery rate of small planets
that may then be searched for transits; (ii) improve the dynamic range
of the massive photometric surveys; (iii) concentrate specifically on
smaller stars, which offer a greater transit depth for a planet of a
given size. Of course these options are not exclusive. Improvements in
velocimetry is discussed elsewhere; here we note only that it is
easier to achieve the high dynamic range needed for photometric
confirmation of transits over a narrow and targeted field of view than
over the wide field of view required for photometric surveys. This
fact is demonstrated by the discovery of the transiting
``super-Neptune'' HD 149026b, with a transit depth of only 0.3\%. This
especially interesting system was first identified as a
radial-velocity variable and was then intensively searched for
transits (the geometric transit probability was about 10\%).  It is
doubtful whether current photometric surveys would have been able to
identify such a low-amplitude variable without also admitting a slew
of statistical false positives.

It might be possible to improve the dynamic range of wide-field
ground-based photometry through the use of larger apertures (thereby
decreasing both the Poisson and scintillation noise) and the careful
control of potential systematic errors due to factors such as guiding
errors, focus and other optical variations, and detector
inhomogeneities.  The typical level of time-correlated noise (or ``red
noise'') that is introduced by atmospheric and detector effects has
been shown to slash discovery rates by an order of magnitude (Pont et
al. 2007). Whether the red noise of ground-based photometry can be
reduced enough to enable the discovery of significantly smaller
transit depths is unknown but deserves further study.

Naturally, it is also appealing to contemplate a space-based survey,
in which the atmosphere is not an issue and at least some aspects of
the environment are extremely stable. The CoRoT mission has provided
the first space-based discovery platform for detecting extrasolar
planets in transit. The CoRoT spacecraft is in a polar circular orbit
at 896 km altitude, which allows it uninterrupted 5-month access to
two regions of the sky.  The spacecraft houses a 27-cm photometric
telescope that illuminates a 4-CCD array of detectors. The standard
mode of operation is to observe $\sim 10,000$ stars in the $11 < V <
16.5$ range at 512 second cadence.

COROT was launched in December 2006.  In May 2007, the COROT team
announced the detection of a short-period Jovian planet transiting a
solar-type star at 500 pc distance.  The planet is anomalously large,
with a $1.78 R_{\rm Jup}$ radius, and a mass approximately 1.3 times
that of Jupiter; it orbits its parent star once every 1.5
days. The signal to noise ratio of the detection is excellent,
indicating that the spacecraft is performing remarkably well. The
photometric error bars on the light curve accompanying the press
release describing the first planet discovery indicate that the
detection of planets of 2$R_{\oplus}$ or less is potentially possible
if red-noise contamination can be reduced via observation of the 
target star during a large number of individual transits.
Indeed, COROT will give important information regarding the
frequency of large terrestrial planets and giant planet cores on
reasonably short period (50 days or less) orbits around a wide variety
of parent stars, and will provide statistical expectations for what
the more sensitive Kepler mission should be able to see.

The Kepler mission is perhaps the most anticipated project in
transit-related exoplanetary science. Kepler, currently planned for
launch in 2009, is a $\sim 1000$kg spacecraft that will be inserted
into an Earth-trailing heliocentric orbit. The spacecraft houses a
95cm aperture photometer and an array of 42 CCDs. Over the nominal
4-year mission lifetime, the spacecraft will photometrically monitor
$\sim100,000$ 9th-15th stars for planets by maintaining vigilance on a
single 105-square degree star field in the galactic plane in the
direction of the constellation Cygnus.

Kepler's photometic sensitivity on a 12th-magnitude star over a 6.5
hour integration is $\sim2\times10^{-5}$. Earth-sized planets on
$\simeq$ 1 AU orbits about solar-type stars produce transit durations of
order 13 hours, with transit depths of order
$1\times10^{-4}$. Assuming that the frequency of Earth-sized planets
is high, Kepler will be able to detect of order 50 $1 R_{\oplus}$
planets in 1-year orbits at the close of the nominal 4-year mission.

In addition to detecting Earth-sized planets in the habitable zones of
solar-type stars, Kepler will have a significant impact on the study
of short and intermediate-period giant planets. It will provide an
excellent platform for the detection and initial characterization of
transiting eccentric giant planets in the ~1 AU zone, and it will have
the photometric sensitivity to detect of order 870 planets with
periods less than one week by observing modulation in the reflected
light.

The clearest lesson of the last few years of transit investigations is
``brighter is better,'' meaning specifically that with brighter target
stars it is easier to rule out false positives; it is easier to detect
the orbit by RV and hence measure the planetary mass; and it
is more rewarding to conduct all of the high-precision follow-up
investigations that make transits so valuable. This sounds rather
obvious but the extreme difficulty of working with fainter target
stars was not fully appreciated at the outset of this subfield.  The
five OGLE planets were confirmed only after many nights' investment
with 8-10m class telescopes, and due to the faintness of their parent
stars those planets have not been the subjects of the most interesting
types of follow-up observations. Likewise, the transit candidates
detected in SWEEPS, a large program using HST, cannot be confirmed
with existing equipment and are therefore of limited interest. 

An exciting opportunity exists to perform a wider and shallower
equivalent of the Kepler survey; that is, a space-based transit
survey of all the brightest stars on the sky.  The main prize would be
to detect small planets that cannot be easily detected from the ground
around the brightest possible stars.  These systems would be the best
possible targets for detailed characteriation with Spitzer, Warm
Spitzer, JWST, and ground-based RV  with large telescopes.
It will be a challenge to design the equipment that will enable
precise photometry over a very wide field, along with an efficient
scan strategy that would permit an all-sky survey in a reasonable
time.  Because the target stars would be bright, the collection area
need not be large and thus the weight requirement for the mission need
not be burdensome and as prohibitively expensive as general purpose
space telescopes.  Depending on the chosen technology it may even be
possible for a nimble wide-field transit mission to be launched within
5 years.

The Task Force received briefings on several such concepts.  LEAVITT is
envisioned as a {\it Hipparcos}-style rotating spacecraft that
performs low-resolution slitless spectroscopy (essentially multiband
photometry) of all stars with $V<14.8$. Each star would be visited for
6-7~hr on a few hundred occasions over 5~yr.  The PLATO consortium,
involving dozens of European institutions, is exploring two different
mission concepts: a battery of 100 small wide-field cameras on a
single platform, or a trio of 1m telescopes spaced by 120 degrees
around a rotating platform. The Transiting Exoplanet Survey Satellite
(TESS) would have 9 small wide-field cameras canted at different
angles, giving a total instantaneous field of view of 3,000 square
degrees.  The scan strategy is to face the antisolar point and allow
stars to drift across the wide field of view over the course of a
year, thereby monitoring a given star continuously for 120 days.  The
photometric accuracy would be a few times $10^{-4}$ per 90 minute
orbit of the spacecraft. Over a 2-year mission it would examine $\sim
10^6$ stars with $V<12$ over the entire sky.  Only planets with
relatively short periods ($<$2 months) would be found, but this is not
as strict a limitation as it may seem, for the reasons described
above: transits with longer periods are of limited practical value
because of the infrequency of follow-up opportunities, and periods of
a few weeks are in the habitable range for low-mass stars.

The latter point brings up the general notion that low-mass stars
offer a number of important advantages to the transit seeker.  The
habitable zone is closer to the star, where the transit probability is
relatively large and the frequency of transits is relatively high.
The transit depth is obviously larger for a smaller star, for a planet
of a given size; indeed, an Earth-sized planet transiting a
main-sequence M-dwarf would produce an easily detectable transit depth
of 0.5-2\%. Likewise, the radial velocity variation of the star would
have a larger amplitude, although this must be weighed against the
time demands of optical Doppler velocimetry on faint red stars. 
The developement of high-precisions Doppler velocimetry in the infrared
would greatly aid the discovery and characterization of low-mass, potentially
habitable planets orbiting M-type stars.

Supposing that low-mass stars are not unexpectedly deficient in
low-mass planets, and that such planets are potentially habitable
despite the unusual characteristics of the host star's radiation field
and strong tidal interactions, it would be very desirable to design a
survey that specifically targets low-mass stars. A difficulty is that
such stars are intrinsically faint and therefore rare in any
magnitude-limited survey. One would want to search the stars with the
brightest apparent magnitudes, but because their surface density on
the sky is low they would need to be observed sequentially rather than
simultaneously with a single detector.  In a budding project titled
"MEarth", a sample of $\sim2000$ nearby M-dwarfs would be observed
individually by one of an armada of 10 ground-based telescopes, each
with an aperture of approximately 0.3~m (Nutzman \& Charbonneau 2007).
If planets larger than $\sim 2$ Earth radii have an occurence rate
of $\sim 20\%$ within the habitable zones of late M-dwarfs, then this
effort will detect several potentially habitable worlds. 

\subsection {Prospects for characterization}

\subsubsection {Spitzer}

In July of 1994, the astronomical community was treated to observations
of the serendipitous collision between Jupiter and
the comet Shoemaker-Levy 9.  The sudden deposition of energy 
into the Jovian atmosphere caused a dramatic response on a planetary scale, 
providing remarkable opportunities for observation and simulation 
Harrington \& Deming 2001, Deming \& Harrington 2001)
and fostering the development of a deeper understanding of the structure 
and behavior of gas-giant atmospheres.   Within our own solar system, 
such an occurrence was a once-in-a-lifetime event.  
However, the discovery of transiting extrasolar 
giant planets on short-period orbits 
(some of which have substantial eccentricity) affords the opportunity to 
study atmospheric dynamics on worlds where both steady and periodic doses 
of energy are imparted on on a continual basis, and
with the advent of Spitzer, direct observation of such planets 
has become feasible.

Indeed, the Spitzer telescope has been an unexpectedly useful tool in the
study of hot Jupiters, and has already been used to make a number of landmark
observations of extrasolar planets. 
These began with the detections of the secondary
eclipses of HD 209458 b (Deming et al. 2005) and TrES-1 (Charbonneau et al. 2005),
which confirmed that hot Jupiters have low albedos and day-side effective
temperatures of order $T_{\rm eff}\sim1100\,{\rm K}$. 
The TrES-1 observations were
carried out at 4.5$\mu{\rm m}$ and 8$\mu{\rm m}$, and thus provided an estimate of that
planet's mid-infrared spectral slope.
High S/N observations at 16 $\mu{\rm m}$ of
the secondary transit of HD 189733 (Deming et al. 2006) produced a similar
($1117\pm42\,{\rm K}$) day-side temperature.

Harrington et al. (2006) detected strong orbital phase-dependent brightness
variations for $\upsilon$ And b at 24 $\mu{\rm m}$, which indicate a strong 
day-night temperature contrast in the far IR, and suggest that heat transfer
between the hemispheres is inefficient at the level in the atmosphere that 
generates the observed $24\mu{\rm m}$ emission. Conversely, 
Cowan et al. (2007) found no significant orbital
phase variations at 8 $\mu{\rm m}$ for 51\,Peg b and HD\,179949 b, indicating that
for those two planets, surface flows {\it are} effective at redistributing heat at
the level probed by the 8 $\mu{\rm m}$ photosphere.

More recently, Knutson et al. (2007) have presented results from a continuous
30-hour Spitzer observational campaign on HD 189733. The
resulting 8$\mu$ photometric time-series contains extraordinary
(part in $10^{-4}$) signal-to-noise and clearly shows phase-dependent
flux variations that indicate a day-night temperature contrast of order 200K.
Of particular interest in the time series is a 
$\delta F/F \sim 0.0005$
increase in planetary flux of $\sim 5$hr duration just past the primary 
transit. This
feature can be interpreted as arising from a hotspot near the dawn terminator
of the planet, and can fairly be ascribed to constitute our first ``weather
map" from an alien solar system. 

The discovery of transits by the Neptune-mass (and Neptune-sized) planet
orbiting the red dwarf Gliese 436 (Gillon et al. 2007) 
was followed up almost immediately with Spitzer.
Deming et al . (2007) report 4.5-micron photometry of the secondary eclipse.
The light curve indicates that this planet has an effective temperature of
order 800K. This temperature is higher than expected from baseline models, and
indicates that the planet is likely experiencing tidal heating in response
to its anomalously eccentric orbit (which was fixed to high accuracy by the
timing of the secondary eclipse to $e=0.15\pm0.01$.)

In addition to providing infrared photometric time-series data,
the Spitzer Space Telescope has also been used to obtain mid-infrared spectra
of transiting extrasolar planets, including
HD\,209458b (Richardson et al. 2007) and HD\,189733b (Grillmair et al. 2007).
In both cases, spectra in the 7-14$\mu$m range were obtained by subtracting 
the stellar spectrum measured
during secondary eclipse from the combined star+planet spectrum obtained
just outside eclipse. Model atmospheres of extrasolar planets under strong
insolation conditions (e.g. Sudarsky, Burrows \& Hubeny 2003, Marley
et al. 2006) predict that ${\rm H_2O}$ absorption features would be
prominent within the observed spectral range. Remarkably, however,
neither planet displayed the predicted signature. This null result can be
interpreted in several ways. First, it is possible that the atmospheres
are dry, perhaps because available oxygen is sequestered in CO. This
interpretation, however would require both planets to have unexpectedly
large C to O ratios. Alternately,
the effectively featureless spectrum could arise from a very clear
atmosphere with an isothermal P-T profile down to large optical 
depth. This
interpretation is favored by the fact that the hot Jupiters have extremely
low albedos (Rowe et al. 2006). As a third possibility, 
suggest that
HD\,209458b may be displaying line emission at 10$\mu$m. They
interpret these features as possibly arising from silicate clouds. Any
such clouds, however, would need to be highly non-reflective in order
to be consistent with the low optical albedo.

At the time of this writing, Spitzer observations have been made of both
the transit and the secondary transit of HAT-P-2b (Bakos et al. 2007). These
data, when analyzed will allow for a direct measurment of the effective
radiative time constant in the atmosphere of this planet, which experiences
a ten-fold increase in insolation between apastron and periastron.

\subsubsection{Warm Spitzer} 

Spectroscopic and photometric work with Spitzer is revolutionizing our
understanding of the atmospheric and dynamical properties of short-period
extrasolar planets. Unfortunately, however, Spitzer's cryogen is scheduled
to run out in February 2009, which will effectively eliminate the spacecraft's ability
to obtain observations at wavelengths longer than 4.5 microns. An 
extended ``Warm Spitzer" mission has been proposed, in which the spacecraft
would continue to be operated at the shortest IR bands. Such a mission
would be of considerable importance for transit-related studies of 
short-period extrasolar planets, and in particular, would allow for
extended, full-phase observations of transiting planets on eccentric
orbits to be carried out. It would also be useful for characterizing the
temperatures and atmospheric properties of very low mass planets that
are expected to be detected around nearby M-dwarfs during the next few
years. Warm Spitzer would allow the community to retain the ability to do
follow-up work in the near IR during the next half-decade. This is an important capability and one that is key to the early years of the M-dwarf track portion of the strategy proposed here. 

\subsection{Spectroscopy during transits}

A transit produces not only a photometric signal but also at least 2
different spectroscopic signals. These spectroscopic signals are
worthy of intensive observations because of the information they
provide about the planetary atmosphere and its orbit.  First, any
strong absorption features in the planetary spectrum would cause the
transit depth to appear larger when viewed within the wavelength band
of the absorption.  Using this methodology, often called
``transmission spectroscopy,'' investigators have claimed evidence for
neutral sodium, neutral hydrogen, and water in planetary atmospheres,
with varying degrees of statistical significance (but none above
4$\sigma$).  Observations with an extremely high dynamic range are
required because the expected variation in the transit depth is of
order $(h/R_\star)^2 \lesssim 10^{-5}$, where $h$ is the atmospheric
scale height.  In addition, detailed atmospheric models are needed for
the interpretation of the signals, since the starlight that is
affected by the planetary atmosphere grazes the surface of the planet
and experiences a wide range of pressures and temperatures while
traversing the planetary atmosphere.

Transmission spectroscopy is perhaps one of the most futuristic
applications that can be brought to bear on transiting extrasolar
planets. The dominant hurdle to this technique is simply the number 
of photons. Integration times are limited to the duration of the
planetary transit, and generally high spectral resolution and 
especially signal-to-noise are required. The advent of the next
generation of 30-meter class ground-based telescopes (expected within
ten years) may allow significant advances to be made in this field. With
10-meter class telescopes, transmission spectroscopy will probably be confined
to a handul of favorable cases involving giant planets.

A second spectroscopic (and from the the standpoint of current observations)
very exciting effect is an ``anomalous'' Doppler shift that
occurs during transits arises because of stellar rotation.  The
emergent spectrum from a given point on the stellar disk is
Doppler-shifted by an amount that depends on the local line-of-sight
velocity. The spread in velocities across the disk broadens the
spectral lines (along with thermal and turbulent broadening). When the
planet hides a portion of the stellar surface, the corresponding
velocity components are missing from the spectral lines.  When the
planet is in front of the approaching (blueshifted) half of the
stellar disk, the starlight appears slightly redshifted. The anomalous
Doppler shift vanishes when the planet is in front of the stellar
rotation axis, and then reverses sign as the planet moves to the
receding (redshifted) half of the stellar disk.  This phenomenon is
called the Rossiter-McLaughlin (RM) effect because it was described by
those two researchers in 1924 (of course, this was in the context of
eclipsing binary stars, rather than exoplanets).

The RM effect scales as the transit depth times the projected rotation
velocity of the parent star, which is often tens of meters per second
for the known transiting planets.  Thus it can be observed with a
high signal-to-noise ratio, which allows one to determine the trajectory of
the planet relative to the (sky-projected) stellar rotation axis.
Specifically, one can measure the angle between the sky projections of
the orbital axis and the stellar rotation axis, and thereby assess
spin-orbit alignment.  Current measurements are all consistent with
well-aligned systems, suggesting that the migration process that
produces close-in gas giant planets apparently preserves the initial
spin-orbit alignment of a planetary system.  It will be interesting to
apply this technique to longer-period planets, which are observed to
often have large eccentricities. One might expect that the mechanism
that perturbs the eccentricity would also perturb the inclination of
the orbit, an expectation that can be tested with RM observations.

It has also been pointed out that the RM effect is an alternative
means of detecting a transit, which may offer advantages in some
cases. Specifically, for small planets with long periods, the
amplitude of the RM effect will sometimes exceed the stellar orbital
velocity. This raises the appealing possibility of using the RM effect
to confirm transits that will be detected by the forthcoming satellite
missions COROT and Kepler. Many of their most interesting candidates
will have small transit depths that cannot be detected photometrically
from the ground and they will be very difficult to confirm by
detecting the star's Doppler spectroscopic orbit.  RM confirmation appears
more feasible for at least some stars that are not too
slowly-rotating, not only because of the relatively large velocity
amplitude but also because the RM velocity variation occurs over the
time scale of the transit duration, which is much shorter than the
time scale of the orbital velocity variation. This relaxes the
requirement for long-term stability of the spectrograph. Details are
given by Gaudi \& Winn (2007).

Another fundamental and potentially valuable quantity that can be
measured during transits is the precise time of the transit.  The
transits will be strictly periodic if the star and the transiting
planet are the only bodies in the system, but a second planet in that
system would introduce gravitational perturbations that would cause
the transit interval to vary. Transit timing variations (TTV) thereby
offer an alternative means for detecting planetary systems and
characterizing their orbits. The niche of this technique seems likely
to be its high sensitivity to bodies that are in mean-motion
resonances with the transiting planet; such bodies produce maximal
TTV, whereas the Doppler and astrometric signals from such bodies are
often hidden because of parameter degeneracies.  And while mean-motion
resonances might sound like an unlikely coincidence, in fact many of
the known planetary systems exhibit some kind of resonance, and some
planet formation scenarios predict a large number of small bodies that
are caught in resonances due to differential migration.  A large-scale
TTV campaign would involve a ground-based, longitudinally distributed
network of small automated telescopes, of generally modest cost. In
the near term, this subfield needs more theoretical attention. Issues
that need work are the best strategies for detection campaigns, the
optimal selection of target systems, and the solution of the ``inverse
problem'' of inferring the orbital elements of any perturbing planet
from a list of observed transit times.  Some work has been done on
these issues but much more work is indicated before a large-scale
campaign should be undertaken.

\subsection{Transit spectroscopy with James Webb Space Telescope}

JWSTÕs science payload is planned to comprise four instruments. NIRCam is a wide
field, deep imaging camera that will also make the wavefront sensing and control
(WFSC) measurements, necessary to phase the telescope. NIRSpec is a multi-object
spectrometer contributed by the European Space Agency (ESA). MIRI provides
mid-infrared (5-28.5 $\mu m$) imaging and spectroscopy. The Tunable Filter Imager (TFI) is a
camera contributed by the Canadian Space Agency. Each of these science instruments
features capabilities that enable either transit imaging or spectroscopy and these are
summarized in figure ~\ref{fig:JWST_Table}.

\begin{sidewaysfigure}
\begin{center}
\includegraphics[angle=0,scale=1.0]{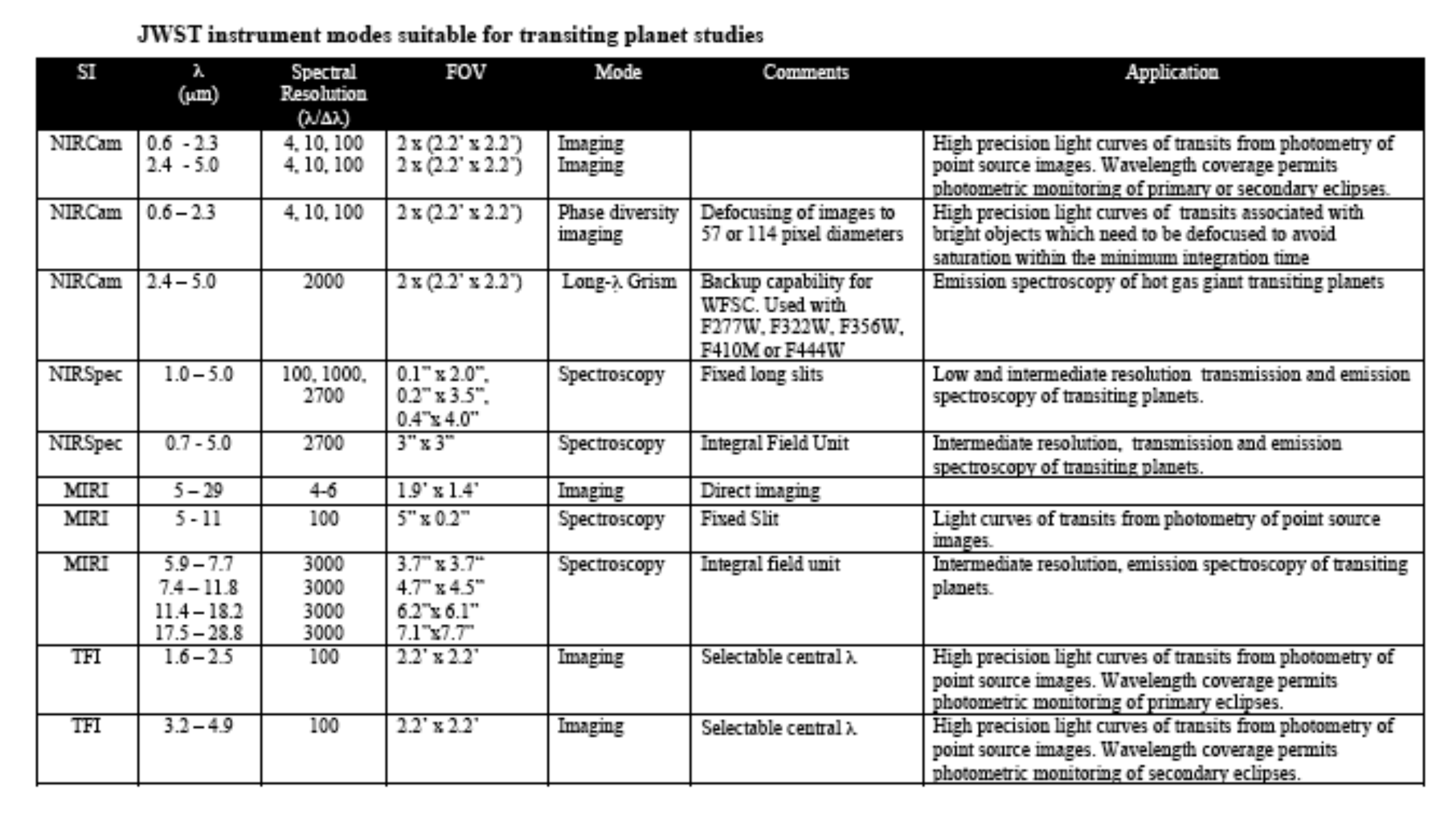}
\label{fig:JWST_Table} 
\end{center}
\end{sidewaysfigure}

Two techniques can be used to probe transiting extrasolar planet atmospheres with
JWST. The absorption spectrum of the planet can be measured by detecting the
signatures imposed on stellar light transmitted through the planet's atmosphere during
transit. The emission spectrum of the planet can also be measured using the secondary eclipse technique.The two techniques are complementary. Emission spectra produce
potentially larger signals than transmission spectra at infrared wavelengths. However,
features in transmission spectra will be present even in the extreme case when the
atmospheric temperature profile of the exoplanet is isothermal - which would produce a
featureless spectrum in emission. Note also that JWST thermal emission spectra can be binned to lower resolution, producing the equivalent of secondary eclipse photometry. In
this case, the dispersion of the light on the detector can help to alleviate potential
saturation for bright systems.

A recent study by Valenti et al. (2007) concludes that the NIRSpec on JWST could be used to detect deep features in the spectrum of an ocean-covered, Earth-sized planet orbiting an M3V planet within 20 pc. Preliminary calculations of sample size for JWST instruments are given in the Figures of Merit section of this report. 

\clearpage
% - - - - - - - - - - - - - - - - - - - - - - - - - - - - - - - - - - - - - - - - - - - - - - - - - - - - - - - - - - - - - - - - - - -
%SECTION 9 MICROLENSING DELETED
% - - - - - - - - - - - - - - - - - - - - - - - - - - - - - - - - - - - - - - - - - - - - - - - - - - - - - - - - - - - - - - - - - - -
\section{Coronagraphy}

\subsection{Introduction to Coronagraphy}

\vspace{1.5mm}

Originally invented to study the sun, a coronagraph is a telescope designed to block 
light coming from the solar disk, in order to see the extremely faint emission from the 
region around the sun, called the corona. Invented in 1930 by Bernard Lyot to 
study the sun's corona at times other than during a solar eclipse, the coronagraph, at 
its simplest, is an occulting disk in the focal plane of a telescope or out in front of the 
entrance aperture that blocks the image of the solar disk, and various other 
features, to reduce stray light so that the corona surrounding the occulting disk can
be studied.

If a pointlike star is observed with a space telescope equipped with perfect
optics, unaffected by the aberrations of the Earth's atmosphere, the stellar
image will still be surrounded by a halo of light diffracted from the edges
of the telescope aperture.  Known as the Airy pattern, this halo is many
orders of magnitude brighter than any extrasolar planet 
(see Fig.~\ref{fig:Airy_Pattern}).

Suppressing this halo is the function of a coronagraph. Over the past decade,
due to increased interest in the direct detection of extrasolar planets and
the possibility of a coronagraphic Terrestrial Planet Finder, coronagraphic
concepts have proliferated
far beyond Lyot's basic vision into a vast interrelated family of 
diffraction-control devices.  Guyon et al. (2006; hereafter G06) 
break these down into four
broad categories: interferometric coronagraphs, pupil-apodization
coronagraphs, amplitude-based Lyot coronagraphs, and phase-based Lyot
coronagraphs.  G06 lists 18 separate coronagraphs within these categories.
We refer the interested reader to this paper, and the references therein,
for the details related to each technique.  In this report, we discuss briefly
a subset of the most promising coronagraph architectures that span a range of 
performance and technical maturity. 

A given coronagraph type can be characterized by several key parameters.
The most fundamental is the Inner Working Angle.
Although this is not normally
a sharp step function, as discussed in G06, the flux from the planet is 
rapidly attenuated and/or the flux from the star rapidly increases as the 
IWA is approached, yielding a sharply plummeting detection sensitivity.  
Thus, the IWA is often treated as a hard lower limit in
modeling of coronagraphic missions.  Currently, the most mature coronagraph
concepts assume an IWA of $\sim\,$4$\,\lambda/D$; some newer, but less-proven, 
concepts suggest that an IWA of $\sim\,$2$\,\lambda/D$ can be achieved, effectively 
allowing the same planets to be resolved from the star with a smaller telescope.

\begin{figure}
\centering
\includegraphics[scale=0.67]{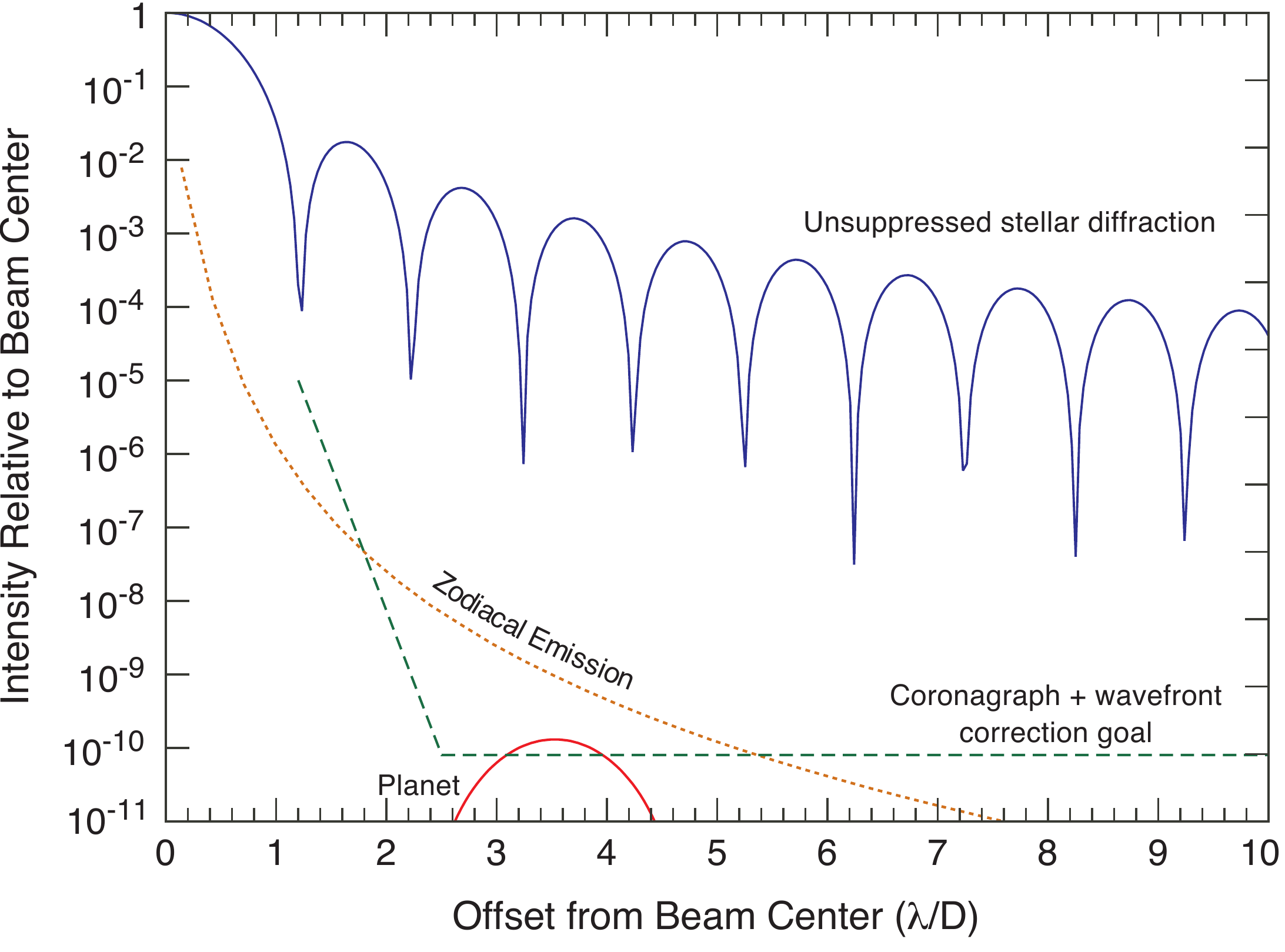}
%\vspace{-1mm}
\caption {Flux contributions in the image plane of the telescope, vs. 
angle from the beam center, measured in $\lambda/D$.
({\em Solid blue curve}):~Airy, or diffraction, pattern resulting from an 
ideal filled circular aperture of 4-meters diameter operating at 5500 \AA\
employing no light suppression techniques.  
({\em Solid red curve}):~Reflected-light flux from an Earth-like exoplanet
relative to the flux from its Sun-like central star.
({\em Dashed green curve}):~Required performance of a coronagraphic 
plus wavefront
correction system designed to detect Earth-sized exoplanets -- i.e., the
Airy rings beyond a few $\times\:\lambda/D$ must be suppressed 
to a level of about 10$^{-10}$.
({\em Dotted orange curve}):~Zodiacal scattered light background due
to our Solar System plus an equivalent background in the target stellar
system.  Detection of an Earth-like exoplanet requires both stability in
the suppressed-light image and spatially smooth extrasolar zodiacal
dust emission to confidently extract the planet signal.}
\label{fig:Airy_Pattern}
\end{figure}

A second key parameter is planet throughput, the fraction of the planet's light that 
survives the process of suppressing the starlight, and reaches the detector.  The more 
mature designs have throughput of 8-30\% while some newer, less-proven designs claim 
throughput of 80\% or higher.  In addition,
most coronagraph designs produce the required stellar light suppression
within only a fraction of the entire field of view, for example only some range of 
azimuth around the star, or angular separation from it.  This area of deep stellar
light suppression is sometimes referred to as the ``discovery space" and varies
for each design.
Thus, use of such coronagraphs to conduct blind surveys for previously
undetected planets would often require that the telescope be rotated through multiple
angles around the line-of-sight in order to sample the full annulus around the
star, effectively reducing the throughput of the system.  This is less of a limitation
for the characterization of a planet whose position has been previously
determined and where it becomes possible to place the area of deep stellar
light suppression -- however small -- directly on the position of the exoplanet.

A third key parameter is the sensitivity of the coronagraph to low-order
wavefront errors.  Although all coronagraphs are sensitive to mid-spatial-frequency aberrations,
some designs have additional sensitivity to low-order aberrations
such as tilt (image position), focus, astigmatism, etc. These aberrations arise easily due 
to drift and misalignment, and their effects can quickly obscure or mimic a planetary signature. 
Many otherwise promising concepts achieve near-perfect
sensitivity only for an unresolved star located exactly on-axis with a perfect telescope; 
however, for realistic conditions, these concepts lose much of their performance.   This is
discussed in some detail in G06. 

\begin{figure}
\centering
\includegraphics[scale=0.2]{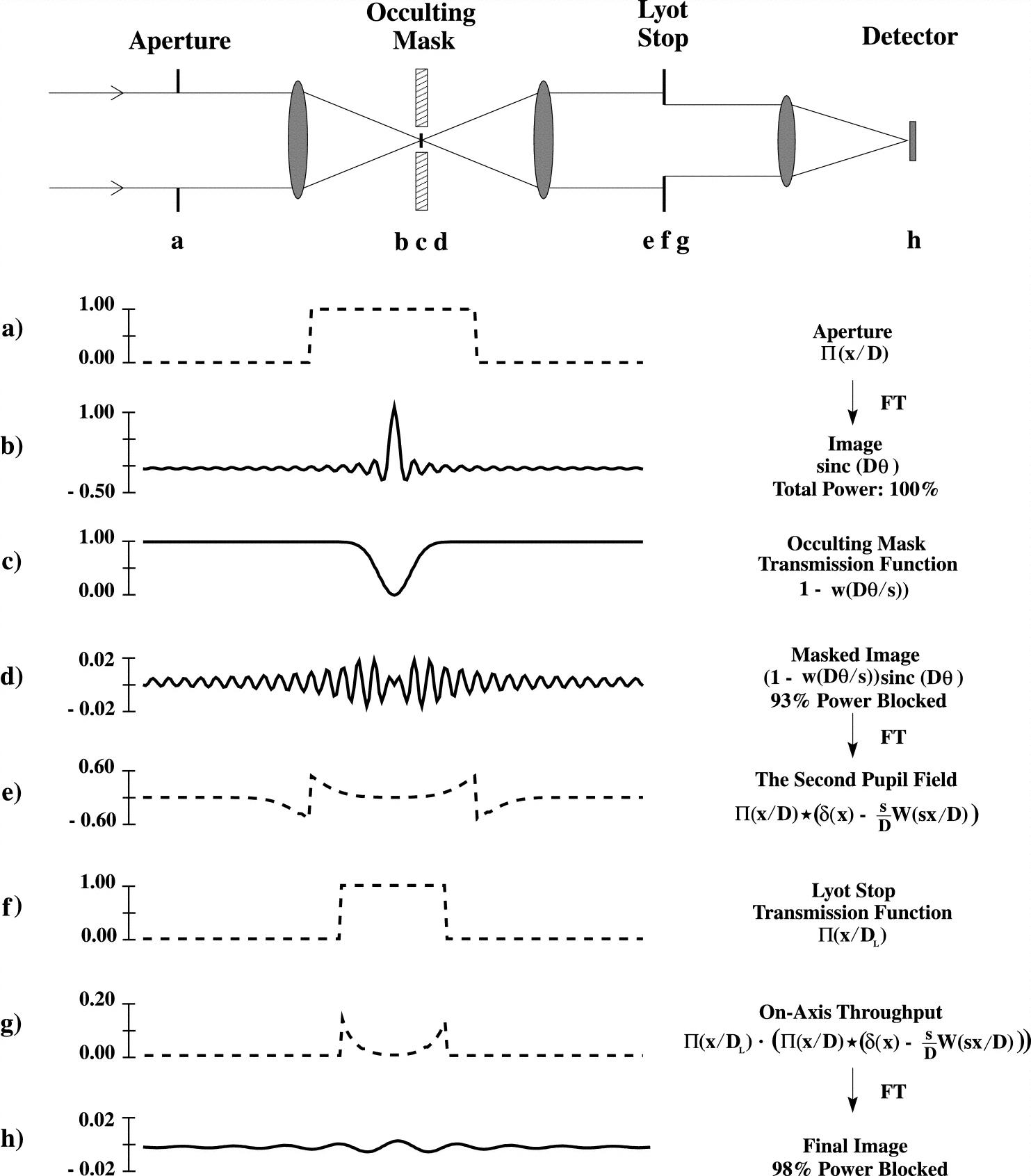}
\vspace{1mm}
\caption{Basic schematic of a coronagraph employing a Lyot mask
(Sivaramakrishnan et al., 2001).}
\label{fig:Coronagraph_Schematic}
\end{figure}

A fourth key factor is the chromaticity of the coronagraph, i.e. its ability 
to suppress starlight across a broad wavelength range.  Low chromaticity allows 
each observation to span a broader passband, allowing a lower total integration time. 
While chromaticity
is often tied to the physical details of the implementation of the coronagraph
and the wavefront control performance, some coronagraphic designs
are intrinsically easier to adapt for broadband operation. These would allow 
simultaneous observations over a much wider passband. Typically the useful bandwidth 
is limited to $\Delta\lambda/\lambda\:\sim$20\% or less; but some concepts exist for 
operating in multiple passbands, 
with parallel coronagraph instruments. These would allow simultaneous observations 
spanning a much wider passband. Such concepts have not yet been demonstrated experimentally. 

\vspace{2mm}

\subsubsection{Amplitude-Based Lyot Coronagraphs}

The most mature family of coronagraphs are those evolved from Lyot's
original solar coronagraph, comprising a mask at an image of the star, 
and another mask at a later image of the entrance aperture (pupil).  As 
illustrated in Fig.~\ref{fig:Coronagraph_Schematic},
the first mask blocks the core of the star, but leaves residual diffracted 
starlight in the re-imaged pupil, where the second mask blocks it.
The most important variant of this concept
is the band-limited coronagraph (BLC), devised by Kuchner \& Traub (2002). 
This approach uses a focal plane mask with carefully tailored transmission profile, to 
almost perfectly confine the residual diffracted light to a finite outer region
of the pupil.  Typically, the BLC has an inner working angle of 
2-4.5$\,\lambda/D$, a throughput of 20-40\% and, for an IWA
$>\,$4$\,\lambda/D$, it has
good robustness against low-order wavefront errors. This type of coronagraph
has produced the best laboratory results to date, achieving a measured
contrast of 6$\times$10$^{-10}$ (Trauger \& Traub 2007),
albeit only in monochromatic light.  Manufacture of the necessary masks for
broadband operation remains a technological challenge, but promising
approaches exist. The BLC has served as the baseline for many TPF-C architecture
studies and currently represents the most conservative coronagraph approach, 
with moderate performance but a relatively high level of maturity. 

Another Lyot variant is the Apodized-Pupil Lyot Coronagraph (APLC), which
instead modifies the entrance pupil with a tapered transmission mask.  This 
coronagraph class is capable of similar performance to the BLC in monochromatic
light but is challenging to manufacture for broadband use. Manufacture of
the necessary mask is a significant technical challenge, and laboratory 
demonstrations to date have only achieved contrast levels in the 10$^{-6}$ range. 
This coronagraph
type is easily adapted to conventional telescopes with on-axis secondary mirrors, 
and has been selected by the major next-generation ground-based coronagraph 
projects.

\vspace{2mm}

\subsubsection{Phase-Based Lyot Coronagraphs}

An alternate approach to the Lyot coronagraph is to induce a phase
shift in part of the starlight in the focal plane, creating
destructive interference for on-axis starlight within the telescope pupil and
effectively blocking the star. The best-known example of this concept
is the 4-quadrant
phase-mask (4QPM) coronagraph, which focuses the starlight onto a mask
that shifts light passing through half the focal plane by half a wavelength.
Such coronagraphs often have near-ideal theoretical performance with 
high throughput and an IWA of 1$\,\lambda/D$.  However,
they are also extremely sensitive to position in the focal plane, with
performance degrading rapidly for even small tip/tilt errors or stars that are
partially resolved.  Since this technique relies on phase shifts, they are also difficult
to manufacture for broadband light. The 4QPM remains popular in various
low-performance European mission concepts (contrast levels of 10$^{-6}$ to 10$^{-8}$), 
but is unlikely to meet future TPF-C requirements.

A recently-proposed variant of the 4QPM is the Optical Vortex 
Coronagraph, which uses a spiral-staircase phase mask 
(see Fig.~\ref{fig:optical_vortex}).
These vortices are quite robust against low-order aberrations and are
extremely promising for small (i.e., 2-3$\,\lambda/D$) IWA coronagraphs
for TPF-C.  However, manufacture of the masks, especially for broadband
use, remains a significant challenge.

\begin{figure}
%\vspace{-2mm}
\begin{center}
\includegraphics[scale=0.7]{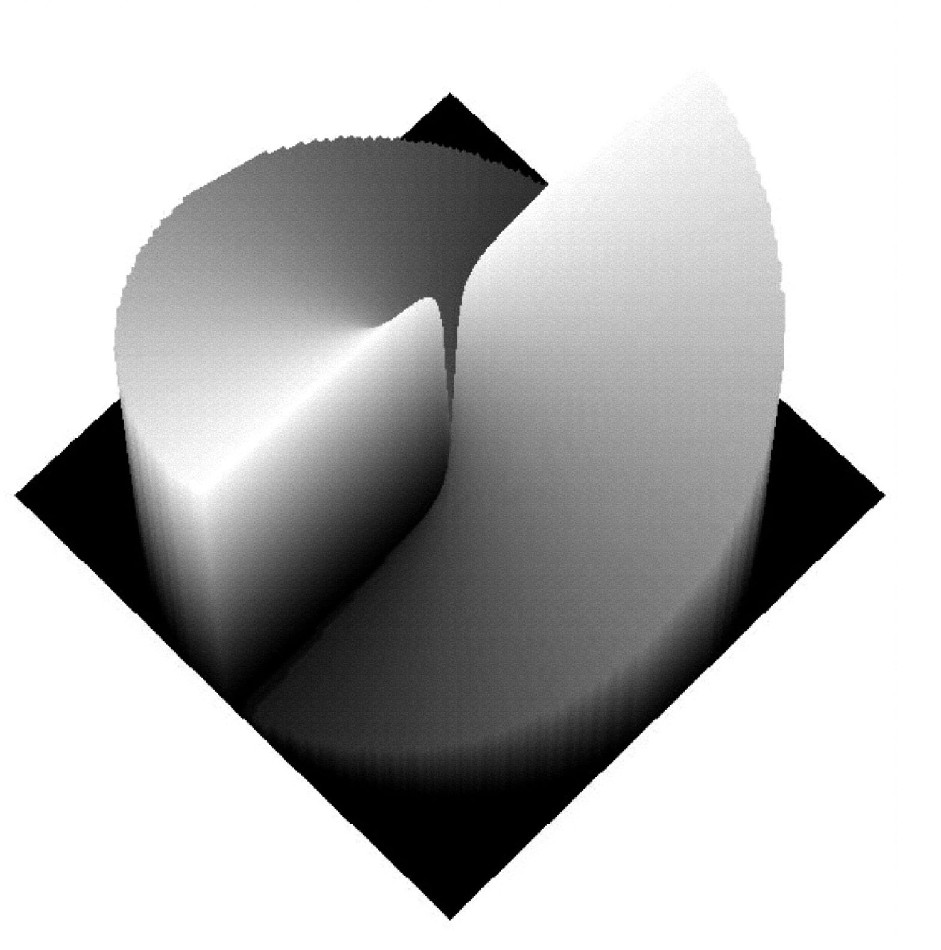}
\caption{Mask for an optical vortex coronagraph 
(after Mawet et al. 2005, Ap.J,. 633, 1119).}
\label{fig:optical_vortex}
\end{center}
\end{figure}

\subsubsection{Interferometric Coronagraphs}

Several coronagraph architectures combine phase-shifted 
copies of the telescope pupil in a way similar to a multi-telescope 
interferometer. Some of these concepts have significant chromaticity and/or
low-order wavefront issues. The most mature concept of this class
is the Visible Nuller Coronagraph
(VNC), which recombines displaced copies of the telescope pupil sheared
laterally to cancel out the on-axis starlight. This can be shown
mathematically to be equivalent to the BLC described above, and hence has
similar performance, though it trades the complexity of manufacturing
masks for cascades of mirrors and beamsplitters.  In principle, though,
the VNC can be dynamically adjusted for planets of different angular
separations from their central star.
Laboratory demonstrations of the basic nulling technology exist but, to date,
end-to-end demonstrations have achieved contrast levels of 
10$^{-6}\,$--$\,$10$^{-7}$ for a single fiber.
Analysis indicates that a system employing $n$ fibers, where $n$ is
typically envisioned to be $>\,$ 1000, would see an
improvement in stellar light suppression by a factor of $n$,
but this scaling has not yet been demonstrated in a full end-to-end test.

\subsubsection{Pupil Apodization}

The diffraction pattern of a telescope is simply the Fourier transform
of its entrance aperture; the sharp edges of a conventional telescope
result in the Fourier ringing that produces the well-known Airy pattern. 
Conceptually, the simplest coronagraph would entail a modification of the
telescope aperture so it lacks these sharp edges. 
Called apodization, this modification can be achieved by tapering
the telescope transmission through a grey mask that falls gradually to 
zero transmission at the edges. 
These masks trade
throughput for inner working angle, with a typical mask designed for an IWA 
of 4$\,\lambda/D$ having a transmission of 8\%.  However,
manufacturing such a mask with graded attenuation methods
is extremely difficult. 

Significant effort has been invested in 
binary approximations to these apodization functions using sharp-edged
metal masks that suppress diffraction over only part of the field of
view (see Fig.~\ref{fig:shaped-pupil}).  Referred to as ``shaped pupils," 
these masks, which can include micron-sized features, must be carefully
fabricated but are feasibly produced with current manufacturing techniques. 
These masks operate in a complex trade space of inner working angle,
throughput, and discovery space.  These designs are inherently achromatic,
and may be attractive for characterization of planets with known positions,
where they can be optimized for throughput over a narrow region of the
focal plane. Shaped-pupil masks have been demonstrated in the laboratory at
the 2$\times$10$^{-9}$ contrast level, even in broadband light. 

\begin{figure}
\centering
\includegraphics[scale=0.7]{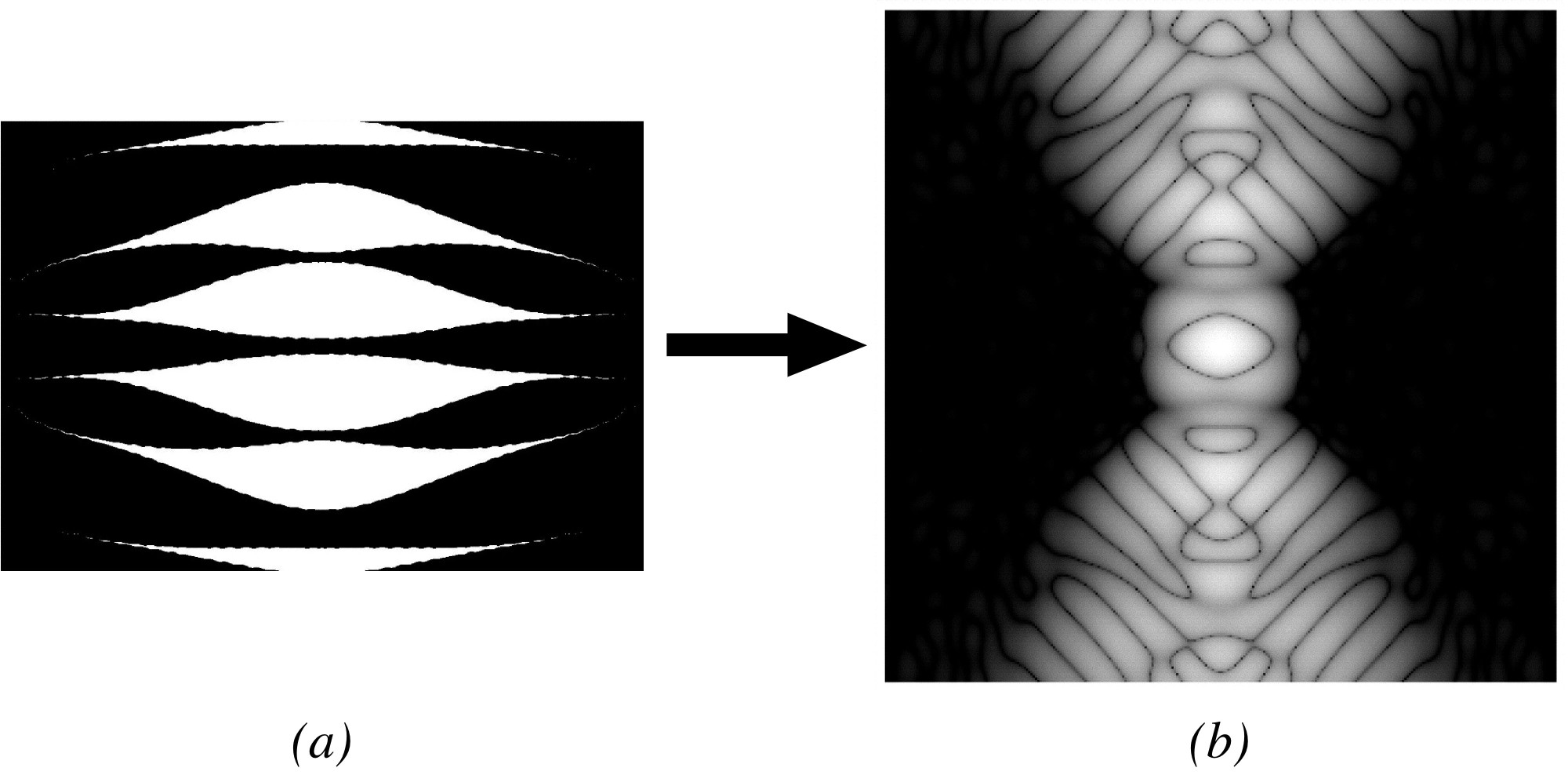}
\vspace{-1mm}
\caption{$(a)$ Example of a shaped-pupil  mask and, $(b)$, the resulting
light suppression pattern in the image plane
(after Kasdin et al. 2003, Ap.J., 582, 1147).  The dark areas in panel $(b)$
represent the regions of greatest diffracted light suppression, and thus the
area within which it is most likely that an exoplanet can be detected.}
\label{fig:shaped-pupil}
\end{figure}

An alternative approach to apodization is to use the phase of the 
light to modify the intensity in the pupil plane. The 
Phase-Induced-Amplitude-Apodization Coronagraph 
(PIAAC; Guyon et al., 2005, see Fig.~\ref{fig:piaa}) takes this approach,
using highly aspheric mirrors to redistribute the uniform beam of light into a 
tapered profile.  (These optics also highly distort off-axis planet
images, but this effect can be compensated.)  In simulations, a PIAAC
can have an inner working angle of 2-3$\,\lambda/D$ and a throughput of 80\% 
or more, which can significantly reduce the telescope size needed for
a given TPF-C-like mission compared to the baseline BLC. However, 
this requires complex optical surfaces located in various conjugate planes. 
The effects of wavefront errors on these and other optical surfaces have
not yet been fully explored. Also, wavefront requirements are dramatically
tighter for such small IWAs, as discussed in Section~2.
Laboratory experiments of this technique are ongoing.

\begin{figure}[h]
\centering
\includegraphics[scale=0.9]{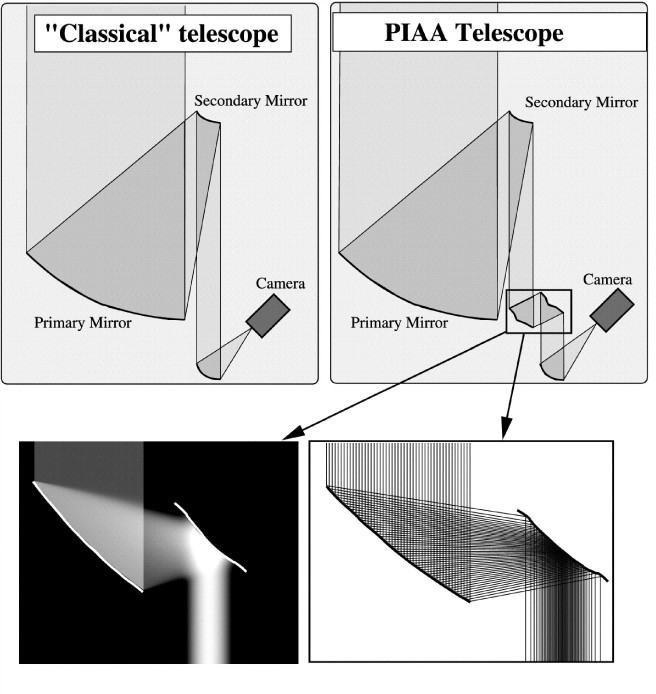}
\vspace{-2mm}
\caption{Schematic representation of the 
Phase-Induced-Amplitude-Apodization
Coronagraph (after Guyon et al., 2005.
Ap.J., 622 744).}
\label{fig:piaa}
\end{figure}

\vspace{2mm}

\subsubsection{External Occulter Coronagraphs}

A different type of starlight suppression is as old as the human hand -- 
blocking the bright star before it reaches the telescope. In 1960, Lyman Spitzer
proposed combining a telescope and starshade in space for discovery of 
planets (Spitzer, 1960).  The size of the shade and the inter-spacecraft separation were 
enormous and thus impractical, but over many years refinements in starshade 
design have reduced the required starshade dimensions and improved the 
level of suppression.  Though some concepts for the starshade have used 
transmitting sheets with graded transmission for apodization, the most recent 
work has focused on optimizing the shapes of serrated-edge binary masks.  Petal shapes have been found (Vanderbei, 2007, for example) that permit operation at
IWA $<\:$100 milli-arcseconds at wavelengths from 0.5-1.1$\mu$m, using a shade
with a nominal diameter of 40$\;$meters at a telescope-starshade separation
as small as 40,000 km. 

This concept yields an IWA that is only weakly dependent upon the telescope
diameter -- in principle, as small as 1$\,\lambda/D$.  The telescope can be an 
ordinary diffraction-limited space telescope, and its diameter is determined
mainly by the integration time required to detect faint planets ($\propto\,D^{-4}$), 
and by the need for planet-star astrometry ($\propto\,\lambda/D$). It can 
observe the planet in the entire passband from 0.5 to 1.1 $\mu$m in a single 
integration, whereas internal coronagraphs typically are limited to 20$\%$ 
passband slices at a time. 
Slewing from one star to the next requires that the starshade travel several 
thousand kilometers.  To accomplish this within a few weeks requires large
starshade velocities and delta-velocities. With conventional thrusters, this 
would take a hefty amount of fuel; advanced electric propulsion eliminates 
this concern, but requires substantial electrical power. 
In sum, agility on the sky is difficult, and a single-occulter system would be 
essentially unable to provide the scheduling flexibility typical of an internal 
coronagraph.  However, a substantial engineering effort has been dedicated to 
minimizing the time between observations and the resources required, and some 
mission scenarios have been found that yield satisfactory efficiencies with one 
occulters, and much better with two occulters. The comparative lack of nimbleness 
also can be offset significantly by having prior knowledge of the planet orbits so that 
a schedule of observations can be optimized. In the absence of such astrometric 
orbit data, the external occulter might use a strategy of immediate spectral 
characterization after a detection, especially if it would take only a few days; 
whereas for the internal coronagraph we conservatively assume a need to take 
multiple confirming observations before investing this integration time.

The deployment of the large starshade and the propulsion required for many 
stellar observations are the major technology issues to be addressed. 
Conventional prelaunch end-to-end testing -- i.e., demonstrating stellar
suppression at typical mission distances -- is impossible. Thus it will be necessary 
to rely on diffraction models validated by subscale testing. This modeling and 
validation is likely within the next year or two.

\vspace{2mm}

\subsection{Wavefront Control}

A problem common to all internal coronagraphs is wavefront accuracy 
and stability.  Wavefront errors produced by imperfections in the telescope
mirrors and coatings cause speckles in the image after the coronagraph masks have 
suppressed the starlight.  The intensity and variability of these speckles 
can easily be high enough to obscure
the faint exoplanet image. It is possible to suppress these speckles using aggressive
wavefront sensing and control (WFSC) with deformable mirrors, limited
only by optical disturbances that cause the speckles to move before the control
system can sense and remove them.  This race between disturbance and 
compensation sets a practical floor to the planet sensitivity of a coronagraphic
telescope.

WFSC performance in turn constrains the choice of IWA.  Reducing 
the IWA from 
4$\,\lambda/D$ to 2$\,\lambda/D$ brings dramatically higher sensitivity to 
small changes in low-order telescope aberrations. If the WFSC system is quick 
and accurate, and can control these aberrations adequately, 
then the IWA of 2$\,\lambda/D$ becomes usable.
If, instead, the WFSC system is slow, the telescope structure must be
designed {\em and demonstrated} to be inherently stable enough to hold these 
low-order aberrations over correspondingly 
long intervals. Thus the performance of the WFSC system and the chosen IWA 
together set the required passive stability of the large telescope. At present, this 
is a major technical challenge facing the construction of a planet-detecting 
internal coronagraph telescope.

A few years ago, slow WFSC update rates drove telescope designs toward 
extraordinary passive stability. Integrated modeling showed these requirements 
might be achievable for on-orbit conditions, but a feasible concept for pre-launch 
testing to verify this stability was never developed. In fact, that testing would 
be thousands of times more challenging than for JWST. With greater WFSC 
speed, the passive stability required of the telescope would be relieved, and 
testing needs with them -- perhaps close to the current state of the art. 
Recent concepts to improve the WFSC update rates offer the possibility of less 
demanding passive stability and/or a smaller IWA.

The profile of the image of a point source is called the point spread function
(PSF), and the relationship between PSF and the wavefront
phase and amplitude is discussed extensively in the literature 
(Malbet et al., 1995; Perrin et al., 2003; 
Give'on et al., 2006; Marois et al., 2006). 
To first order, the PSF of a coronagraphic telescope resembles the
power spectrum (modulus squared of the Fourier transform) of the
wavefront phase.  Achieving a PSF with Airy rings suppressed
sufficiently to allow detection
of Earth-sized planets would require wavefront phase errors of less than an
angstrom (10$^{-10}$ meters) -- well beyond current polishing
capabilities for telescope-sized mirrors.  However, use of a deformable
mirror (DM) in the optical train allows irregularities in the wavefront to be
corrected by feedback, up to a spatial frequency set by the deformable mirror's actuator
spacing. With good DM performance, this allows low- and mid-spatial-frequency 
errors to be removed from the wavefront, producing a PSF with 
a characteristic ``dark hole" region. 

This approach drove early designs for the Terrestrial Planet Finder coronagraph,
but is now recognized as significantly incomplete.  At contrast levels of
10$^{-10}$, several higher-order effects become important, such as wavefront
amplitude (intensity) variations caused by optical coating non-uniformities 
or by Fresnel propagation of surface irregularities
on mirrors in the optical train, polarization effects, and high-frequency wavefront
errors ``folding back" by aliasing into the inner parts of the PSF. These effects require 
more sophisticated approaches to wavefront sensing
and correction; for example, using additional phase corrections
introduced by multiple deformable mirrors to cancel PSF artifacts caused by 
(uncontrollable) amplitude errors. 
Such corrections have been demonstrated using a single deformable
mirror to control the wavefront over only half of the focal plane at almost the TPF-C
levels of performance, but only in monochromatic or narrow-band light. 
Many of these higher-order effects are inherently wavelength dependent,
so correction degrades over broader wavelength ranges. Detailed modeling
(e.g., Shaklan \& Green 2006) shows that for state-of-the-art optics,
the necessary correction can be achieved for some coronagraphs
over scientifically useful bandwidths (typically up to $\sim$20\%); but 
this must be verified experimentally
and extended to the more advanced coronagraph types discussed above.

Determining the necessary DM corrections in a real operational scenario
is also challenging.  The ``speckle nulling" algorithm that was used for
the most successful demonstrations to date is extremely time-consuming,
potentially requiring hours of observations to remove the speckles iteratively. 
This in turn requires that the telescope and optical system remain stable 
at the angstrom level over the same timescale.  New algorithms have
been proposed, and partially tested, that reduce the required time
significantly, potentially down to a few seconds.  The trade between wavefront 
sensing time, algorithm accuracy and PSF stability is critical to determining
the practicality of a TPF-C mission.  For instance, designing and verifying a
telescope that must remain stable on day-long timescales probably would require
full-scale testing, whereas demonstrating stability on minute-long timescales
may be practical with only subsystem-level laboratory testing combined with
simulations and analysis.  The interrelation of coronagraphic technique,
wavefront control strategy, system stability, and ground verification underscores that modeling
of TPF-C must be done in an integrated fashion to predict the ultimate
performance.  Such modeling is partially complete for the baseline 
(IWA $\sim\,$4$\,\lambda/D$ BLC on a 6-8m telescope) TPF, but 
remains incomplete for more advanced and lower-cost concepts. 

\vspace{2mm}

\subsection{Coronagraph Summary}

Coronagraphy as an area has shown enormous progress in the
past decade.  Assuming commensurate progress in wavefront control
(discussed below), there are currently at least two internal 
coronagraph architectures --
the band-limited coronagraph and the shaped-pupil -- that appear likely to
meet the requirements for detecting terrestrial planets at 
IWA $>$ 3.5$\,\lambda/D$
with throughput of 10$\,$--$\,$40\%.  Other concepts, most notably the PIAAC
and the optical vortex, could significantly improve on this, but are presently at
lower levels of technical maturity.  In all cases, 
the interrelated problems of wavefront control and system testing must be solved.  Modeling of
the coronagraph + telescope + wavefront control is relatively complete 
for the BLC architecture, but remains a significant area of uncertainty
for other architectures.  An approach employing an external occulter
sidesteps these issues; however, shade-diffraction modeling and the engineering 
issues of shade deployment and long-distance maneuvering remain a concern.

\vspace{2mm}

\subsection{State of Ground-Testing Capabilities}

A number of laboratory setups are capable of testing coronagraphic and
wavefront performance to contrast levels of 10$^{-7}$.  Testing
performance to a contrast level of 10$^{-10}$ requires that atmospheric
and thermal fluctuations be effectively eliminated and that sources
of vibration be minimized.  The High-Contrast Imaging Testbed (HCIT)
at JPL has been designed to provide an extremely stable environment
for testing to the 10$^{-10}$ contrast level.  Unfortunately, the testing time
for advanced concepts within the HCIT has been limited -- a situation
that can only be remedied with greater support for the HCIT and the
construction of additional comparable facilities. 

Presently, these testbeds are only applicable to the coronagraph instrument; no test 
facility yet has the capability to demonstrate stability of a full-size 
telescope at the required level -- of order 1 angstrom RMS per day as presently 
understood. Advances in WFSC and in telescope testing concepts 
could dramatically ease these requirements, but we 
emphasize that a substantial technology development need remains in this area.

For external coronagraphs, the verification approach is also far from historical 
practice. Assuming that diffraction modeling can be used to prove the diffraction performance 
of the shade as {\em designed}, the final verification of the shade {\em as built} will entail 
measurements of the edge shape to show it matches the design within tolerances. 
The tolerances are currently believed to be of order a millimeter in petal 
width and several centimeters in lateral position.
Testing is also needed for thrusters and alignment sensing, likely done at the 
subsystem level. Then the most challenging remaining test is of deployment 
accuracy and reliability.

\vspace{3mm}

\subsection{Current Status of Coronagraphy Science}

Coronagraphic instruments on the Hubble Space Telescope and ground-based
adaptive optics (AO) coronagraphs
have produced many significant results in the past decade, but have yet
to achieve detections 
of even giant planets orbiting normal stars. 

HST has had several coronagraphic capabilities. The most significant for
purposes
of exoplanet detection is the coronagraph in the NICMOS instrument. 
This produces a moderate
level of coronagraphic suppression of scattered starlight, but optical
errors
still scatter light into a residual halo. However, this halo is very
stable. 
This allows successive images taken of different stars or the same star
at different
orientations to be subtracted, enhancing contrast by a factor of 5--10.
The final contrast
is then limited by the time evolution of optical errors as the
temperature of
the telescope changes. HST also had a powerful visible-light coronagraph
on the 
(currently defunct) Advanced Camera for Surveys instrument. At visible
wavelengths this had little
sensitivity to warm planets but excellent sensitivity to scattered light
from
circumstellar dust, particularly using PSF subtraction. Restoration of
this capability during
the next HST servicing mission is crucial to studies of such debris
disks. Both major
HST coronagraphs have relatively large inner working angles (400 to 800
milliarcseconds). 

Ground-based AO cameras (some with simple coronagraphs)
are operational on 8-10m class telescopes. These produce
near-diffraction-limited images
in the near-infrared (1.0 -- 5.0 $\mu$m). The AO images are still
limited in
dynamic range, however. First, imperfect correction of atmospheric
wavefront errors
with current AO hardware leaves a broad diffuse halo containing 30-70\%
of the target
star's light. The ability to detect point source companions such as
extrasolar planets
is limited not by this diffuse halo but by image artifacts caused by
quasi-static 
optical aberrations
within the telescope, AO system, coronagraph, or camera that are not
correctly calibrated
or corrected. These aberrations produce slowly-evolving speckle
patterns. Several
techniques have been developed to overcome these speckle patterns,
including
comparison of simultaneous images at different wavelengths or images of
the same
star at different orientations. Current ground-based AO coronagraphs, even with
advanced image processing, are limited by their internal aberrations
and atmospheric image artifacts (``speckles''). These limit the achievable contrast of both Keck telescopes to
$10^{-4}$ to $10^{-6}$ at separations
of 500 to 5000 milliarcseconds. This is insufficient to detect a mature
Jupiter-sized planet,
but could potentially detect very young planets ($\leq$ 100 Myr), still hot
($>$1000 K)
from their initial formation, through near-infrared radiation.   

Several surveys of 50-100 young stars in the solar neighborhood have
been conducted
using HST or ground-based AO coronagraphs (Figure \ref{fig:lafrfigure}). Since such young stars are generally
distant (30-100 pc),
these surveys are sensitive to planets in wide orbits. 
The only unambiguously planet-mass object detected is a 
5--8 Jupiter-mass companion to the very low mass brown dwarf 2M1207J,
first discovered
with an AO coronagraph at the European Southern Observatory. No
planetary-mass
companion to a solar-type star has been proven to be detected. (Several 
candidate companions have been proposed, but have generally proven to
have higher
mass than initially estimated or to be unrelated background stars.)
These surveys
have produced upper limits on massive planets ($\geq$ 2 \mjup) in orbits beyond 50
AU, but
only probe solar-system-like scales around a handful of the youngest,
closest stars.

\begin{figure}
%\plotone{Thermal_IR_Venus_Earth_Mars}
\includegraphics[scale=1.]{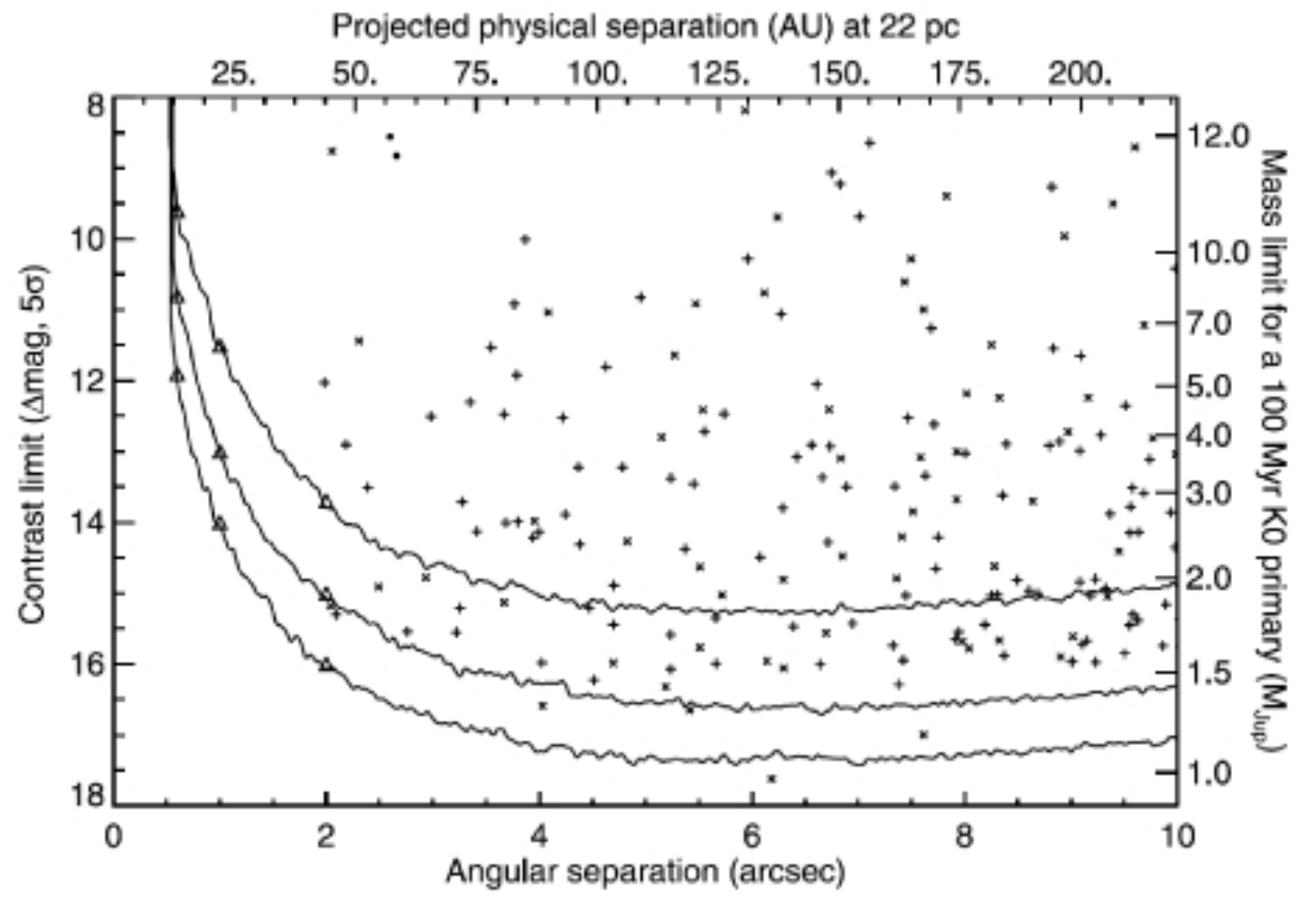}
\vspace{0.1 in}
\caption{Detectable planet contrast vs. angle for  
ground-based adaptive optics images from Lafreniere \etal\ (2007),
reporting on the Gemini Deep Planet Survey of young stars. 
The $y$-axis plots contrast in delta-magnitudes, where 2.5 magnitudes
corresponds to a factor of 10, so delta-magnitude of 10 is a planet:star
contrast of 10$^4$, and delta magnitude of 15 is 10$^6$.  The three curves
show the worst, median, and best performance on the 85 stars surveyed.
Crosses and {\small $\times$}'s show background stars discovered in the survey.
Although a mature Jupiter analog would be undetectable (delta-Magnitude
of 22.5), younger planets would be detectable through their
self-luminosity; the right and top axis show the detectable mass and
angular separation respectively for a typical target in this survey.
Next-generation high-contrast AO systems, such as the Gemini Planet
Imager (under construction for 2011),
should reach delta-magnitude of 17.5 down to angular separations
of 0.2 arcseconds.}
\label{fig:lafrfigure}
\end{figure}

Current-generation instruments have been highly successful in the study
of massive
debris disks surrounding young stars, allowing studies of the dust
distribution (including
asymmetries that may be induced by unseen planets), dust composition
(through photometry
or polarimetry), etc. 

To improve performance, both the European Southern Observatory and
the Gemini Observatory are building next generation (so-called ``Extreme") AO coronagraphs dedicated to extrasolar planet science.  Known
respectively as SPHERE (Kjetil et al., 2006) and the Gemini Planet Imager 
(GPI; Macintosh et al., 2006), these incorporate advanced AO, precision optics and calibration,
and near-infrared spectrographs and polarimeters
for study of extrasolar planets.  Modeling predicts that these systems will achieve
contrasts on the order of 10$^{-7}$ at separations of 0.1-1 arcsec for typical science targets 
(I$\,<\,$8$^{\rm th}$ mag.). 
Although this is insufficient for the detection of Earth-sized or even mature
Jovian planets, these high contrast levels and IWAs nonetheless enable 
significant science studies.  In particular, it may be possible to
detect ``warm" Jovian planets which retain infrared self-luminosity for
perhaps 100~Myr following their formation; more massive planets
remain self-luminous for a billion years, creating a large
sample of potential targets. 

If these systems perform as predicted, they will probe the architectures of other solar 
systems in the 5$\,$--$\,$50 AU radius range, beyond the limit of current Doppler 
searches, and allow spectroscopic
characterization of giant planet temperatures, radii, and composition at spectral
resolving powers of $\sim\,$40$\:$--$\:$50. These systems will also be capable 
of imaging extrasolar debris disks
in polarized light at much lower dust densities than current HST capabilities,
tracing the presence of otherwise-undetectable planets orbiting nearby dusty stars.

In addition to its science reach, ground-based AO coronagraphy provides a unique 
opportunity to test all aspects of direct planet detection in an operational
scientific setting.   These systems will use coronagraphs in the field rather 
than laboratory conditions, operating on starlight over broad wavelength ranges. 
They will incorporate deformable mirrors  similar to those needed for TPF-C.
(Current TPF-C mirrors are in fact derived from 
technology developed for civilian and military AO applications.)
The same observing techniques must be applied to extract planet signals 
from noisy speckle backgrounds and, in fact, many such techniques have been developed or perfected for ground-based astronomy. 

\subsection{Role of Coronagraphy in the 5-, 10-, and 15-Year Time Frames}

\subsubsection{Ground-Based}

Both the European Southern Observatory and the Gemini Observatory
coronagraphic efforts
are predicting first light in early 2011, and the observatories plan to allocate 
substantial telescope time -- 100 to 300 nights -- to
large-scale exoplanet surveys. Facilities such as the Large Binocular Telescope
and the Subaru Telescope may have specialized instruments such as thermal-IR
nullers or PIAAC-type coronagraphs with significant planet detection capability
as well, albeit on a PI level rather than facility level.

On longer timescales, several proposals exist for 20$\,$--$\,$40$\:$m Extremely 
Large Telescopes (ELTs), including Extreme AO coronagraphs.  Although even
these instruments will be unable to detect Earth-sized planets, it is predicted that
they can achieve contrast levels in the 10$^{-8}$ -- 10$^{-9}$  range and inner 
working angles as small as 0.03 arcseconds -- allowing them to see closer to 
stars than even the most ambitious variation of TPF-C.  The combination of small
IWA and higher contrast would allow spectral characterization of mature Jovian
or ice-giant planets in reflected starlight at $\sim\,$1--3 AU
scales and studies of zodiacal circumstellar dust in inner parts of solar systems. 
An IWA of 0.03 arcseconds corresponds to $\sim\,$4$\,$--$\,$5 AU in nearby 
star-forming regions, such as the Taurus and Ophiucus
associations, making possible the direct study of the formation of planetary 
systems in their first few million years.

Overall, though, ground-based AO coronagraphs are unlikely ever to image Earth-sized planets, 
such telescopes will be capable of imaging giant planets in the moderately 
near (2011) future, producing both the first census of giant planets in the 
5--20 AU range and the first spectra of such planets. By 2020, ELT systems will 
push this into the 10--20 earth mass range. 

%\vspace{3mm}

\subsubsection{Space-Based}

The technologies required for a coronagraphic space mission capable of
characterizing Earth-sized exoplanets are not sufficiently mature
to allow such a mission to be undertaken today.  Thus, it is unlikely 
that such a mission can be launched within the next 5 years.  With
immediate and ongoing investment in technology development, the light suppression
and wavefront control performance
necessary to enable an internal coronagraph mission can likely be demonstrated
during the next 3 to 6 years, allowing the launch of such a mission sometime
in the next 10 to 15 years. A similar investment in understanding external 
occulter technologies would likewise position it for a launch in perhaps the 
next 7-15 years, if funding were available.

However, in addition to technological readiness, for a coronagraphic space mission
to be realized in the next 10--15 years, such a mission must be affordable.  
Coronagraphic missions designed both to conduct extensive blind searches for
planets (entailing
many revisits to a given star) and to characterize any discovered planets are currently
not affordable.  To minimize costs, it is therefore
important that such a mission be designed to accomplish {\em only}
those goals which are deemed to be critical
scientifically and unique to a visible-light direct-imaging mission.
In addition, the mission efficiency should be high -- i.e., most of the time
should be spent characterizing exoplanets with minimal time spent
on blind searches.

Thus, a {\em minimum cost} coronagraphic mission capable of
directly imaging and characterizing Earth-sized exoplanets can only be designed
after the locations of nearby exoplanets have been determined, as these data
will establish the minimum required telescope diameter.  A maximally efficient
mission would rely on knowing the orbital elements of nearby candidate
planets, as these data can determine when a given exoplanet is visible at a 
sufficient angle from its central star.  For these reasons, a coronagraphic mission
should be preceded by an astrometric mission capable of providing these data.

\clearpage

\section{Mid-Infrared Interferometry}

\subsection{Why Interferometry?}

The mid-infrared spectral range at wavelengths between 6 and 20\,$\mu$m is rich
in signatures of key atmospheric constituents; the continuum windows between
these absorption bands can be used to determine the surface temperature (see
Fig.~\ref{Mid-IR}). A mission designed to perform such an analysis for
potentially habitable planets must be able to separate planets in the habitable
zone from their parent stars. This translates to an angular resolution of $\sim
35$\,mas, corresponding to 0.35\,AU at 10\,pc, or 0.7\,AU at 20\,pc. This
``inner working angle'', determines the ``natural'' implementation of
spectroscopic missions at different wavelengths: whether via monolithic telescopes in the
visible as discussed in the previous chapter, or via space interferometers with
baselines of $\sim 100$\,m in the mid-infrared.

Once it has been separated from the bright star, the planetary light can be
analyzed spectroscopically. In principle, much information about the chemistry
and physics of planetary atmospheres is available at high spectral resolution,
both in the visible and in the mid-infrared. However, because of limitations in
the achievable signal-to-noise ratio, a first-generation mission which could be
considered for implementation in the 2015-2025 time frame will likely be
equipped with a low-resolution ($R \equiv \lambda / \Delta \lambda \approx 50$)
spectrograph. With such an instrument it will be possible to detect the most
important spectral bands (see Fig.~\ref{Mid-IR}); the detailed determination of
their shapes and the detection of weaker features will have to be left for
future more capable missions.

\begin{figure}[h]
 \begin{center}
 %\begin{tabular}{c}
\includegraphics[width=\textwidth]{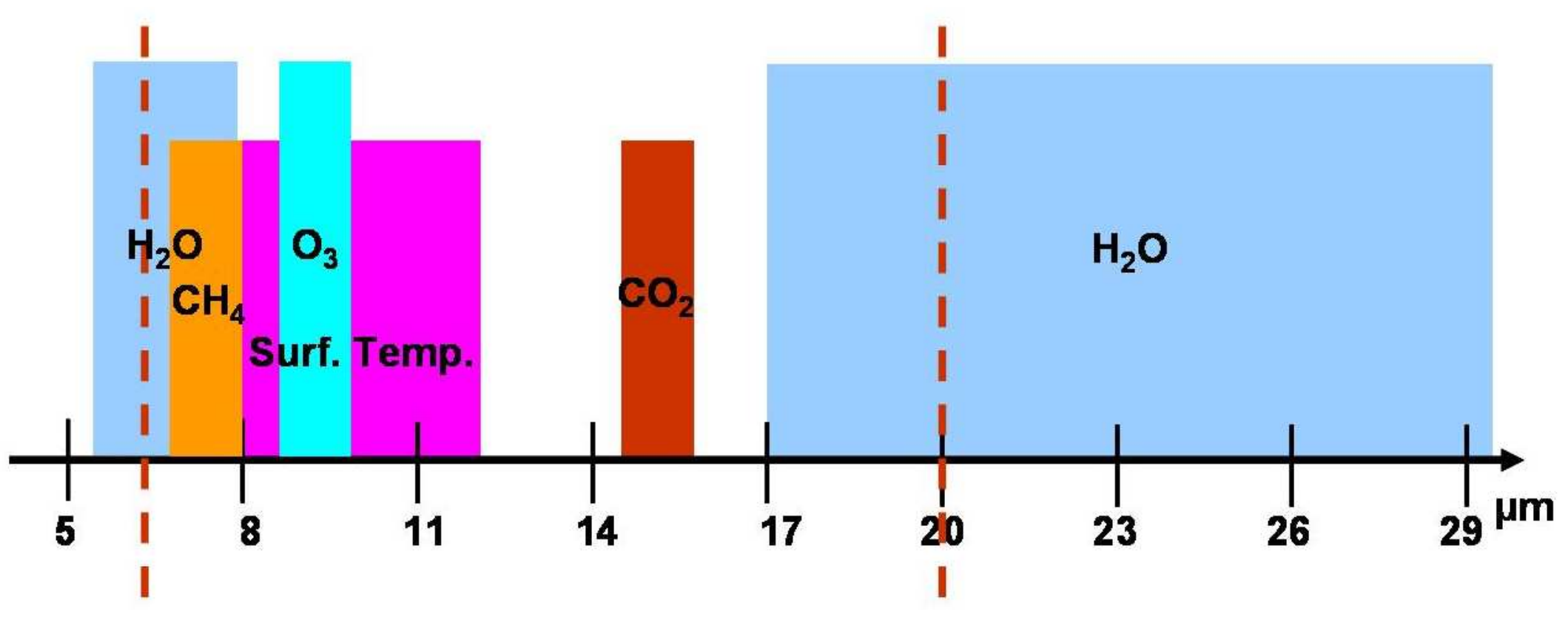}
 %\end{tabular}
 \end{center}
\vspace*{-3mm}
\caption
%>>>> use \label inside caption to get Fig. number with \ref{}
{\label{Mid-IR} Important diagnostic bands in the mid-infrared wavelength
range. The dashed vertical lines indicate the possible wavelength coverage of
an interferometric mission capable of performing an analysis of planetary
atmospheres.}
 \end{figure}

\subsection{Introduction to Nulling Interferometry}

In an infrared interferometer the starlight can be suppressed by introducing an
achromatic $\pi$ phase shift in one arm of the interferometer, so that there is
destructive interference for light arriving on-axis; this concept was conceived
by Bracewell (Bracewell, 1978). The most important characteristic of such a nulling
interferometer is the quality of this destructive output, called the null depth
$N$, defined as the ratio of the intensity of leakage through the null, $I_{null}$, divided by the unnulled intensity $I_0$: $I_{null}/I_0 = (\pi   \sigma_{OPD}/\lambda)^2$. Here $\sigma_{OPD}$ is the optical path difference error between the two beams.

Very precise control of wavefront defects and
precise matching of the interferometer arms is required to achieve a null depth
of $\sim 10^{-5}$, as required for the spectroscopy of Earth-sized planets. In a single-baseline interferometer one would also have to know the absolute
null depth and other sources of background photons very precisely -- otherwise
a planet could not be distinguished from a slightly larger contribution from
the stellar and background photons leaking through the ``destructive'' output
of the interferometer. This is almost impossible to achieve in practice;
therefore the configurations considered for a planet characterization mission
are based on at least three telescopes forming separate nulling
interferometers, whose outputs can be combined with time-variable phases. With this ``phase chopping'' technique it is possible to produce a modulated signal proportional in strength to the brightness of the planet.

\subsection{Ground-Based Nulling Interferrometry}

The zodiacal light from the target planetary system is among the most important
contributions to the background hampering the detection and characterization of
Earth-sized planets within the HZ. It is currently unknown whether the level of zodiacal light
in our Solar System is typical for Sun-like stars, or whether the majority of
such stars are surrounded by much brighter zodiacal clouds. For the
optimization of missions aimed at characterizing Earth-sized planets in the relevant AU distance range,  it is
necessary to obtain statistical information on the expected exo-zodiacal light
levels, or at least upper limits for this quantity. This can be done by
ground-based two-telescope nulling interferometers such as the Keck
Interferometer and the Large Binocular Telescope; nulling instruments are
currently under development at both facilities.

\subsection{Nulling Techniques, Performance, and Progress}

The core of a nulling interferometer consists of a beam combiner with a phase
shifter, which must introduce the required $\pi$ phase shift for all
wavelengths within the observing band. This achromatic phase shift can be
introduced either by geometric methods (e.g., through-focus field-inversion or
field counter-rotation with pairs of periscopes) or with pairs of dispersive
glasses. In any case, one has to maintain very precise symmetry in all other
parts of the system in order to avoid spurious additional phase shifts.

Based on the expected star/planet contrast ($1.5 \cdot 10^{-7}$ at 10\,$\mu$m
and $10^{-6}$ at 18\,$\mu$m for an Sun-Earth analog) and on an evaluation of
instrumental noise, the null depth must be $\sim 10^{-5}$ on average, and it
must be sufficiently stable on the timescale of days so that the signal to
noise ratio improves essentially as the square root of time. This stability
requirement translates into tight instrument control specifications, which can
however be relaxed by means of noise mitigation techniques.

A deep null also requires extremely flat wavefronts, much beyond any feasible
surface quality of the optical elements. It is therefore necessary to flatten
the wavefront with modal filters such as single-mode fibers or wave guides
realized with integrated optics. While these technologies are readily available
at near-infrared wavelengths, materials issues pose serious challenges to the
development of suitable components for the mid-infrared (6 to 20\,$\mu$m)
wavelength range.

A nulling beam combiner is a complex optical system, with considerable mass and
very stringent alignment tolerances that must be maintained during launch,
cool-down and operations. It would therefore be very appealing to employ
integrated optics, which holds the promise of providing compact and rugged beam
combiners that are amenable to inexpensive replication. In principle, with this
technology the entire nulling combiner may be fabricated in a thin plate
comparable in size to a microscope slide. This technology may potentially
reduce the risk inherent in the complexity of the beam-combining subsystem of
the interferometer.

These issues are all subject to ongoing research and development efforts in the
US as well as in Europe. Key technological developments in recent years
include:

\begin{itemize}
\item{Different techniques to build achromatic phase shifters, which allow
broadband destructive interference between beams, have been developed.
Comparative studies should identify the preferred approach.}
\item{Space-qualified delay lines to balance the different optical paths to
nanometer accuracy have been demonstrated by European industry at a temperature
of 40\,K.}
\item{Single mode fibers, and integrated optics modal filters that enable
broadband nulling are under development. Chalcogenide fibres have demonstrated
the required performance of 40\% throughput and 30\,dB rejection of
higher-order spatial modes. Ongoing work is emphasizing silver halide
single-mode filters, which will operate in the 12-20\,$\mu$m band. Photonic
crystal fibers that can cover the whole spectral domain in a single optical
channel are under consideration.}
\item{Considerable progress has been made towards detector arrays with
appropriate read noise and dark current. The Si:As impurity band conductor
arrays developed for JWST appear to be fully compliant with the requirements of
nulling interferometry in space. A reduced-size version of the JWST $1024
\times 1024$ detector, e.g.\ $512 \times 8$, could be read out at the required
rate with a dissipation of a few tens to hundreds of $\mu$W. These devices
exhibit high quantum efficiency (80\%), low read noise (19\,e$^-$), and minimal
dark current (0.03\,e$^{-}\!$/s at 6.7\,K). Such performance permits sensitive
observations, even at moderately high spectral resolution ($R \approx 300$).}
\item{Low vibration cryo-coolers for the detector system are under development.
A European program has led to a prototype absorption cooler providing 5\,mW of
cooling power at 4.5\,K. JPL scientists have demonstrated a system with 30\,mW
of cooling at 6\,K.}
\end{itemize}

\subsection{Nulling Interferometer Configurations}

\begin{figure}
 \begin{center}
 %\begin{tabular}{c}
\includegraphics[width=\textwidth]{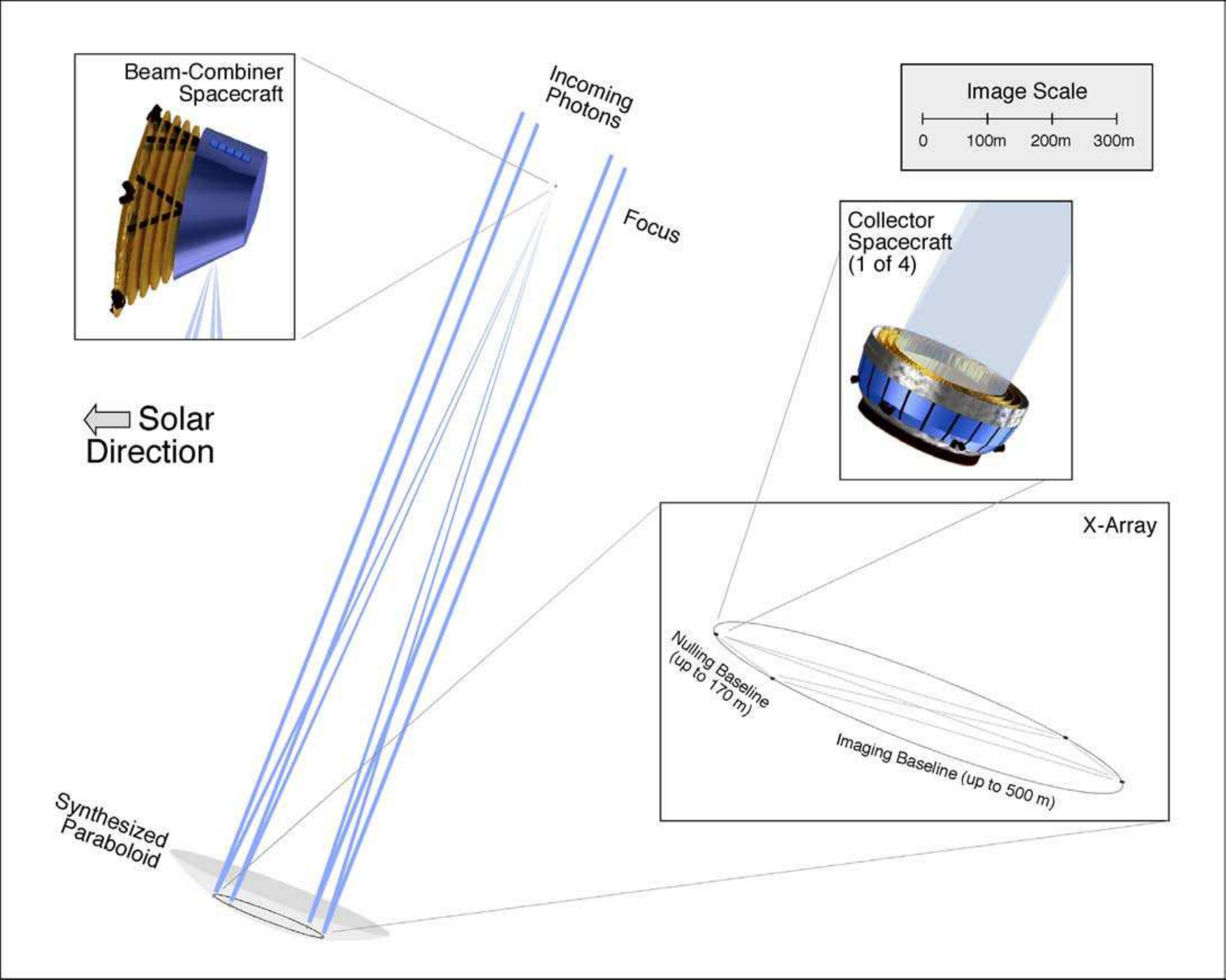}
 %\end{tabular}
 \end{center}
\vspace*{-3mm}
\caption
%>>>> use \label inside caption to get Fig. number with \ref{}
{\label{Emma} ``Emma'' X-array configuration of a nulling interferometer. The
four light collectors are spherical mirrors arranged on the surface of a
paraboloid; the beam combiner spacecraft is located at its focus. (From the
Darwin proposal to ESA.)}
 \end{figure}

A dozen array configurations using phase chopping have been proposed and
studied by ESA and NASA during the past decade. In 2004, common figures of
merit to evaluate their performance were agreed upon. The most important
criteria are the modulation efficiency of the beam combination scheme, the
structure of the point spread function and its associated ability to handle
multiple planets, the overall complexity of beam routing and combination, and
finally, the number of stars that can be surveyed during the mission lifetime.
Among the many configurations studied, the X-array has been identified as the
most promising.

The X-array configuration (see Fig.~\ref{Emma}) separates the nulling and
imaging functions, thus allowing independent optimal tuning of the shorter
dimension of the array for starlight suppression, and that of the longer
dimension for resolving the planet. Most other configurations are partially
degenerate for these functions. The X-array also lends itself naturally to
techniques for removing instability noise, a key limit to the sensitivity of a
nulling interferometer. The assessment studies settled on an imaging to nulling
baseline ratio of 3:1, based on scientific and instrument design constraints. A
somewhat larger ratio of 6:1 may improve performance by simplifying noise
reduction in the post-processing of science data.

The X-array geometry consists of five spacecraft: four telescopes or light
collectors and one beam combiner spacecraft. It can be implemented in two
different ways, either as a planar array, or as an out-of-plane configuration
nicknamed ``Emma''.

In a planar array, all five spacecraft are kept in a plane orthogonal to the
line of sight. The light is collected with four telescopes, which send
collimated beams to the central beam combiner. This arrangement is highly
symmetric, which facilitates achieving a high null depth. The main
disadvantages of planar arrays are the relatively complex telescopes, which
have to be deployed after launch, and limited accessibility of the sky
restricted at any given time to a cone around the anti-Sun direction.

The Emma configuration consists of four collector spacecraft carrying spherical
mirrors, which are arranged on a paraboloidal surface and a beam combiner
located at the focus of the paraboloid (see Fig.~\ref{Emma}). This arrangement
is less symmetric than a planar array, which makes it more difficult to achieve
a deep null. On the other hand, the light collectors are relatively simple and
do not need on-orbit deployment, and a larger fraction of the sky is accessible
at any given time. The Emma configuration is therefore currently considered as
the baseline concept for a planet characterization mission.

\subsection{Formation Flying}

To realize the baseline lengths of $\sim 100$\,m or more needed for a
space-based nulling interferometer, one cannot mount all elements on rigid
connected structure, and tethered satellites appear to be impractical, too.
Therefore one has to place each collector and the beam combiner on a separate
spacecraft, as shown in Fig.~\ref{Emma}, and manoeuver the formation safely and
with cm-level precision.

Simulations of algorithms for interferometer constellation deployment at L2,
and a 2D robotic breadboard at JPL have shown the feasibility of formation
flying. The European PRISMA mission intends to demonstrate formation flying and
rendezvous technologies in space. PRISMA comprises two spacecraft and will be
launched in autumn 2008 into a low, Sun-synchronous orbit (600-1000\,km) with a
mission lifetime of about 8 months. The main objectives are to carry out
technological flight demonstrations and manoeuvring experiments, including
guidance, navigation, control, and sensor techniques. The positioning of the
spacecraft relies on a relative GPS technology, which should have an accuracy
of $\sim 10$\,cm. For inter-satellite distances less than 6\,m, additional
optical metrology should improve this accuracy. Micro-thrusters needed for
station keeping based on field effect electric propulsion (FEEP) or cold gas
have also been developed over the past few years. It thus appears that the
implementation of an interferometric formation will be a complex and
challenging engineering task, but the technologies required are under development both for exoplanet and for gravitational wave searches (LISA Pathfinder). 

\subsection{Role of Nulling Interferometry in the 5-, 10-, and 15-Year Time
Frame}

In the five-year time frame, ground-based nulling interferometry can obtain
statistical information on the brightness of the exo-zodiacal light from
Sun-like stars, or place relevant upper limits on this quantity if the typical
level is not much higher than in the Solar System. This will enable an
optimization of space-based missions aimed at characterizing terrestrial
planets, and will put the predictions of the expected performance of such
missions on a more secure footing.

A space-based nulling interferometer for the spectroscopic characterization of
Earth-sized planets could be implemented in the 10- to 15-year time frame. It is
important, however, to develop the enabling technologies for such a mission
rapidly to a readiness level that is sufficient to begin implementation from a
technical point of view. This will mitigate the inherent risks, support
realistic cost estimates, and enable a trade-off between visible-light
coronography and mid-infrared interferometry for the first terrestrial planet
characterization mission.

\subsection{Detailed recommendations on nulling interferometry}

Nulling interferometry represents an attractive way of obtaining fundamental
data on Earth-sized planets, including the composition of their atmospheres.
This information is complementary to that obtainable at visible wavelengths.
NASA should therefore continue the preparations for an infrared nulling
interferometer. Since a visible-light coronograph and an infrared
interferometer are equally appealing scientifically, the decision which one to
fly first should be based on cost and technology readiness at a time when one
of the two concepts is ready for implementation.

We therefore recommend the following actions:

\begin{itemize}
\item{The ongoing efforts to characterize the typical level of exo-zodiacal
light around Sun-like stars with ground-based nulling interferometry should be
continued.}
\item{A vigorous technology program, including component development,
integrated testbeds, and end-to-end modeling, should be carried out in the
areas of formation flying and mid-infrared nulling, with the goal of enabling a
nulling interferometry mission around the end of the next decade.}
\item{The fruitful collaboration with European groups on mission concepts and
relevant technologies should be continued.}
\end{itemize}

\clearpage

\section{Microlensing} 

The only information from general relativity that is required in the
analysis of a microlensing search for planets is the
small angle deflection of a light ray passing 
close to a spherically symmetric mass, $\alpha= 4GMD_{OL}D_{LS}/
(rD_{OS}c^2)$, where the parameters are defined in
Fig. \ref{fig:lensgeom} with the symbols $O,L,S$  referring to observer,
lens and source respectively, $M$ is the mass of the lens, $c$ is the
speed of light and $G$ is the gravitational constant.
 Microlensing occurs
when an intervening star (lens) passes almost directly between a more
distant star (source) and the observer, and the light from the source
is gravitationally focused by the lens into two images on either side
of the lens as shown in Fig. \ref{fig:lensgeom}. The location of the
images is a transformation of source coordinates to image coordinates
via the lens equation shown in Fig. \ref{fig:lensgeom}.
These images spread into a ring (Einstein ring) of radius $R_{
E}=\sqrt{4GMD_{OL}D_{LS}/(c^2D_{OS})}$ if the lens is
directly in front of the source ($r_0=0$ in Fig. \ref{fig:lensgeom}).
Generally, the image separation is on the order of 
milliarcseconds, and the images usually cannot  be
resolved. But the light  received by the observer from the two images
is magnified achromatically. Surface brightness ($\rm 
ergs\,cm^{-2}sec^{-1}sterad^{-1}$) is conserved along a ray path, so
the magnification $A$ is the transformation of an element of area of
the source projected onto the lens plane to an element of area of the
images (like $\vec{dx}\times\vec{dy}\rightarrow rdrd\theta$) with
\begin{eqnarray}
A=A_1+A_2&=&\frac{(u^2+2)}{u\sqrt{u^2+4}}, \nonumber\\
u(t)&=&\sqrt{u_{min}^2+[(t-t_0)/t_{E}]^2},
\label{eq:amplification}
\end{eqnarray}
where $u=\theta_s/\theta_E$, the 
ratio of the angular separation of source and lens to the angular Einstein
ring angular radius, $u_{min}$ is the normalized impact parameter, and
$t_0$ is the time of closest approach.  The magnification for each
image is most easily 
represented as the inverse of the Jacobian determinant of the
transformation from the image coordinates to the source
coordinates. The magnification increases as the source-lens separation
decreases, such that a microlensing event is characterized by a bell
shaped light curve.  The singularity in $A$ at $u=0$ is called a
caustic point, which singularity is removed when the source is not a
point source. The time scale of an event $t_{E}=R_{E}/v
\propto \sqrt{M}$ is the time it takes the source to move an Einstein ring
radius relative to the lens with $v$ being the relative velocity
projected onto the lens plane. $t_{E}=\theta_{E}/\mu_{rel}$ in
terms of the relative proper motion $\mu_{rel}$. 

\begin{figure}[h]
\begin{center}
\includegraphics*[scale=0.5]{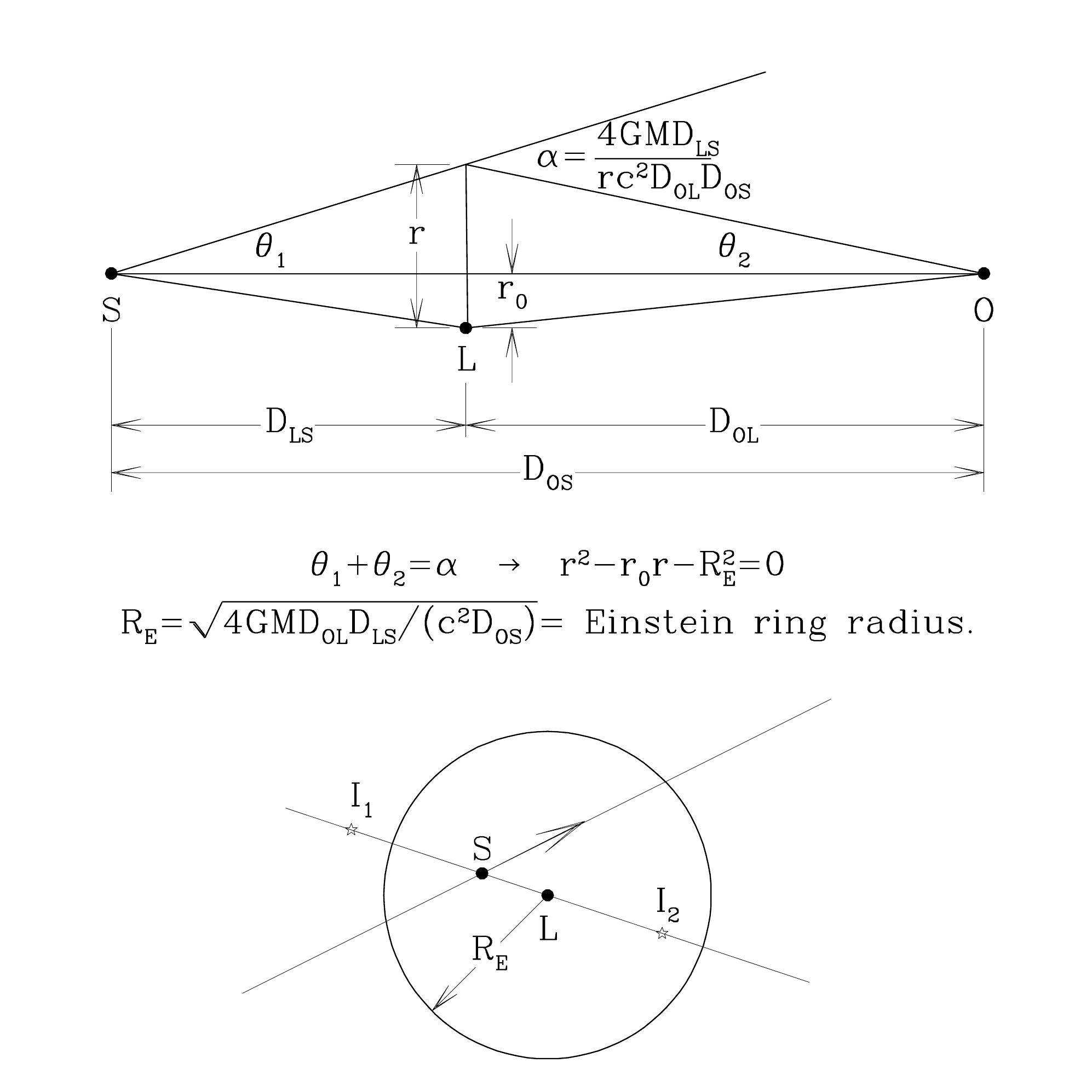}
\caption{Geometry for microlensing with the lens equation
defined. \label{fig:lensgeom}} 
\end{center}
\end{figure} 

The analysis is more complicated if the lens is binary, but it is
still straightforward. The lens equation relating coordinates of an
image in the lens plane to the coordinates of the source in the source
plane is now a vector equation. The quadratic equation for a single
lens goes to two 5th order equations, or more conveniently, to a single
5th order equation in the complex plane ({\it e.g.} Witt, 1990) where
the solution yields either three or five image positions instead of
two for the single lens. Determination of the magnification is the
same transformation of areas according to the coordinate transformation
between source and image coordinates,  $A=A_1+A_2+A_3$ or
$A_1+A_2+A_3+A_4+A_5$. What was a caustic point for a single lens has
become one, two or three closed caustic curves in the source plane which
transform to critical curves in the lens plane. (The Einstein ring is
the critical curve for the caustic point for a single lens, so when
the source is on a caustic, the 
image is spread along the corresponding critical curve.) There are
three images if the source is outside the closed caustic curves, and
five images if inside. There is thereby a sharp peak in the light
curve on a caustic crossing (somewhat rounded by finite source
effects), with increased magnification while the source is inside the
caustic curve.  

If a planet is the second mass in a binary lens, it significantly
perturbs the light curve  when it is near one of the unperturbed
image positions corresponding to the source being near a caustic
curve.  This explains why a puny planet can have a much
larger effect on the light curve than simply adding its lens effect on
the source to that of the star; it focuses or defocuses the light
already focused by the star.   Contours of a given magnification can
be drawn in the lens plane for fixed source position or in the source
plane for fixed planet position. The planetary perturbations can be
quite large, but also quite short.  The widths of 5\% perturbation
contours in the source plane are comparable to the projected Einstein ring
radius of the planet. The time scale of a planetary perturbation will
thus be comparable to the crossing time of the planetary Einstein
ring, $t_{Ep}=R_{Ep}/v\propto \sqrt{m}$. For $D_{OL}=4$ kpc, and
$v=200$ km/sec, $t_{Ep}=1.1$ day for a Jupiter mass object and 1.5
hours for an Earth mass object. Fig. \ref{fig:perturblc} shows the
perturbation of a microlensing light curve by a Jupiter mass planet. 
Frequent photometric measurements are necessary to catch the
perturbations from the smaller planets.   

\begin{figure}
\begin{center}
\includegraphics*[scale=0.3]{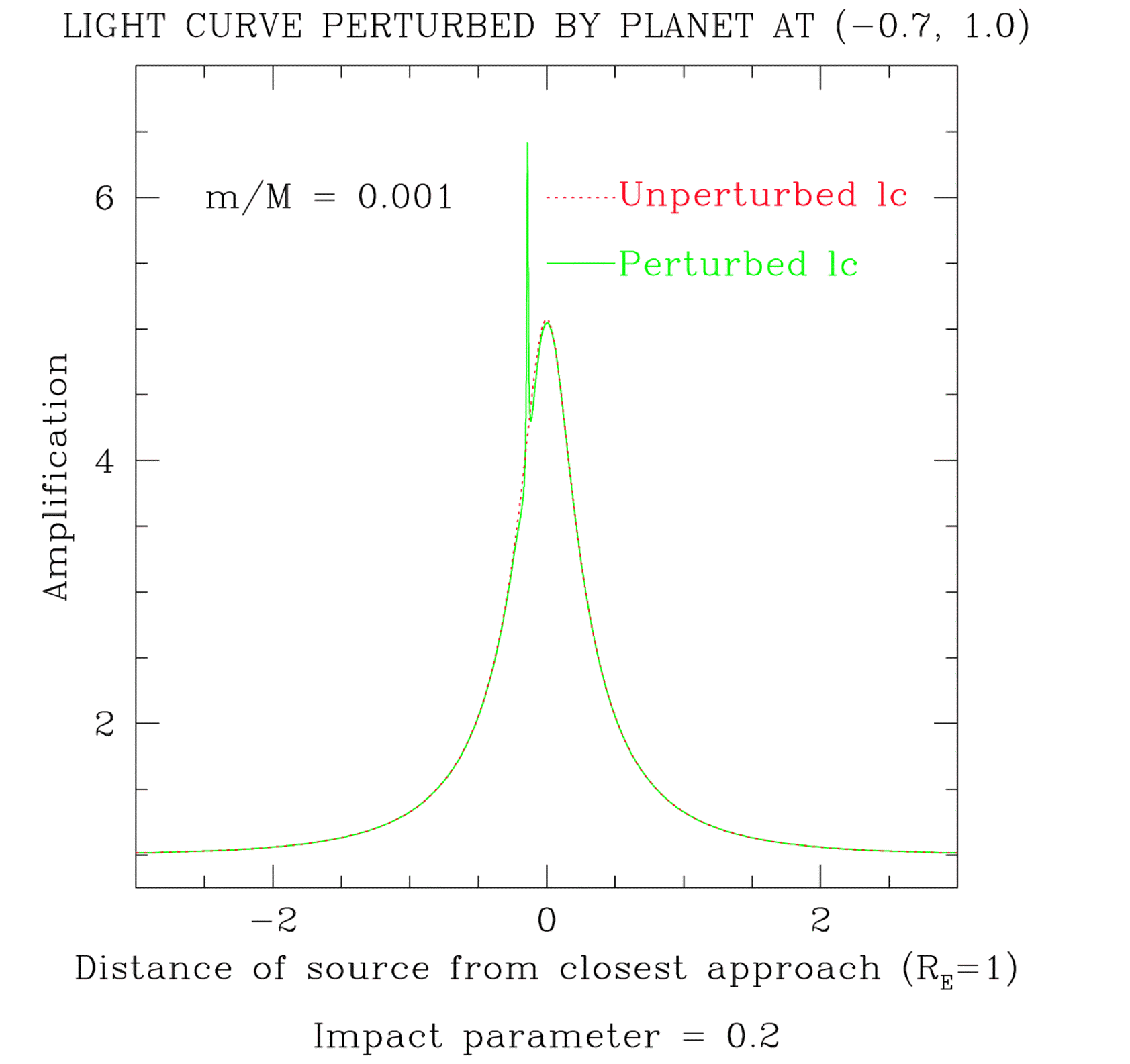}
\caption{Perturbation of a light curve for a planet with $q=m/M=0.001$
located in the lens plane at $x,y=-0.7,\,1.0$ in units of $R_{\s
E}$. The origin of coordinates is at the lens  with the $x$ axis
parallel to the source trajectory. \label{fig:perturblc}}  
\end{center}
\end{figure}

For a star with a planetary companion, there is usually a closed caustic
curve at the lens position as well as one or two additional closed
caustics symmetric about the 
lens-planet line depending on whether the planet is outside or inside
the Einstein ring radius.  A single caustic results at the lens
position if the planet is exactly on the Einstein ring or if the
planet mass is sufficiently large. The central caustic means that for
a high magnification event, where the source passes very close to the
lens, there is a good chance for the source to cross the central caustic
near the peak magnification and reveal the planet's presence without
the planet being near one of the images. The probability of detecting
a planet in a high magnification event (above a specified
minimum peak magnification, $A_{crit}\approx 1/u_{th}$), is determined
empirically by calculating the light curves for trajectories of the
source across a magnification map in the source plane for a particular
value of $q=m/M$ and star-planet separation $x_p=r_p/R_{E}$, where
the trajectories are distributed uniformly in 
$u_{min}<u_{th}$ and angle relative to the star-planet line.  A
representative criteria for detection in these events is a 5\%
perturbation of the microlensing light curve.  For $q=m/M=0.003$ the
probability of detection is unity for a wide range of star-planet
separations in the ``lensing zone'' spanning the Einstein ring radius
(Griest and Safizadeh, 1998), although there is an $r_p$ {\it vs.}
$1/r_p$ degeneracy in the lens-planet separation with modest
differences in $m/M$ for the two solutions (Udalski {\it et al.} 2005;
Bennett, 2007b).
The probability of detection decreases with decreasing minimum peak 
magnification (increasing $u_{th}$) and with decreasing $q$, such that
for $q=10^{-4}$, the region of high probability of detection is
limited to approximately the range $x_p=1.0\pm 0.2$. The high
probability of planet detection in the high magnification events has
motivated the MicroFun group to limit their high time resolution, high
precision photometric followup coverage of the light curves to only a
time interval around the peaks of the relatively rare high
magnification events found by the OGLE and MOA surveys described
in the Appendix. 

Generally, from the light curve alone, one can obtain $m/M$ from the
ratio of the time scales of the planetary perturbation to that of
the total event, and the projected displacement of the planet from the
lens in units of $R_{E}$ and the angle the star-planet line makes
with the source trajectory from the location on the light curve
of the perturbation relative to the maximum magnification. From the
ground the source will almost always be blended with other nearby
stars, but the intrinsic brightness of the source can often be
obtained if the light curve is sufficiently well sampled. 
At any instant the measured flux density is  $F_i=F_0 A_i + B$, where
$F_0$ is the flux density from the source (intrinsic brightness), $A_i$ 
is the magnification and $B$ is the flux density from the blended
stars. The only thing measured is $F_i$. There are many determinations
of $F_i$ over the event. $A_i$ is a function of $t-t_0$, $t_{E}$ and
$u_{min}$, where $t_0$ is the time of maximum magnification.  There are
five constant parameters, $t_0,t_{E}, F_0, u_{min}$ and $B$ that can
be solved for in a least squares fit to the light curve, where $F_0$
is the desired flux density from the unmagnified source. The
uncertainty in the parameters depends on how well the light curve is
sampled and the photometric uncertainties in the measurements.

The point source representation of a microlensing event is
an adequate approximation if the angular Einstein radius of the planet
$\theta_{Ep}$ is much larger than the angular size of the star
$\theta_*$. For $1M_{\odot}$ at 4 kpc with a source at 8 kpc,
$\theta_{E}\approx 1$ mas, $\theta_{Ep}\approx
30 \mu$as for Jupiter with the same geometry but only 1.8 $\mu$as for
Earth. The angular radius $\theta_*$ for a star with radius
$1\,R_{\odot}$ at 8 kpc is 0.6 $\mu$as, so Jupiters and perhaps
Neptunes can be treated with the point source  analysis if
the source is a main sequence star, but not Earths. 

Magnification of a finite source is determined in principle by
integrating the magnification over the area of a limb darkened
source. However, singularities when the source overlaps a caustic make
such an integration tricky.  Bennett and Rhie (1996) solve this
problem by integrating in the image plane, and Wambsganss (1997)
creates a magnification map in the source plane by ray tracing photons
at the image plane back to their source positions for a given lens-planet
configuration. The density of source photons is proportional to the
ratio of image to source areas and hence to the magnification. A
round, limb darkened source is convolved with the magnification map to
obtain the true magnification of the source. The source trajectory
through the magnification map thereby determines the light curve.
In practice, some variation of this procedure must be adopted in a
least squares fit to the observed light curve to obtain the properties
of small planets. The angular size of the source can be determined
from the measured baseline brightness and its color (flux density
$=\theta_*^2\sigma T_e^4$). The problem of determining the
magnification for a finite source for arbitrary planet sizes is
solved. 

Fundamental to estimating the yield of planetary detections from a
microlensing survey is the probability of detecting a planet during an
event under the assumption that the lens has a planet with particular
$m/M$ and projected separation. Several schemes have been proposed to
determine whether or not a planet is detected. Gould and Loeb (1992),
in an analytic analysis, assume a planet has been detected if the
perturbation of the light curve exceeds 5\% of the single lens light
curve at the point of the perturbation. Peale (2001) assumes a planet
is detectable if the perturbation of the single lens light curve
exceeds $2/(S/N)$ sometime during the event, where $S/N$ is the
instantaneous photometric signal-to-noise ratio for the magnified source
and where it is assumed that at least 20 consecutive photometric
points during the perturbation are necessary for a detection. In
addition, 2 m telescopes with 60 s integrations in I band with high
time resolution 
photometry throughout the event are assumed. The probabilities are
derived as a function of the semimajor axis $a$ of the planet, where the
peak probability occurs where $a$ is approximately the mean Einstein
ring radius of the distribution of lenses along the line-of-sight (LOS).
Because of the numerous degeneracies in the analysis of a microlensing
event, it is necessary to average the probabilities over the
distribution of the projected position of the planet on the lens plane
for each semimajor axis, over the lens mass function, over the
distribution of impact parameters, over the distribution of the lenses
and sources along the LOS, and over the I band luminosity functions of
the sources adjusted for source distance and extinction.  Both the
Gould and Loeb and Peale analyses assume a point source 
with $q=0.001$.  Probabilities for other values of $q$ vary
approximately as $q^{1/2}$ in these analyses but vary somewhat with
semimajor axis ({\it e.g.} Peale, 2001). 

The maximum
probability of detecting a planet with $q=0.001$ was near 17\% for at
a projected separation near the Einstein ring radius in Gould and Loeb
analysis and about 10\% in the Peale analysis, although the latter was
raised to about 20\% if the distribution of impact parameters out to
$u_{min}=1$ was according to empirical detection efficiencies rather
than being uniform. It was the relatively high probability of planet
detection found by Gould and Loeb (1992) that motivated the
development of the microlensing searches for planets. Detection
probabilities for low mass planets, where finite source size must be
accounted for were first determined by Bennett and Rhie (1996).

More accurate, semi-empirical schemes for determining the likelihood of
planet detection during an event, which are described below, include
finite source size and simulated light curve data analysis.  With a
detection criterion of a minimum reduction in $\chi^2$ of 160 in going
from a single lens fit to the simulated data to a binary lens-planet
fit to the same data, the probability of detection scales as $q^{1/2}$
for large $q$, like the analytic result, but closer to $q$ for small
$q$ (Gaudi, 2007a; Bennett, private communication, 2007). 

\subsection{Description of current ground based microlensing searches
for planets}
Two groups are performing microlensing
surveys from the ground, OGLE (Optical Gravitational Lensing
Experiment) and MOA (Microlensing Observations in Astrophysics).
Two additional groups, PLANET (Probing Lensing Anomalies Network) and
MicroFun (Microlensing Follow-up Network) are performing high time
resolution, high precision photometry on selected events from alerts
by the survey groups. 
 
The OGLE group uses a dedicated 1.3 meter telescope at Las Campanas
Observatory in Chili (field of view (FOV) 35' x 35') to monitor 120 million
stars toward the Galactic bulge.  Difference Image Analysis (DIA)
photometry (Alard and Lupton, 1998; Alard, 2000; Wozniak, 2000) leads
to better quality photometry in dense stellar fields. 
Candidate microlensing events are posted on the web, and there is also
a mailing list providing on line notification of events for follow-up
by high time resolution high precision photometry in search of the
short time scale planetary perturbations of the microlensing light
curve. In 2006, the OGLE group detected about 550 events, which is a
typical yearly output.
 
MOA has a new (2004) 1.8 meter telescope at the Mt. John Observatory
in New Zealand that is dedicated until 2010 (field of view 2 square
degrees). More than 100 million stars are monitored toward the Galactic
bulge with two fields containing perhaps $10^8$ stars including
unresolved stars being imaged every 10 minutes. The remaining MOA fields
are imaged every 50 minutes.   MOA also uses DIA and
does more frequent sampling of selected high magnification events.
Microlensing events are posted on the web, and there is on
line notification of events for follow-up. During 2007, the MOA group
detected about 500 events, and this should rise slightly in future years.

The PLANET Group consists of 32 team members affiliated with 20
institutions in 10 countries. Currently 5 telescopes at 4 different
longitudes are listed on the PLANET web site. (Boyden 1.5 meter,
S. Africa; SAAO 1 meter; La Silla Danish 1.54 meter; Canopus 1.0
meter, Tasmania; Perth 0.6 meter) PLANET attempts to follow all of the
events that it can, and that practice led to the detection of the
first sub-Neptune mass planet in a low magnification event (Beaulieu {\it
et al.} 2006).  In addition to the above telescopes, the PLANET group
is now collaborating with Robotnet-1.0, a collection of three 2 meter
class, UK-operated robotic telescopes. Two of the latter three
telescopes in Hawaii and Australia have been purchased and refurbished
by the Las Cumbres Observatory, where microlensing events continue to
be monitored (T. Brown, private communication 2007).

The MicroFun Group lists 25 active members observing on 16 telescopes
on 5 continents with apertures ranging from 0.35 to 2.4
meters. Amateur astronomers are participating in the high 
time resolution photometry, and they seem to be increasingly important
in establishing a dense longitudinal distribution of observers to
maximize the probability of complete coverage of the perturbed light
curves. Because of much higher probability of detecting a planet near
the peak of a high magnification event (Griest and Safizadeh, 1998),
MicroFun only follows these relative rare events (usually with
A of order or larger than 50 with an all-out effort on those events with $A>100$) that are
captured before the peak. The telescope at CTIO follows the
alert until it is clear that it will be high magnification at which
time the other observatories are alerted. We note that the
detection probability for the high magnification events is unity for a
range of planetary positions spanning the Einstein ring, and only a
relatively short time span around the peak of light curve need be
covered with high time resolution observations leading to an efficient
allocation of scarce observing time.  

\subsection{Simulation of the next generation ground based program}
The simulated event rate is determined from the empirical microlensing
optical depth toward the Galactic center (GC), which is a function of the
line of sight. The optical depth $\tau$ is defined as the probability of a
random single source being microlensed at any instant with a
magnification $A>1.34$, meaning  that the source is inside an
Einstein ring.  Empirically, this can be represented as the fraction
of the total observing time with $A>1.34$ per source or $\tau=\Sigma_i
[t_i/\epsilon(t_i)]/(T_oN_*)$, where $t_i$ is the time event $i$ has
$A>1.34$ weighted by the efficiency $\epsilon_i$ of detecting events
with this value of $t_i$, $T_o$ is the total time interval of
observation, and $N_*$ is the number of sources being monitored. To
express this in terms of $t_{E,i}=\theta_{E,i}/\mu_i$, the time to
cross an Einstein ring radius, an average over a uniform distribution
of impact parameters is assumed for a particular event, with the result
\begin{equation}
\tau=\frac{\pi}{2N_*T_o}\Sigma_i\frac{t_{E,i}}{\epsilon_i}.
\label{eq:opticaldepth}   
\end{equation}
Typical empirical values of $\tau=1\,{\rm to}\,3\times 10^{-6}$ ({\it
e.g.} Sumi {\it et al.} 2006). Theoretical estimates of $\tau$ from
plausible Galactic models are consistent with the empirical values
({\it e.g.} Peale, 1998). If $N_*=10^8$, and the average time
scale for an event $\langle t_{E}\rangle\approx 20$ days, there would
be about 10 new microlensing events/day with about 190 ongoing at any 
time.  

Some assumptions and procedures are similar to both this simulation of
a ground based program and that of a space based program treated by
Bennett and Rhie (2002). The luminosity function of the
source stars is that of Holtzman {\it et al.} (1998). Star densities
are assigned relative to Holtzman {\it et al.} densities in Baade's
window. I-band extinctions in two fields are $A_I=1.6,\,1.5$. An
optical depth of $\tau=\sim 3\times 10^{-6}$
from several references before 2002 is comparable to more recent
determinations ({\it e.g.} Sumi {\it et al.} 2006). The mass function
of the lenses (Zoccali, 2000) is given by
$f(m)=m^{-\alpha}$, $\alpha=2.3$ for $m>0.8M_{\odot}$, 1.33 for
$0.15<m<0.8M_{\s\odot}$, 0.3 for $.05<m<0.15M_{\odot}$. Stellar
remnants include: white dwarfs $\sim 13\%$; neutron stars and black
holes contribute less than 1\% and 0.1\% respectively.  

Rather than average over the distributions as was done for the
analytic probabilities of detection ({\it e.g.} Peale, 2001), a Monte
Carlo analysis is used, where a stellar lensing event is selected for
each star in the frame with lens parameters selected at random from
the mass function, and density and velocity distribution and
dispersion from a standard galaxy model. If the
selected lens is not a main sequence star, the event is ignored. All
lensing events have $u_{min} <3$ with source stars at 0.5 kpc behind
the GC at 8 kpc. The orientation of each planetary orbit plane is
selected at random and the planet put at random phase in the orbit with
the planets being on circular orbits between .25 and 30 AU distributed
logarithmically. Mass fractions range  from $3\times 10^{-7}$ to
$1\times 10^{-3}$ distributed uniformly. Light curves are constructed
using a 10 minute cadence. Finite source effects are included assuming
a mass radius relation from Bertelli {\it et al.} (1994), and limb
darkening effects are included. The CCD camera is assumed to collect 13
photons/sec for I=22.  High mag events $A_{max}> 200$ are excluded
because of the ambiguity in determining the lens-planet
separation for planets detected near the peak magnification.

Seeing and weather statistics at each site are used to randomly select
these conditions for each observing night, where standard air mass and wave
length correction formulae (Walker, 1987) are used to convert the
seeing data to seeing estimates for bulge observations in the I-band.
Sky brightness was calculated following Krisciunas and Schaefer
(1991). Systematic photometric errors of 0.7\% arising from blending
of sources in the crowded field are assigned based on the set of
photometry data for stars varying in a predictable way from
microlensing (Udalski, 2003). Photon noise from all sources is added
to the seeing and systematic photometric errors to determine the
uncertainty of the points on the light curve, where parts of the light
curve are sometimes missing because of weather, limited coverage off
the central part of the the bulge season, or seeing
$>2^{\prime\prime}$. If the seeing and weather at Mt. John in New
Zealand is as poor as that at Siding Springs in Australia and Las
Campanas has similar conditions to Paranal in Chile, the probability
of observing the GC on a given day for three telescopes
at Mt. John in New Zealand, Las Campanas in Chile, and SAAO in South
Africa reaches a maximum close to 60\% in the middle of the bulge
season and falls to zero when the Sun is close to the GC (Peale,
2003).  A probability of 1 would indicate that the GC is observable at
all times during the day from one or more observatories. The GC is
considered observable if the zenith angle of the GC is less than
$70^\circ$, the zenith angle of the Sun is greater than $105^\circ$,
and the night is usable. 
 
With light curves constructed as above for each simulated event with
uncertainties appropriate for each observation, a single lens,
point source light curve is fitted to the event. The planet is assumed
detected if a second light curve fitted to the binary lens with the
same data set reduces $\chi^2$ by more than 160 from that of the single
lens fit. Bennett (2004) shows light curves constructed as
above, which satisfy the criteria for a detected terrestrial mass
planet, but which are too noisy, or lack sufficient coverage to allow
the determination of the planet-star mass ratio or its projected
normalized separation from the star.  For this reason, he defines a
second class of ``discovered'' planets for which the planetary
parameters can be determined. Criteria for a ``discovered''
planet include at least 60\% coveraged of the total planetary
perturbation of the light curve, at least 40\% coverage of each of the
beginning, middle and end regions of the perturbation, where the first
and last regions begin and end when the deviation reaches 10\% of the
maximum planetary deviation or 0.3\% of single lens magnification. $S/N$
criteria require at least one measurement that detected a stellar
microlensing event by $\geq 10\sigma$ and one measurement that detects
the planetary deviation by $\geq 5\sigma$. 

Bennett assumes different star densities in different fields, and
considers different observing strategies, which vary the number of
fields observed with the integration times on each field.  The optimum
strategy to detect the maximum number of planets depends on the mass
of the planet. Detection efficiency is highest in sparse fields but
there are many more events and planetary detections in dense fields,
since both optical depth and the number of stars go as $N_*$.

The unpublished ground based simulations by Bennett {\it et al.}
(2007a) and separately by Gaudi {\it et al.} (2007a), found
considerably higher planetary detection rates than found 
for the next generation ground based program shown in Fig.
\ref{fig:bennett1}. However, these latter simulations
hypothesized the existence of 3 new 2
meter telescopes with 2 square degree FOV (Bennett) or 4 new 2 meter
telescopes with 4 square degrees FOV (Gaudi). Since funding for such
new arrays is unlikely, we chose to describe the more conservative
next generation ground based program simulated by Bennett
(2007b), which program could be implemented with modest funding for
one new telescope, especially with the likely foreign participation.  
The red, hatched curve in Fig. \ref{fig:bennett2} shows the sensitivity for
``discovered'' planets from this conservative simulation for the next
generation ground based program.  The
microlensing sensitivity criterion is that more than 10 planets are
discovered in a four year program for all points above the curve. The
pink curve in Fig \ref{fig:bennett2} shows the sensitivity for the
current ground based program. The red points indicate the currently
published planets that have been detected by microlensing. The solar
system planets are indicated by letters. Regions of
sensitivity for various techniques and planned or notional missions are also indicated for
comparison, where the planets detected by Doppler are indicated as
lower bounds on the masses, and the transiting planets are indicated by
blue squares.  

\begin{figure}
\begin{center}
\includegraphics*[scale=0.5]{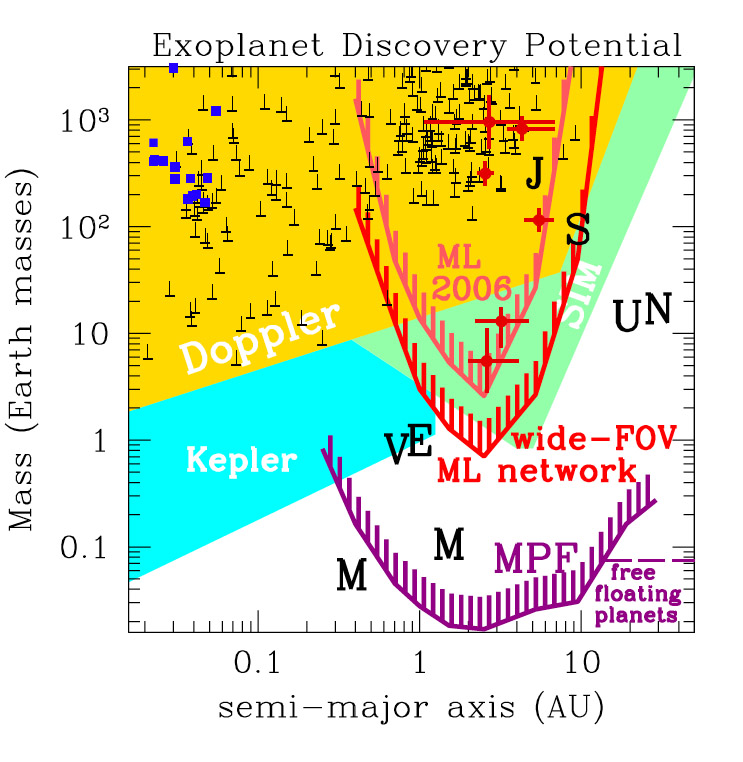}
\vspace{1.5in}
\caption{Regions of sensitivity for current ground based, next
generation ground based, and space based microlensing programs for
planet discovery along with the regions of sensitivity for 
various techniques and systems. 
The microlensing sensitivity criterion is that more than 10 planets are
discovered in a four year program for all points above the curves. See
text for details.
(From Bennett, 2007b)
\label{fig:bennett2}}  
\end{center}
\end{figure}

\subsection{Antarctica}
Microlensing followup and survey telescopes have been proposed for
Antarctica as a way to provide continuous coverage of events from the
ground  during the winter months (Yock, 2007). Atmospheric seeing at
the south pole itself is poor, averaging $1.8^{\prime\prime}$
(Travouillon {\it et al.} 2002), but seeing is phenomenally good at
Dome C at $74^\circ$ south and 3233 m altitude, with a median of
$0.27^{\prime\prime}$ 
and a best measurement during the winter of 2003-4 of
$0.07^{\prime\prime}$ at 500 nm (Lawrence, {\it et al.}
2004). However, these statistics were determined by unattended
scintillation detectors, which are insensitive to near surface
turbulence (B. MacIntosh, private communication, 2007). Further
testing of the site indicates that a telescope would have to 
be mounted more than 30 m above the surface to approach these seeing
statistics, where median seeing is only $1.3\pm0.8$ arcsec on an 8.5 m
tower. (Aristidi, {\it et al.} 2005).  Wind
velocities are low and the site is cloud free 75\% of the time with
10-100 times less IR sky emission than any mid latitude site (Ashley
{\it et al.} 2005). The atmosphere is so dry that the IR absorption bands
blocking the spectrum at mid latitudes are largely absent, allowing
observations in a wide range of IR bands (Burton, {\it et al.} 2005).  
Dome A at $81^\circ$ south and at an altitude of 4200 m should have
perhaps even superior characteristics, but the complete lack of
infrastructure at this site may make Dome C preferable (Lawrence {\it
et al.} 2004). Still, the Chinese plan to build a scientific base on
Dome A, and Don York of the University of Chicago is planning a giant
array of 0.5 meter telescopes to continuously monitor 8000 square
degrees of sky with half at Dome C and half at Dome A (D. Bennett,
private communication, 2007).  

The high latitude has a disadvantage in that the galactic bulge,
toward which most microlensing observations are directed, has zenith
angles varying from $44^\circ$ to $80^\circ$ from Dome C and
$51^\circ$ to $73^\circ$ at Dome A. The largest of these zenith angles
corresponds to 5.6 air masses,  so absorption, differential refraction
and a degradation of the seeing are going to be a problem.  There have
been no determinations of the seeing at large zenith angles from Dome
C and no site evaluation at all of Dome A. Transportation to either
site is a major obstacle to any of these developments, and none are
likely to become functional in the next decade (D. Bennett, private
communication, 2007) There have been no
simulations of microlensing planet detections accounting for all
constraints from either site. Until such is complete, the actual
advantages {\it vs} disadvantages of a microlensing survey telescope
located in Antarctica over the next generation ground based program
described above cannot be ascertained. 

\subsection{Simulation of a space based program}
The simulation of expected planetary discoveries from space in
Fig. \ref{fig:bennett1} is modified somewhat from the MPF proposal 
in that a PSF of $0.32^{\prime\prime}$ in the IR instead of
$0.24^{\prime\prime}$ in the visible. Still, this PSF is sufficient to
resolve most of the sources, so blending is much less than in the
ground based program.  Assumptions similar to those of  Bennett and
Rhie (2002) apply, except the survey is in the IR, where more lens
stars can detected than in a visible survey. Seeing, weather, night
sky brightness, lack of light curve 
coverage and other problems that hinder the detection of planets from
the ground are eliminated.  The total field assumed is the 2.6 square
degrees of MPF split up into four 0.65 square degree fields of the
telescope. The number of source stars that are monitored is
approximately $2\times 10^8$. Light curve error bars are generated
under the assumption that 
the photometric accuracy is limited by photon statistics for noise
levels down to 0.3\%.  In addition to the source star, the lens
star and nearby blended stars are assumed to contribute to the
photon noise. Light curves for lenses with a single planet are
generated by the Monte Carlo scheme described above with parameters
selected randomly from appropriate distributions as in the ground
based simulation. A point source
single lens light curve is fitted to the data, and a planet is assumed
detected if a binary light curve fitted to the same data reduces
$\chi^2$ by more than 160. Unlike the ground based observations,
essentially all of the ``detected'' planets in the space based program
will be ``discovered,'' and the  planet-star mass ratios and separations
in terms of $R_{E}$ will be determined. 

The space based program can collect enough additional data to find a
complete gravitational lens solution and determine the masses of the
planets and their host stars and the displacement of the planet in
AU (Bennett, Anderson and Gaudi, 2007). This information depends entirely on
the detection of the lens star through the stability of the PSF 
in a space borne telescope, where the focal plane is dithered between
images to reduce the pixel undersampling of the field.  If the source
star brightness is 
determined from the shape of the light curve, it can be subtracted
from the total brightness of the blended source-lens image after the
event to obtain the brightness of the lens in two or more colors. The
color of the lens and its brightness determines $D_{OL}$ and allows
a complete determination of the lens mass, planetary mass and
separation of the planet from the star.  The precision of this
determination can be checked by several alternative procedures
discussed below involving determining the relative proper motion when
the lens is detected. 

The $0.32^{\prime\prime}$ FWHM
PSF used in the construction of the space curves in
Fig. \ref{fig:bennett1} will resolve most of the sources, and 
the PSF stability allows the relative proper motion of the lens and
source to be determined. In a typical event, the relative proper motion
$\mu_{rel}$ will be $\sim 5$ mas/yr, so 3 years after the event, the
separation would be increased from $<1$ milliarcsecond during the event to 15
milliarcseconds or about 0.06 pixel in the MPF example. The observable effects are
that the source will be slightly elongated in the direction of motion,
and if the lens and source are different colors, the blue and red
centroids will be displaced. The elongation of the PSF due to the
relative proper motion depends on both the source-lens separation
$\Delta x$ and their relative brightness. But the brightness of the
source is found from the shape of the well sampled light curve as
described above, and the brightness of the lens can be determined by
subtracting the flux density from the source from the total flux
density. $\Delta x$ from the elongation and $\mu_{rel}=\Delta x/T$ can be
found, $T$ being the time after the magnification maximum that the
images are taken. If $\mu_{rel}$ is 
determined by another procedure, the image elongation can be used as a
measure of the brightness ratio of the source and lens, which can be
compared to the brightness of the source from the light curve and
thereby test for the presence of a bright binary companion of the
source. The Einstein ring crossing time $t_{E}$ follows directly
from the event light curve, so if the 
proper motion is available, we have the Einstein ring angular radius
$\mu_{rel}t_{E}=\theta_{E}$. A relation between the mass of the
lens and the distance $D_{OL}$ follows from the Einstein ring radius.
\begin{equation}
M=\frac{c^2}{4G}\theta_{E}^2\frac{D_{OS}D_{OL}}{D_{\s
OS}-D_{OL}}, \label{eq:massdistance}
\end{equation}  
where $D_{OS}$ is approximately known from the source brightness and
color. The brightness of the lens gives the luminosity as a function
of $D_{OL}$, and if the lens is a main sequence star, the
mass-luminosity relation allows a solution for $M$, $D_{OL}$ and the
mass of the planet from the light-curve-determined planet-lens mass
ratio. The value of $R_{E}=D_{OL}\theta_{E}$ gives the
separation of the planet from 
the star in physical units. If the spectral type of the lens can be
determined, the brightness allows $D_{OL}$ to be solved for
directly in the above mass-distance relation. Redundant analysis in
several pass bands can be used to constrain the possibility that the
source or lens has a bright binary companion.  Methods of determining
the uncertainties in the determinations of $\mu_{rel}$ are described
in Bennett, Anderson and Gaudi (2007).

For a source-lens angular separation of $\Delta x$, the image of the
source is a blend of two peaks characterized by the PSFs for each
star, which differ for V-band and I-band. If the source and lens are
different colors, the 
centroid of the combined images in I-band will be displaced from that
in V-band by $\Delta x_{\rm{V-I}}$, which displacement can be measured
with the stable PSFs that are attainable from space (Bennett {\it et
al.} 2006). The displacement
of the blue from the red centroids is related to the actual
displacement of source and lens $\Delta x$ by calculating the
combination of source and lens in the V-band and separately in the
I-band PSFs as a function of $\Delta x$.  
This requires knowing the relative intensities of the source and lens
separately in the two pass bands. The source brightness in the two
bands can be determined by the shape of the light curve as discussed
above, and the total brightness source plus lens determined after the
event. The brightness of the lens in two or more colors is then
available by subtracting the source brightness from the total
brightness.  The solution for $\Delta x$ from $\Delta x_{\rm V-I}$
yields the relative proper motion, which can be used in the total
solution for the system when there is no caustic crossing during the
event to determine this same relative proper motion. 

For many planetary perturbations of a microlensing light curve, the
source comes close to or crosses a caustic.  The singularity is rounded
off by the effect of the finite source, and the width of the rounded
peak in the light curve is a measure of the caustic crossing time by
the source $t_*$, where $t_*$ is now solved for in the least squares
fit to the light curve. Since the angular radius of the source
$\theta_*$ can be estimated from the brightness and color, the
relative proper motion $\mu_{rel}=\theta_*/t_*$ follows. Then $\theta_{
E}=\mu_{rel}t_{E}$ and the mass-distance relation  along with
the brightness of the lens and the mass-luminosity relation can be
used for the complete solution of the lens equation as above. A
caustic crossing thus provides a check on $\mu_{rel}$ by the imaging
methods described above. The event OGLE-2003-BLG-235/MOA-2003-BLG-53
is an example of a caustic crossing event where the host lens was
identified using the Hubble telescope and the offsets of the centroids
of the PSFs in B, V and I were determined (Bennett {\it et al.} 2006). 

A final procedure for constraining planet and lens properties involves
the parallax effect evident in long duration events. The point source
single lens magnification given in Eq. (\ref{eq:amplification}) assumed
that source, lens and observer all had constant velocities during the
event. But the observer sitting on the Earth is accelerated during the
orbital motion, which results in the light curve being
asymmetric. Detection of this asymmetry and the gain in information
about the lens system that results depends on the precision of the
light curve and the duration of the event.  This parallax affect has
been detected and used to improve the fit to a light curve  (Alcock
{\it et al.} 1995).  In the parallax analysis, $u(t)$ for the constant
velocity determination of $A(t)$ is replaced with a constant velocity
solution relative to the Sun with the Earth's motion relative to the
Sun being added. In that procedure the relative velocity between lens and
source is projected to the Sun's position, and $\tilde{v}=v/(1-z),\,(z=D_{
OL}/D_{OS})$ is solved for in the least squares solution for the
asymmetric light curve. The projection of the Einstein ring radius to
the Sun $\tilde{r}_{E}=\tilde{v}t_{E}$ is then known. If $R_{
E}=vt_{E}= \tilde{v}(1-z)$ is substituted into the expression for
$R_{E}$, we can write
\begin{equation}
M=\frac{c^2}{4GD_{OS}}\frac{1-z}{z}\tilde{v}^2t_{
E}^2=\frac{c^2}{4GD_{OS}}\frac{1-z}{z}\tilde{r}_{
E}^2=\frac{c^2}{4G}\tilde{r}_{E}\theta_{E}, \label{eq:massparallax}
\end{equation}
where the last form determines the mass unambiguously if $\theta_{
E}$ is determined from a measured $\mu_{rel}$ (Bennett Anderson and Gaudi,
2007; Gould, 1992).  

It is evident that there is considerable redundancy of methods in the
determinations of the lens star and planet masses by detecting the
lens star from space, where such redundancy can reduce the uncertainty
in lens-planet parameters and detect binary companions of the source.
An example of the application of these procedures to a simulation of a
space based mission is shown in Fig. \ref{fig:bennett3} (Bennett
Anderson and Gaudi, 2007). This figure shows the distribution of
stellar masses 
and the distributions of the uncertainties in both the projected
separation and in the stellar and planetary masses for lenses with
detected terrestrial planets, where Fig. \ref{fig:bennett3}a shows in
red that 69\% of the discoveries allow a complete solution yielding
the planet and stellar masses to better than 20\%. The probability of
a complete solution in a space based microlensing search for planets
for lens and planet mass and separation is 100\% for F and G stars but
steadily drops for K and M-dwarfs.  Most of the planetary discoveries
are completely characterized, and the complete statistics for the
occurrence of planets as a function of their mass, stellar type of
host star, semimajor axis and distribution between here and the
galactic center as advertised earlier are realized for a
space based microlensing search for planets.

The current paradigm of planetary system formation recognizes frequent
planetary encounters that will result in the ejection of a significant
number of planets into the interstellar medium.  Only a space based
mission has the capability of detecting enough of these free floating
planets ($\sim 60$ by MPF, if every star loses a planet) to determine
whether close planetary encounters are the rule or the exception in
the evolution of planetary systems. There are a number of additional
discovery and characterization capabilities of a space based
microlensing program, including $\sim 50,000$ transiting Jupiters,
planetary moons exceeding a lunar mass, and 600 multiplanet systems if each
star has a distribution of planets like that of the solar system. 

\bibliographystyle{plainnat}

\bibliography{planeto-biblio}

\chapter{References}
\parindent=0pt

Alard, C. (2000) Image subtraction using a space-varying kernel, {\it 
Astron. Astrophys. Suppl. Ser.} {\bf 144}, 363-370. 

Alard, C. and R.H. Lupton (1998) A Method for Optimal Image Subtraction
{\it Astrophys. J.}  {\bf 503}, 325-331. 

Alcock, C. R.A. Allsman, D. Alves, T.S. Axelrod, D.P. Bennett + 12
coauthors (1995) First observation of parallax in a gravitational
microlensing event, {\it Astrophys. J.} {\bf 454}, L125-L128.

Aristidi, E., A Agabi, E. Fossat, M. Azouit, J. Vernin, T. Sadibekova,
T. Travouillon, A. Ziad, F. Martin (2005) Site testing in winter at
Dome C, {\it SF2A-2005: Semaine de l'Astrophysique Francaise}, meeting
held in Strasbourg, France, June 27 - July 1, 2005, Edited by
F. Casoli, T. Contini, J.M. Hameury and L. Pagani. Published by
EdP-Sciences, Conference Series, 2005, p. 45. 

Ashley, M.C.B., M.G. Burton, P.G. Calisse, A. Phillips and J.W.V
Storey (2005) Site testing at Dome C - Cloud statistics from the
ICECAM experiment, {\it Highlights of Astronomy} (ASP
Conf. ser. Vol. {\bf 13} Astron. Soc. Pacific), Ed. O. Engvold, p932.

Basri, G., Marcy, G., Oppenheimer, B., Kulkarni, S.R., and Nakajima, T. (1996) . Rotation and activity in the coolest stars. {\it ASP Conf Ser.} {\bf109}, 587

Basri, G. \& Marcy, G. (1995), A surprise at the bottom of the main sequence: Rapid rotation and NO H-alpha emission. {\it AJ} {\bf 109}, 762.

Beaulieu, J.-P., D.P. Bennett,  P. Fouqu\'e, A. Williams and 69
coauthors (2006) Discovery of a cool planet of 5.5 Earth masses
through gravitational microlensing, {\it Nature} {\bf 437}, 439-440.  

Bennett, D.P. and S.H. Rhie (1996) Detecting Earth-Mass Planets with Gravitational Microlensing
 {\it Astrophys. J.} {\bf 472}, 660.

Bennett, D.P. and S.H. Rhie (2002) Simulation of a space-based
microlensing survey for terrestrial extrasolar planets, {\it
Astrophys. J.} {\bf 574}, 985-1003.

Bennett, D.P. (2004) The detection of terrestrial
planets via gravitational microlensing: space vs. ground-based surveys
{\it Astro-ph}/0404075. 

Bennett, D.P. (2007a) Simulation of a next generation ground based
search for planets with microlensing, In preparation.
 
Bennett, D.P. (2007b) Detection of extrasolar planets by gravitational
microlensing, Chapter in exoplanet book.

Bennett, D.P., I. Bond, E. Cheng, S. Friedman, S. Gaudi and 18
coauthors (2004) The Microlensing Planet Finder: completing the census
of extrasolar planets in the Milky Way, {\it. Proc. SPIE} {\bf 5487},
1453-1464. 

Bennett, D.P., J. Anderson, I.A. Bond, A. UIdalski and A. Gould (2006)
Identification of the OGL-2003-BLG-253/MOA-2003-BLG-53 planetary host
star, {\it Astrophys. J.} {\bf 647}, L171-L174.

Bennett, D.P., J. Anderson, and S. Gaudi (2007a) Characterization of
gravitational microlensing planetary host stars., {\it Astrophys. J.}
{\bf 660}, 781-790.

Bennett, D.P. {\it et al.} (2007b) Characterization of the
OGLE-2006-BLG-109Lb,c two planet system, In preparation. 

Bertelli, G., A. Bressan, C. Chiosi, F. Fagotto, E. Nasi (1994)
Theoretical isochrones from models with new radiative opacities, {\it
Astron. Astrophys. Suppl.} {\bf 106}, 275-302.    

Bond, I.A., A. Udalski, M. Jaroszy\'nski, N.J. Rattenbury,
B. Paczy\'nski, I. Soszy\'nski, L. Wyrzykowski, and 25 coauthors
(2004) OGLE 2003-BLG-235/MOA 2003-BLG-53: A Planetary Microlensing
Event, {\it Astrophys. J.} {\bf 606}, L155-L158.  

Bracewell, R. (1978). {\it The Fourier Transform and its Applications}. McGraw-Hill, New York. 

Brown, M. et al. (2003, ApJ 593, L125) {\it Astrophys. J.} {\bf 593}, L125

Brown, R. A. (2008) ``Keplerian Mission Studies," $http://sco.stsci.edu/studies /{\it keplerian_mission_studies.pdf}$

Burton, M.G., J.S. Lawrence, M.C.B. Ashley, J.A. Bailey and 23
coauthors (2005) Science Programs for a 2-m Class Telescope at Dome C,
Antarctica: PILOT, the Pathfinder for an International Large Optical
Telescope, {\it Pub. Astron. Soc. Austral.} {\bf 22}, 199-235. 

Catling, D. C. (2006) Comment on "A hydrogen-rich early Earth atmosphere". Science 311, 38a-38b.

 Cox, A.~N.\ 2000, Allen's 
 astrophysical quantities, 4th ed.~Publisher: New York: AIP Press;
 Springer, 2000.~Edited by Arthur N.~Cox.~ ISBN: 0387987460.
 
Delfosse, X., Forveille, T., Perrier, C., and Mayor, M.  (1998). Rotation and chromospheric activity in field M-dwarfs A\&A, 331, 581

Di Stefano, R. (1999) Microlensing and the search for extraterrestrial
life, {\it Astrophys. J.} {\bf 512}, 558.

Docobo, J., Tamazian, V., Balega, Y., Andrade, M., Schertl, D., Weigelt, G., Campo, P. and Palacios, M. (2008) A methodology for studying physical and dynamical properties of mutiple stars. {\it Astronomy and Astrophysics}, submitted.

Dole, S. H. (1964). {\it Habitable Planets for Man}. New York: Blaisdell Publishing.

Erskine, D.J.  (2003).  An Externally Dispersed Interferometer Prototype for Sensitive Radial Velocimetry: Theory and Demonstration on Sunlight, [\it PASP}  {\bf 115}, 255.

Forget, F. and Pierrehumbert, R.T. (1997) Warming early Mars with carbon dioxide clouds that scatter infrared radiation. {\it Science} {\bf 278}, 1273-1276.

Franck, S., Block, A., von Bloh, W., Bounama, C., Schellnhuber, H. J., and Svirezhev, Y. (2000). Habitable zone for Earth-like planets in the solar system. {\it Planetary and Space Science} {\bf 48}, 1099-1105.

Gaudi, S., C. Han, and A. Gould (2007a) Simulations of a ground-based
next generation survey for exoplanets with microlensing, In preparation.

Gaudi, S. {\it et al.} (2007b) Discovery of a Jupiter/Saturn Analog with Gravitational Microlensing. {\it Science} {\bf 319}, 927-930.

Give'on et al (2006)  Closed-loop Wavefront Correction for High-contrast Imaging: The "Peak-A-Boo" Algorithm. {\it BAAS} abst. 164.15.

Gough, D.O. (1981) Solar interior structure and luminosity variations. {\it Solar Phys.} {\it 74}, 21-34.

Gould, A. (1992) Extending the MACHO search to $\sim 10^6
M_{\s\odot}$, {\it Astrophys. J.} {\bf 392}, 442-451.

Gould, A., and A. Loeb (1992) Discovering planetary systems through
gravitational microlenses, {\it Astrophys. J.} {\bf 396}, 104-114.

Gould, A. A. Udalski, D. An, D.P. Bennett, A.-Y Zhou, and 31 coauthors
(2006) Microlens OGLE-2005-BLG-169 Implies That Cool Neptune-like
Planets Are Common, {\it Astrophys. J.} {\bf 644}, L37-L40.  

Griest, K. and N. Safizadeh (1998) The use of High-magnification
microlensing events in discovering extrasolar planets, {\it
Astrophys. J.} {\bf 500}, 37-50.

Han, C. and H. Chang (2003) Direct lens imaging of Galactic bulge
microlensing events, {\it Mon. Not. Roy. Ast. Soc.} {\bf 338}, 637.

Hart, M. H. (1979) Habitable zones around main sequence stars. {\it Icarus} {\bf 37}, 351-357.

Henry, T.~J., Kirkpatrick, J.~D., \& Simons, D.~A.\ 1994, \aj, 108, 1437.
 
Holtzman, A. A.M. Watson, W.A. Baum, C.J. Grillmair, E.J. Groth,
R.M. Light, R. Lynds, and E.J. O'Neil Jr. (1998) The Luminosity Function
and Initial Mass Function in the Galactic Bulge, {\it Astron. J.} {\bf
115}, 1946-1957. 

Ingersoll, A. P. (1969). The runaway greenhouse: A history of water on Venus. {\it J. Atmos. Sci.} {\bf 26}, 1191-1198.

Joshi, M. (2003).  Climate model studies of synchronously rotating planets. {\it Astrobiology} {\bf 3}, 415-427.

Kasting, J. F. (1988). Runaway and moist greenhouse atmospheres and the evolution of Earth and Venus. {\it Icarus} {\bf 74}, 472-94.

Kasting, J. F. (1991). CO$_2$ condensation and the climate of early Mars. {\it Icarus} {\bf 94}, 1-13.

Kasting, J.F., Whitmire, D.P., and Reynolds, R.T. (1993) {\it Icarus},
{\bf 101}, 108.

Kay, W. (1958) A child's introduction to outer space. {\it Golden Records} (audio recording). 

Khodachenko, M. L., Ribas, I., Lammer, H., et al. (2007) Coronal Mass Ejection (CME) activity of low mass M-dwarfs as an important factor for the habitability of terrestrial exoplanets. I. CME impact on expected magnetospheres of earth-like exoplanets in close-in habitable zones. {\it Astrobiology} {\bf 7}, 167-184.

Krisciunas, K. and B. Schaefer (1991) A model of the brightness of
moonlight, {\it Pub. Astron. Soc. Pac.} {\bf 103}, 1033-1039.

Langton, J. and Laughlin, G. 2008, "Hydrodynamic Simulations of Unevenly Irradiated Jovian Planets" {\it ApJ} {\bf 674}, 1106-1116 

Lawrence, J.S., M.C.B. Ashley, A. Tokovinin, T. Travouillon (2004)
Exceptional astronomincal seeing conditions above Dome C in
Antarctica, {\it Nature} {\bf 431}, 278-281.

Nearby Stars from the LSPM-North Proper-Motion Catalog. I. Main-Sequence Dwarfs and Giants within 33 Parsecs of the Sun. {\it AJ} {\bf 130}, 1247. 

Lovelock, J. E. (1965) A physical basis for life detection experiments. {\it Nature} {\bf 207}, 568-570 (1965).

Malbet, F., Liu, D.T., Yu, J.W., and Saho, M.  (1995).  Space adaptive optics coronography. {\it Proc. SPIE} {\bf 478}, 230-238. 

Marois, C., Lafrenire, D., Macintosh, B. Doyon, R. (2006). Accurate Astrometry and Photometry of Saturated and Coronagraphic Point Spread Functions. {\it Ap.J.} {\bf 647}, 612. 

Mawet, D., Riaud, P., Absil, O., Surdej, J. 2005, Annular groove phase mask coronagraph. {\it Ap.J.} {\bf 633}, 1119

Mayor, M. and Queloz, D. (1995). A Jupiter-Mass Companion to a Solar-Type Star, {\it Nature} {\bf 378}, 355. 

McLean, I.S., Prato, L., McGovern, M.R., Burgasser, A.J., Kirkpatrick, J.D., Rice, E.L., Kim, S.S. (2007) The NIRSPEC Brown Dwarf Spectroscopic Survey. II. High-Resolution J-Band Spectra of M, L, and T Dwarfs {\it ApJ} {\bf 658}, 1217

Mischna, M.M., Kasting, J.F., Pavlov, A.A., and Freedman, R. (2000) Influence of carbon dioxide clouds on early martian climate. {\it Icarus} {\bf 145}, 546-554.

National Research Council. (2007). {\it  The Limits of Organic Life in Planetary Systems}, National Academies Press, Washington, DC.

Peale, S.J. (1998) On microlensing event rates and optical depth
toward the Galactic center, {\it Astrophys. J.} {\bf 509}, 177-191.

Peale, S.J. (2001) Probability of detecting a planetary companion
during a microlensing event, {\it Astrophys. J.} {\bf 552}, 889-911.

Perrin, M.D., Sivaramakrishnan, A., Makidon, R.B., Oppenheimer, B.R., Graham, J.R. (2003).The Structure of High Strehl Ratio Point-Spread Functions {\it ApJ} {\bf 596}, 702.  

Pollack, J. B., Kasting, J. F., Richardson, S. M., and Poliakoff, K. (1987) The case for a wet, warm climate on early Mars. {\it Icarus} {\bf 71}, 203-224.

Rasool, S.I. and DeBergh, C. (1970) The runaway greenhouse and the accumulation of CO$_2$ in the Venus atmosphere. {\it Nature} {\bf 226}, 1037-1039.

 Reid, N., \& Hawley, 
 S.~L.\ 2000, New light on dark stars : red dwarfs, low mass stars,
 brown dwarfs (New York : Springer)

Reipurth, B., Jewitt, D., and Keil, K. (eds.) 2006. {\it Protostars and Planets V}, University of Arizona Press, Tucson. 

Sivaramakrishnan, A., Koresko, C., Makidon, R B.. Berkefeld, T. Kuchner, M. J. 2001.Ground-based Coronagraphy with High-order Adaptive Optics
{\it Ap.J.} {\bf 552}, 397-408. 

Solomon, S. C. and Head, J. W. (1991). Fundamental issues in the geology and geophysics of Venus. {\it Science} {\bf 252}, 252-260.

Spitzer, L. (1960).Space telescopes and components. {\it AJ} {65}, 242. 

Sumi, T., P.R. Wo\'zniak, A. Udalski, M. Szyma\'nski, M
Kubiak,G. Pietrzy\'nhski, I. Soszy\'nski,K. Zegru\'n, O. Szewczyk,
L. Wyrzykowski, and B. Paczy\'nski (2006) Microlensing optical depth
toward the galactic bulge using bright sources from OGLE-II, {\it
Astrophys. J.} {\bf 636}, 240-260.

Tian, F., Toon, O. B., Pavlov, A. A., and De Sterck, H. (2005) A hydrogen rich early Earth atmosphere. {\it Science} {\bf 308}, 1014-1017.

Travouillon, T., M.C.B. Ashley, M.G. Burton, J.W.V. Storey, and R.F
Lowenstein (2002) Atmospheric turbulence at the South Pole and its
implications for astronomy, {\it Astron. Astrophys.} {\bf 400},
1163-1172. 

Udalski, A. (2003) The Optical Gravitational Lensing Experiment. Real
Time Data Analysis Systems in the OGLE-III Survey, {\it Acta Astron.}
{\bf 53}, 291-305.

Udalski, A. M. Jaroszy\'nski, B. Paczy\'nski, M. Kubiak, and 32
coauthors (2006) The Optical Gravitational Lensing
Experiment. OGLE-III Long Term Monitoring of the Gravitational Lens
QSO 2237+0305, {\it Astrophys. J.} {\bf 628}, L109-L112. 

Valencia, D., O'Connell, R.J., Sasselov, D.D.  (2007). Inevitability of Plate Tectonics on Super-Earths {\it Ap.J.} {\bf 670}, 45. 

Vanderbei, R.J., Cady, E., Kasdin, N.J.  (2007). Optimal Occulter Design for Finding Extrasolar Planets {\it ApJ} {\bf 665} 794. 

Walker, G.A.H. (1987) {\it Astronomical Observations: An Optical
Perspective} Cambridge University Press, Cambridge.

Wambsganss, J. (1997)  Discovering Galactic planets by gravitational
microlensing: magnification patterns and light curves {\it
Mon. Not. Roy. Astron. Soc.} {\bf 284}, 172-188. 

Witt, H.J. (1990) Investigation of high amplification events in light
curves of gravitationally lensed quasars, {\it Astron. Astrophys.}
{\bf 236}, 311-322.

Wolf, S., Moro-Mart'n, A, D'Angelo, G.  2007, Signatures of planets in protoplanetary and debris disks. {\it Pl. and Sp. Sci} {\bf 55} 569. 

Wolf, S.  \& Klahr, H.  2005,  Observing early stages of planet formation with ALMA: large-scale vortices in protoplanetary disks {\it Proceedings of the dusty and molecular universe: a prelude to Herschel and ALMA}  Ed. by A. Wilson. ESA SP-577, Noordwijk, Netherlands: ESA Publications Division, ISBN 92-9092-855-7, 2005, p. 473 - 474.

Wolszczan, A. \& Frail, D.A. (1992). A planetary system around the millisecond pulsar PSR1257 + 12, {\it Nature} {\bf 355} 145. 

Wozniak, P.R. (2000) Difference Image Analysis of the OGLE-II Bulge
Data. I. The Method, {\it Acta Astron.} 50, 421-450. 

Yock, P. (2007) Detecting Earth-like extra-solar planets from
Antarctica by Gravitational microlensing, {\it Chinese
Astron. Astrophys} {\bf 31}, 101-108.

Zoccali, M., S. Cassisi, J.A.Frogel, A. Gould, S. Ortolani,
A. Renzini, R.M. Rich, and A.W. Stephens (2000) The Initial Mass
Function of the Galactic Bulge down to ~0.15 Msolar, {\it
Astrophys. J.} {\bf 530}, 418-428. 

\chapter{Selected acronyms and abbreviations}
\begin{acronym}[TDMA]
 \acro{2MASS}{Two-micron All Sky Survey}
 \acro{4QPM}{4-quadrant phase-mask}
 \acro{AAAC}{Astronomy and Astrophysics Advisory Committee}
 \acro{AAS}{American Astronomical Society}
 \acro{AAT}{Anglo-Australian Telescope}
 \acro{ALMA}{Atacoma Large Millimeter Array}
 \acro{AO}{Adaptive optics}
 \acro{APLC}{Apodized-Pupil Lyot Coronagraph}
 \acro{arXiv}{electronic archive}
 \acro{AU}{Astronomical Unit}
 \acro{BLC}{Band-Limited Coronagraph}
 \acro{CARMA}{Combined Array for Research in Millimeter-wave Astronomy}
 \acro{CCD}{Charge couple device}
 \acro{Cen}{Centaurus}
 \acro{CHZ}{Continuously Habitable Zone}
 \acro{Cnc}{Cancri}
 \acro{COROT}{Convection, Rotation, and Planetary Transits}
 \acro{CTIO}{Cerro Tololo Inter-American Observatory}
 \acro{DIA}{Difference image analysis}
 \acro{DM}{Deformable mirror}
 \acro{EDI}{Externally Dispersed Interferometer}
 \acro{ELT}{Extremely Large Telescope}
 \acro{ESA}{European Space Agency}
 \acro{ESO}{European Southern Observatory}
 \acro{ExAO}{Extreme Adaptive Optics}
 \acro{ExoPTF}{Exo-Planet Task Force}
 \acro{FAME}{Full-sky Astrometric Mapping Explorer}
 \acro{FEEP}{Field Effect Electric Propulsion}
 \acro{FOV}{Field-of-view}
 \acro{GC}{Galactic center}
 \acro{GJ}{Gliese-Jahreiss}
 \acro{GPI}{Gemini Planet Imager}
 \acro{GSMT}{Giant Magellan Telescope}
 \acro{Gyr}{billion years}
 \acro{HARPS}{High Accuracy Radial velocity Planet Searcher}
 \acro{HAT}{Hungarian Automated Telescope Network}
 \acro{HCIT}{High Contrast Imaging Testbed}
 \acro{HD}{Henry-Draper (catalog)}
 \acro{HDS}{(Subaru) High Dispersion Spectrograph}
 \acro{HET}{Hobby-Eberly Telescope}
 \acro{HIRES}{High Resolution Echelle Spectrometer}
 \acro{HITRAN}{High Resolution TRANsmission (database)}
 \acro{HST}{Hubble Space Telescope}
 \acro{HZ}{Habitable Zone; I=Inner, M=Middle, O=Outer}
 \acro{IR}{Infrared}
 \acro{IRAC}{Infrared Array Camera}
 \acro{IWA}{Inner working angle}
 \acro{JPL}{Jet Propulsion Laboratory}
 \acro{JWST}{James Webb Space Telescope}
 \acro{KI}{Keck Interferometer}
 \acro{LBT}{Large Binocular Telescope}
 \acro{LBTI}{Large Binocular Telescope Interferometer}
 \acro{LISA}{Laser Interferometer Space Antenna}
 \acro{LOS}{Line of sight}
 \acro{LSST}{Large Synoptic Survey Telescope}
 \acro{MIDEX}{Mid-level Explorer satellite}
 \acro{MIKE}{Magellan Inamori Kyocera Echelle}
 \acro{MIRI}{Mid-infrared Instrument}
 \acro{MMT}{Monolithic (formerly Multiple)Mirror Telescope}
 \acro{MOA}{Microwave Observations in Astrophysics (group)}
 \acro{MOST}{Microvariability and Oscillations of STars}
 \acro{MPF}{Microlensing Planet Finder}
 \acro{NASA}{National Aeronautics and Space Administration}
 \acro{NICMOS}{Near-Infrared Camera and Multi-Object Spectrometer}
 \acro{NIRCam}{Near-infrared Camera}
 \acro{NIRSpec}{Near-infrared Spectrograph}
 \acro{NSF}{National Science Foundation}
 \acro{OGLE}{Optical Gravitational Lensing Experiment}
 \acro{OPD}{optical path difference}
 \acro{PACS}{Photodetector Array Camera and Spectrometer }
 \acro{PanSTARSS}{Panoramic Survey Telescope and Rapid Response System}
 \acro{Peg}{Pegasi}
 \acro{PI}{Principal Investigator}
 \acro{PIAAC}{Phase-induced Amplitude Apodization Coronagraph}
 \acro{PLANET}{Probing Lensing Anomalies Network}
 \acro{PPV}{Protostars and Planets V}
 \acro{PRIMA}{Phase-Referenced Imaging and Micro-arcsecond Astrometry}
 \acro{PSF}{Point spread function}
 \acro{RV}{Radial Velocity}
 \acro{SCUBA}{Sub-millimetre Common-User Bolometer Array}
 \acro{SIM}{Space Interferometry Mission}
 \acro{SMA}{Sub-millimeter Array}
 \acro{SNR}{Signal-to-noise ratio}
 \acro{SOFIA}{Stratospheric Observatory for Infrared Astronomy}
 \acro{SPARTA}{Standard Platform for Adaptive Optics Real Time Applications}
 \acro{SPHERE}{Spectro-Polarimetric High-contrast Exoplanet Research}
 \acro{SPIRE}{Spectral and Photometric Imaging Receiver}
 \acro{SPIRIT}{Proposed Space Infrared Interferometer Telescope}
 \acro{STIS}{Space Telescope Interferometric Spectrometer}
 \acro{TMT}{Thirty-Meter Telescope}
 \acro{TPF}{Terrestrial Planet Finder; C = coronagraph, I interferometer, O occulter}
 \acro{TrES}{Transatlantic Exoplanet Survey}
 \acro{UCLES}{University College London Spectrograph}
 \acro{UK}{United Kingdom}
 \acro{US}{United States}
 \acro{USNO}{United States Naval Observatory}
 \acro{UVES}{UV-Visual Echelle Spectrograph}
 \acro{VLT}{Very Large Telescope}
 \acro{VLTI}{Very Large Telescope Interferometer}
 \acro{WASP}{Wide Angle Search for Planets}
 \acro{WFS}{Wavefront sensing}
 \acro{WFSC}{Wavefront sensing and control}
\end{acronym}

\clearpage

\chapter{Charge Letter to the Task Force}

\begin{figure}
\centering
\includegraphics[scale=1.1]{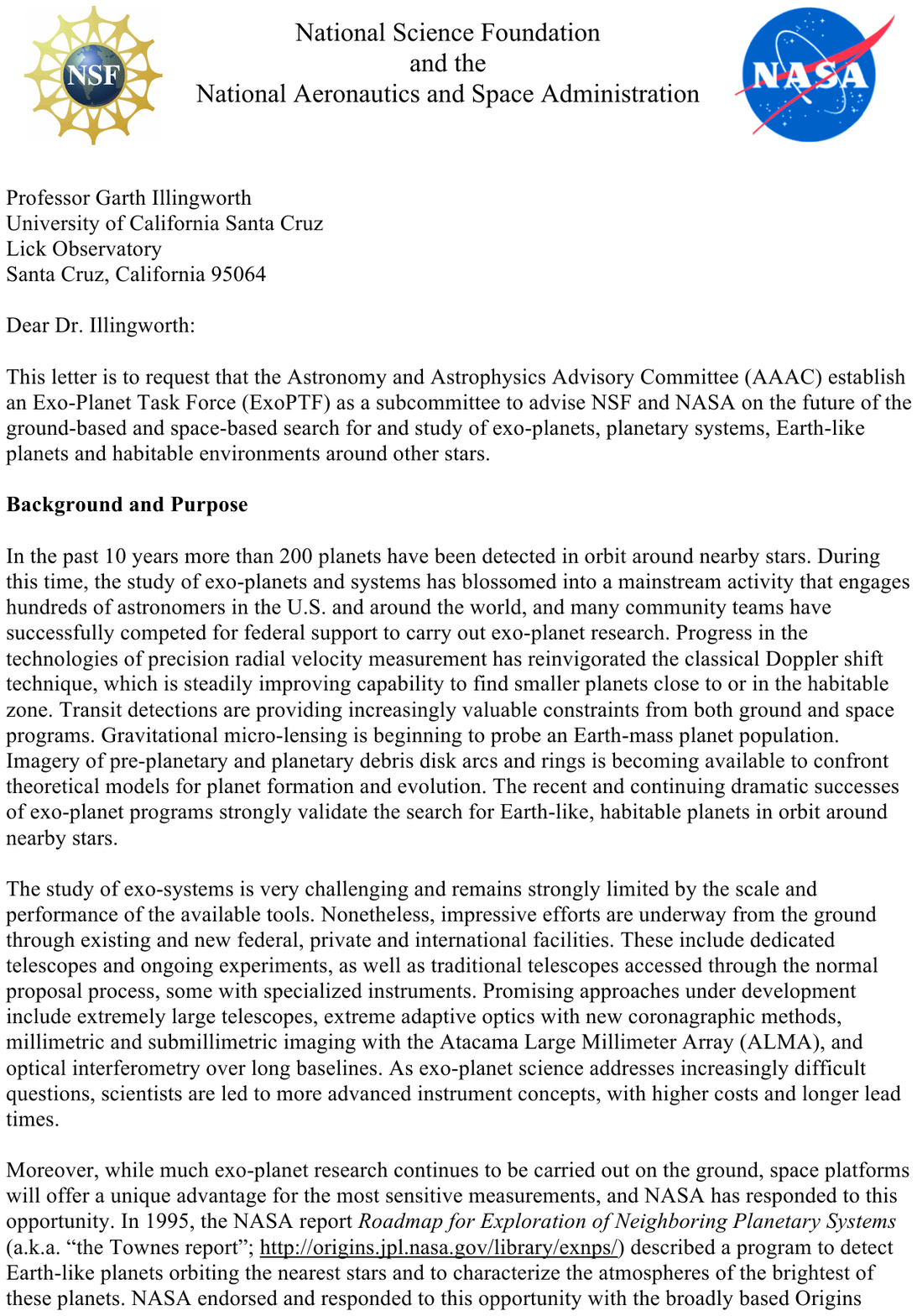}
\vspace{-1cm}
\label{fig:charge}
\end{figure}

\begin{figure}
\centering
\includegraphics[scale=1.1]{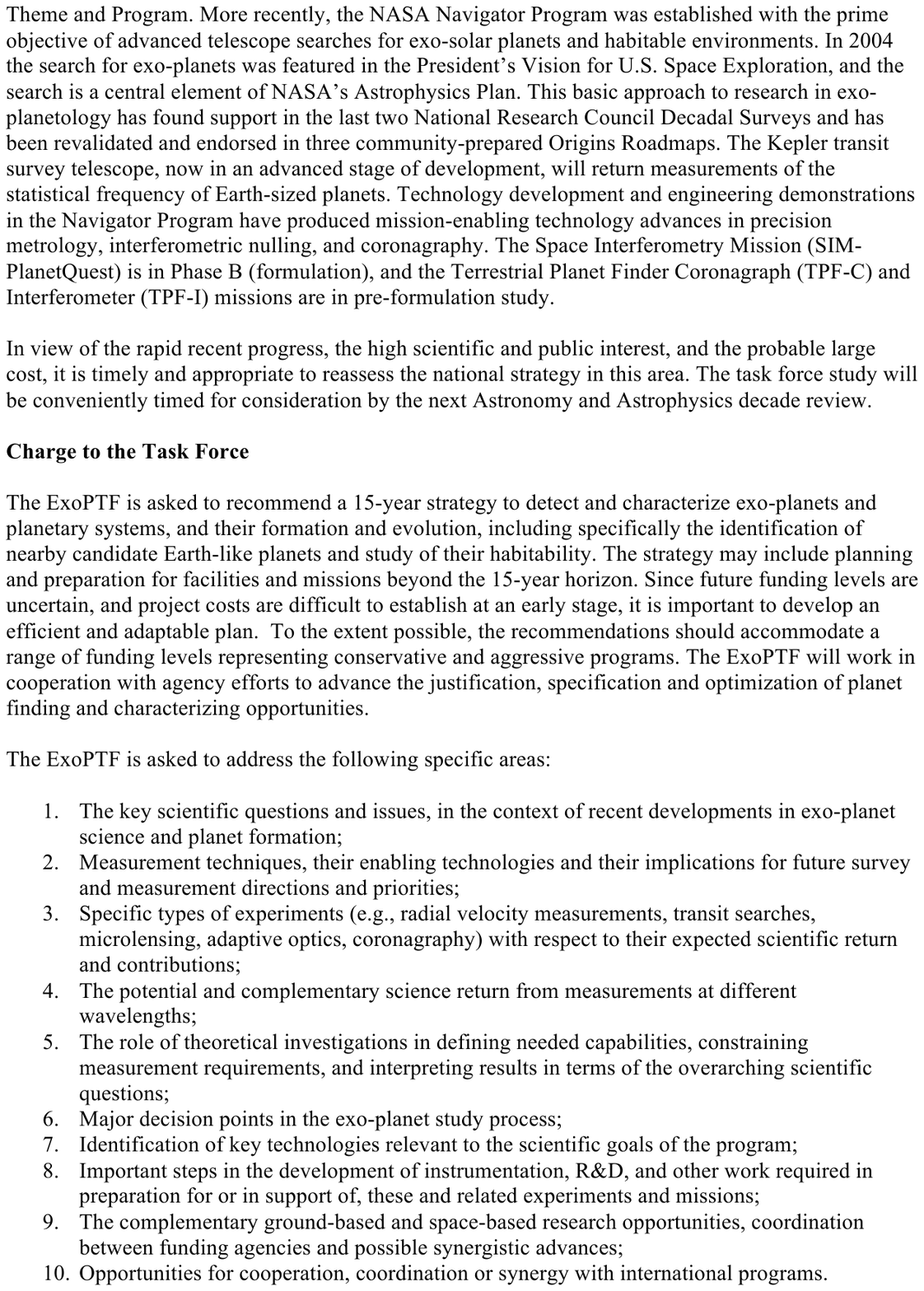}
\vspace{-1cm}
\label{fig:charge2}
\end{figure} 

\begin{figure}
\centering
\includegraphics[scale=1.0]{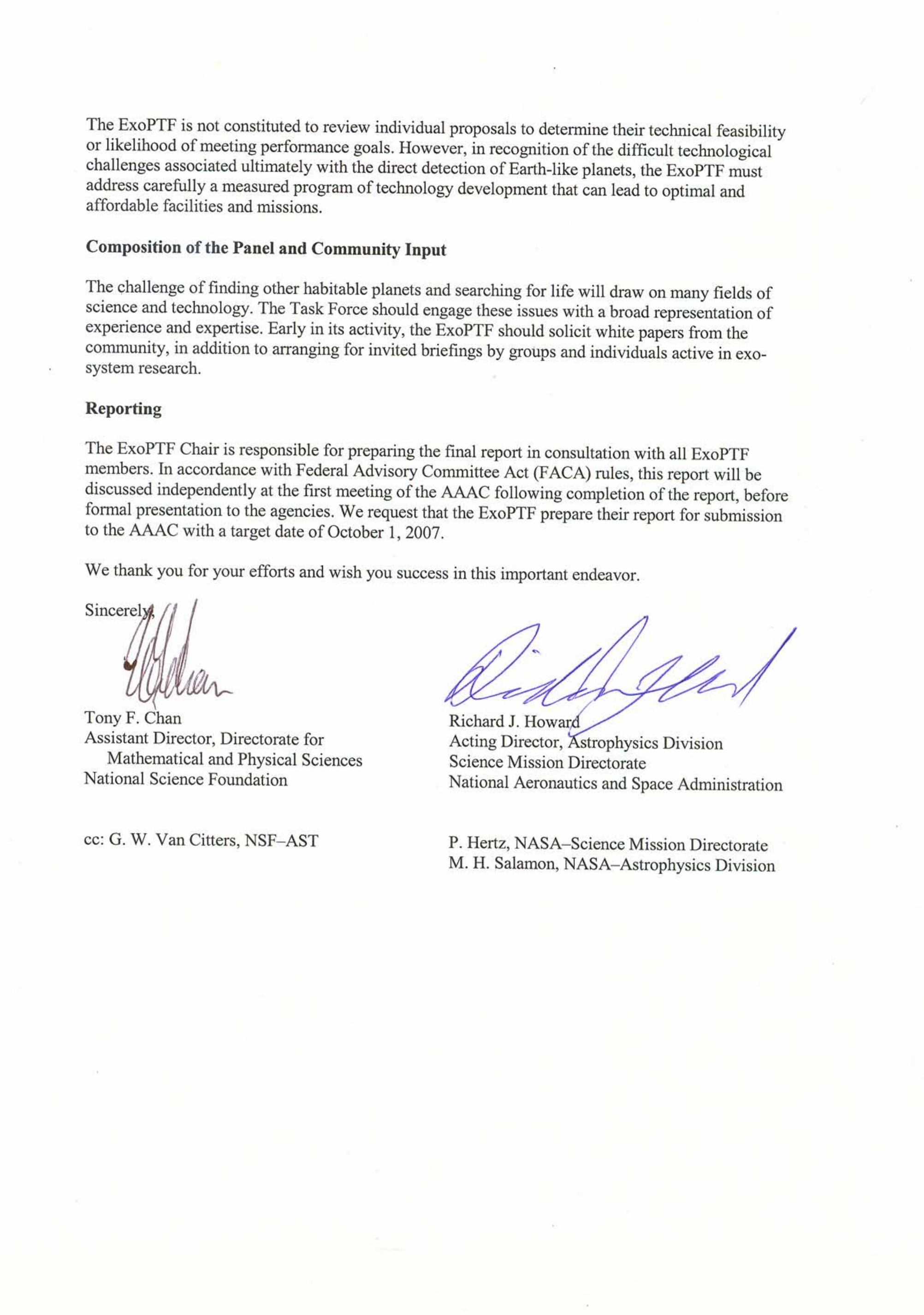}
\vspace{-1mm}
\label{fig:charge3}
\end{figure} 

\end{document}